\pdfoutput=1
\documentclass[11pt]{article}

\usepackage{jheppub}

\usepackage[english]{babel}

\usepackage{subfigure}
\usepackage{placeins}

\usepackage{braket}   % for matrix elements
\usepackage{bbold}
\usepackage{appendix}

\newcommand{\sect}[1]{section~#1}

\newcommand{\appx}[1]{appendix~#1}
\newcommand{\Fig}[1]{Figure~#1}
\newcommand{\fig}[1]{figure~#1}
\newcommand{\figs}[1]{figures~#1}
\newcommand{\tab}[1]{table~#1}

\newcommand{\half}{{\textstyle\frac{1}{2}}}

\newcommand{\re}{\operatorname{Re}}
\newcommand{\im}{\operatorname{Im}}

\newcommand{\mev}{\operatorname{MeV}}

\newcommand{\fm}{\operatorname{fm}}

\newcommand{\ms}{\mskip 1.5mu}

\newcommand{\tr}{\operatorname{tr}}

\newcommand{\tvec}[1]{\boldsymbol{#1}}

\newcommand{\mvec}[1]{\vec{\mskip 0.5mu #1}\mskip 1.5mu}

\newcommand{\etac}{\eta_{\textit{\tiny C}}}
\newcommand{\etapt}{\eta_{\textit{\tiny PT}}}
\newcommand{\etaf}{\eta_{\textrm{\tiny 4}}}

\newcommand{\one}{\mathbb{1}}

\newcommand{\ie}{i.e.\ }
\newcommand{\iex}{i.e.:}
\newcommand{\eg}{e.g.\ }

\newcommand{\wrt}{w.r.t.\ }
\newcommand{\rhs}{r.h.s.\ }
\newcommand{\lhs}{l.h.s.\ }
\newcommand{\D}{\mathcal{D}}
\newcommand{\Op}{\mathcal{O}}

\newcommand{\pha}{\mathcal{E}}
\newcommand{\srcph}{\vphantom{\left( \overline{S}_j^{123} \right)}}
\newcommand{\dd}{\mathrm{d}}
\newcommand{\avsum}{\sideset{}{^\prime}\sum}

\graphicspath{{plots/},{graphs/}}

\allowdisplaybreaks[2]

\title{Double parton distributions in the nucleon from lattice QCD}

\abstract{
We evaluate nucleon four-point functions in the framework of lattice QCD in order to extract the first Mellin moment of double parton distributions (DPDs) in the unpolarized proton. In this first study, we employ an $n_f = 2+1$ ensemble with pseudoscalar masses of $m_\pi = 355~\mathrm{MeV}$ and $m_K = 441~\mathrm{MeV}$. The results are converted to the scale $\mu = 2~\mathrm{GeV}$. Our calculation includes all Wick contractions, and for almost all of them a good statistical signal is obtained. We analyze the dependence of the DPD Mellin moments on the quark flavor and the quark polarization. Furthermore, the validity of frequently used factorization assumptions is investigated.
}

\preprint{\vbox{
\hbox{DESY-21-066}}}

\author[a,b]{Gunnar S. Bali}
\author[c]{Markus Diehl}
\author[d]{Benjamin Gl{\"a}{\ss}le}
\author[a]{Andreas Sch{\"a}fer}
\author[a]{and Christian Zimmermann}

\affiliation[a]{Institute for Theoretical Physics, University of Regensburg, 93040 Regensburg, Germany}
\affiliation[b]{Department of Theoretical Physics, Tata Institute of Fundamental Research, Homi Bhabha Road, Mumbai 400005, India}
\affiliation[c]{Deutsches Elektronen-Synchrotron DESY, Notkestr.~85, 22607 Hamburg, Germany}
\affiliation[d]{Zentrum f\"ur Datenverarbeitung, Universit\"at T\"ubingen, W\"achterstr.\ 76, 72074 T\"ubingen, Germany}

\emailAdd{gunnar.bali@ur.de}
\emailAdd{markus.diehl@desy.de}
\emailAdd{benjamin.glaessle@uni-tuebingen.de}
\emailAdd{andreas.schaefer@physik.uni-regensburg.de}
\emailAdd{christian.zimmermann@ur.de}

\begin{document}

\maketitle

\section{Introduction}
\label{sec:intro}

The high-luminosity upgrade of the LHC will substantially improve its potential for discovering physics beyond the Standard Model. In parallel to the expected decrease of statistical errors, theoretical uncertainties for standard model processes must be reduced as much as possible to optimize the physics output. A particular challenge is the description of multiple hard scattering, which means that several hard parton-level interactions occur within the same proton-proton collision. Contributions from multiple scattering generically increase with the collision energy. They can be substantial for final states with high multiplicity. Many discovery channels for new physics are of this type. In this work, we focus on double parton scattering (DPS), which is the least complex and often most important representative of multiple hard scattering.

The discussion of DPS started already in the late 1970s and produced a remarkable amount of theoretical insight \cite{Landshoff:1978fq, Kirschner:1979im, Politzer:1980me, Paver:1982yp, Shelest:1982dg, Mekhfi:1983az, Sjostrand:1986ep}.  During the last decade, there has been a considerable effort to develop a full description of DPS from first principles in QCD \cite{Blok:2011bu, Diehl:2011tt, Gaunt:2011xd, Ryskin:2011kk, Blok:2010ge, Diehl:2011yj, Manohar:2012jr, Manohar:2012pe, Ryskin:2012qx, Gaunt:2012dd, Blok:2013bpa, Diehl:2017kgu}.  Experimental searches for DPS contributions to various final states also started long ago \cite{Akesson:1986iv,Alitti:1991rd} and were greatly intensified at the Tevatron and the LHC, see \cite{Abe:1997xk,Abazov:2015nnn,Aaij:2016bqq,Aaboud:2018tiq,Sirunyan:2019zox} and references therein.  At LHC energies, it is possible to study double Drell-Yan-type reactions, in particular like-sign $W$ pair production \cite{Kulesza:1999zh,Gaunt:2010pi,Sirunyan:2019zox,Ceccopieri:2017oqe, Cotogno:2018mfv,Cotogno:2020iio}, which is particularly clean at the theoretical level. For a comprehensive and recent overview of multiparton interactions, we refer to~\cite{Bartalini:2017jkk}.

Whilst the two hard-scattering processes in DPS proceed independently of each other, the partons that initiate them can be correlated.  These correlations are quantified by double parton distributions (DPDs), which extend the familiar concept of parton distribution functions (PDFs) to the case of two partons extracted from one hadron.  To date, little is known about DPDs, apart from constraints from sum rules that reflect quark number and momentum conservation \cite{Gaunt:2009re, Golec-Biernat:2014bva, Golec-Biernat:2015aza, Diehl:2020xyg} and from their behavior in the limit of small inter-parton distances \cite{Diehl:2011tt, Diehl:2011yj, Diehl:2019rdh}.  Beyond this, a considerable number of papers have investigated DPDs in quark models \cite{Chang:2012nw, Rinaldi:2013vpa, Broniowski:2013xba, Rinaldi:2014ddl, Broniowski:2016trx, Kasemets:2016nio, Rinaldi:2016jvu, Rinaldi:2016mlk, Rinaldi:2018zng, Courtoy:2019cxq, Broniowski:2019rmu, Rinaldi:2020ybv,Courtoy:2020tkd}.

A complementary approach is to study correlations inside a hadron using lattice QCD.  This has long been pursued at the level of two-current correlation functions, with a focus on various physics aspects \cite{Barad:1984px, Barad:1985qd, Wilcox:1986ge, Wilcox:1986dk, Wilcox:1990zc, Chu:1990ps, Lissia:1991gv, Burkardt:1994pw, Alexandrou:2002nn, Alexandrou:2003qt, Alexandrou:2008ru, Bali:2018nde}.  In the short-distance limit, such correlation functions can be used to extract parton distributions, which has recently been done in \cite{Sufian:2019bol}.

As was observed in \cite{Diehl:2011tt, Diehl:2011yj}, correlation functions of two currents in a hadron can also be related to the Mellin moments of DPDs.  This generalizes the well-known relation between single-current matrix elements and the Mellin moments of PDFs, which has been extensively exploited in lattice calculations as reviewed in \cite{Hagler:2009ni, Lin:2017snn, Lin:2020rut}.  We recently presented a corresponding calculation for DPDs in the pion \cite{Bali:2020mij}, using $n_f = 2$ ensembles with a pion mass around $300 \mev$.  The results have rather high statistical precision, and they reveal a number of interesting patterns.   For unpolarized quarks, they agree rather well with the quark model results in \cite{Courtoy:2019cxq, Broniowski:2019rmu}.  Encouraged by these findings, we went on to study the DPDs in an unpolarized proton, which are relevant to collider experiments.  We report on this in the present work; preliminary results have been published in \cite{Zimmermann:2019quf}.

Our paper is organized as follows. In \sect\ref{sec:theory}, we review the different theoretical objects relevant to our study and explain how they are related to each other.  Details of the lattice setup and ensembles we used are given in \sect\ref{sec:lattice}, with more technical information being collected in an appendix.  Sections \ref{sec:mellin} and \ref{sec:factorization} contain the results of our calculation.  In \sect\ref{sec:mellin}, the Mellin moments for different flavor and polarization combinations are presented, while the results of various factorization tests are discussed in \sect\ref{sec:factorization}. Such tests are especially important in view of the fact that many phenomenological models of DPS use similar factorization assumptions.  We summarize our findings in \sect\ref{sec:summary}.

\section{Theory}
\label{sec:theory}

In this section we review certain basics on double parton distributions (DPDs) and their relevance in the context of double parton scattering.

\subsection{Double parton distributions}
\label{sec:dps}

The DPD of a given hadron parameterizes the joint probability of finding two partons with certain polarization and momentum fractions $x_i$ at a given relative transverse distance $\tvec{y}$. For the case of the unpolarized proton, DPDs of quarks and antiquarks are defined by a proton matrix element of two operators:

\begin{align}
\label{eq:dpd-def}
F_{a_1 a_2}(x_1,x_2,\tvec{y})
= 2p^+ \int \dd y^- \int \frac{\dd z^-_1}{2\pi}\, \frac{\dd z^-_2}{2\pi}\,
&      e^{i\ms ( x_1^{} z_1^- + x_2^{} z_2^-)\ms p^+}
\nonumber \\
& \times
 \avsum_\lambda \bra{p,\lambda} \mathcal{O}_{a_1}(y,z_1)\, \mathcal{O}_{a_2}(0,z_2) \ket{p,\lambda}
\,,
\end{align}
where $\sum^\prime$ indicates the average over the proton helicity states. In \eqref{eq:dpd-def} we use light-cone coordinates $v^\pm = (v^0 \pm v^3)/\sqrt{2}$, $\tvec{v} = (v^1,v^2)$ for a given four-vector $v^\mu$. Moreover, we work in a frame where the transverse proton momentum vanishes, \ie $\tvec{p} = \tvec{0}$. The light-cone operators appearing in \eqref{eq:dpd-def} are defined as:

\begin{align}
\label{eq:quark-ops}
\mathcal{O}_{a}(y,z)
&= \bar{q}\bigl( y - \half z \bigr)\, \Gamma_{a} \, q\bigl( y + \half z \bigr)
   \Big|_{z^+ = y^+_{} = 0,\, \tvec{z} = \tvec{0}}\,,
\end{align}
where $a$ specifies the quark flavor and polarization, which is determined by the spin projections

\begin{align}
  \label{eq:quark-proj}
\Gamma_q & = \half \gamma^+ \,, &
\Gamma_{\Delta q} &= \half \gamma^+\gamma_5 \,, &
\Gamma_{\delta q}^j = \half i \sigma^{j+}_{} \gamma_5  \quad (j=1,2) \,.
\end{align}
In this notation, $q$ refers to the sum over all quark polarizations. $\Delta q$ denotes the difference between positive and negative helicity contributions and, therefore, corresponds to the longitudinal quark polarization, whereas $\delta q$ is the analogue for the case of transverse polarization. The expression of the light cone operators given in \eqref{eq:quark-ops} is only valid in light cone gauge, otherwise a Wilson line has to be inserted. Notice that the light cone operators have to be renormalized. This leads to a scale dependence of the operators and, consequently, of the DPDs. For brevity, we do not indicate the scale.\

Because of momentum conservation, the momentum fractions can take only values satisfying $|x_1| + |x_2| \le 1$. Negative momentum fractions are associated with antiquarks, \iex

\begin{align}
\label{eq:dpd-neg-x}
F_{a_1 a_2}(-x_1,x_2,\tvec{y}) = \etac^{a_1} F_{\bar{a}_1 a_2}(x_1,x_2,\tvec{y})\ ,
\qquad
F_{a_1 a_2}(x_1,-x_2,\tvec{y}) = \etac^{a_2} F_{a_1 \bar{a}_2}(x_1,x_2,\tvec{y})\ ,
\end{align} 
where

\begin{align}
\label{eq:dpd-eta-c}
\etac^{a} = -1 ~~~\text{for}~ a = q, \delta q 
\ , \qquad
\etac^{a} = +1 ~~~\text{for}~ a = \Delta q\,.
\end{align}
DPDs fulfill certain sum rules, which have been proposed in \cite{Gaunt:2009re} and proven in \cite{Gaunt:2012tfk,Diehl:2018kgr}. In this paper, we consider the number sum rule. In position space this can be formulated as:

\begin{align}
   \label{eq:dpd-sr}
&\int_{-1+|x_1|}^{1-|x_1|} \dd x_2 \int_{|\tvec{y}|>y_{\text{cut}}}^\infty \!
   \dd^2 \tvec{y}\, F_{q q^\prime}(x_1,x_2,\tvec{y}) \nonumber \\
& \qquad = \left( N_{q^\prime} + \delta_{q\bar{q}^\prime} - \delta_{qq^\prime} \right) f_q(x_1) + \mathcal{O}(\Lambda^2 y_{\text{cut}}^2)
        + \mathcal{O}\bigl( \alpha_s \bigr) \,,
\end{align}
where $f_{q}(x)$ is an ordinary PDF for an unpolarized quark with flavor $q$ and satisfies $f_{\bar{q}}(x) = f_{q}(-x)$. $N_{q^\prime}$ is the number of valence quarks with flavor $q^\prime$. The lower cutoff in the integral over $\tvec{y}$ is necessary, because DPDs have a singular $1/\tvec{y}^2$ behavior for $\tvec{y}^2 \rightarrow 0$. This is caused by perturbative splitting processes, which are of $\Op(\alpha_s(\mu))$. For more details see \cite{Diehl:2017wew}. A common choice for the lower cutoff is $y_{\mathrm{cut}} = b_0/\mu$, where $\mu$ is the renormalization scale and $b_0 = 2e^{-\gamma} \approx 1.12$ with the Euler-Mascheroni constant $\gamma$.

The double parton distributions defined in \eqref{eq:dpd-def} are needed to compute double parton scattering processes. The corresponding cross section can be written in terms of two DPDs, integrated over the transverse parton distance:

\begin{align}
\int \dd^2\tvec{y} \;
    F_{a_1 a_2}(x_1, x_2, \tvec{y}) \, F_{b_1 b_2}(x_1', x_2', \tvec{y}) \,.
\end{align}
Hence, the dependence of DPDs on the transverse distance is not directly accessible in experiments. 

DPDs are often simplified and expressed in terms of single parton distributions within certain factorization approaches. The first procedure in this context is based on the insertion of a complete set of states between the operators in the matrix element in \eqref{eq:dpd-def}. Then it is assumed that nucleon states dominate, such that all other states can be neglected. This leads to an expression of DPDs in terms of impact parameter distributions $f_a(x,\tvec{b})$:

\begin{align}
   \label{eq:dpd-fact}
F_{a_1 a_2}(x_1, x_2, \tvec{y})
& \overset{?}{=} \int \dd^2\tvec{b}\; f_{a_1}(x_1, \tvec{b} + \tvec{y})\,
                      f_{a_2}(x_2, \tvec{b}) \,.
\end{align}
This kind of factorization has been investigated on the lattice for the case of the pion. Significant differences between the \rhs and \lhs of \eqref{eq:dpd-fact} have been found, while the orders of magnitude are consistent with each other \cite{Bali:2020mij}. Similar observations have been made in quark model studies \cite{Rinaldi:2020ybv}.  We shall perform analogous investigations for the case of the nucleon in \sect\ref{sec:factorization}. 

The other factorization approach frequently used assumes a complete factorization \wrt all arguments:

\begin{align}
   \label{eq:dpd-pocket}
F_{a_1 a_2}(x_1, x_2, \tvec{y})
& \overset{?}{=} f_{a_1}(x_1)\, f_{a_2}(x_2) \, G(\tvec{y}) \,.
\end{align}
This leads to the so-called "pocket formula", where the DPS cross section is written as a product of two SPS cross sections \cite{Bartalini:2011jp}:

\begin{align}
\sigma_{\mathrm{DPS},ij} = \frac{1}{C} \frac{\sigma_{\mathrm{SPS},i}\  \sigma_{\mathrm{SPS},j}}{\sigma_{\mathrm{eff}}} \,,
\label{eq:dpd-pocket-formula}
\end{align}
where $i$ and $j$ indicate the final states of the two hard scattering processes. $C$ is a combinatoric factor, which is $2$ if $i=j$ and $1$, otherwise. The effective cross section $\sigma_{\mathrm{eff}}$ is defined by $\sigma_{\mathrm{eff}}^{-1} = \int \dd^2 \tvec{y} [G(\tvec{y})]^2$. The function $G(\tvec{y})$ must be independent of the quark flavor, which leads to the prediction that $\sigma_{\mathrm{eff}}$ should be a universal constant. Since we are not able to resolve the $x_i$ dependence of DPDs in lattice studies, we cannot investigate to what extent factorization approaches \wrt the momentum fractions are valid. However, we shall perform the evaluation of DPD Mellin moments for different quark flavor combinations, such that we are able to check the universality of the function $G(\tvec{y})$.

\subsection{Skewed double parton distributions}
\label{sec:sdps}

\begin{figure}
\begin{center}
\includegraphics[width=0.48\textwidth]{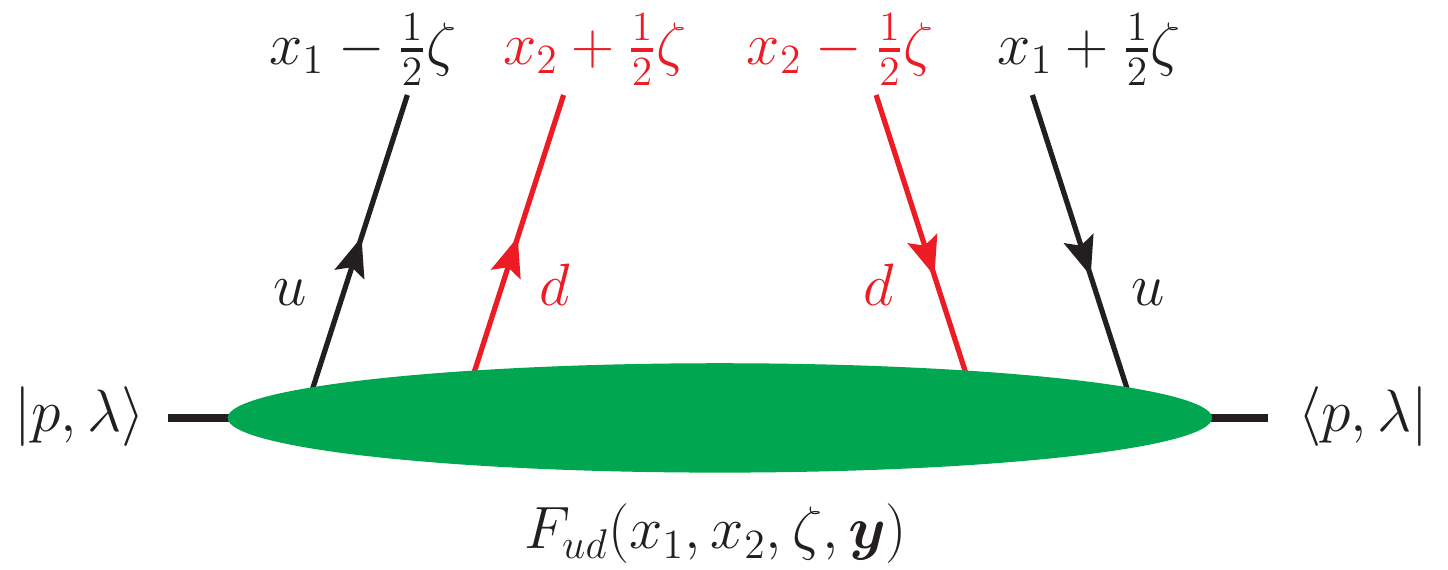}
\end{center}
\caption{Illustration of a skewed DPD of a $u$ and a $d$ quark for the case where all fractions $x_i \pm \zeta/2$ are positive. \label{fig:zeta-distrib}}
\end{figure}

\begin{figure*}
\begin{center}
\includegraphics[width=0.99\textwidth]{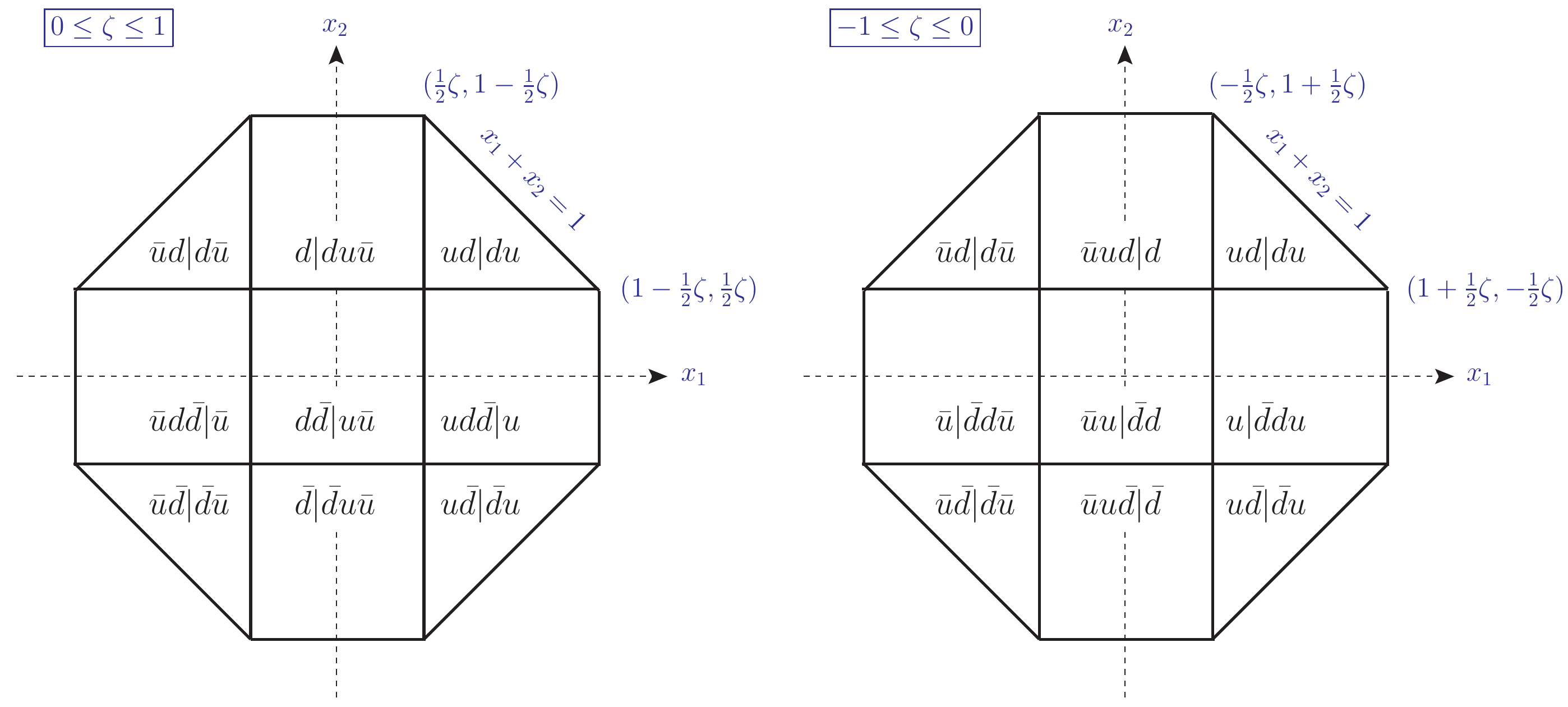}
\end{center}
\caption{Support regions of $F_{ud}(x_i,\zeta,\tvec{y})$ \wrt the arguments $x_i$ and $\zeta$. This is shown for positive (left) and negative (right) skewness parameter $\zeta$. For each sub-region we indicate the (anti-)quark content of the wave function and its complex conjugate. The notation $u|\bar{d}du$ means that we have a $u$-quark in the proton wave function and $\bar{d}du$ in its complex conjugate. \label{fig:dpd-support}}
\end{figure*}

The DPDs defined in \eqref{eq:dpd-def} can be extended by introducing an additional phase in the definition. This causes a difference between the momenta of the emitted and absorbed partons, respectively. We call the resulting functions skewed DPDs, which additionally depend on the skewness parameter $\zeta$:

\begin{align}
\label{eq:dpd-skew-def}
F_{a_1 a_2}(x_1,x_2,\zeta, \tvec{y})
= 2p^+ \int \dd y^- e^{-i \zeta y^- p^+}
 &  \int \frac{\dd z^-_1}{2\pi}\, \frac{\dd z^-_2}{2\pi}\,
          e^{i\ms ( x_1^{} z_1^- + x_2^{} z_2^-)\ms p^+}
\nonumber \\
& \quad \times
    \avsum_\lambda \bra{p,\lambda} \mathcal{O}_{a_1}(y,z_1)\, \mathcal{O}_{a_2}(0,z_2)
    \ket{p,\lambda} \,.
\end{align}
The partons have momentum fractions $x_i \pm \half \zeta$. The sign of the fractions determines whether there is a quark (antiquark) in the proton wave function or an antiquark (quark) in its complex conjugate. An overview of the corresponding regions is given in \fig\ref{fig:dpd-support}. If all fractions are positive, we have two quarks with momentum fractions $x_1 - \half \zeta$ and $x_2 + \half \zeta$ in the proton wave function and two quarks with $x_1 + \half \zeta$, $x_2 - \half \zeta$ in its complex conjugate. This is sketched in \fig\ref{fig:zeta-distrib} for the case of a $u$ and a $d$ quark.
Because of momentum conservation, the region in the $(x_1,x_2,\zeta)$-parameter space where the skewed DPDs are non-zero is restricted by

\begin{align}
\label{eq:support-region}
|x_i \pm \half \zeta| \le 1\,, \qquad |x_1|+|x_2| \le 1\,, \qquad |\zeta| \le 1\,.
\end{align}
The corresponding support region is also indicated in Figure~\ref{fig:dpd-support}.
From $PT$ invariance it follows that:

\begin{align}
  \label{eq:reverse-zeta}
F_{a_1 a_2}(x_1,x_2,\zeta, \tvec{y})
&= \etapt^{a_1} \, \etapt^{a_2} \, F_{a_1 a_2}(x_1,x_2,-\zeta, -\tvec{y}) \,,
\end{align}
with

\begin{align}
\label{eq:dpd-eta-pt}
\etapt^{a} = -1 ~~~\text{for}~ a = \Delta q, \delta q 
\ , \qquad
\etapt^{a} = +1 ~~~\text{for}~ a =  q \,.
\end{align}
Moreover, one can give a decomposition of the skewed DPDs in terms of functions that are rotationally invariant in the transverse plane.  $F_{q_1 q_2}$ and $F_{\Delta q_1 \Delta q_2}$ have even parity and are scalar quantities, therefore they are already rotationally invariant. By contrast $F_{\Delta q_1 q_2}$ and $F_{q_1 \Delta q_2}$ are parity-odd, which implies that they have to vanish. From invariance under time reflection, it also follows that the $T$-odd quantities $F^{j_1}_{\delta q_1 \Delta q_2}$ and $F^{j_2}_{\Delta q_1 \delta q_2}$ are zero. The remaining DPDs can be decomposed in terms of transverse vectors as follows: 

\begin{align} \label{eq:invar-dpds}
F_{q_1 q_2}(x_1,x_2,\zeta, \tvec{y}) &= f_{q_1 q_2}(x_1,x_2,\zeta, y^2) \,,
\nonumber \\
F_{\Delta q_1 \Delta q_2}(x_1,x_2,\zeta, \tvec{y})
&= f_{\Delta q_1 \Delta q_2}(x_1,x_2,\zeta, y^2) \,,
\nonumber \\
F_{\delta q_1 q_2}^{j_1}(x_1,x_2,\zeta, \tvec{y})
&= \epsilon^{j_1 k} \tvec{y}^k\, m f_{\delta q_1 q_2}(x_1,x_2,\zeta, y^2) \,,
\nonumber \\
F_{q_1 \delta q_2}^{j_2}(x_1,x_2,\zeta, \tvec{y})
&= \epsilon^{j_2 k} \tvec{y}^k\, m f_{q_1 \delta q_2}(x_1,x_2,\zeta, y^2) \,,
\nonumber \\
F_{\delta q_1 \delta q_2}^{j_1 j_2}(x_1,x_2,\zeta, \tvec{y})
&= \delta^{j_1 j_2} f^{}_{\delta q_1 \delta q_2}(x_1,x_2,\zeta, y^2)
\nonumber \\
&\quad  + \bigl( 2 \tvec{y}^{j_1} \tvec{y}^{j_2}
         - \delta^{j_1 j_2} \tvec{y}^2 \bigr)\ms
   m^2 f^{\ms t}_{\delta q_1 \delta q_2}(x_1,x_2,\zeta, y^2) \,,
\end{align}
where $m$ is the proton mass and $\epsilon^{ij}$ is the antisymmetric tensor in two dimensions, with $\epsilon^{12}=1$. Notice that $y^2 = y^\mu y_\mu$ denotes the Lorentz invariant scalar product. In our case we have $y^+ = 0$, \ie $y^2 = -\tvec{y}^2$.
For $\zeta = 0$ the functions on the \rhs of \eqref{eq:invar-dpds} have the following physical interpretation:
\begin{itemize}
\item $f_{q_1 q_2}$ describes the probability of finding two quarks with momentum fractions $x_1$ and $x_2$ at a transverse distance $\tvec{y}$. It contains a sum over all quark polarization states, \ie the quarks are unpolarized.
\item $f_{\Delta q_1 \Delta q_2}$ describes the difference between the probabilities of finding the two quarks with aligned or anti-aligned spins in the longitudinal direction. This gives a measure for the longitudinal quark polarization.
\item $f_{\delta q_1 \delta q_2}$ is the analogue of $f_{\Delta q_1 \Delta q_2}$ for polarization in the transverse direction.
\item $f_{\delta q_1 q_2}$ describes the correlation between the transverse polarization of the first quark and the transverse distance $\tvec{y}$ between the two quarks. $f_{q_1 \delta q_2}$ can be interpreted in analogy, where the second quark is polarized and the first unpolarized.
\item $f_{\delta q_1 \delta q_2}^t$ gives the correlation between the transverse distance $\tvec{y}$ of the quarks and their transverse polarizations.
\end{itemize}
The functions $f_{\delta q_1 q_2}$, $f_{q_1 \delta q_2}$, and $f_{\delta q_1 \delta q_2}^t$ describe spin-orbit correlations, whereas $f_{\delta q_1 \delta q_2}$ and $f_{\Delta q_1 \Delta q_2}$ quantify spin-spin correlations. Combining \eqref{eq:invar-dpds} and \eqref{eq:reverse-zeta} we find:

\begin{align}
  \label{eq:even-in-zeta}
f_{a_1 a_2}(x_1,x_2,\zeta, y^2)
&= f_{a_1 a_2}(x_1,x_2,-\zeta, y^2)\,.
\end{align}
The matrix elements in the definitions \eqref{eq:dpd-def} and \eqref{eq:dpd-skew-def} are not directly accessible on a Euclidean lattice, since they involve light-like distances. A way to circumvent this obstacle is to consider Mellin moments of skewed DPDs.  The lowest Mellin moment is defined as:

\begin{align}
  \label{eq:skewed-inv-mellin-mom-def}
I_{a_1 a_2}(\zeta, y^2)
&= \int_{-1}^{1} \dd x_1^{} \int_{-1}^{1} \dd x_2^{} \;
   f_{a_1 a_2}(x_1,x_2,\zeta, y^2) 
\nonumber \\
& = \int_0^1 \dd x_1 \int_0^1 \dd x_2 \left[ f_{a_1 a_2}(x_1,x_2,\zeta,y^2) + \etac^{a_1} f_{\bar{a}_1 a_2}(x_1,x_2,\zeta,y^2) \right.
\nonumber \\ 
&\qquad\left. + \etac^{a_2} f_{a_1 \bar{a}_2}(x_1,x_2,\zeta,y^2) + \etac^{a_1}\etac^{a_2} f_{\bar{a}_1 \bar{a}_2}(x_1,x_2,\zeta,y^2)
 \right]\,.
\end{align}
The integrals over $x_1$ and $x_2$ in \eqref{eq:skewed-inv-mellin-mom-def} together with the exponentials $e^{ix_i p^+ z_i^-}$ in \eqref{eq:dpd-def} and \eqref{eq:dpd-skew-def} set $z_i^-$ to zero, which is what we intended. The resulting matrix elements involve only local quark bilinears, which can be evaluated in lattice simulations.

\subsection{Two-current matrix elements}
\label{sec:tcme_def}

We define two-current matrix elements of the proton with momentum $p$ as:

\begin{align}
  \label{eq:mat-els}
M^{\mu_1 \cdots \mu_2 \cdots}_{q_1 q_2, i_1 i_2}(p,y)
&:= \avsum_\lambda \bra{p,\lambda} J^{\mu_1 \cdots}_{q_1, i_1}(y)\,
              J^{\mu_2 \cdots}_{q_2, i_2}(0) \ket{p,\lambda} \,,
\end{align}
where we take the average of the proton spin. The currents $J^{\mu\dots}_{q,i}$ are local quark bilinear operators. In this work we focus on three types of currents, which are defined as:

\begin{align}
  \label{eq:local-ops}
J_{q, V}^\mu(y) &= \bar{q}(y) \ms \gamma^\mu\ms q(y) \,,
&
J_{q, A}^\mu(y) &= \bar{q}(y) \ms \gamma^\mu \gamma_5\, q(y) \,,
&
J_{q, T}^{\mu\nu}(y) &= \bar{q}(y) \ms \sigma^{\mu\nu} \ms q(y) \,.
\end{align}
These operators commute if the distance vector $y$ is space-like, in particular for $y^0=0$. A consequence of this property is the relation:

\begin{align}
   \label{eq:exchange-J}
M^{\mu_1 \cdots \mu_2 \cdots}_{q_1 q_2, i_1 i_2}(p,y)
   &= M^{\mu_2 \cdots \mu_1 \cdots}_{q_2 q_1, i_2 i_1}(p,-y) \,.
\end{align}
Moreover, the currents have definite transformation behavior under charge conjugation and the combination of parity and time reflection: 

\begin{align}
J_{q, i}^{\mu \cdots}(y)
 & \underset{C}{\to} \etac^i\, J_{q, i}^{\mu \cdots}(y) \,,
&
J_{q, i}^{\mu \cdots}(y)
 & \underset{PT}{\to} \etapt^i\, J_{q, i}^{\mu \cdots}(-y)\,,
\end{align}
where in analogy to \eqref{eq:dpd-eta-c} and \eqref{eq:dpd-eta-pt} the sign factors $\etac$ and $\etapt$ are defined as:

\begin{align}
  \label{eq:cur-eta-c}
\etac^{i}
	&= -1  ~~~\text{for}~ i = V, T \,,
&
\etac^{i}
  	&= +1  ~~~\text{for}~ i = A 
\intertext{and}
\label{eq:cur-eta-pt}
\etapt^{i}
  &= -1  ~~~\text{for}~ i = A, T \,,
&
\etapt^{i}
  &= +1  ~~~\text{for}~ i = V \,.
\end{align}
The combined $PT$ symmetry implies for the two-current matrix elements:

\begin{align}
\label{eq:tcme-pt}
M^{\mu_1 \cdots \mu_2 \cdots}_{q_1 q_2, i_1 i_2}(p,y)
  &= \etapt^{i_1}\, \etapt^{i_2}\,
     M^{\mu_1 \cdots \mu_2 \cdots}_{q_1 q_2, i_1 i_2}(p,-y)\,.
\end{align}
In the context of DPDs we have to consider the current combinations $VV$, $AA$, $VT$, $TV$, and $TT$. The corresponding two-current matrix elements are by definition Lorentz tensors of a certain rank, which is determined by the involved currents. Therefore, the matrix elements can be decomposed in terms of Lorentz invariant functions and Lorentz tensors constructed from the four-vectors $p$ and $y$. In order to reduce the number of independent quantities, we subtract trace contributions and consider symmetric combinations. For brevity we skip the arguments $y$ and $p$ of the matrix elements $M$:

\begin{align}
  \label{eq:tensor-decomp}
M^{\{\mu\nu\}}_{q_1 q_2, V V} 
	- \tfrac{1}{4} g^{\mu\nu} g_{\alpha\beta}
		M^{\alpha\beta}_{q_1 q_2, V V}
 & = u_{V V,A}^{\mu\nu}\, A_{q_1 q_2}^{}
   + u_{V V,B}^{\mu\nu}\, m^2\ms B_{q_1 q_2}^{}
   + u_{V V,C}^{\mu\nu}\, m^4\ms C_{q_1 q_2}^{} \,,
\nonumber \\[0.1em]
M^{\mu\nu\rho}_{q_1 q_2, T V} + \tfrac{2}{3} g_{\vphantom{q_1T}}^{\rho[\mu} M^{\nu]\alpha\beta}_{q_1 q_2, T V} g_{\alpha\beta}
 & = u_{T V,A}^{\mu\nu\rho}\, m\ms A_{\delta q_1 q_2}^{}
   + u_{T V,B}^{\mu\nu\rho}\,  m^3\ms B_{\delta q_1 q_2}^{} \,,
\nonumber \\[0.3em]
\tfrac{1}{2} \, \bigl[ M^{\mu\nu\rho\sigma}_{q_1 q_2, T T}
     + M^{\rho\sigma\mu\nu}_{q_1 q_2, T T} \bigr]
 &= \tilde{u}_{T T,A}^{\mu\nu\rho\sigma}\, A_{\delta q_1 \delta q_2}^{}
    + \tilde{u}_{T T,B}^{\mu\nu\rho\sigma}\, m^2\ms B_{\delta q_1 \delta q_2}^{}
    + \tilde{u}_{T T,C}^{\mu\nu\rho\sigma}\, m^2\ms C_{\delta q_1 \delta q_2}^{}
\nonumber \\[0.3em]
 & \quad
    + \tilde{u}_{T T,D}^{\mu\nu\rho\sigma}\, m^4\ms D_{\delta q_1 \delta q_2}^{}
    + u_{T T,E}^{\mu\nu\rho\sigma}\, m^2\ms \widetilde{E}_{\delta q_1 \delta q_2}^{} \,.
\end{align}
Here we write $t^{\{\mu\nu\}} = (t^{\mu\nu}+t^{\nu\mu})/2$ and $t^{[\mu\nu]} = (t^{\mu\nu}-t^{\nu\mu})/2$ for an arbitrary tensor $t^{\mu\nu}$. The quantities $A$, $B$,...\ are Lorentz invariant functions, \ie they only depend on $py = p^\mu y_\mu$ and $y^2 = y^\mu y_\mu$. The decomposition of $M_{q_1 q_2, A A}^{\mu\nu}$, which is not explicitly given in \eqref{eq:tensor-decomp}, has the same form as the one for $M_{q_1 q_2, V V}^{\mu\nu}$ and introduces the functions $A_{\Delta q_1 \Delta q_2}$, $B_{\Delta q_1 \Delta q_2}$, and $C_{\Delta q_1 \Delta q_2}$. Decomposing $M_{q_1 q_2, V T}^{\mu\nu\rho}$ works in analogy to $M_{q_1 q_2, T V}^{\mu\nu\rho}$ with the Lorentz indices interchanged appropriately. The basis tensors are given by: 

\begin{align}
  \label{basis-tensors}
u_{V V,A}^{\mu\nu} &= 2 p^\mu p^\nu - \half\ms g^{\mu\nu} p^2 \,,
\nonumber \\
u_{V V,B}^{\mu\nu} &= 2 p^{\{\mu} y^{\nu\}} - \half\ms g^{\mu\nu} py \,,
\nonumber \\
u_{V V,C}^{\mu\nu} &= 2 y^\mu y^\nu - \half\ms g^{\mu\nu} y^2 \,,
\nonumber \\[0.3em]
u_{T V,A}^{\mu\nu\rho} &= 4y^{[\mu} p^{\nu]} p^\rho
      + \tfrac{4}{3}\ms g^{\rho[\mu} y^{\nu]} p^2
      - \tfrac{4}{3}\ms g^{\rho[\mu} p^{\nu]} py \,,
\nonumber \\
u_{T V,B}^{\mu\nu\rho} &= 4y^{[\mu} p^{\nu]} y^\rho
      + \tfrac{4}{3}\ms g^{\rho[\mu} y^{\nu]} py
      - \tfrac{4}{3}\ms g^{\rho[\mu} p^{\nu]} y^2 \,,
\nonumber \\[0.3em]
\tilde{u}_{T T,A}^{\mu\nu\rho\sigma} &= - 8\ms p^{[\nu} g^{\mu][\rho}  p^{\sigma]} \,,
\nonumber \\
\tilde{u}_{T T,B}^{\mu\nu\rho\sigma} &= - y^2\, u_{T T,A}^{\mu\nu\rho\sigma}
    - 16\ms y^{[\mu} p^{\nu]} y^{[\rho} p^{\sigma]} \,,
\nonumber \\
\tilde{u}_{T T,C}^{\mu\nu\rho\sigma} &= - 4 p^{[\nu} g^{\mu][\rho} y^{\sigma]} 
    - 4 y^{[\nu} g^{\mu][\rho} p^{\sigma]} \,,
\nonumber \\
\tilde{u}_{T T,D}^{\mu\nu\rho\sigma} &= - 8 y^{[\nu} g^{\mu][\rho}  y^{\sigma]} \,,
\nonumber \\
u_{T T,E}^{\mu\nu\rho\sigma} &=  2 g^{\mu[\rho} g^{\sigma]\nu} \,.
\end{align}
Notice that the tensors $\tilde{u}_{TT,A}\dots\tilde{u}_{TT,D}$ are \emph{not} trace-subtracted, which is in contrast to the analogous tensors $u_{TT,A}\dots u_{TT,D}$ defined in \cite{Bali:2020mij}. For this reason, the last term in \eqref{eq:tensor-decomp}, which is proportional to the trace, involves a modified invariant function $\widetilde{E}_{\delta q_1 \delta q_2}$ rather than the original function $E_{\delta q_1 \delta q_2}$. The remaining functions $A_{\delta q_1 \delta q_2}\dots D_{\delta q_1 \delta q_2}$ are the same as in \cite{Bali:2020mij}. Using the decomposition \eqref{eq:invar-dpds}, we can relate the two-current matrix elements \eqref{eq:mat-els} to the DPD Mellin moments \eqref{eq:skewed-inv-mellin-mom-def}:

\begin{align}
  \label{eq:t2-mat-els}
\int_{-\infty}^{\infty} \dd y^-\, e^{-i\zeta y^- p^+} M_{q_1 q_2, V V}^{++}(p,y)
   \, \Big|_{y^+ = 0,\, \tvec{p} = \tvec{0}}
  &= 2 p^+\ms I^{}_{q_1 q_2}(\zeta,y^2) \,,
\nonumber \\
\int_{-\infty}^{\infty} \dd y^-\ms e^{-i\zeta y^- p^+} M_{q_1 q_2, A A}^{++}(p,y)
   \, \Big|_{y^+ = 0,\, \tvec{p} = \tvec{0}}
  &= 2 p^+\ms I^{}_{\Delta q_1 \Delta q_2}(\zeta,y^2) \,,
\nonumber \\
\int_{-\infty}^{\infty} \dd y^-\ms e^{-i\zeta y^- p^+} M_{q_1 q_2, T V}^{k_1 ++}(p,y)
   \, \Big|_{y^+ = 0,\, \tvec{p} = \tvec{0}}
  &= 2 p^+\ms \tvec{y}^{k_1} m I^{}_{\delta q_1 q_2}(\zeta,y^2) \,,
\nonumber \\
\int_{-\infty}^{\infty} \dd y^-\ms e^{-i\zeta y^- p^+} M_{q_1 q_2, V T}^{+ k_2 +}(p,y)
   \, \Big|_{y^+ = 0,\, \tvec{p} = \tvec{0}}
  &= 2 p^+\ms \tvec{y}^{k_2} m I^{}_{q_1 \delta q_2}(\zeta,y^2) \,,
\nonumber \\
\int_{-\infty}^{\infty} \dd y^-\ms e^{-i\zeta y^- p^+} M_{q_1 q_2, T T}^{k_1 + k_2 +}(p,y)
   \, \Big|_{y^+ = 0,\, \tvec{p} = \tvec{0}} &
\nonumber \\
  & \!\!\!\!\!\!\!\!\!\!\!\!\!\!\!\!\!\!\!\!\!\!\!\!\!\!\!\!\!\!\!\!\!\!\!\!\!\!\!\!\!\!\!\!\!\!\!\!\!\!\!\!\!\!\!\!\!\!\!\!\!\!\!\!\!\!\!\!\!\!\!\!\!\!\!\!\!\!\!\!\!\!\!\!
    = 2 p^+\ms \bigl[\ms
      \delta^{k_1 k_2}\, I^{}_{\delta q_1 \delta q_2}(\zeta,y^2) - \bigl( 2 \tvec{y}^{k_1} \tvec{y}^{k_2} - \delta^{k_1 k_2} \tvec{y}^2 \bigr)
     m^2 \ms I^t_{\delta q_1 \delta q_2}(\zeta,y^2) \bigr]\,.
\end{align}
Notice that the Dirac structure in the local tensor operator $J_{q,T}^{\mu\nu}$ differs from that in the spin projection $\Gamma_{\delta q}^j$ by an extra $\gamma_5$, see \eqref{eq:quark-proj} and \eqref{eq:local-ops}. This corresponds to a rotation by 90° in the transverse plane, which follows from the relation $i\sigma^{j+}\gamma_5 = \epsilon^{jk}\sigma^{k+}$ and has been taken into account in \eqref{eq:t2-mat-els}.

Comparing \eqref{eq:t2-mat-els} with \eqref{eq:tensor-decomp} we find the following relations between the DPD Mellin moments and the invariant functions:

\begin{align}
\label{eq:skewed-mellin-inv-fct}
I_{a_1 a_2}(\zeta, y^2)
&= \int_{-\infty}^{\infty} \dd(py)\, e^{-i\zeta py}\, A_{a_1 a_2}(py,y^2) \,,
\nonumber \\
I^t_{\delta q \delta q^\prime}(\zeta, y^2)
&= \int_{-\infty}^{\infty} \dd(py)\, e^{-i\zeta py}\, B_{\delta q \delta q^\prime}(py,y^2) \,,
\end{align}
\ie the Mellin moments are Fourier transforms of the invariant functions $A_{a_1 a_2}$ and $B_{\delta q \delta q^\prime}$. We refer to this subset of invariant functions as twist-two functions throughout this paper. Since the Mellin moments are symmetric in $\zeta$, which follows from \eqref{eq:reverse-zeta}, the inverse Fourier transform at $py=0$ can be written as:

\begin{align}
\label{eq:py-zero-fct}
A_{a_1 a_2}(py=0, y^2) = \frac{1}{\pi}
  \int_{0}^1 \dd \zeta \, I_{a_1 a_2}(\zeta, y^2) \,.
\end{align}
We define even $\zeta$-moments of the Mellin moments:

\begin{align}
  \label{eq:zeta-moments}
\langle \zeta^{2 m} \rangle_{a_1 a_2}^{}(y^2) &= \frac{%
    \int_{-1}^1 \dd \zeta \; \zeta^{2 m}\, I_{a_1 a_2}(\zeta, y^2)}{%
    \int_{-1}^1 \dd \zeta \, I_{a_1 a_2}(\zeta, y^2)}
 = \Biggl[\ms \frac{(-1)^m}{A_{a_1 a_2}(py, y^2)}
      \frac{\partial^{2 m} A_{a_1 a_2}(py, y^2)}{(\partial\ms py)^{2 m}}
   \ms\Biggr]_{py=0} \,,
\end{align}
whereas odd $\zeta$-moments vanish because of parity. The last expression in \eqref{eq:zeta-moments} follows from inserting \eqref{eq:skewed-mellin-inv-fct} and performing an integration by parts. Hence, the $2m$-th moment in $\zeta$ is directly related to the $2m$-th derivative in $py$ of the corresponding twist-two function at $py=0$.

\section{Two-current matrix elements on the lattice}
\label{sec:lattice}

In order to perform lattice simulations, we switch to Euclidean spacetime in this section. The corresponding time component of a four vector $x^\mu$ is denoted by $x^4$ instead of $x^0$. In Euclidean spacetime, the matrix elements given in \eqref{eq:mat-els} can be directly calculated in lattice QCD if the distance between the two insertion operators is purely spatial, \ie $y^4 = y^0 = 0$. In this section we describe the relation between the matrix elements and the lattice four-point functions defined below for the case of the nucleon and explain the techniques we use for the evaluation of the latter.

\subsection{Four-point functions and matrix elements}
\label{sec:fourpt_def}

\paragraph{Definition:} We define the proton four-point function $C^{ij,\mvec{p}}_{\mathrm{4pt}}$ as the correlator of a proton creation operator $\overline{\mathcal{P}}$ (source), the corresponding annihilation operator $\mathcal{P}$ (sink), and the two local currents $J_i$ defined in \eqref{eq:local-ops}:

\begin{align}
C^{ij,\mvec{p}}_{\mathrm{4pt}}(\mvec{y},t,\tau) 
&:= 
	a^6 \sum_{\mvec{z}^\prime,\mvec{z}} 
	e^{-i\mvec{p}(\mvec{z}^\prime-\mvec{z})}\  
	\left\langle \tr \left\{
		P_+ \mathcal{P}(\mvec{z}^\prime,t)\ J_i(\mvec{y},\tau)\ 
		J_j(\mvec{0},\tau)\ \overline{\mathcal{P}}(\mvec{z},0) 
	\right\} \right\rangle\,, 
\label{eq:4ptdef}
\end{align}
where the sum over $\mvec{z}$, $\mvec{z}^\prime$ combined with the exponential injects a total proton momentum, and the operator

\begin{align}
P_+ = \frac{1}{2} \left( \one + \gamma_4 \right)\,
\label{eq:parity-op}
\end{align} 
projects onto positive parity. The proton creation and annihilation operators, which we also refer to as interpolators, are given by tri-quark operators matching the proton's spin $J=1/2$ and isospin $I=1/2$:

\begin{align}
\overline{\mathcal{P}}(\mvec{x},t) &:= 
	\left.\epsilon_{abc}\ 
		\left[ 
			\bar{u}_a(x)\ C \gamma_5\ \bar{d}_b^{\,T}(x) 
		\right] \bar{u}_c(x) 
	\right|_{x^4=t} \,, 
	\nonumber\\
\mathcal{P}(\mvec{x},t) &:= 
	\left.\epsilon_{abc}\ u_a(x) 
	\left[ 
		u_b^T(x)\ C \gamma_5\ d_c(x) 
	\right] \right|_{x^4=t} \,,
\label{eq:interpdef}
\end{align}
where $C$ is the charge conjugation matrix in spinor space, and $[.]$ indicates a scalar quantity \wrt spinor indices. The traces in \eqref{eq:4ptdef} are taken \wrt the open spinor indices introduced by the quark fields $u_a$ and $\bar{u}_c$, respectively. Furthermore, we define the two-point function:

\begin{align}
C^{\mvec{p}}_{\mathrm{2pt}}(t) &:= 
	a^6 \sum_{\mvec{z}^\prime,\mvec{z}} 
	e^{-i\mvec{p}(\mvec{z}^\prime-\mvec{z})}\  
	\left\langle \tr \left\{ 
		P_+ \mathcal{P}(\mvec{z}^\prime,t)\ 
		\overline{\mathcal{P}}(\mvec{z},0) \right\} 
	\right\rangle\,.
\label{eq:2ptdef}
\end{align}
We denote the separation in Euclidean time direction between the source and the current insertions by $\tau$, and the separation between the source and the sink by $t$.

\begin{figure}
\begin{center}
\includegraphics[scale=1]{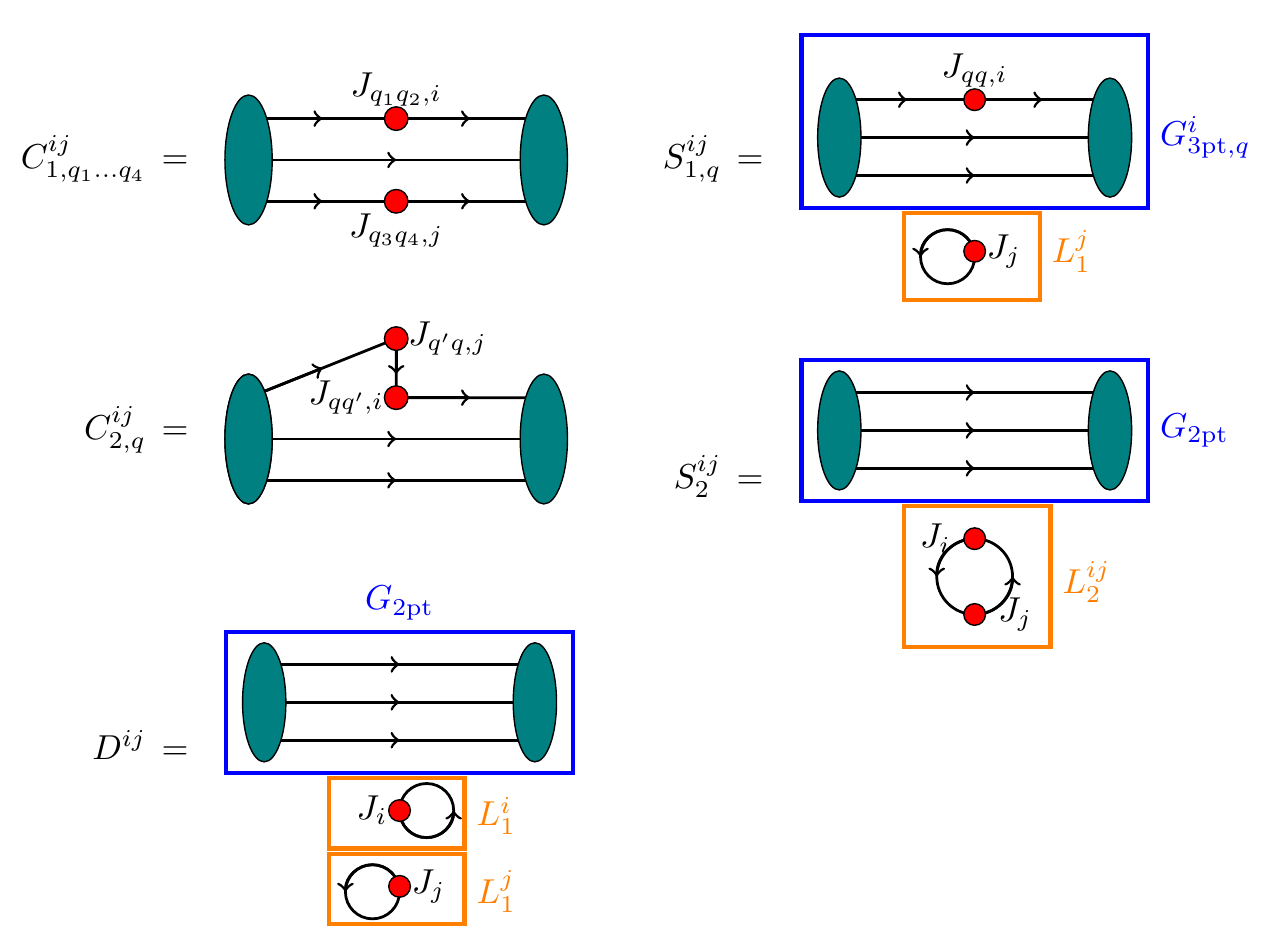}
\end{center}
\caption{Illustration of the five kinds of Wick contractions (graphs) contributing to a four-point function of a baryon. The explicit contributions for the graphs $C_1$, $C_2$ and $S_1$ depend on the quark flavor of the current insertions (red points). In the case where all quark flavors have the same mass, $C_2$ only depends on the flavors of the two propagators connected to the source or the sink. These flavors have to be the same for proton-proton matrix elements. For the graphs $S_1$, $S_2$ and $D$ we also indicate the parts connected to the proton source and sink, \ie $G_{3\mathrm{pt}}$ and $G_{2\mathrm{pt}}$ (blue), as well as the disconnected loops $L_1$ and $L_2$ (orange). \label{fig:graphs}}
\end{figure}

\paragraph{Wick contractions:} 
The evaluation of the correlation functions \eqref{eq:4ptdef} and \eqref{eq:2ptdef} leads to a definite set of Wick contractions. While there are only two contractions arising from permutations of $u\bar{u}$-pairs in the two-point function, there is a multitude of possible contractions in the case of the four-point functions, which can be grouped into five types. Following the notation of \cite{Bali:2018nde} we call them $C_1$, $C_2$, $S_1$, $S_2$ and $D$. They can be represented by the graphs illustrated in \fig{\ref{fig:graphs}}. $S_1$, $S_2$ and $D$ are disconnected contractions involving the sub-graphs $G_{3\mathrm{pt}}$ and $G_{2\mathrm{pt}}$, as well as the loops $L_1$ and $L_2$. Explicit expressions are given in \appx\ref{sec:fptcontr}. The explicit contribution of a given type depends on the flavor content of the inserted operators $J_{q q^\prime,i}$, which in general can be flavor changing. In the case of the graph $C_1$ this is indicated by the four flavors $q_1 \dots q_4$ of the quark lines connected to the current insertions, where the first two indices correspond to the flavor of the first operator $J_{q_1 q_2,i}$ and the last two flavor indices are those of the second operator $J_{q_3 q_4,j}$. For the proton there are three independent contributions called $C_{1,uuuu}$, $C_{1,uudd}$ and $C_{1,uddu}$, where the latter is not considered in this work, since we restrict ourselves to flavor conserving currents $J_{q q,i} = J_{q,i}$, see definition \eqref{eq:local-ops}. If all considered quarks have the same mass, the graphs $C_{2,q}$ and $S_{1,q}$ depend only on the flavor $q$ of the two propagators connecting the source or sink with one of the current insertions. Therefore, in the case of proton-proton matrix elements there are two possibilities for each contraction: $C_{2,u}$, $C_{2,d}$, $S_{1,u}$, $S_{1,d}$. For each of the contractions $S_2$ and $D$ there is only one contribution, which is flavor independent. Notice that we define the quantities $C_{1,uuuu}$, $C_{1,uudd}$, $C_{1,uddu}$, $C_{2,u}$, $C_{2,d}$, $S_{1,u}$, $S_{1,d}$, $S_2$, and $D$ as a sum of all quark permutations that share the same quark line topology. In particular, this includes permutations of the two $u$-quarks of the proton itself (see the definitions in \eqref{eq:graph_C1}, \eqref{eq:graph_C2S1}, and \eqref{eq:graph_S2D}).

In addition to the desired nucleon ground state, the interpolators $\mathcal{P}$ and $\overline{\mathcal{P}}$ also create and annihilate excited states. In order to relate the four-point functions to physical matrix elements of the nucleon ground state, we have to ensure that these excited states are sufficiently suppressed. This can be achieved by taking large Euclidean time separations $t$ and $t-\tau$. In this context we define:

\begin{align}
C^{ij,\mvec{p}}_{\mathrm{4pt}}(\mvec{y}) := 
	2V \sqrt{m^2 + \mvec{p}^2} 
	\left. 
		\frac{C^{ij,\mvec{p}}_{\mathrm{4pt}}(\mvec{y},t,\tau)}
		{C^{\mvec{p}}_\mathrm{2pt}(t)} 
	\right|_{0 \ll \tau \ll t} \,,
\label{eq:4pt-2pt-ratio}
\end{align}
where $V = L^3 a^3$ denotes the spatial volume. The factor $2V \sqrt{m^2 + \mvec{p}^2}$ on the \rhs of \eqref{eq:4pt-2pt-ratio} ensures the correct normalization of states. In a similar manner, we define:

\begin{align}
C^{ij,\mvec{p}}_{1,uudd}(\mvec{y}) = 
	2V \sqrt{m^2+\mvec{p}^2} 
	\left. 
		\frac{
			C^{ij,\mvec{p}}_{1,uudd}(\mvec{y},t,\tau)
		}{
			C_{2\mathrm{pt}}^{\mvec{p}}(t)
		} 
	\right|_{0 \ll \tau \ll t}\,,
\end{align}
and likewise for the other contractions $C_{1,uuuu}$, $C_{2,u}$, \ldots that contribute to $C_{\mathrm{4pt}}(\mvec{y},t,\tau)$.
 
Let us now point out some properties of these contractions: Using \eqref{eq:K_sym} and \eqref{eq:Gijk_pt} and $PT$ invariance, as well as invariance under translations in the time direction we are able to deduce the relations

\begin{align}
G^{ij,\mvec{p}}(\mvec{y}) &= 
	\etapt^{ij}\ G^{ij,\mvec{p}}(-\mvec{y}) 
&\qquad 
\mathrm{for}\ G &= 
	C_{1,uudd}, C_{1,uuuu}, S_{1,u}, S_{1,d}, S_2, D\ ,\nonumber \\
G^{ij,\mvec{p}}(\mvec{y}) &= 
	\etapt^{ij}\ G^{ji,\mvec{p}}(\mvec{y}) 
&\qquad 
\mathrm{for}\ G &= 
	C_{2,u}, C_{2,d}\,,
\label{eq:graph_pt}
\end{align}
where $\etapt^{ij}$ is defined in \eqref{eq:cur-eta-pt}. Notice that strictly speaking \eqref{eq:graph_pt} is exactly fulfilled only if $\tau=t/2$. In the limit of large Euclidean time separations $0 \ll \tau\ll t$ we consider $G^{ij,\mvec{p}}$ to be constant \wrt $\tau$ so that the $PT$ relations can also be applied to the ratio \eqref{eq:4pt-2pt-ratio}. Invariance under $CP$ transformations, together with the relations \eqref{eq:K_sym} and \eqref{eq:Gijk_ccp}, implies for all contractions:

\begin{align}
\left[ G^{ij,\mvec{p}}(\mvec{y}) \right]^{*} = 
	\etaf^{ij} \etapt^{ij}\ G^{ij,\mvec{p}}(-\mvec{y})\,,
\label{eq:graph_ccp}
\end{align}
with

\begin{align}
\etaf^i &= +1  ~~~\text{for}~ i = V,T \,, 
& 
\etaf^i &= -1  ~~~\text{for}~ i = A \,, 
& 
\etaf^{ij} = \etaf^i \etaf^j\,.
\label{eq:eta_4}
\end{align}
If $\etaf^{ij} = 1$, which is the case for the matrix elements we consider in this work, the relations \eqref{eq:graph_pt} and \eqref{eq:graph_ccp} imply that $C_{1,uudd}$, $C_{1,uuuu}$, $S_{1,u}$, $S_{1,d}$, $S_2$, $D$ are real-valued, whereas $C_{2,u}, C_{2,d}$ can have non-vanishing imaginary parts. For these contractions we find

\begin{align}
2\re \left\{ C_{2,q}^{ij,\mvec{p}}(\mvec{y}) \right\} 
&= 
	C_{2,q}^{ij,\mvec{p}}(\mvec{y}) + 
	C_{2,q}^{ji,\mvec{p}}(-\mvec{y}) \,,
\nonumber\\
2i\im \left\{ C_{2,q}^{ij,\mvec{p}}(\mvec{y}) \right\} 
&= 
	C_{2,q}^{ij,\mvec{p}}(\mvec{y}) - 
	C_{2,q}^{ji,\mvec{p}}(-\mvec{y}) \,.
\label{eq:C2_re_im}
\end{align}
Moreover, translational invariance implies that

\begin{align}
G^{ij,\mvec{p}}(\mvec{y}) = 
	G^{ji,\mvec{p}}(-\mvec{y})
~~~\mathrm{for}\ G = S_2, D\,.
\end{align}

\paragraph{Contribution to physical matrix elements:}

Inserting a complete set of states between the interpolators and the current insertions and taking the limit $0\ll \tau\ll t$ (see \eqref{eq:4pt-2pt-ratio}), we find:

\begin{align}
C^{ij,\mvec{p}}_{\mathrm{4pt}}(\mvec{y}) = 
	\left. 
	\frac{
		\sum_{\lambda\lambda^\prime} 
		\bar{u}^{\lambda^\prime}(p) P_+ u^{\lambda}(p)\  
		\bra{p,\lambda} J_i(y)\ J_j(0) \ket{p,\lambda^\prime}
	}{
		\sum_{\lambda} 
		\bar{u}^{\lambda}(p) P_+ u^{\lambda}(p)
	}\right|_{y^0 = 0} \,,
\label{eq:4pt-spin-sum}
\end{align}
where $u^\lambda(p)$ is the usual spinor solution of the Dirac equation for the nucleon. Again we note that we set $y^0 = y^4 = 0$ so that the translation to Minkowski spacetime is trivial. By writing $y^0$ instead of $y^4$ on the \rhs of \eqref{eq:4pt-spin-sum} we refer to the matrix elements in Minkowski spacetime, which we are actually interested in. For the parity projection $P_+$ defined in \eqref{eq:parity-op} the \rhs of \eqref{eq:4pt-spin-sum} turns into the desired spin averaged proton matrix element. Considering the currents defined in \eqref{eq:local-ops} (we omit Lorentz indices for brevity), we can write:

\begin{align}
\label{eq:4pt-to-me}
C_{4\mathrm{pt}}^{ij,\mvec{p}}(\mvec{y}) = 
\left. \vphantom{\sum} 
	M_{q_1 q_2, i_1 i_2}(p,y) 
\right|_{y^0 = 0} \,,
\end{align}
where $M_{q_1 q_2, i_1 i_2}$ is the two-current matrix element \eqref{eq:mat-els} to be investigated. For the currents containing only the light quarks $u$ and $d$, we find for the proton matrix elements:

\begin{align}
\left. M_{ud, ij}(p,y)\right|_{y^0 = 0} 
&= 
	C^{ij,\mvec{p}}_{1,uudd}(\mvec{y}) + 
	S^{ij,\mvec{p}}_{1,u}(\mvec{y}) + 
	S^{ji,\mvec{p}}_{1,d}(-\mvec{y}) + 
	D^{ij,\mvec{p}}(\mvec{y})\,,
\nonumber \\
\left. M_{uu, ij}(p,y)\right|_{y^0 = 0} 
&= 
	C^{ij,\mvec{p}}_{1,uuuu}(\mvec{y}) + 
	C^{ij,\mvec{p}}_{2,u}(\mvec{y}) + 
	C^{ji,\mvec{p}}_{2,u}(-\mvec{y}) + 
	S^{ij,\mvec{p}}_{1,u}(\mvec{y}) + 
	S^{ji,\mvec{p}}_{1,u}(-\mvec{y})
\nonumber \\
&\quad + 
	S_{2}^{ij,\mvec{p}}(\mvec{y}) + 
	D^{ij,\mvec{p}}(\mvec{y})\,,
\nonumber \\
\left. M_{dd, ij}(p,y)\right|_{y^0 = 0} 
&= 
	C^{ij,\mvec{p}}_{2,d}(\mvec{y}) + 
	C^{ji,\mvec{p}}_{2,d}(-\mvec{y}) + 
	S^{ij,\mvec{p}}_{1,d}(\mvec{y}) + 
	S^{ji,\mvec{p}}_{1,d}(-\mvec{y}) 
\nonumber \\
&\quad +
	S_{2}^{ij,\mvec{p}}(\mvec{y}) + 
	D^{ij,\mvec{p}}(\mvec{y})\,.
\label{eq:phys_me_decomp}
\end{align}
According to \eqref{eq:C2_re_im}, we can identify the combination $C^{ij,\mvec{p}}_{2,q}(\mvec{y}) + C^{ji,\mvec{p}}_{2,q}(-\mvec{y})$ in \eqref{eq:phys_me_decomp} with the real part $2\re \{ C_{2,q}^{ij,\mvec{p}}(y) \}$. Since we consider the $u$ and $d$ quarks to have the same mass, the quantities we calculate exhibit an exact isospin symmetry. Therefore, we can relate our results for the proton matrix elements to those of the neutron:

\begin{align}
\left. M_{du, ij}\right|_n = \left. M_{ud, ij}\right|_p\,, 
\qquad 
\left. M_{dd, ij}\right|_n = \left. M_{uu, ij}\right|_p\,, 
\qquad 
\left. M_{uu, ij}\right|_n &= \left. M_{dd, ij}\right|_p\,.
\end{align}

\paragraph{Renormalization:} The operators $J_i(y)$ have to be renormalized multiplicatively, \ie :

\begin{align}
\label{eq:latt_op_ren}
J_i^{\overline{\mathrm{MS}}}(y) = Z_i J_i^{\mathrm{latt}}(y)\,,
\end{align}
where $J_i^{\mathrm{latt}}(y)$ are the bare lattice operators. The renormalization factors $Z_A$ and $Z_V$ for the axial and vector currents do not depend on the renormalization scale, because the associated anomalous dimensions vanish. By contrast, $Z_T$ refers to the scale

\begin{align}
\label{eq:the_scale}
\mu = 2~\mathrm{GeV}\,.
\end{align}
The renormalization constants $Z_i$ specific to our lattice setup with $\beta = 3.4$ have been determined in \cite{Bali:2020lwx} (see table X therein) using the $\mathrm{RI}^\prime$-MOM scheme. They include the conversion to the $\overline{\mathrm{MS}}$-scheme at 3-loop accuracy. We summarize the corresponding values in \tab{\ref{tab:renorm}}. 
\begin{table}
\begin{center}
\begin{tabular}{ccc}
\hline\hline
$Z_V$ & $Z_A$ & $Z_T$ \\
\hline
$0.7128$ & $0.7525$ & $0.8335$ \\
\hline\hline
\end{tabular}
\end{center}
\caption{Renormalization constants for the local currents $J_V$, $J_A$ and $J_T$ for $\beta = 3.4$ and the renormalization scale $\mu = 2~\mathrm{GeV}$ \cite{Bali:2020lwx} (table X therein). \label{tab:renorm}}
\end{table}

The matrix elements we are interested in contain two local operators. Hence, the two-current matrix element renormalized in the $\overline{\mathrm{MS}}$ scheme is given by:

\begin{align}
M^{\overline{\mathrm{MS}}}_{q_1 q_2, i_1 i_2} = 
Z_{i_1} Z_{i_2} M^{\mathrm{latt}}_{q_1 q_2, i_1 i_2}\,.
\end{align}
In other words, the product of renormalized operators $J_i^{\overline{\mathrm{MS}}}(y) \, J_j^{\overline{\mathrm{MS}}}(0)$ requires no additional renormalization, because we always consider a finite spacelike distance $y$ between the two currents.

\vspace*{\baselineskip}

\subsection{Technical details on Wick contractions}
\label{sec:tcme_wick}

\begin{figure}[ht]
\includegraphics[scale=1]{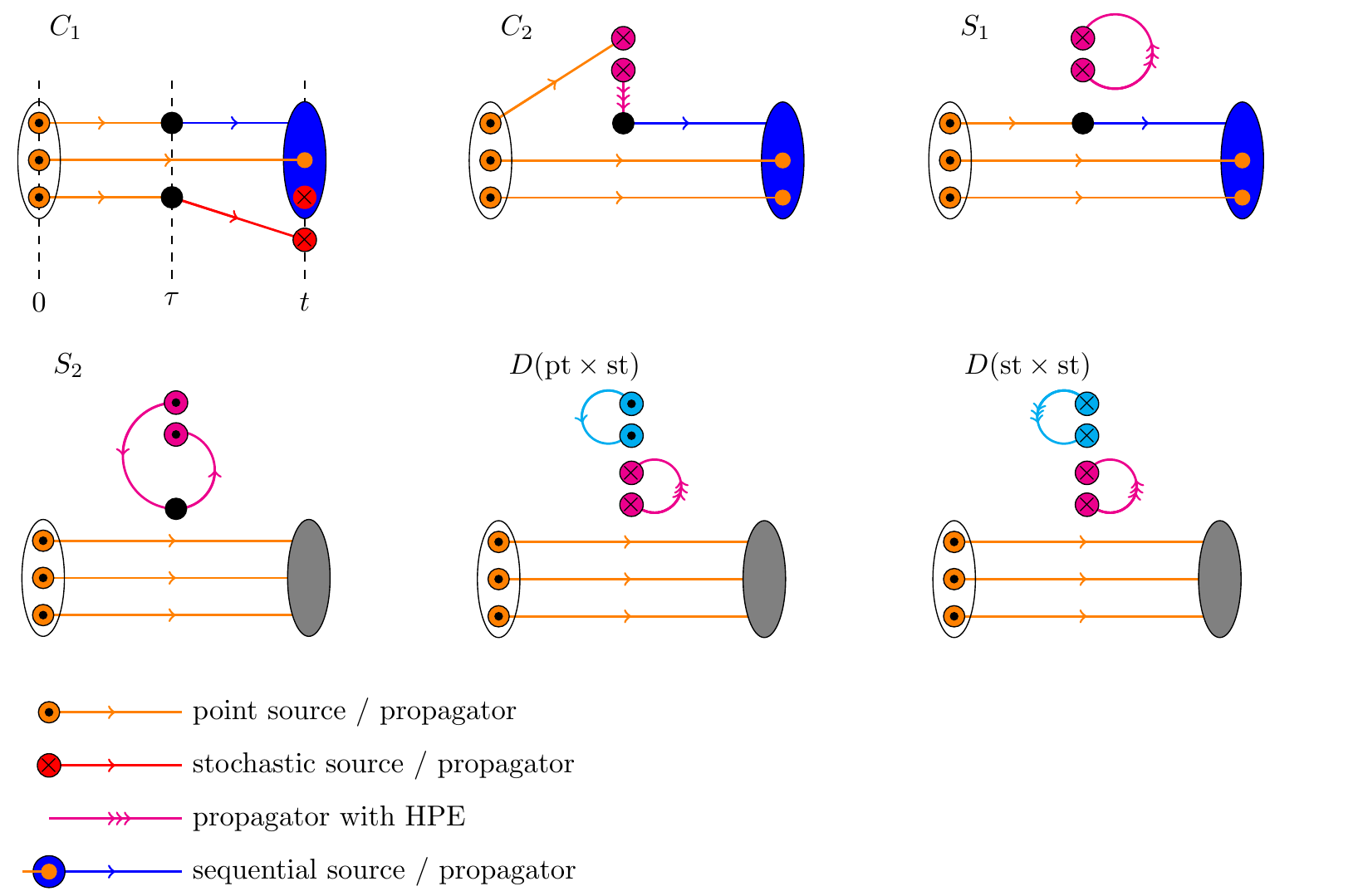}
\caption{Sketch of all Wick contractions including the techniques used for the evaluation of each piece on the lattice. We use one color for each involved quark source, \ie if two or more pieces within a graph share the same color, they involve the same quark source. Colors have no meaning regarding the evaluation technique. There are two versions of the $D$ graph: In the first we use two stochastic loops (bottom right), whereas in the second (bottom center) one stochastic loop is replaced by a point source loop. \label{fig:graphs_tech}}
\end{figure}

In the following, we discuss the technical details regarding the evaluation of each Wick contraction we have previously defined. A technical sketch of all graphs is shown in \fig{\ref{fig:graphs_tech}}. Each contraction is calculated on a smeared quark point source $S_z^{\Phi,\mvec{p}} = \Phi^{\mvec{p}} S_z$. It is a diagonal spinor-color matrix located at position $z$, \ie $(S_z)_{\alpha\beta}^{ab}(y) = \delta_{zy} \delta_{\alpha\beta} \delta_{ab}$, where $z^4$ is the nucleon source timeslice. Notice that here and in the following spinor indices are denoted by Greek letters $\alpha, \beta, \dots$, whereas Latin letters $a, b, \dots$ denote color indices of the fundamental representation. More details and explanations on the notation can be found in \appx\ref{sec:notation}

\paragraph{Smearing:} 

As already mentioned above, we apply a smearing function $\Phi^{\mvec{p}}$ to the corresponding sources and propagators, in order to increase the overlap of the proton interpolators with the proton ground state. $\Phi^{\mvec{p}}$ includes a phase injecting a momentum $b\mvec{p}$ to each of the quarks, where $\mvec{p}$ denotes the proton momentum. The method is known as momentum smearing \cite{Bali:2016lva}, which is based on the Wuppertal smearing technique \cite{Gusken:1989ad}. Explicitly, the smearing function $\Phi^{\mvec{p}}$ reads:

\begin{align}
(\Phi_0^{\mvec{p}})(x|y) = 
\frac{1}{1+6\epsilon} 
\left[ 
	\delta_{x,y} + 
	\epsilon \sum_{j=1}^3 
	\left( 
		\delta_{x+\hat{\jmath},y} U^{\mathrm{sm}}_{j}(x)\ 
		e^{ib\mvec{p}\hat{\jmath}} + 
		\delta_{y+\hat{\jmath},x} U^{\mathrm{sm},\dagger}_{j}(y)\ 
		e^{-ib\mvec{p}\hat{\jmath}} 
	\right) 
\right]\ ,
\label{eq:mom_smearing}
\end{align}
where we set $\epsilon = 0.25$ and $b = 0.45$, in order to obtain a maximal overlap with the ground state. The value of the latter parameter is specific to our setup. The smearing function is applied $n$ times, which is denoted by $\Phi = \Phi_0^n$. The gauge links $U^{\mathrm{sm}}$ appearing in \eqref{eq:mom_smearing} are obtained from the original gauge links by applying spatial APE-smearing \cite{Falcioni:1984ei}, which reduces unphysical short-distance fluctuations. $M^{\Phi,\mvec{p}}_z(y)$ denotes the source-smeared point-to-all quark propagator, which is obtained by solving:

\begin{align}
\D M^{\Phi,\mvec{p}}_z = S_z^{\Phi,\mvec{p}} := \Phi^{\mvec{p}} S_z\,,
\label{eq:point_to_all_prop}
\end{align}
where $\D$ is the Dirac operator. This propagator is used for the construction of each of the contractions relevant for the four-point function $C_{4\mathrm{pt}}$.

\paragraph{Stochastic propagators and improvements:} 

The all-to-all propagators required for the evaluation of most of the four-point graphs are estimated by use of time-local stochastic sources $\eta_t^{(\ell)}$. In this context the spatial unit matrix is approximated in the following way: 

\begin{align}
\frac{1}{N_{\mathrm{st}}} 
\sum_{\ell}^{N_{\mathrm{st}}} 
\eta_t^{(\ell)} \otimes \eta_t^{\dagger (\ell)} 
\xrightarrow{N_{\mathrm{st}} \rightarrow \infty} \one_t\,.
\label{eq:stoch_one}
\end{align}
In the present study we employ $Z_2 \otimes Z_2$ wall sources defined for a specific timeslice $t$, \ie the entries can take the values

\begin{align}
\left(\eta_t^{(\ell)}\right)_{\alpha a}(x) 
&= 
\frac{1}{\sqrt{2}} (\pm 1 \pm i)\ \delta_{x^4 t}\,.
\end{align}
The propagated stochastic source $\psi_t^{(\ell)}$, which we call "stochastic propagator" in the remainder of this work, is obtained by solving:

\begin{align}
\D \psi_t^{(\ell)} &= \eta_t^{(\ell)}\,.
\label{eq:stoch_prop}
\end{align}
It describes the propagation from any spatial position on timeslice $t$ to any other site on the lattice.

The off-diagonal components in \eqref{eq:stoch_one} are pure noise. This noise is particularly large for near-diagonal terms, where the propagator takes large values. Quantities involving these terms can be improved by a method that has also been used in \cite{Bali:2009hu}, where one exploits ultra-locality of the action. The method consists of applying the hopping parameter expansion (HPE), where the Dirac operator is rewritten as $\D = (\one - H)/(2a\kappa)$. Subsequently, the corresponding propagator can be expanded in terms of powers $H^n$ using the geometric series. Depending on the situation there exists a maximal order $N$ in the series, up to which the corresponding terms vanish exactly in the stochastic limit or, equivalently, for the exact solution of the propagator. This enables us to rewrite the propagator as:

\begin{align}
\D^{-1} 
&= 
	2a\kappa \sum_{n=0}^\infty H^n = 
	2a\kappa \sum_{n=0}^{N-1} H^n + 
		2a\kappa \sum_{n=N}^\infty H^n 
\nonumber \\
&\to 
	0 + 2a\kappa \sum_{n=N}^\infty H^n = 
	H^N 2a\kappa \sum_{n=0}^\infty H^n = 
	H^N \D^{-1}\,.
\label{eq:hpe_def}
\end{align}
The replacement $\D^{-1} \rightarrow H^N \D^{-1}$ removes the first $N$ terms in the expansion.

In the context of our calculations this method is used in two different places. The first one is a propagator connecting two sites on the same timeslice, which is needed for the calculation of the $C_2$ graph. Since the hopping term $H$ connects only nearest neighbors, we set $aN(\mvec{y}) = |y_1| + |y_2| + |y_3|$. Taking into account the periodicity of the lattice, the exact definition of $N(\mvec{y})$ is:

\begin{align}
N(\mvec{y}) = 
	\sum_{i=1}^3 \min 
	\left( \frac{|y_i|}{a}, L - \frac{|y_i|}{a} \right)\,.
\label{eq:nhop}
\end{align}
In the expression of the graph to be evaluated, we then have to replace $\psi_\tau^{(\ell)}$ by

\begin{align}
\label{eq:stoch_prop_hpe}
\xi_\tau^{(\ell),N} = H^{N} \psi_\tau^{(\ell)}\,.
\end{align}
If the propagator is contracted with a Dirac matrix, \eg in loops containing only one current, there is also a certain number of terms in the hopping parameter expansion that cancel. The number of terms that vanish depends on the Dirac matrix, see \tab{\ref{tab:loop_hpe}}.

\begin{table}
\begin{center}
\begin{tabular}{c|ccccc}
\hline\hline
$\Gamma$ & $V^\mu$ & $A^\mu$ & $T^{\mu\nu}$ \\
\hline
$N$ & $3$ & $4$ & $1$ \\
\hline\hline 
\end{tabular}
\end{center}
\caption{Number $N$ of omitted hopping terms in the $L_1$ contraction for each considered operator insertion type. \label{tab:loop_hpe}}
\end{table}

\paragraph{Interpolator kernels:} 

Before we continue to define the expressions to be evaluated, we introduce the compact notation:

\begin{align}
\label{eq:latt-bar-abbrev}
\left( E^a \right)^{bc}_{\beta\gamma} 
&:= 
	\epsilon^{abc} 
	\left( C \gamma_5 \right)_{\beta\gamma}\,,
\nonumber \\
\pha^{\mvec{p}}(x) &:= e^{-i\mvec{x}\mvec{p}}\,.
\end{align}
We then define the annihilation operator kernel:

\begin{align}
\label{eq:bar-ann-kern}
{}_{\sigma}\Op^{abc}_{\alpha\beta\gamma} = 
	\left(P_+\right)_{\sigma\alpha} 
	\left( E^a \right)^{bc}_{\beta\gamma}\,.
\end{align}
Contracting with the quark field operators, this yields the baryon annihilation operator \eqref{eq:interpdef} itself. In analogy, the baryon creation operator kernel is defined as:

\begin{align}
\label{eq:bar-cre-kern}
{}_{\sigma}\overline{\Op}^{abc}_{\alpha\beta\gamma} = 
	\left( E^c \right)^{ab}_{\alpha\beta} 
	\left(P_+\right)_{\gamma\sigma}\,.
\end{align}
In both cases, the index $\sigma$ corresponds to the open fermion index, which is consistent with the fermionic nature of baryons. $P_+$ again denotes the parity projection operator \eqref{eq:parity-op}.

\paragraph{The two point function $C_{2\mathrm{pt}}$:}

The proton two-point function involves two Wick contractions arising from permutations of the two $u$-quarks. In terms of the smeared point-to-all propagator \eqref{eq:point_to_all_prop} evaluated at the source at position $z$, the total contribution for momentum $\mvec{p}$ is given by:

\begin{align}
G^{\mvec{p}}_{2\mathrm{pt}}(z,z^\prime) 
&= 
\pha^{\mvec{p}}(z^\prime - z)
\left[ 
	\vphantom{
		\left[
			\left(M_z^{\Phi,\mvec{z}}\right)^T
		\right]_{\gamma\beta}^{cc}
	} %square bracket size
	\left(
		P_+ \Phi^{\mvec{p}} M_z^{\Phi,\mvec{p}} (z^\prime)
	\right)_{\alpha\alpha}^{ab}\ 
	\tr\left\{ 
		\left( 
			\Phi^{\mvec{p}} M_z^{\Phi,\mvec{p}}(z^\prime) E^b 	
		\right)^T 
		E^a \Phi^{\mvec{p}} M_z^{\Phi,\mvec{p}}(z^\prime) 
	\right\} 
\right.\nonumber \\
&\quad 
+\left. 
	\left(
		P_+ \Phi^{\mvec{p}} M_z^{\Phi,\mvec{p}} (z^\prime) 
	\right)_{\beta\gamma}^{ab}
	\left[ 
		\left( 
			\Phi^{\mvec{p}} M_z^{\Phi,\mvec{p}}(z^\prime) E^b 
		\right)^T 
		E^a \Phi^{\mvec{p}} M_z^{\Phi,\mvec{p}}(z^\prime) 
	\right]^{cc}_{\gamma\beta} 
\right]\,, 
\label{eq:def_g_2pt}
\end{align}
where $z^\prime$ denotes the sink position. Together with the phase introduced by the factor $\pha^{\mvec{p}}(z^\prime - z)$, a sum over $z^\prime$ at the sink timeslice projects onto the proton momentum $\mvec{p}$. The two-point function itself is defined as the average over all gauge fields, which is indicated by the $\langle . \rangle$-notation:

\begin{align}
C^{\mvec{p}}_{2\mathrm{pt}}(t) = 
	\left\langle 
		\tilde{G}^{\mvec{p}}_{2\mathrm{pt}}(t)
	\right\rangle\,, 
\qquad 
\widetilde{G}^{\mvec{p}}_{2\mathrm{pt}}(t) = 
	a^3 V \sum_{\mvec{z}^\prime} \left.  
		G^{\mvec{p}}_{2\mathrm{pt}}(z,z^\prime)  
	\right|_{(z^\prime)^4 = z^4 + t}\,,
\label{eq:def_c_2pt}
\end{align}
where $t$ is the source-sink separation in the time direction. A momentum projecting sum at the source is not necessary because of translational invariance, \ie there is no dependence on the source position. In the second expression in \eqref{eq:def_c_2pt}, the omitted sum over $z$ has been compensated for by a factor $V$.

As previously discussed, the two-point function $C_{2\mathrm{pt}}$ is needed to normalize the two-current matrix element, see \eqref{eq:4pt-2pt-ratio}. Furthermore, the expression $\tilde{G}_{2\mathrm{pt}}$ is part of the contractions $S_2$ and $D$, which will be discussed later.

\paragraph{Graph $C_1$:}

\begin{figure}[ht]
\includegraphics[scale=1]{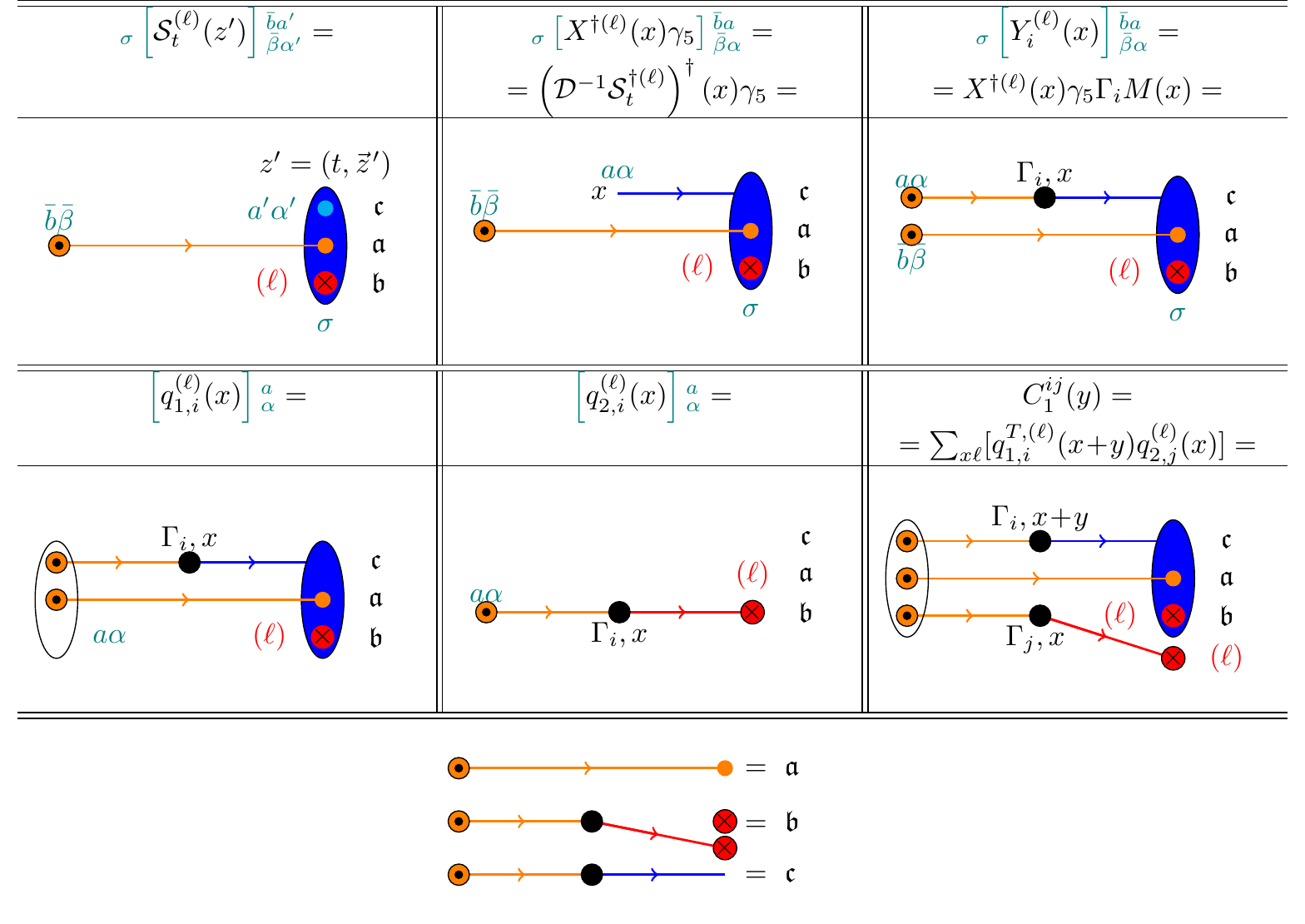}
\caption{Detailed illustration of the different parts involved in the evaluation of the $C_1$ graph. The symbols have the same meaning as in \fig{\ref{fig:graphs_tech}}. For clarity, we also write down the spinor, color and stochastic indices and spacetime arguments. We also indicate the quark lines $\mathfrak{a}$, $\mathfrak{b}$ or $\mathfrak{c}$, which are defined at the bottom. Upper panels: Left: The sequential source $\mathcal{S}_t$ at timeslice $t$, which is a sum of the expressions defined in \eqref{eq:seq_c1_src}. The sequential source already incorporates parts of each quark line. The light blue dot indicates the open spinor and color indices that are used for the inversion of the Dirac operator. Center: Sequential propagator $X^\dagger \gamma_5$ including Hermitian conjugation. Right: The combination of the sequential propagator, the current insertion $\Gamma_i$, and the forward propagator $M$, which defines the quantity $Y$. Lower panels: Left: A linear combination of the contractions \eqref{eq:c1_src}, which is called $q_1$. This is constructed from the quantity $Y$. The open baryon spinor index $\sigma$ of the sink is contracted with the baryon spinor index of the source, which is why it does not appear anymore in $q_1$. Center: A part of the stochastic quark line $\mathfrak{b}$, which is called $q_2$. Right: The complete $C_1$ graph, which is constructed from $q_{1,i}$ and $q_{2,j}$ (or in some cases $q_{1,j}$ and $q_{2,i}$) \label{fig:c1_tech}}
\end{figure}

The evaluation procedure of the $C_1$ graph is shown in \fig{\ref{fig:c1_tech}}. In total, this contraction involves five propagators, where three of them correspond to the smeared point-to-all propagator \eqref{eq:point_to_all_prop}, which we refer to as forward propagator in the following. The two propagators connecting the current insertions and the sink are calculated from sources placed at the sink. Both propagators have to be Hermitian conjugated and multiplied by $\gamma_5$ on both sides in order to obtain the desired propagator in the forward direction. For these two propagators, we use two different methods: The first propagator is obtained from an inversion on a stochastic wall source $\eta_t^{(\ell)}$, which is placed at the sink timeslice, see \eqref{eq:stoch_prop}. This stochastic propagator is denoted by $\psi^{(\ell)}_t$. From the smeared stochastic source, \ie $\Phi^{\mvec{p}}\gamma_5\eta_t^{(\ell)}$ (the $\gamma_5$ is needed to reverse the propagator), and the smeared forward propagator $\Phi^{\mvec{p}} M_z^{\Phi,\mvec{p}}$, both contracted with the baryon annihilation kernel \eqref{eq:bar-ann-kern}, we create a sequential source $\mathcal{S}^{\mvec{p},(\ell)}_t$, where:

\begin{align}
\mathcal{S}^{\mvec{p},(\ell)}_t (z^\prime) = 
	\pha^{\mvec{p}}(z^\prime)\ 
	\mathcal{S}^{(\ell)}(z^\prime)\ \delta_{z^{\prime 4} t} \,.
\label{eq:c1_seq_t}
\end{align}
The exact contraction with the annihilation kernel, \ie which index is contracted with which part, depends on the baryon type and the quark flavor of the local currents. This is discussed in detail in \appx\ref{sec:seqsources}, where all possible expressions for $\mathcal{S}^{(\ell)}$ are listed, see \eqref{eq:seq_c1_src}.

An inversion on the momentum smeared sequential source yields a sequential propagator $X^{\mvec{p},(\ell)}_t$:

\begin{align}
\D X_t^{\Phi,\mvec{p},(\ell)} = 
	\Phi^{\mvec{p}} \mathcal{S}^{\dagger,\mvec{p},(\ell)}_t\,.
\label{eq:c1_seq_inv}
\end{align}
The sequential source technique has been invented in \cite{Maiani:1987by}. The sequential propagator is connected to the second current insertion. In this technical context, the three quark lines between the proton source and sink can be distinguished \wrt the evaluation method of the involved propagators. We shall use the following labels:
\begin{itemize}
\setlength\itemsep{0em}
\item[$\mathfrak{a}$] forward propagator connecting the baryon operators
\item[$\mathfrak{b}$] quark line with the stochastic source, the stochastic propagator, and one current insertion
\item[$\mathfrak{c}$] quark line with the sequential propagator and the other current insertion.
\end{itemize}
Furthermore, we define:

\begin{align}
Y^{\mvec{p},(\ell)}_{t,j}(y) := 
	X_t^{\dagger,\Phi,\mvec{p},(\ell)}(y)\ 
	\gamma_5 \Gamma_j\ M_z^{\Phi,\mvec{p}}(y)\,,
\label{eq:c1_Y}
\end{align}
which represents the quark lines $\mathfrak{a}$ and $\mathfrak{c}$ and the stochastic source $\eta^{(\ell)}_t$ belonging to quark line $\mathfrak{b}$. The contraction of $Y^{\mvec{p},(\ell)}_{t,j}(y)$ with the baryon creation operator kernel \eqref{eq:bar-cre-kern} is denoted by $q_{1,j}^{(\ell)}(y)$.
Like for the sequential source $\mathcal{S}^{\dagger,\mvec{p},(\ell)}_t$, there are multiple possibilities to contract $Y$ with the creation kernel, which again depend on the flavor. All possible terms are summarized in \eqref{eq:c1_src}.
The remaining part of quark line $\mathfrak{b}$ is given by the quantity $q_{2,i}^{(\ell)}$:

\begin{align}
\label{eq:c1_q2}
\left(q_{2,t,i}^{\mvec{p},(\ell)}\right)^a_{\alpha}(y) &:= 
	\left[ 
		\psi_t^{\dagger,(\ell)}(y)\ 
		\gamma_5 \Gamma_i\ M^{\Phi ,\mvec{p}}_z(y) 
	\right]^{a}_{\alpha}\,.
\end{align}
The $\gamma_5$ in \eqref{eq:c1_Y} and \eqref{eq:c1_q2} again appears from reversing the sequential or stochastic propagator, respectively.
The $C_1$ graph itself is obtained by calculating

\begin{align}
C_{1}^{ij,\mvec{p}}(\mvec{y},t,\tau) &= 
	\frac{a^3}{ N_{\mathrm{st}}} 
	\left. 
		\sum_{\mvec{x}} \sum_{\ell}^{N_\mathrm{st}} 
		\left\langle \left[ 
			q_{2,t,i}^{T,\mvec{p},(\ell)}(x+y)\ 
			q_{1,t,j}^{\mvec{p},(\ell)}(x) 
		\right] \right\rangle 
	\right|_{x^4 = \tau, y^4 = 0}\,.
\label{eq:def_c1}
\end{align}
Here $x$ is the position of the operator $\Op_j$. In order to increase statistics, we perform a sum over all spatial positions at the insertion timeslice (volume average), exploiting spatial translational invariance.

Depending on the quark flavors of the baryon and the insertion operators, there are several terms that have to be summed up to obtain the full $C_1$ contribution. This is explained in detail in \appx\ref{sec:c1_contr} for the proton case and the operators $\Op_{u,i} \Op_{d,j}$ and, $\Op_{u,i} \Op_{u,j}$ \ie the graphs $C^{ij}_{1,uudd}$ and $C^{ij}_{1,uuuu}$.
%For each momentum and each flavor combination we discussed so far we require $12+49N$ inversions, where $N$ is the number of stochastic sources being used. These numbers include the inversion of the spinor-color explicit forward propagator $M_z^{\Phi,\mvec{p}}$, which requires $12$ inversions, and the sequential and stochastic propagator. For the sequential propagator $X^{\Phi,\mvec{p},(\ell)}_t$ we need to invert $12\times 4 = 48$ times, \ie for each spinor and color index value at the source and for each value of the open fermion spinor index at the sink. However, this is still requires much less inversions than spinor-color explicit all-to-all propagators ($48 \ll 12L^3$). The stochastic propagator $\psi_t^{(\ell)}$ requires the inversion of only one column.

\paragraph{Loops $L_1$ and $L_2$:}

We implement two methods to calculate the loop $L_1$, which is needed for the evaluation of the $S_1$ and $D$ graphs. The first method involves the stochastic wall source $\eta_\tau^{(\ell)}$ at the insertion timeslice $\tau$ and the corresponding propagator. Fluctuations introduced by the stochastic noise vectors are reduced by employing the hopping parameter expansion trick, which we have discussed previously. The number of omitted terms $N$ depends on the inserted Dirac structure $\Gamma_i$, see \tab{\ref{tab:loop_hpe}}. With the accordingly improved stochastic propagator $\xi_\tau^{(\ell),N}$ (see \eqref{eq:stoch_prop_hpe}), we define: 

\begin{align}
L_{1,\mathrm{st}}^j(\mvec{y},\tau) := 
	\left. 
		\frac{1}{N_{\mathrm{st}}} \sum_{\ell}^{N_{\mathrm{st}}} 
		\left[ 
			\eta_\tau^{\dagger (\ell)}(y)\ 
			\Gamma_j\ \xi_\tau^{(\ell),N}(y) 
		\right] 
	\right|_{y^4 = \tau}\,.
\label{eq:def_l1st}
\end{align}
Alternatively, we compute the loop for fixed spatial positions using point sources. A disadvantage is that the calculation has to be repeated for each loop position we want to consider. This version is only employed for one of the two loops in the $D$ graph:

\begin{align}
L_{1,\mathrm{pt}}^j(\mvec{y},\tau) := 
	\left. 
		\tr \left\{ \Gamma_j\ M_y(y) \right\} 
	\right|_{y^4 = \tau}\,.
\label{def:def_l1pt}
\end{align}
Furthermore, we define the volume average:

\begin{align}
\left\langle \left\langle L_1^j(\tau) \right\rangle \right\rangle 
:= 
	\frac{a^3}{V} 
	\sum_{\mvec{y}} \left\langle L_1^j(\mvec{y},\tau) \right\rangle\,.
\end{align}
The second kind of loop, $L_2$, appears in the $S_2$ graph. It contains the two spatially separated current insertions, which are connected by two propagators. Using stochastic noise vectors is not feasible in this case. Thus, the loop is calculated from point-to-all propagators only:

\begin{align}
L_2^{ij}(\mvec{y},\tau) = 
	\left. \tr \left\{ 
		\gamma_5\ M^\dagger_{x}(x+y)\ \gamma_5 \Gamma_i\ 
		M_{x}(x+y)\ \Gamma_j 
	\right\} \right|_{x^4 = \tau , y^4 = 0}\,.
\label{eq:def_l2}
\end{align}
Statistics can be enhanced by averaging over several spatial positions $\mvec{x}$. For each position the calculation has to be repeated.

\paragraph{Graphs $C_2$ and $S_1$:}

The $C_2$ and $S_1$ graphs are both constructed from a sequential source. For this we use the same source as for usual three-point functions, see \eqref{eq:3pt_seq}. The corresponding sequential propagator $X_t$ is obtained by inverting:

\begin{align}
\D X^{\Phi,\mvec{p}}_{t,3\mathrm{pt}} = 
	\Phi^{\mvec{p}} \gamma_5 S^{\dagger,\mvec{p}}_{t,3\mathrm{pt}}\,,
\label{eq:seq_3pt_inv}
\end{align}
where again

\begin{align}
S^{\mvec{p}}_{t,3\mathrm{pt}}(z^\prime) = 
	\pha^{\mvec{p}}(z^\prime)\ 
	S^{\mvec{p}}_{3\mathrm{pt}}(z^\prime)\ \delta_{z^{\prime 4} t}\,.
\end{align}
In the case of the $C_2$ contraction, the sequential propagator is connected to one current insertion. The other current insertion is contracted with the forward propagator. Both current insertions are connected by a stochastic propagator, which is improved by the HPE trick we have discussed earlier. Explicitly, we find for the $C_2$ graph:

\begin{align}
C_2^{ij,\mvec{p}}(\mvec{y},t,\tau) &= 
	\frac{a^3}{N_{\mathrm{st}}} \pha^{-\mvec{p}}(z) 
	\sum_\ell^{N_{\mathrm{st}}} \sum_{\mvec{x}} 
	\left\langle \left[ 
		X^{\dagger,\Phi ,\mvec{p}}_{t,3\mathrm{pt}} 
			(\mvec{x}+\mvec{y},\tau)\ 
		\gamma_5 \Gamma_i\ 
		\xi_\tau^{(\ell),n(\mvec{y})} (\mvec{x}+\mvec{y},\tau) 
	\right] \right.\nonumber \\
&\times 
	\left. \left. \left[ 
		\eta^{\dagger (\ell)} (\mvec{x},\tau)\ 
		\Gamma_j\ M_z^{\Phi,\mvec{p}} (\mvec{x},\tau) 
	\right]  \right\rangle \right|_{z^4 = 0}\,.
\label{eq:def_c2}
\end{align}
The $S_1$ graph consists of two disconnected pieces. The first one has the same structure as a usual three-point function calculated from a sequential source:

\begin{align}
G^{i,\mvec{p}}_{3\mathrm{pt}}(\mvec{x},\tau,t) = 
	\left. \pha^{-\mvec{p}}(z) 
		\left[ 
			X^{\dagger,\Phi ,\mvec{p}}_{t,3\mathrm{pt}} 
				(\mvec{x},\tau)\ 
			\gamma_5 \Gamma_i\ M_z^{\Phi,\mvec{p}} (\mvec{x},\tau) 
		\right] 
	\right|_{z^4 = 0}\,.
\label{eq:def_g_3pt}
\end{align}
The second part is given by the previously defined loop $L_1$. If the quantum numbers permit, there are disconnected contributions from the vacuum expectation values of $G_{3\mathrm{pt}}$ and $L_1$. These must be subtracted, in order to obtain the $S_1$ contribution we wish to calculate:

\begin{align}
S_1^{ij,\mvec{p}}(\mvec{y},t,\tau) &= 
	-a^3 \sum_{\mvec{x}} \left\langle 
		G^{i,\mvec{p}}_{3\mathrm{pt}}(\mvec{x}+\mvec{y},\tau,t)\ 
		L_1^j(\mvec{x},\tau) 
	\right\rangle\nonumber \\
&+ 
	a^3 \sum_{\mvec{x}} \left\langle 
		G^{i,\mvec{p}}_{3\mathrm{pt}}(\mvec{x},\tau,t)
	\right\rangle 
	\left\langle \left\langle 
		L_1^j(\tau) 
	\right\rangle \right\rangle\,.
\label{eq:def_s1}
\end{align}
Notice that the global sign corresponds to the permutation sign of the Wick contraction.

\paragraph{Graphs $S_2$ and $D$:} 

The graph $S_2$ consists of a two-point contraction and the loop $L_2$, whereas $D$ consists of a two-point contraction and two $L_1$ loops. As for the $S_1$ graph, we have to consider vacuum contributions of the disconnected parts, which have to be subtracted. Notice that we defined the loop $L_2$ at a fixed spatial position. Hence, we are not able to perform a volume average like in the previous cases:

\begin{align}
S_2^{ij,\mvec{p}}(\mvec{y},t,\tau) = 
	- \left\langle 
		\widetilde{G}_{2\mathrm{pt}}^{\mvec{p}}(t)\ 
		L_2^{ij}(\mvec{y},\tau) 
	\right\rangle 
	+ \left\langle 
		\widetilde{G}_{2\mathrm{pt}}^{\mvec{p}}(t) 
	\right\rangle
	\left\langle L_2^{ij}(\mvec{y},\tau) \right\rangle\,.
\label{eq:def_s2}
\end{align}
We use two methods to evaluate the $D$ graph: The first employs two stochastic loops $L_{1,\mathrm{st}}$, which allows us to perform a volume average. In the second method, we replace one stochastic loop by a loop attached to a point source $L_{1,\mathrm{pt}}$. This might reduce the stochastic noise but precludes the possibility to perform a volume average. For the doubly stochastic case, the $D$ graph reads:

\begin{align}
D^{ij,\mvec{p}}(\mvec{y},t,\tau) &= 
	a^3 \sum_{\mvec{x}} \left\{ 
		\left\langle 
			\widetilde{G}_{2\mathrm{pt}}^{\mvec{p}}(t)\ 
			L_{1,\mathrm{st}}^i(\mvec{x}+\mvec{y},\tau)\ 
			L_{1,\mathrm{st}}^j(\mvec{x},\tau) 
		\right\rangle 
	\right.\nonumber \\
&\qquad\qquad - 
	\left\langle 
		\widetilde{G}_{2\mathrm{pt}}^{\mvec{p}}(t) 
	\right\rangle 
	\left\langle 
		L_{1,\mathrm{st}}^i(\mvec{x}+\mvec{y},\tau)\ 
		L_{1,\mathrm{st}}^j(\mvec{x},\tau) 
	\right\rangle \nonumber \\
&\qquad\qquad -
	\left\langle 
		\widetilde{G}_{2\mathrm{pt}}^{\mvec{p}}(t)\ 
		L_{1,\mathrm{st}}^i(\mvec{x},\tau) 
	\right\rangle 
	\left\langle \left\langle 
		L_{1,\mathrm{st}}^j(\tau) 
	\right\rangle \right\rangle \nonumber \\
&\qquad\qquad - 
	\left. 
		\left\langle 
			\widetilde{G}_{2\mathrm{pt}}^{\mvec{p}}(t)\ 
			L_{1,\mathrm{st}}^j(\mvec{x},\tau) 
		\right\rangle 
		\left\langle \left\langle 
			\vphantom{L_{1,\mathrm{st}}^j} 
			L_{1,\mathrm{st}}^i(\tau) 
		\right\rangle \right\rangle 
	\right\} \nonumber \\
&\quad+2 
	\left\langle 
		\widetilde{G}_{2\mathrm{pt}}^{\mvec{p}}(t) 
	\right\rangle 
	\left\langle \left\langle 
		\vphantom{L_{1,\mathrm{st}}^j} L_{1,\mathrm{st}}^i(\tau) 
	\right\rangle \right\rangle 
	\left\langle \left\langle 
		L_{1,\mathrm{st}}^j(\tau) 
	\right\rangle \right\rangle\,.
\label{eq:def_d}
\end{align}
Note that we use two different sets of stochastic sources for the two disconnected loops. Equation \eqref{eq:def_d} is valid for the first method. For the second method $L_{1,\mathrm{st}}$ has to be replaced by $L_{1,\mathrm{pt}}$. Furthermore, one has to replace the sum $a^3 \sum_{\mvec{x}}$ by a volume factor $V$.

\subsection{Lattice setup}
\label{sec:latt_setup}

The simulation is performed on the gauge ensemble H102 of the CLS collaboration \cite{Bruno:2014jqa}. It includes $n_f = 2+1$ dynamical Sheikholeslami-Wohlert fermions and the tree-level improved Lüscher-Weisz gauge action. The extension is $32^3 \times 96$ with open boundary conditions in the time direction. The pseudoscalar masses are $m_\pi = 355~\mathrm{MeV}$ and $m_K = 441~\mathrm{MeV}$, and the lattice spacing is $a=0.0856~\mathrm{fm}$, which corresponds to the inverse lattice coupling $\beta=3.4$. More information can be found in \tab{\ref{tab:cls}}. From this ensemble we use 990 configurations.

\begin{table}
\begin{center}
\begin{tabular}{ccccccccccc}
\hline
\hline
id & $\beta$ & $a[\mathrm{fm}]$  & $L^3 \times T$ & $\kappa_{l}$ & $\kappa_{s}$ & $m_{\pi,K}[\mathrm{MeV}]$ & $m_\pi L a$ & configs \\
\hline
H102 & $3.4$ & $0.0856$ & $32^3 \times 96$ & $0.136865$ & $0.136549339$ & $355$, $441$ & $4.9$ & $2037$ \\
\hline
\hline
\end{tabular}
\end{center}
\caption{Details of the CLS ensemble which we use for the evaluation of the two-current matrix elements. Our simulation includes 990 configurations.\label{tab:cls}}
\end{table}

For the calculation of the ratio \eqref{eq:4pt-2pt-ratio} we need to know the value of the nucleon energy $E_{\mvec{p}} = \sqrt{m^2 + \mvec{p}^2}$ in the given lattice setup. We obtain the corresponding value from an exponential fit to the two-point function data for each momentum. Moreover, the proton mass is needed in the decompositions \eqref{eq:tensor-decomp} and \eqref{eq:t2-mat-els}. From our fits, we obtain $m = 1.1296(75)~\mathrm{GeV}$.

Our analysis requires a wide range of proton momenta. Explicitly, calculations are performed for the momenta 

\begin{align}
\mvec{p}= \frac{2\pi \mvec{P}}{La}
\end{align}
with $\mvec{P} = (0,0,0)$, $(-1,-1,-1)$, $(-2,-2,-2)$, $(2,2,-2)$, $(2,-2,2)$, $(-2,2,2)$. Thus, the largest momentum has the absolute value $|\mvec{p}| \approx 1.57~\mathrm{GeV}$.

To avoid artifacts possibly caused by the open boundary conditions, we place the source at $t_\mathrm{src} = T/2 = 48a$. The spatial position is chosen randomly for each configuration. The distance to the sink in time direction is $t = t_\mathrm{snk} - t_\mathrm{src} = 12a$ for the case $\mvec{p} = \mvec{0}$ and $t=10a$ otherwise. We evaluate the $C_1$ graph for all intermediate insertion times $0 < \tau < t$. A value for $C_1(\mvec{y})$ is then given by a fit \wrt $\tau$ to a constant including a certain region around $t/2$, where excited states are seen to be sufficiently small. The remaining graphs are calculated for $\tau = t/2$, \ie $\tau=6a$ for $\mvec{p}=\mvec{0}$ and $\tau=5a$ for $\mvec{p}\neq\mvec{0}$. The disconnected parts $\langle L_2(\tau) \rangle$ and $\langle L_1(\tau) L_1(\tau) \rangle$ do not depend on the proton momentum. Hence, the corresponding calculations can be combined, which increases statistics. Consequently, the average insertion time for the contractions $S_2$ and $D$ is slightly different from $t/2$, which should not be a problem as long as excited state contributions are small.

We perform the calculations for multiple proton sources located at different source positions, which further enhances statistics. The number of proton sources, as well as the number of stochastic noise vectors being used for each contraction is summarized in \tab{\ref{tab:statnum}}. The propagators are smeared at the proton source and sink by $n=250$ smearing iterations \eqref{eq:mom_smearing}.

\begin{table}
\renewcommand{\arraystretch}{1.1}
\begin{center}
\begin{tabular}{c|c|cccccccc}
\hline
\hline
 & $\mvec{p}$ & $C_1$ & $C_2$ & $S_1$(st) & $S_2$ & $D$(st, st) & $D$(st, pt) & $3\mathrm{pt}$ & $2\mathrm{pt}$ \\
\hline
$N_\mathrm{src}$ & $=\mvec{0}$ & $1$ & $2$ & $4$ & $25$ & $25$ & $25$ & $4$ & $25$  \\
 & $\neq\mvec{0}$ & $1$ & $1$ & $1$ & $21$ & $21$ & $21$ & $1$ & $21$ \\
$N_{\mathrm{st}/\mathrm{pt}}$ & all & $2$ & $96$ & $120$ & $480$ & $16(60, 60)$ & $4(120, 120)$ & - & - \\ 
vol. average & all & y & y & y & n & y & n & y & n \\
\hline
\hline
\end{tabular}
\end{center}
\caption{Overview of the statistics of our simulation for each Wick contraction. If the contractions involve the loop $L_1$, we indicate by (st) or (pt) which version is employed. $N_\mathrm{src}$ refers to the number of proton sources for which each graph is evaluated. $N_{\mathrm{st}/\mathrm{pt}}$ is the number of stochastic sources used for the calculation of the stochastic propagators. For the graphs involving loops where multiple point sources are used ($S_2$ and $D$(st, pt)), $N_{\mathrm{st}/\mathrm{pt}}$ refers to the corresponding number of point sources. In the last line we indicate whether volume averaging is possible. \label{tab:statnum}}
\end{table}

\subsection{Data quality}
\label{sec:latt_results}

\begin{figure}
\subfigure[$\tau$-dependence, $\langle V^0 V^0 \rangle$ \label{fig:tau-VV}]{
\includegraphics[scale=0.25,trim={0.5cm 1.2cm 0.5cm 2.8cm},clip]{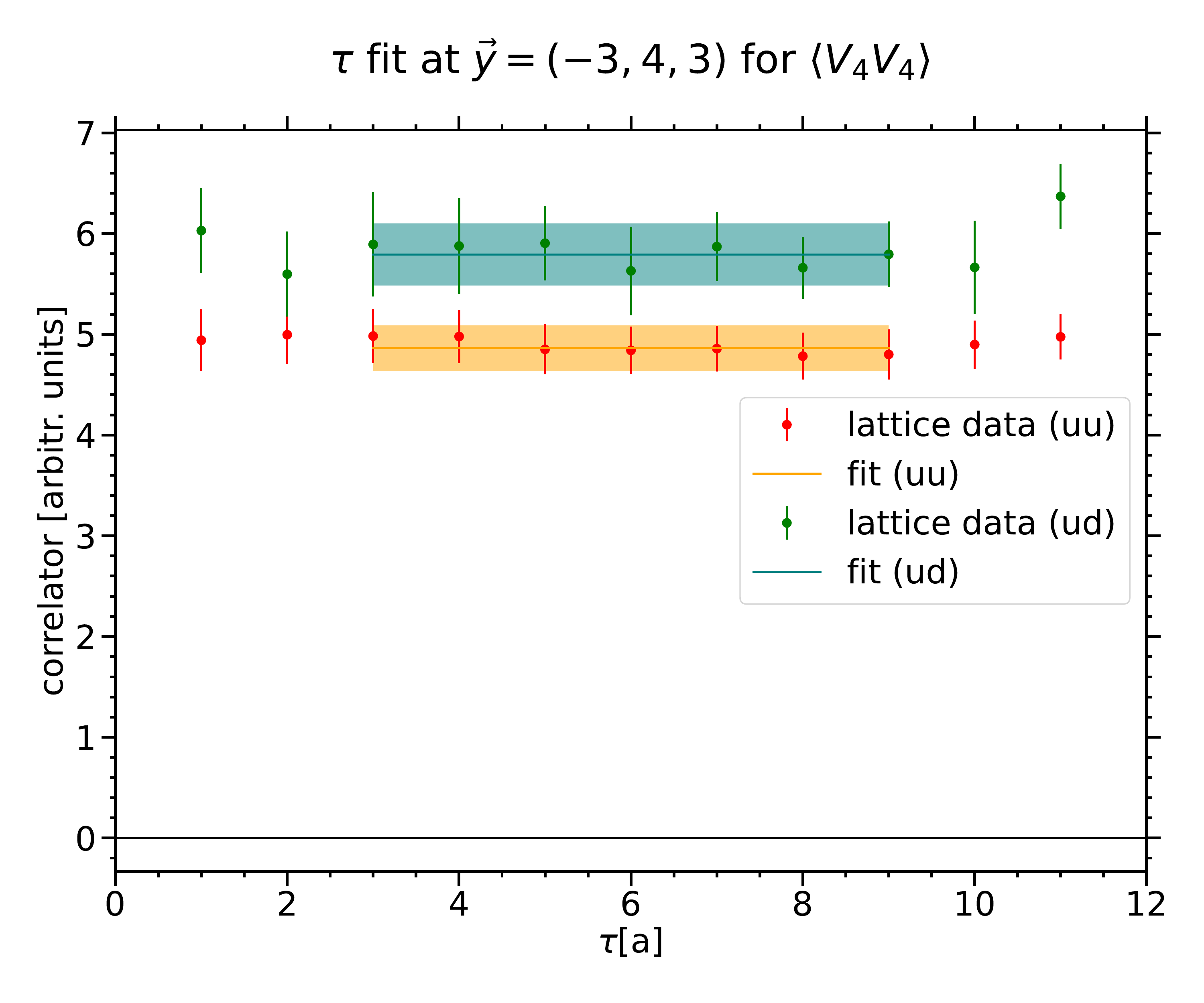}}
\subfigure[$\tau$-dependence, $\langle A^0 A^0 \rangle$ \label{fig:tau-AA}]{
\includegraphics[scale=0.25,trim={0.5cm 1.2cm 0.5cm 2.8cm},clip]{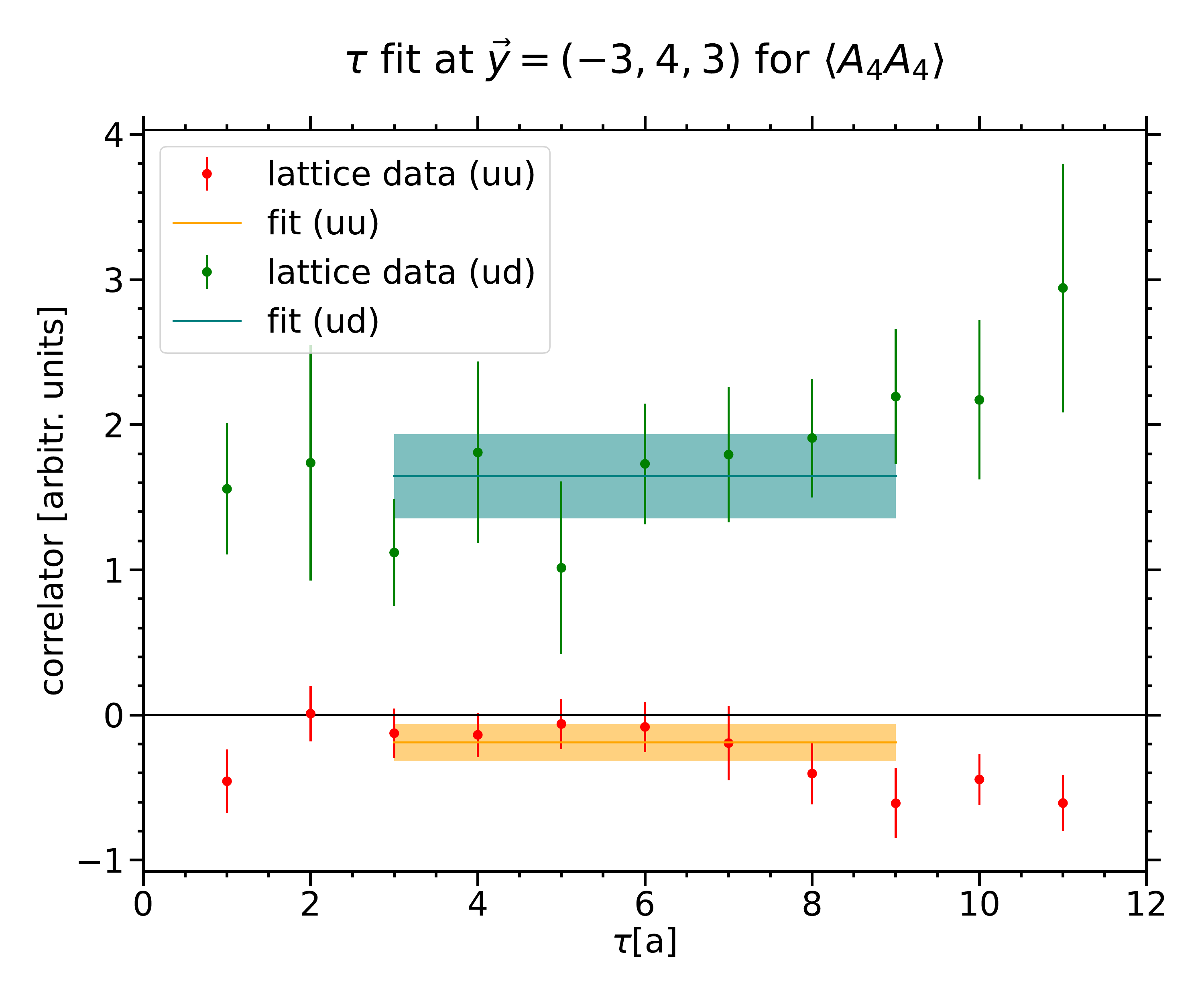}} \\
\caption{$\tau$-dependence of the $C_1$ contraction for the two flavor combinations $uu$ and $ud$. This is plotted for $\langle V^0 V^0 \rangle$ (a) and $\langle A^0 A^0 \rangle$ (b) for fixed $\mvec{y} = (-3,4,3)$ at momentum $\mvec{p}=\mvec{0}$.\label{fig:tau-dep}}
\end{figure}

In the following we want to consider the matrix elements $\langle V^0 V^0 \rangle$ and $\langle A^0 A^0 \rangle$ and discuss a number of artifacts. For the remainder of this paper we shall use the following notation for absolute values of 3-vectors:

\begin{align}
p := |\mvec{p}|
\,, \quad 
P := |\mvec{P}|
\,, \quad 
y := |\mvec{y}|\,.
\end{align}
Nevertheless, we denote the usual 4-vector scalar product by $y^2 = y^\mu y_\mu$. Since $y^0 = 0$, one has $|\mvec{y}|^2 = -y^2$. To avoid confusion, the $n$-th power of $y=|\mvec{y}|$ is denoted by $\sqrt{-y^2}^n$. For details on our notation, see \appx\ref{sec:notation}. At the moment, we consider the data for single contractions instead of the complete four-point functions and, moreover, we restrict ourselves to zero momentum, \ie $p=0$ or, equivalently, $P=0$. In our study, we are interested in the dependence on the current distance $y$. For the $C_1$ graph we are able to investigate the dependence on the insertion time $\tau$, which is plotted in \fig\ref{fig:tau-dep} for $\langle V^0 V^0 \rangle$ and $\langle A^0 A^0 \rangle$ at $\mvec{y} = (-3,4,3)$. We observe a reasonable quality of the data and plateaus around $t/2$. The values for $C_1(\mvec{y})$ are obtained by a fit to a constant \wrt the insertion time $\tau$, where we take into account the timeslices $\tau\in[t/2-3a,t/2+3a]$. The corresponding fit bands are also plotted in \fig\ref{fig:tau-dep}. For all remaining contractions, the insertion time is fixed at $\tau=t/2$ in our simulation, as discussed in the previous section.

For the remainder of this paper, we concentrate on the contributions $C_1$, $C_2$, $S_1$ and $S_2$. For both versions of $\langle L_1 L_1 \rangle$ we have presented in \sect\ref{sec:tcme_wick}, and consequently for the $D$ graph itself, we obtain statistical errors that are much larger than the signals of the remaining graphs. In contrast to our study \cite{Bali:2018nde} for the pion, this is already the case before carrying out the vacuum subtraction. As a consequence, we shall not consider contributions of the $D$ graph in subsequent analysis steps.

\begin{figure}
\subfigure[anisotropy, $\langle V_u^0 V_u^0 \rangle$, $C_1$ \label{fig:aniso-C1uuuu-VV}]{
\includegraphics[scale=0.25,trim={0.5cm 1.2cm 0.5cm 2.8cm},clip]{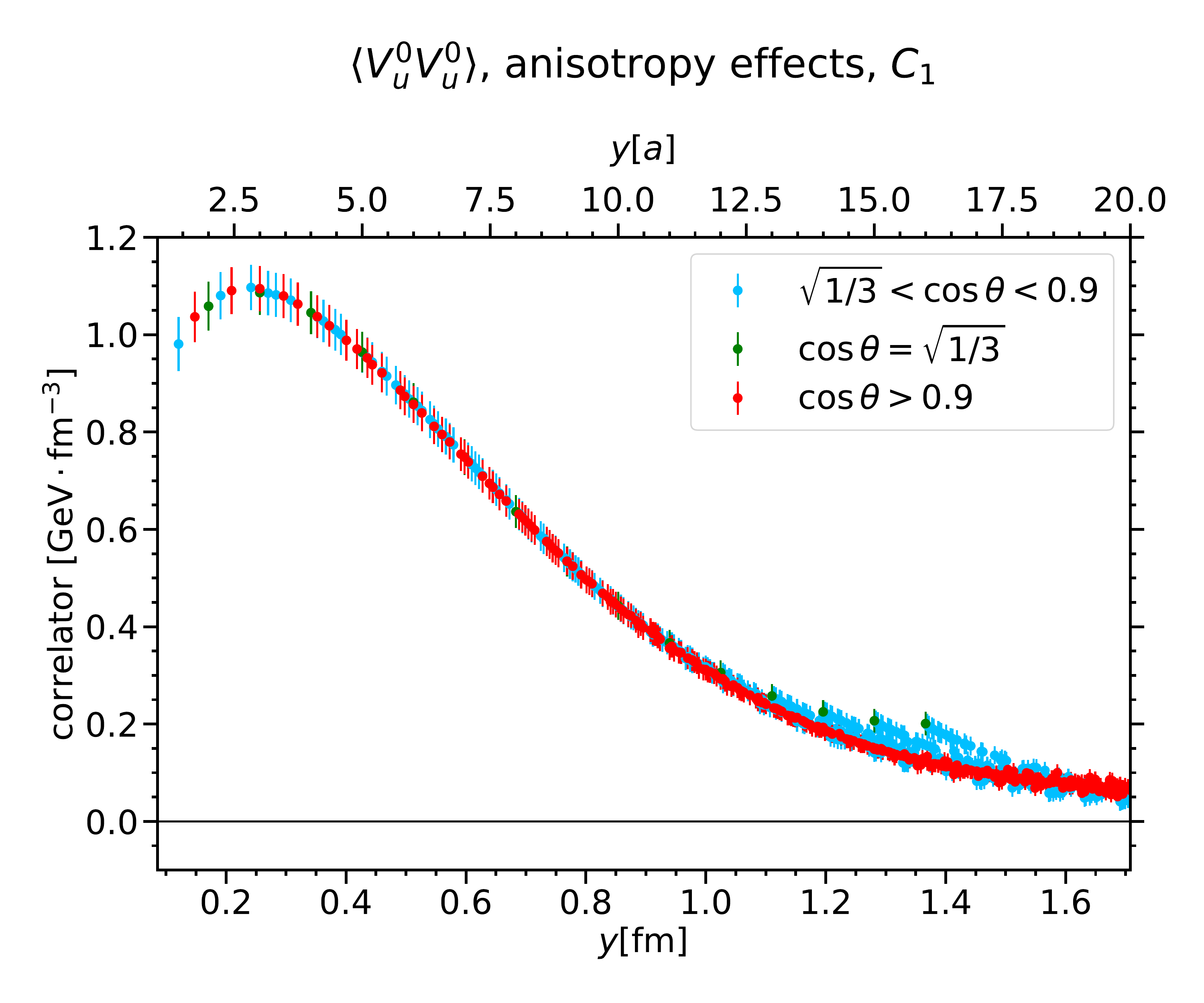}} 
\subfigure[anisotropy, $\langle V_u^0 V_u^0 \rangle$, $C_2$ \label{fig:aniso-C2u-VV}]{
\includegraphics[scale=0.25,trim={0.5cm 1.2cm 0.5cm 2.8cm},clip]{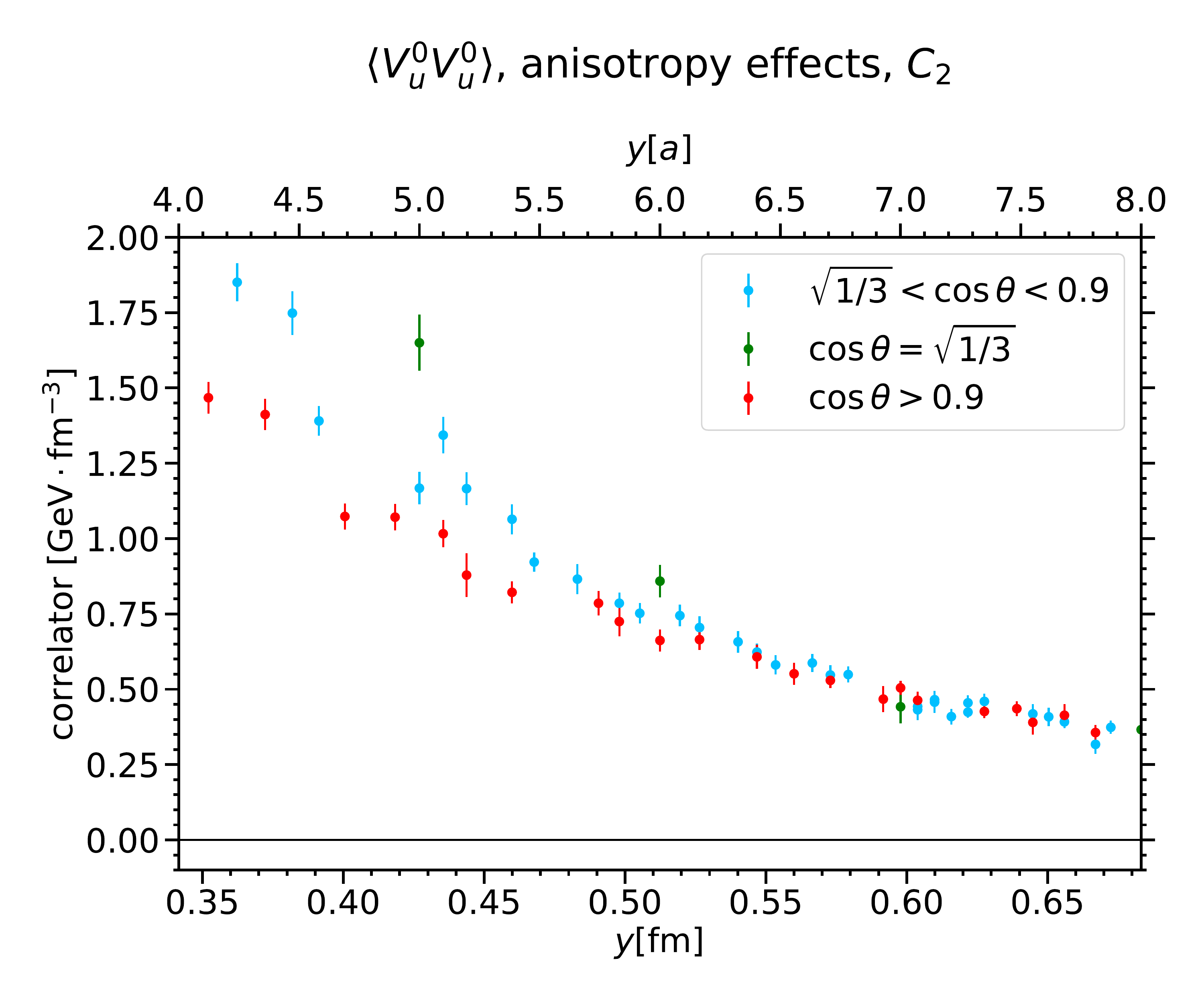}} \\
\subfigure[anisotropy, $\langle V_d^0 V_d^0 \rangle$, $C_2$ \label{fig:aniso-C2d-VV}]{
\includegraphics[scale=0.25,trim={0.5cm 1.2cm 0.5cm 2.8cm},clip]{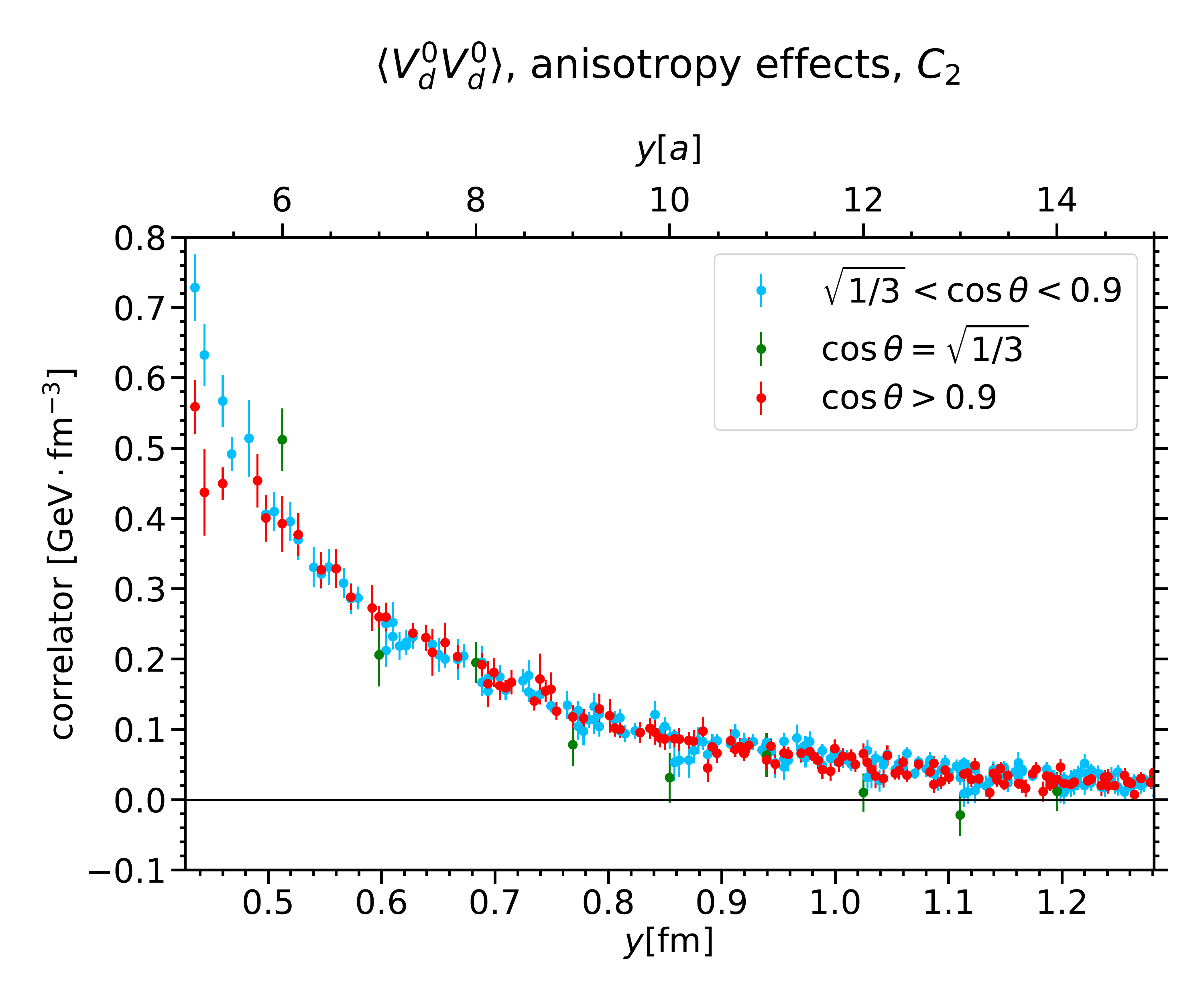}} 
\subfigure[anisotropy, $\langle V_q^0 V_q^0 \rangle$, $S_2$ \label{fig:aniso-S2-VV}]{
\includegraphics[scale=0.25,trim={0.5cm 1.2cm 0.5cm 2.8cm},clip]{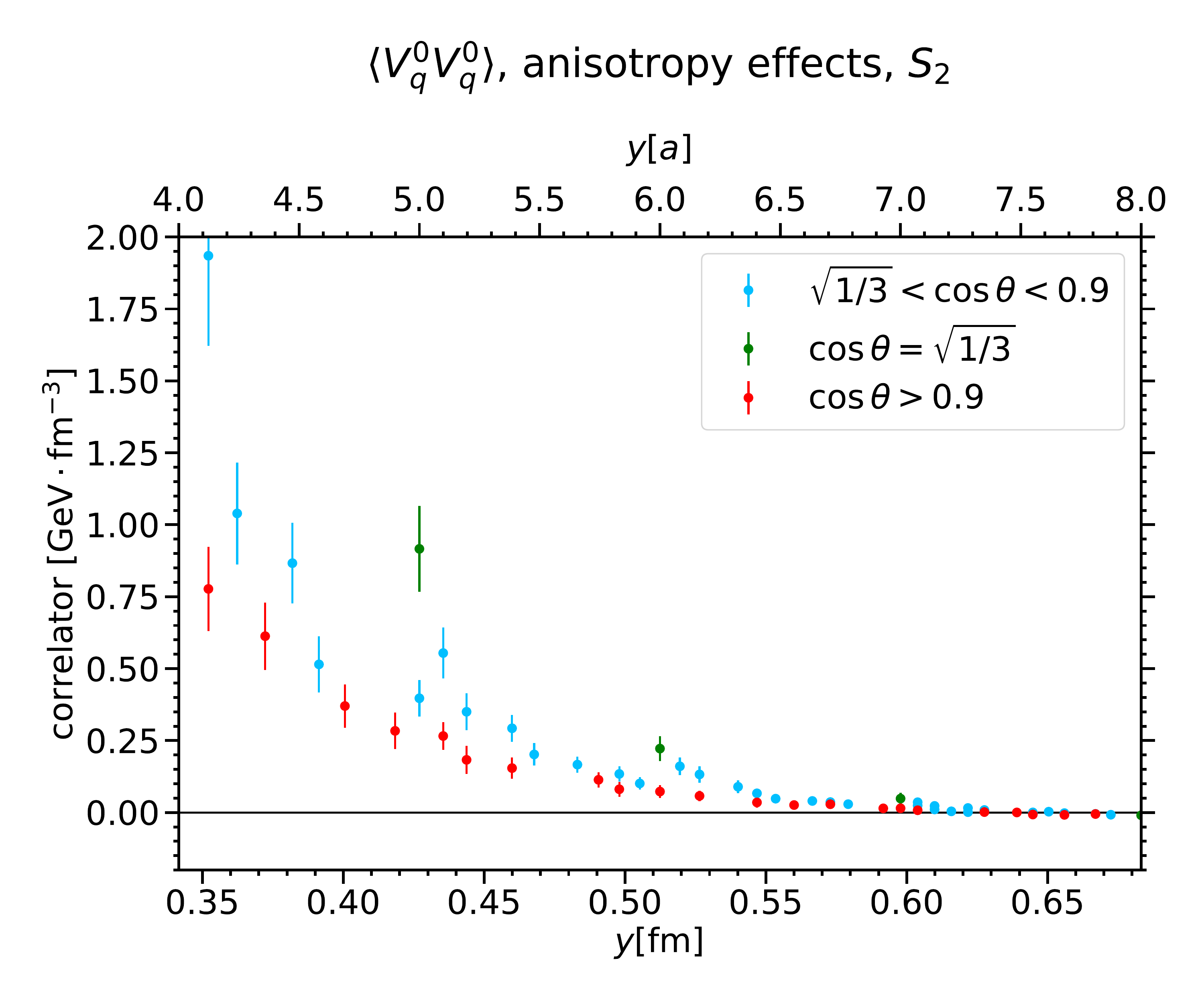}} \\
\caption{Visualization of anisotropies found in the four-point data. The data points are separated \wrt to the angle $\theta$ between the distance vector $\mvec{y}$ and the next nearest space diagonal (see the text). These plots show the results for the $C_1$ contribution to $\langle V^0_u V^0_u \rangle$ (a), the $C_2$ contribution to $\langle V^0_u V^0_u \rangle$ for small $y$ (b) and $\langle V^0_d V^0_d \rangle$ for intermediate $y$ (c), as well as the $S_2$ graph for small $y$ (d).\label{fig:aniso}}
\end{figure}

In order to investigate possible anisotropy effects, we distinguish three sets of data points characterized by the angle $\theta$ between the distance vector $y$ and the nearest lattice space diagonal:
\begin{itemize}
\item $\cos \theta = \sqrt{1/3}$: data points placed on one of the lattice axes
\item $\cos \theta > 0.9$: data points in the vicinity of one lattice space diagonal
\item $\sqrt{1/3} < \cos \theta < 0.9$: all other data points\footnote{A vector with $\cos \theta = 0.9$ does not exist in our lattice setup.}
\end{itemize}
In \fig\ref{fig:aniso} we show some selected results. The first kind of anisotropy effects observed in the lattice data is caused by mirror charges originating from the periodic boundary conditions in the spatial directions, which is explained in detail in \cite{Burkardt:1994pw}. These are stronger along the lattice axes, since the mirror charges lie closer together in this case. This artifact can be observed in \fig\ref{fig:aniso-C1uuuu-VV} at distances $y>12a$, where the data with $\cos\theta < 0.9$ clearly lie above the data for $y$ close to the lattice diagonals. The resulting "saw-tooth" pattern can be seen in each channel in the $C_1$ data.

Another anisotropy effect is caused by the anisotropy of the lattice propagator and is present in all contractions involving at least one propagator directly connecting the two currents, \ie the graphs $C_2$ and $S_2$. Examples are plotted \fig\ref{fig:aniso-C2u-VV} and \ref{fig:aniso-C2d-VV} for the $C_2$ graph and in \ref{fig:aniso-S2-VV} for $S_2$. In these plots, we see a significantly different behavior of the data close to the lattice space diagonals and the remaining data points.

The lattice propagator anisotropy has been studied in detail, \eg in \cite{Bali:2018spj,Cichy:2012is}, where it was found that lattice artifacts are most pronounced along the lattice axes, whereas they are moderate close to the lattice diagonals.

\section{Mellin moments of DPDs}
\label{sec:mellin}

\subsection{Extraction of twist-two functions}
\label{sec:twist2f}

According to \eqref{eq:tensor-decomp}, the two-current matrix elements we obtain in our lattice simulation can be decomposed in terms of Lorentz invariant functions. The twist-two components, which are relevant in the DPD context, are parameterized by a certain subset of these invariant functions. We refer to these functions as twist-two functions. Explicitly, the twist-two functions are $A_{qq^\prime}$, $A_{\Delta q \Delta q^\prime}$, $A_{\delta qq^\prime}$, $A_{q \delta q^\prime}$, $A_{\delta q \delta q^\prime}$, and $B_{\delta q \delta q^\prime}$. Since our calculation includes only light-quark operators, we can extract the twist-two functions for $qq^\prime = uu, ud, dd$. For proton DPDs, which we consider in this paper, the latter probes at least one sea quark.

The twist-two functions are obtained by solving the overdetermined system of equations given by \eqref{eq:tensor-decomp}. This we do by $\chi^2$ minimization. Before we go into physics interpretation, we discuss possible lattice artifacts seen in the data. If Lorentz invariance were intact, the extracted data points of the invariant functions would be boost- and rotationally invariant, \ie for a given $py$ they would be independent of the momentum $\mvec{p}$ and the direction of $\mvec{y}$. In order to check this, the system of equations is solved separately for each graph and for each accessible direction of the distance vector $\mvec{y}$, \ie we obtain one data point for each $y^2$, $py$ and $\theta$, where $\theta$ is the angle between $\mvec{y}$ and the nearest space diagonal on the lattice. We use the same classification of the data points \wrt $\theta$ as in \sect\ref{sec:latt_results}.
\begin{figure}
\subfigure[anisotropy, $A_{ud}$, $C_1$, $\mvec{P} = \mvec{0}$, $py=0$ \label{fig:tw2f_aniso-C1uudd-A_VV-p000}]{
\includegraphics[scale=0.25,trim={0.5cm 1.2cm 0.5cm 2.8cm},clip]{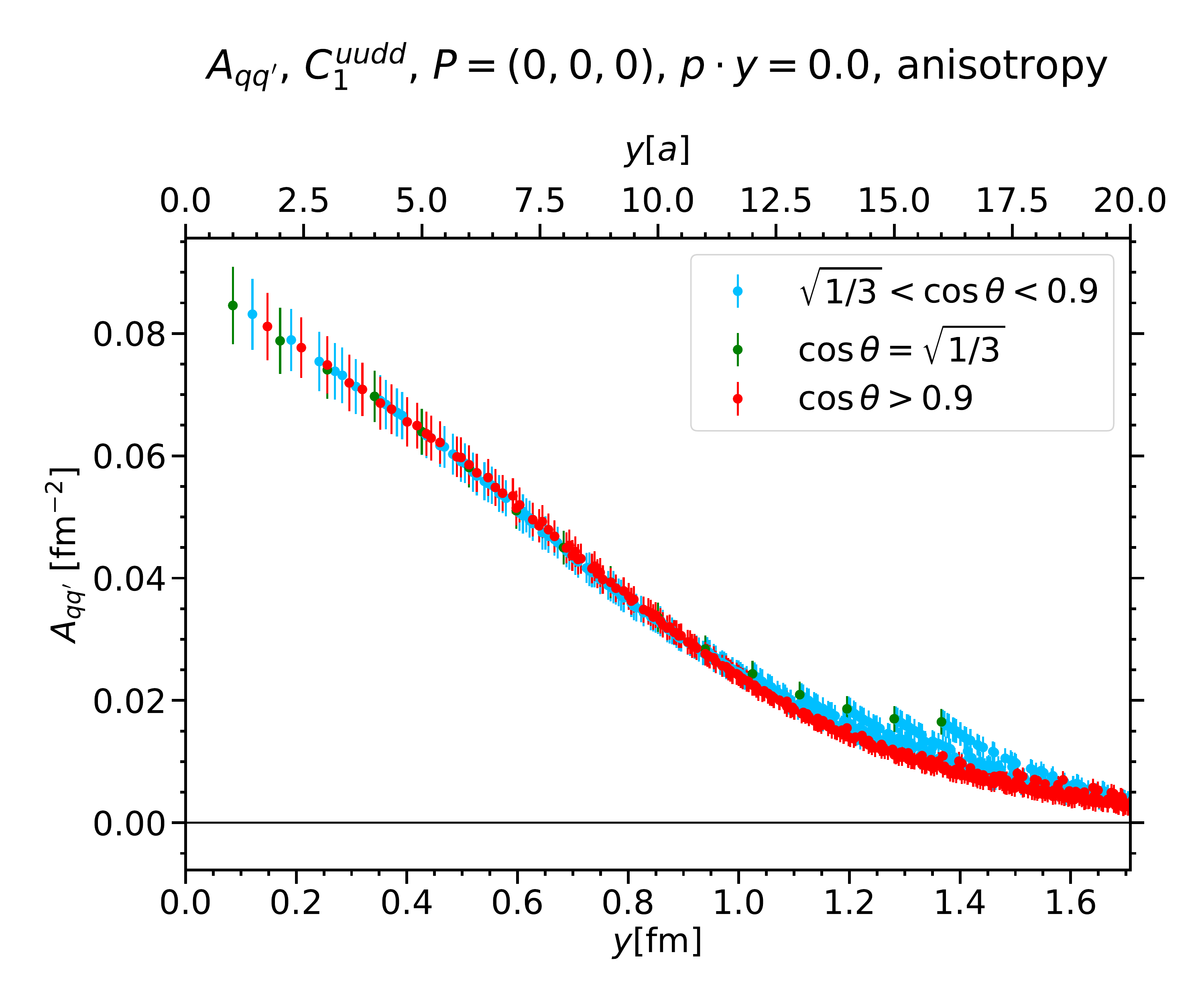}} \hfill 
\subfigure[anisotropy, $A_{uu}$, $C_1$, $\mvec{P} = -(1,1,1)$, $py=1.6$ \label{fig:tw2f_aniso-C1uudd-A_VV-p-1-1-1}]{
\includegraphics[scale=0.25,trim={0.5cm 1.2cm 0.5cm 2.8cm},clip]{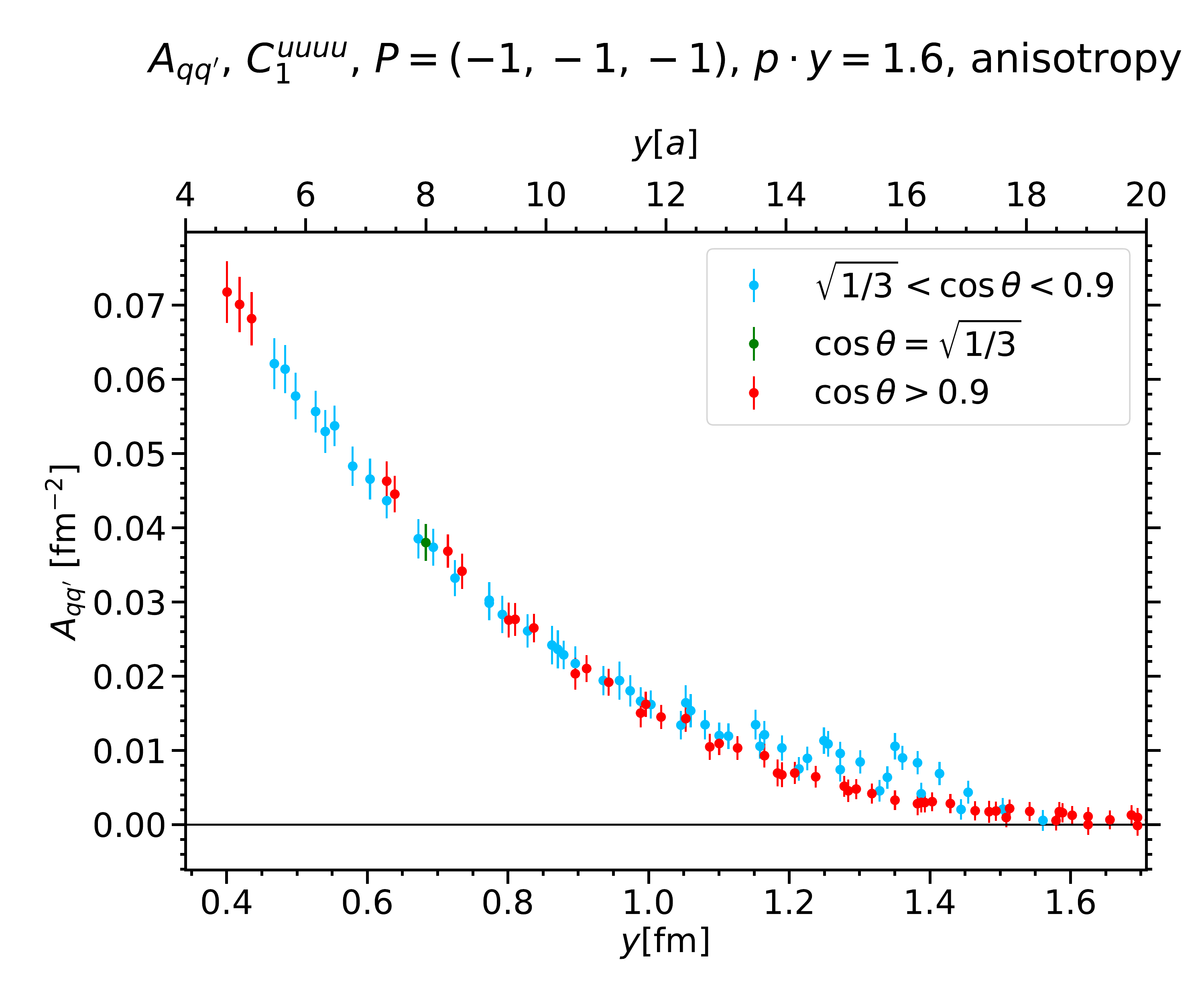}} \\ 
\subfigure[anisotropy, $A_{uu}$, $C_2$, $\mvec{P} = \mvec{0}$, $py=0$ \label{fig:tw2f_aniso-C2u-A_VV-p000-small}]{
\includegraphics[scale=0.25,trim={0.5cm 1.2cm 0.5cm 2.8cm},clip]{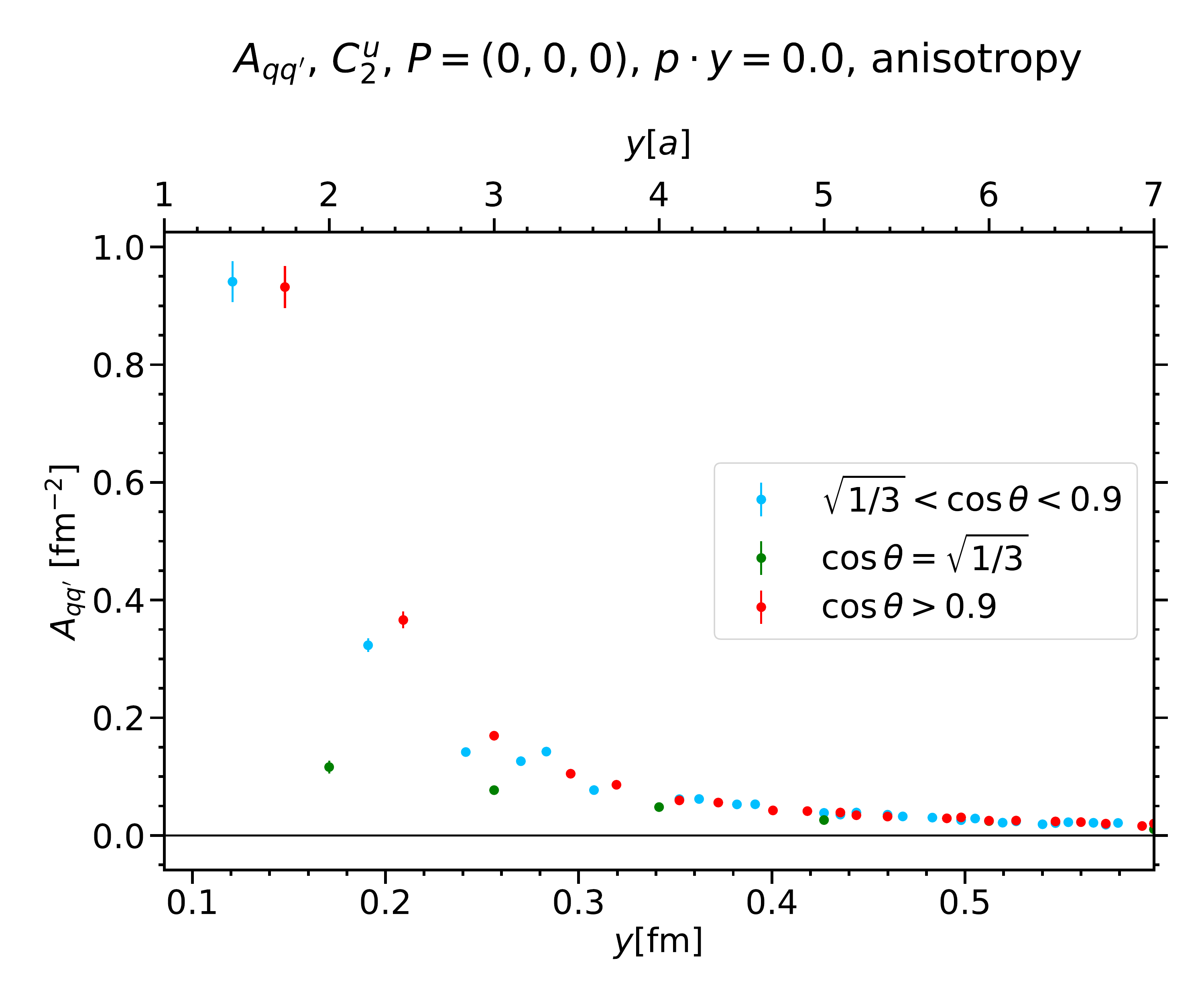}}  \hfill 
\subfigure[anisotropy, $A_{uu}$, $C_2$, $\mvec{P} = \mvec{0}$, $py=0$ \label{fig:tw2f_aniso-C2u-A_VV-p000-mid}]{
\includegraphics[scale=0.25,trim={0.5cm 1.2cm 0.5cm 2.8cm},clip]{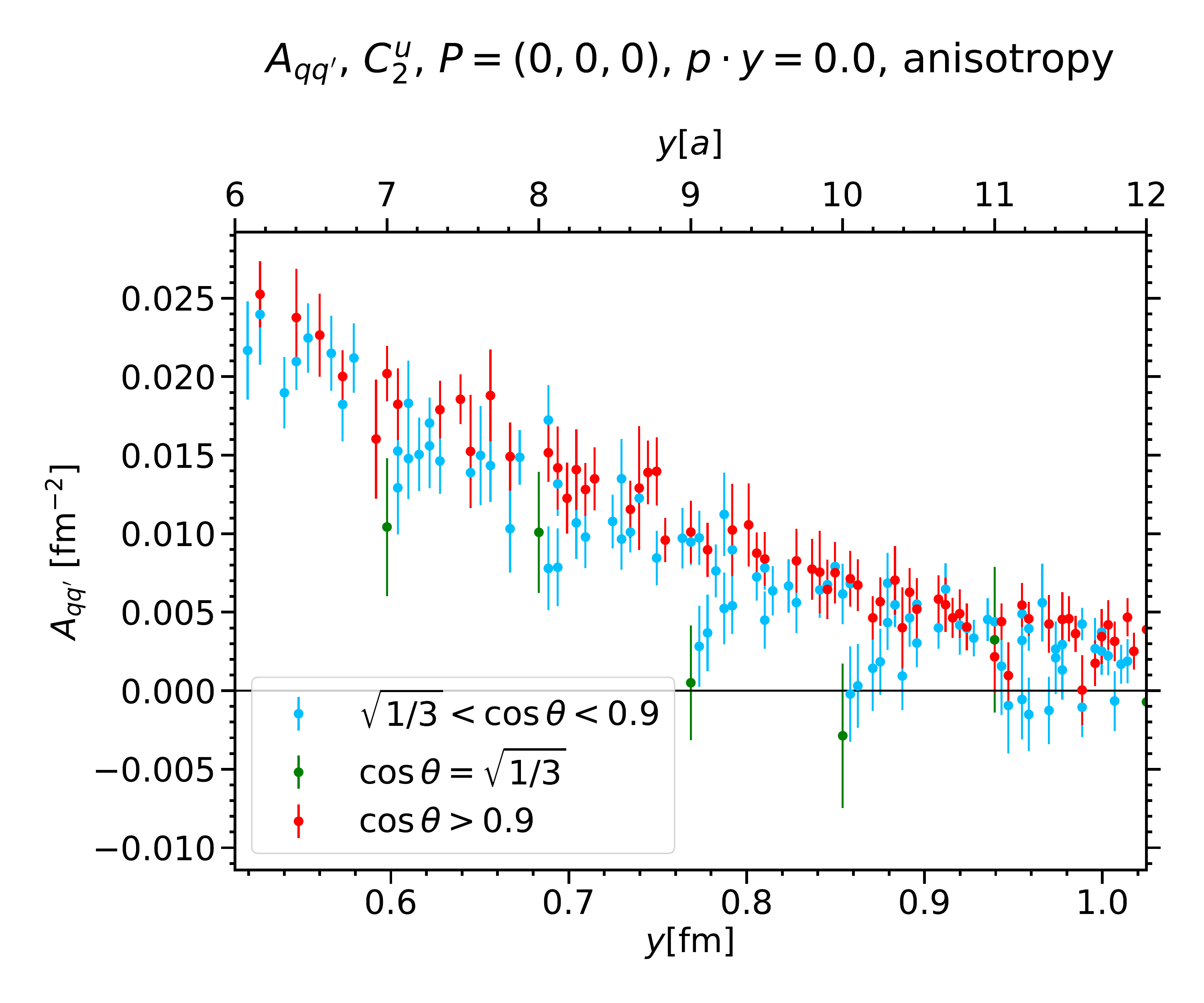}} \\ 
\subfigure[anisotropy, $B_{\delta u \delta d}$, $C_1$, $py=0$ \label{fig:tw2f_anisoBTT-C1uudd}]{
\includegraphics[scale=0.25,trim={0.5cm 1.2cm 0.5cm 2.8cm},clip]{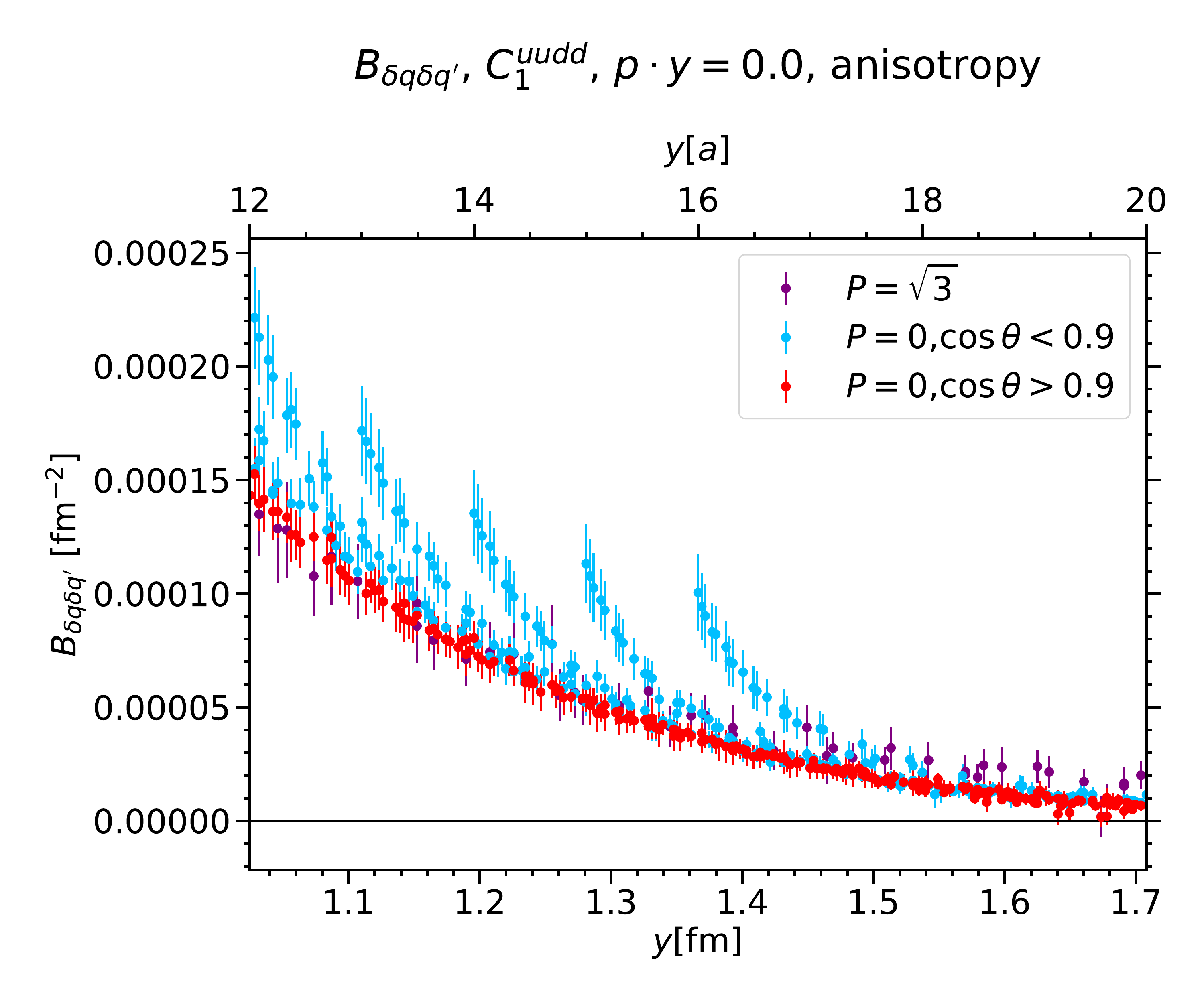}}\hfill 
\subfigure[anisotropy, $B_{\delta u \delta u}$, $C_1$, $py=0$ \label{fig:tw2f_anisoBTT-C1uuuu}]{
\includegraphics[scale=0.25,trim={0.5cm 1.2cm 0.5cm 2.8cm},clip]{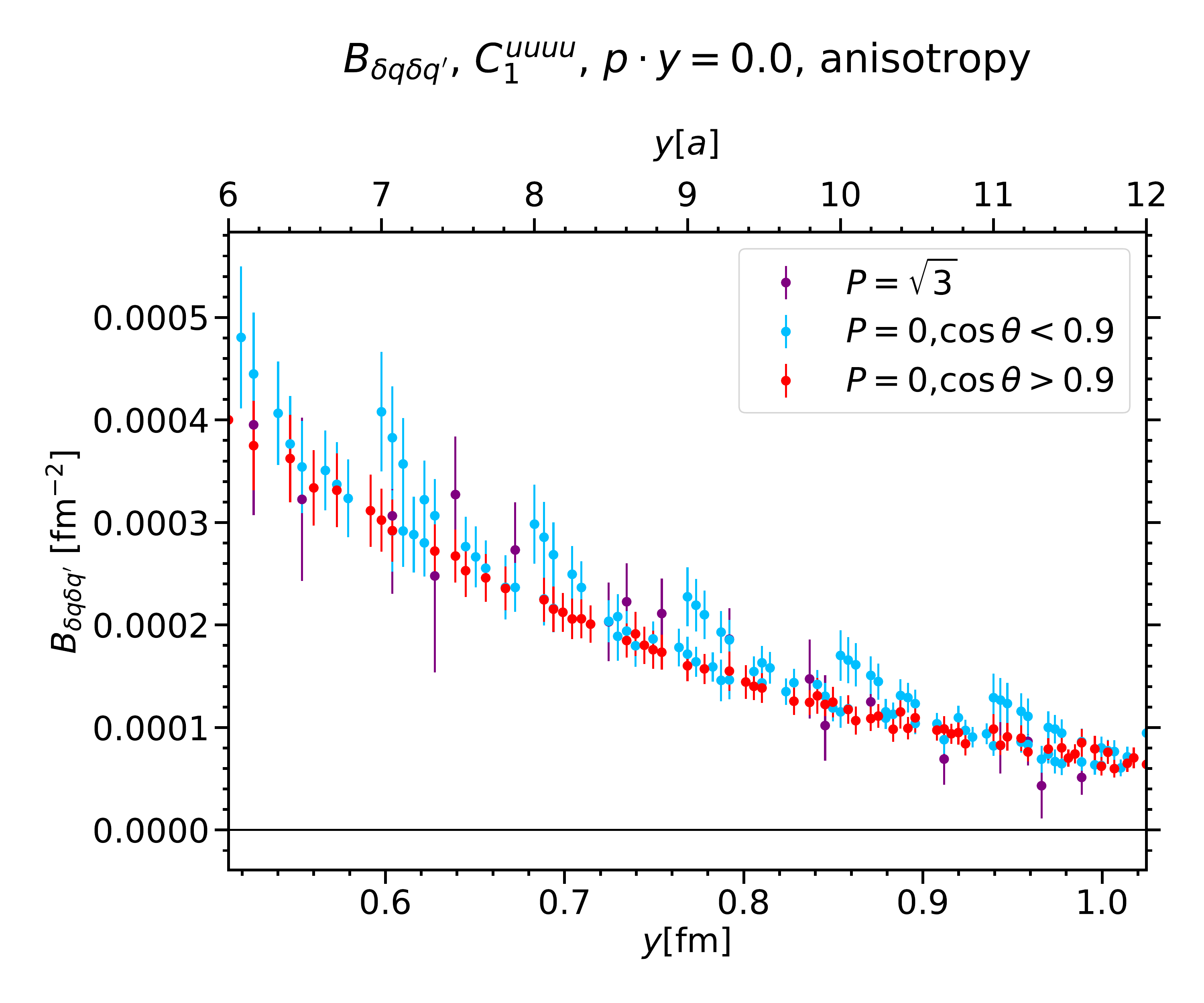}} \\ 
\caption{Visualization of anisotropies in the data of twist-two functions. We separate the data points \wrt the angle between $\mvec{y}$ and the nearest diagonal in the same manner as in \fig\ref{fig:aniso}. This figure shows the corresponding results for the $C_1$ contributions to $A_{ud}$ (a) and $A_{uu}$ (b), as well as the $C_2$ contribution to $A_{uu}$ for small $y$ (c) and large $y$ (d). In panel (b) we plot the data for non-zero momentum and $py=1.6$, whereas in the remaining plots $\mvec{P} = \mvec{0}$. In panels (e) and (f) we show the data for $B_{\delta u \delta d}$ and $B_{\delta u \delta d}$, respectively, where we distinguish only between $\cos \theta < 0.9$ and $\cos \theta > 0.9$. This is compared to the data for $P=\sqrt{3}$ at $py=0$. \label{fig:tw2f_aniso}}
\end{figure}
\Fig{\ref{fig:tw2f_aniso}} shows the data obtained for the twist-two functions separated according to this scheme for some selected channels. As in the data of the bare two-current matrix elements, we observe the saw-tooth pattern in the $C_1$ data for large distances, which originates from mirror charges due to the periodic spatial boundary conditions in our lattice setup. This is plotted in \fig{\ref{fig:tw2f_aniso-C1uudd-A_VV-p000}} for $\mvec{P} = \mvec{0}$ and in \fig\ref{fig:tw2f_aniso-C1uudd-A_VV-p-1-1-1} for $\mvec{P} = -(1,1,1)$ and $py = 1.6$. The data corresponding to distance vectors along one of the lattice diagonals are less affected by mirror charges. In \fig\ref{fig:tw2f_aniso-C2u-A_VV-p000-mid} and \ref{fig:tw2f_aniso-C2u-A_VV-p000-small} we again observe the anisotropy of the lattice propagators in the data of the $C_2$ graph. As discussed in the previous section, the propagator is less affected by this for distance vectors close to one lattice diagonal.

Beside the patterns already discussed, we find an anisotropic behavior of the twist-two function $B_{\delta u \delta d}$ for $\mvec{P} = \mvec{0}$, which can be seen for all regions in $y$. The data points along a lattice axis have a significantly larger value than those corresponding to distance vectors in the vicinity of a space diagonal. This is shown in \fig{\ref{fig:tw2f_anisoBTT-C1uudd}} and \fig{\ref{fig:tw2f_anisoBTT-C1uuuu}}, where we compare these data with those for $\mvec{P} = -(1,1,1)$. The data for non-zero momentum are consistent with the data for zero momentum if again $\mvec{y}$ is close to a space diagonal. Therefore, we regard those data points as more reliable. 

\begin{figure}
\begin{center}
\subfigure[Lorentz invariance, $A_{ud}$, $C_1$, $py=0$ \label{fig:tw2f_linvC1-C1uudd-A_VV}]{
\includegraphics[scale=0.25,trim={0.5cm 1.2cm 0.5cm 2.8cm},clip]{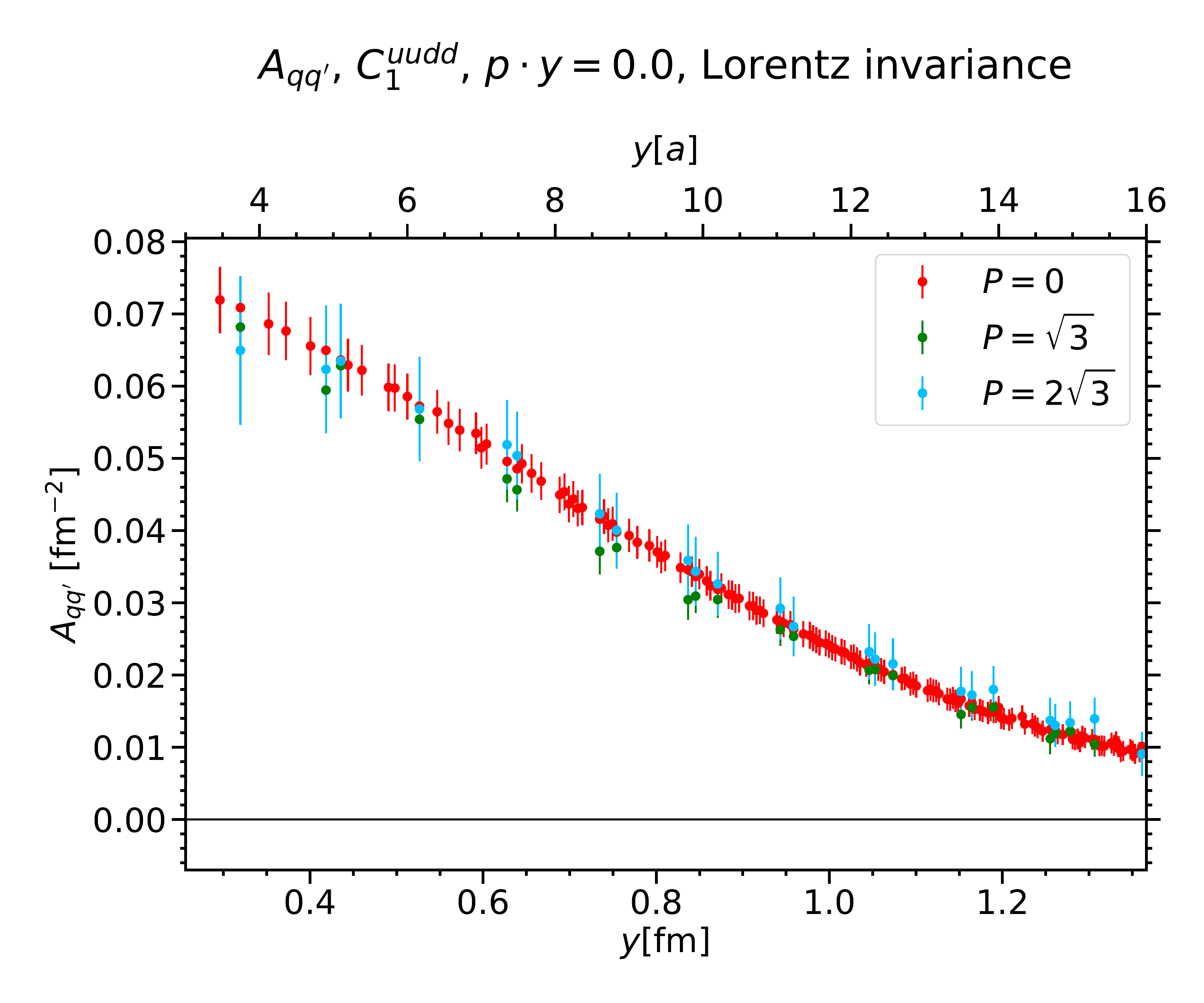}}\hfill 
\subfigure[Lorentz invariance, $A_{\delta d u}$, $C_1$, $py=0$ \label{fig:tw2f_linvC1-C1uuuu-A_VT}]{
\includegraphics[scale=0.25,trim={0.5cm 1.2cm 0.5cm 2.8cm},clip]{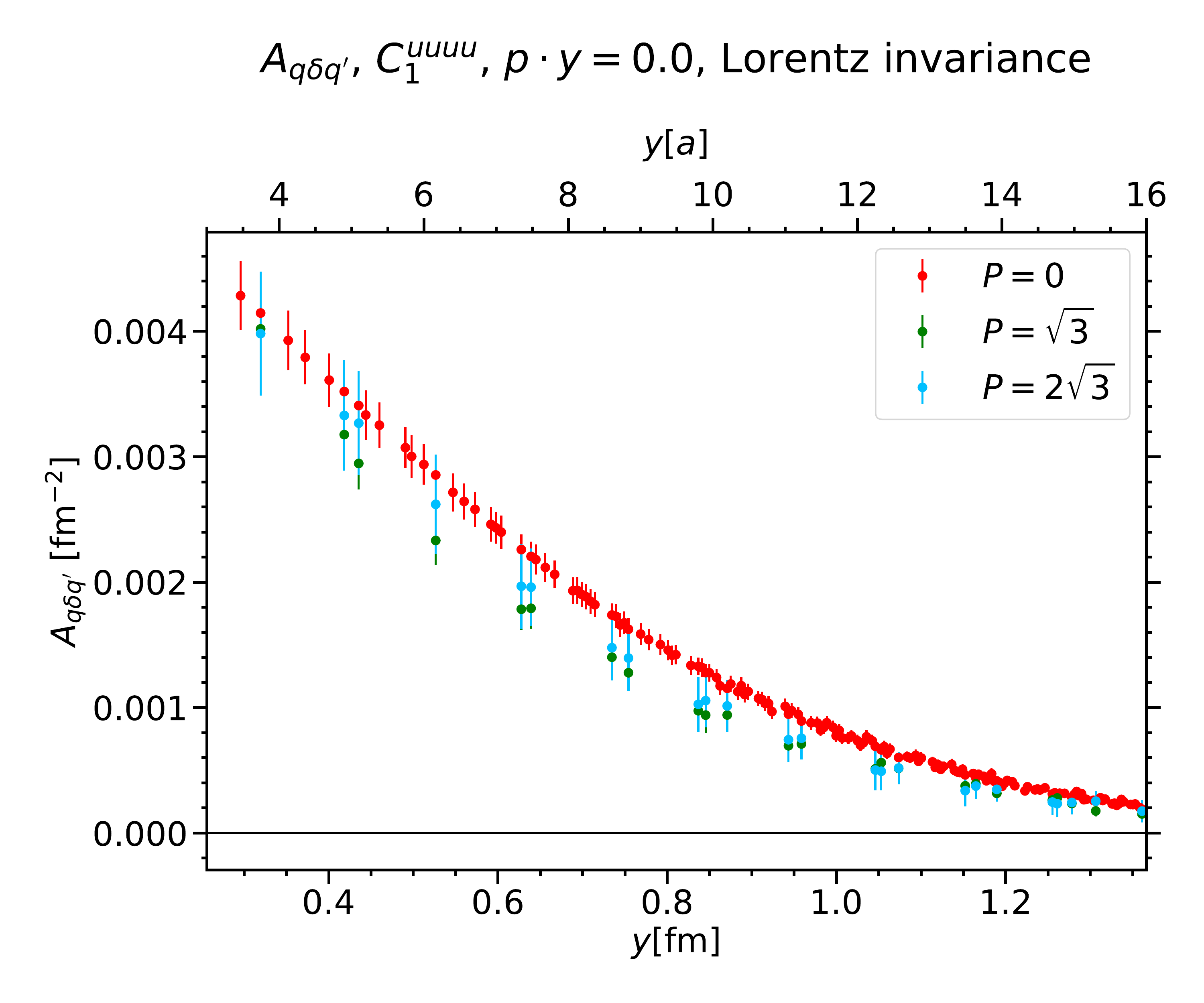}} \\ 
\subfigure[Lorentz invariance, $A_{uu}$, $C_2$, $py=0$ \label{fig:tw2f_linvC2S2-C2u-A_VV}]{
\includegraphics[scale=0.25,trim={0.5cm 1.2cm 0.5cm 2.8cm},clip]{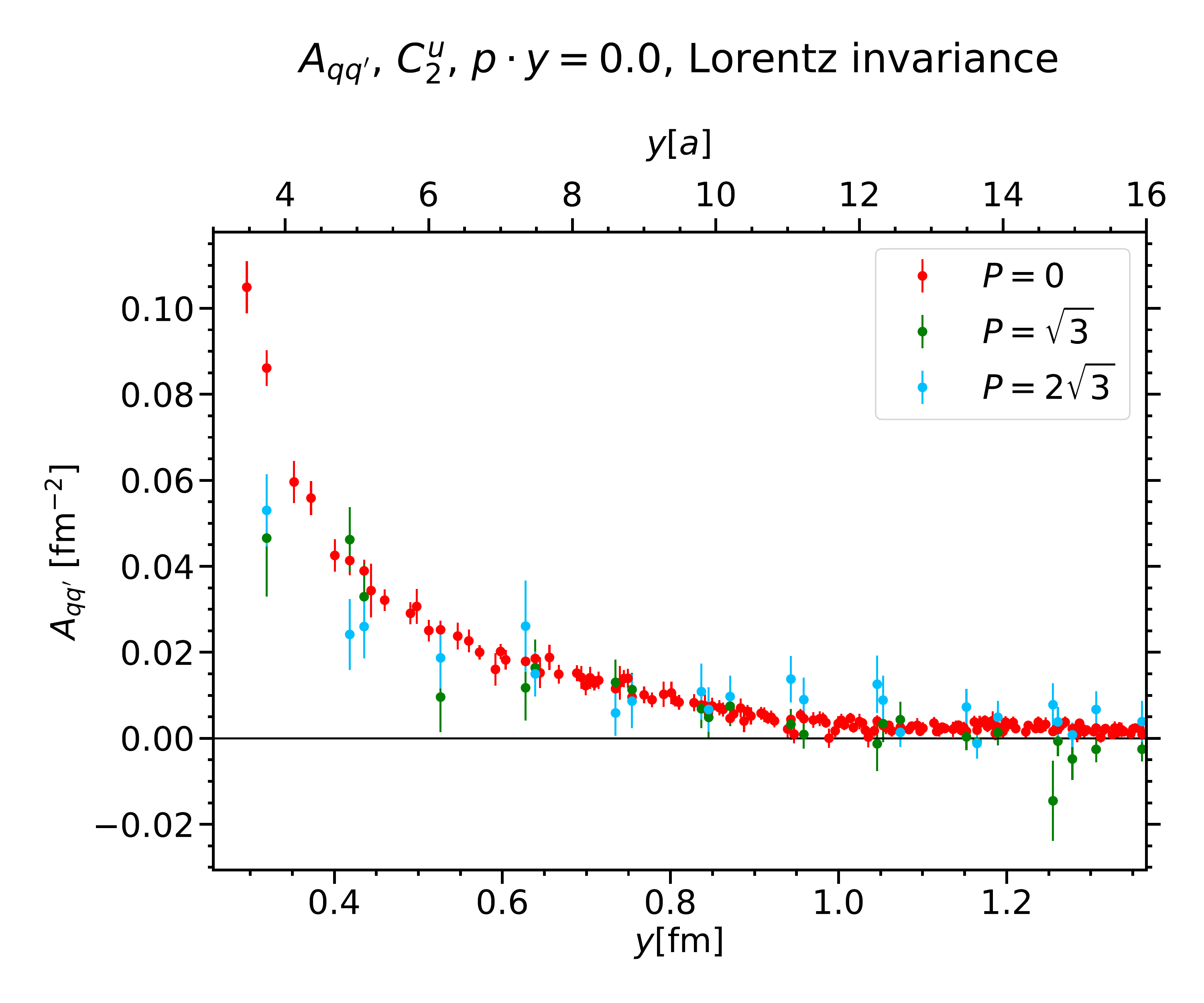}} \hfill 
\subfigure[Lorentz invariance, $A_{\delta u u}$, $C_2$, $py=0$ \label{fig:tw2f_linvC2S2-C2u-A_VT}]{
\includegraphics[scale=0.25,trim={0.5cm 1.2cm 0.5cm 2.8cm},clip]{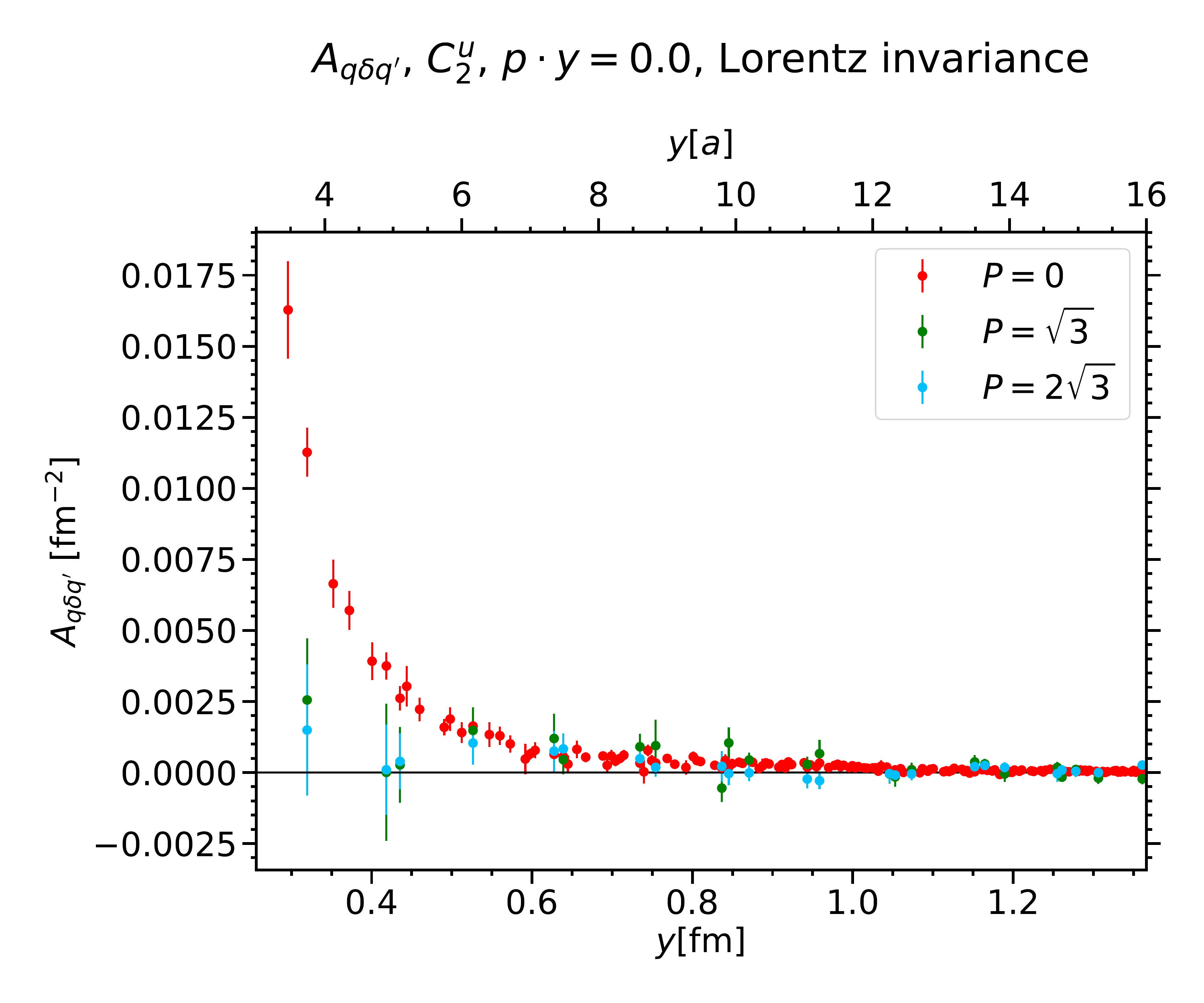}} \\ 
\subfigure[Lorentz invariance, $A_{dd}$, $C_2$, $py=0$ \label{fig:tw2f_linvC2S2-C2d-A_VV}]{
\includegraphics[scale=0.25,trim={0.5cm 1.2cm 0.5cm 2.8cm},clip]{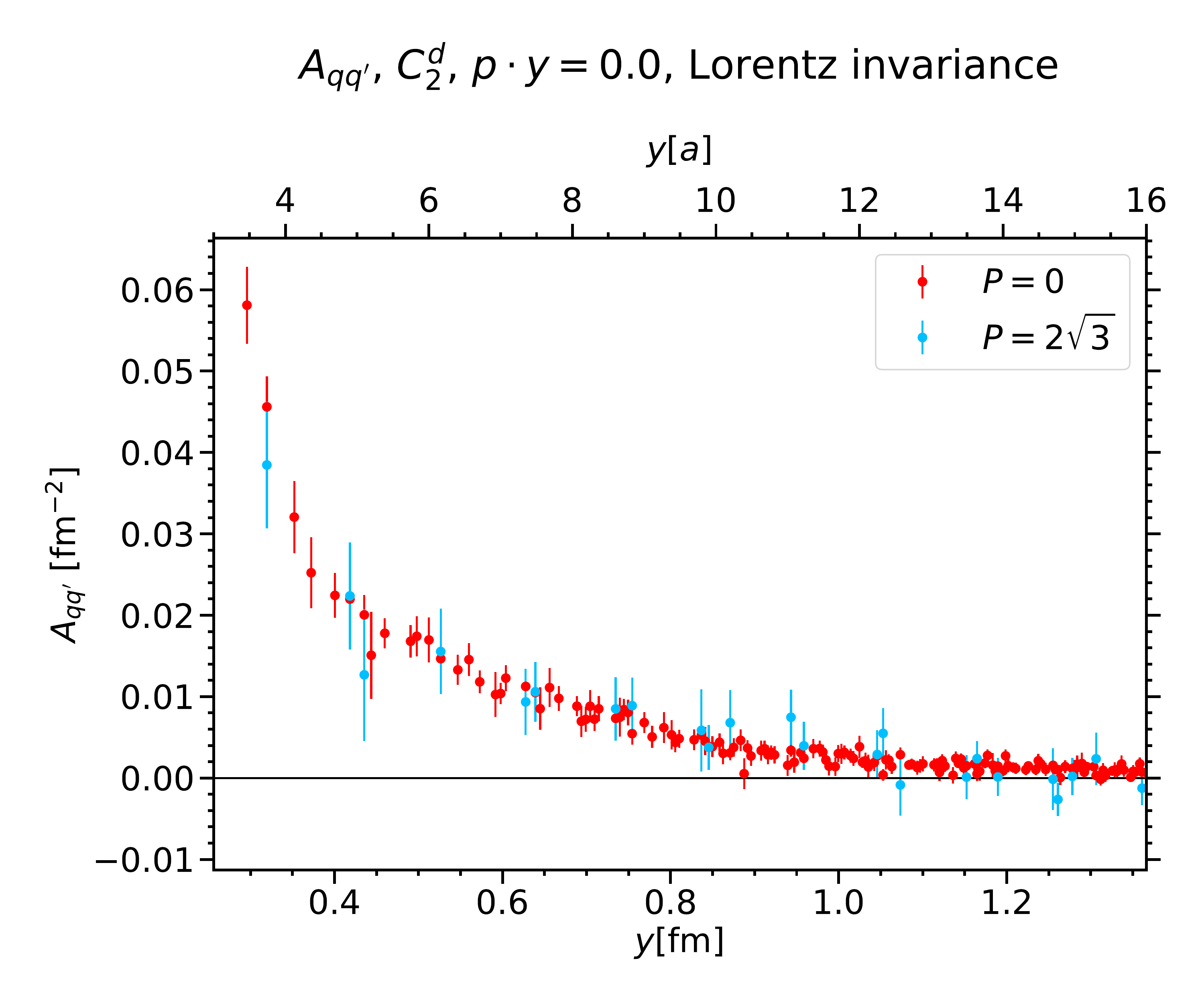}} \hfill 
\end{center}
\caption{Comparison of the twist-two data for $P=0$ (red), $P=\sqrt{3}$ (green) and $P=2\sqrt{3}$ (blue). This is shown for the $C_1$ contributions to $A_{ud}$ (a) and $A_{\delta d u}$ (b), the $C_2$ contributions to $A_{uu}$ (c), $A_{\delta u u}$ (d) and $A_{dd}$ (e). In the latter case we leave out the data for $P=\sqrt{3}$ for clarity, since they have large statistical errors. \label{fig:tw2f_linv}}
\end{figure}

Based on this discussion, we will keep only data corresponding to distances $\mvec{y}$ that satisfy

\begin{align}
\label{eq:cos-cut}
\cos \theta > 0.9\,,
\end{align} 
when discussing physical results. %!!!Also used in \cite{Cichy:2012is} \\
As a further check of the reliability of our data, we compare the twist-two functions obtained for different proton momenta at $py=0$. Because of Lorentz invariance, these should yield the same result within statistical errors. In \fig{\ref{fig:tw2f_linv}} we compare the twist-two data obtained for the momenta $P=0$, $P=\sqrt{3}$ and $P=2\sqrt{3}$. For each value of $P$, $y^2$ and $py$ the data are extracted separately, taking into account all distances $\mvec{y}$ satisfying \eqref{eq:cos-cut} and all contributing momenta $\mvec{P}$.
\begin{figure}
\subfigure[Lorentz invariance, $A_{qq}$, $S_2$, $py=0$ \label{fig:tw2f_linvC2S2-S2-A_VV-smally}]{
\includegraphics[scale=0.25,trim={0.5cm 1.2cm 0.5cm 2.8cm},clip]{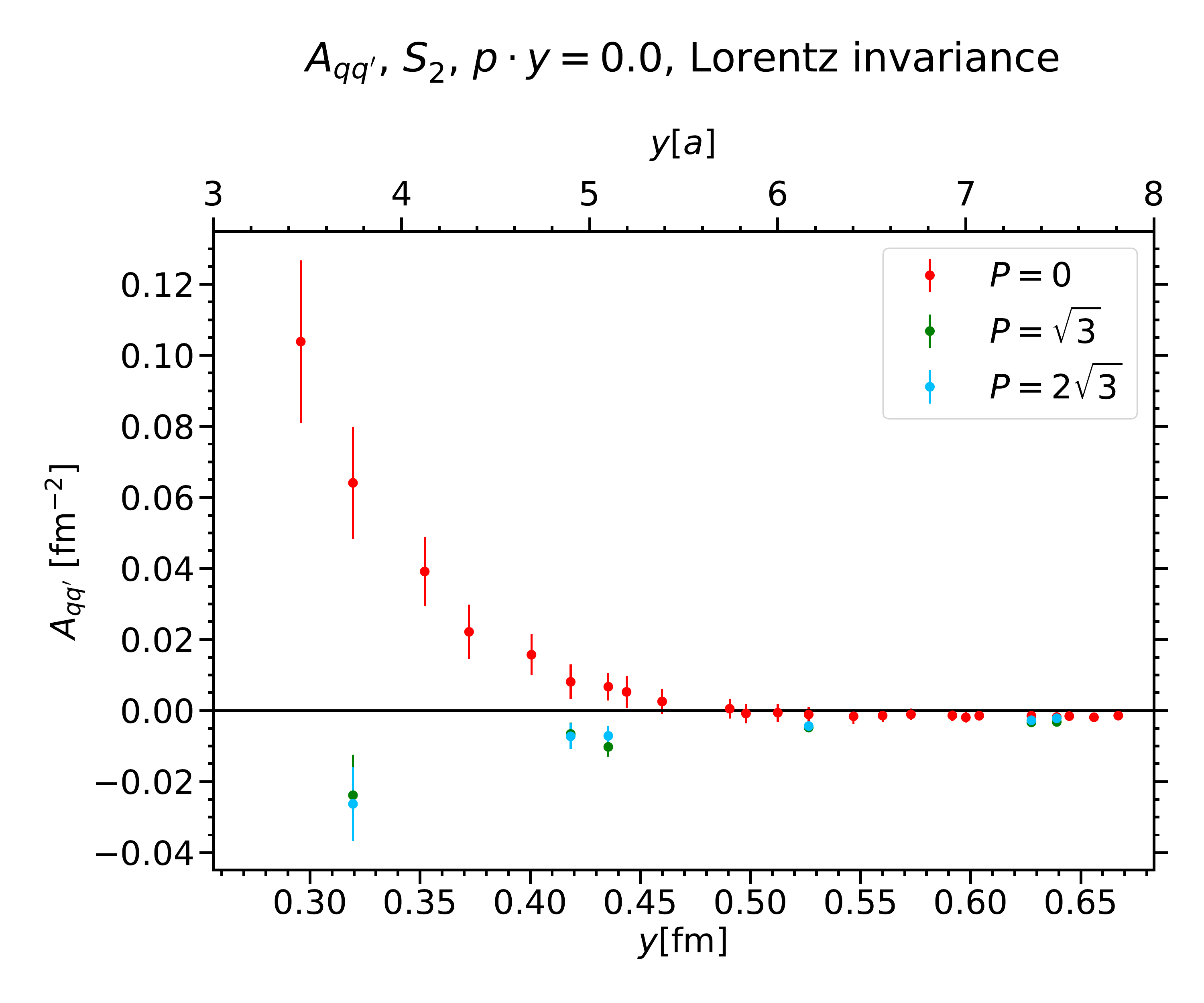}} \hfill
\subfigure[Lorentz invariance, $A_{qq}$, $S_2$, $py=0$ \label{fig:tw2f_linvC2S2-S2-A_VV-largey}]{
\includegraphics[scale=0.25,trim={0.5cm 1.2cm 0.5cm 2.8cm},clip]{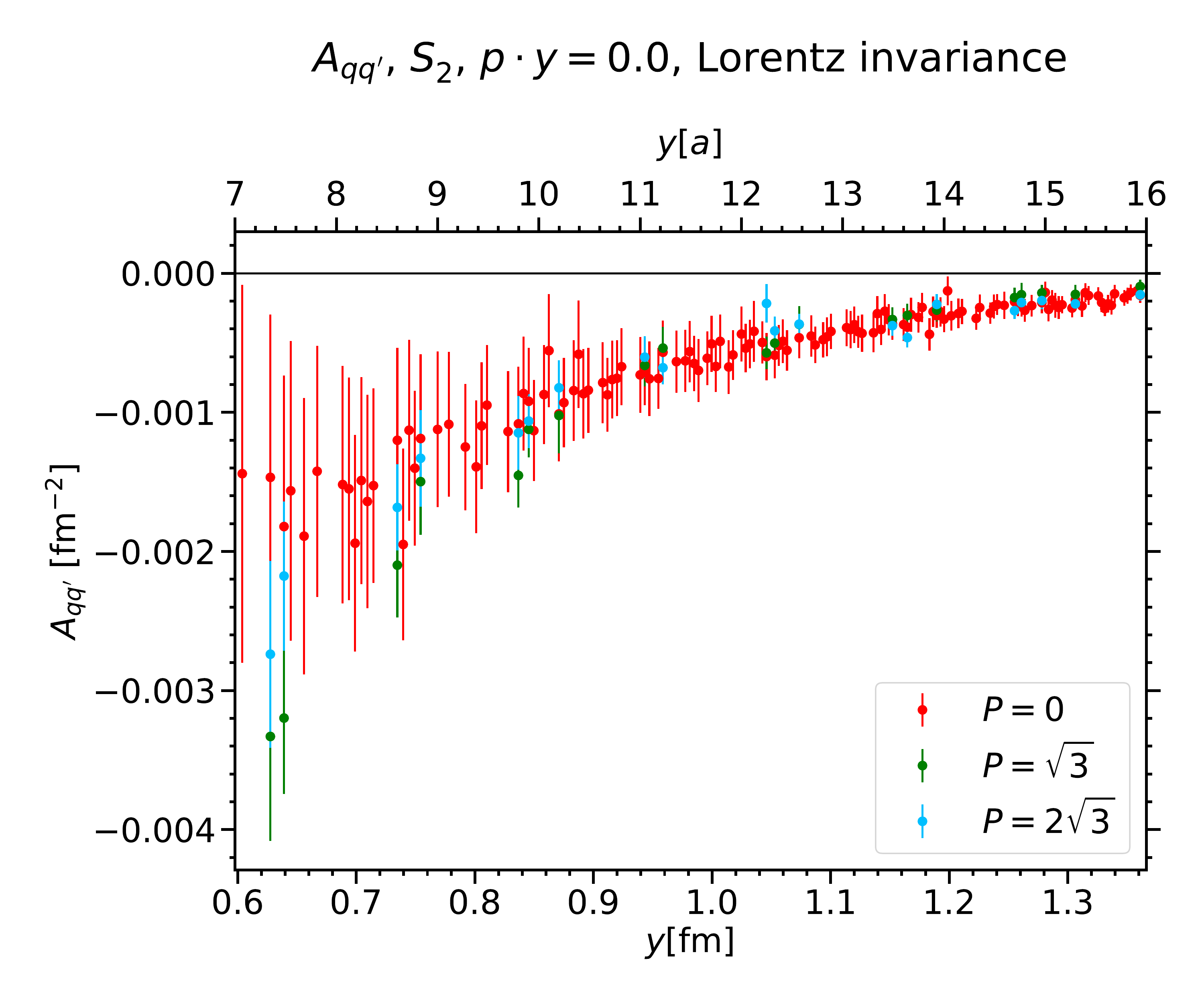}}
\caption{The same as \fig\ref{fig:tw2f_linv} for the $S_2$ contribution to $A_{qq}$. This is shown for small distances (a), where breaking of Lorentz invariance is observed, and for large $y$ (b).\label{fig:tw2f_linv_S2}}
\end{figure}

In the case of $C_1$, see \eg \fig{\ref{fig:tw2f_linvC1-C1uudd-A_VV}}, we observe consistency with Lorentz symmetry. In some cases small deviations are visible, as for $A_{\delta d u}$ shown in \fig{\ref{fig:tw2f_linvC1-C1uuuu-A_VT}}. In this case, the difference occurs between the data for $P=0$ and $P\neq 0$. Notice that we used different source-sink separations for these two cases, hence, the discrepancy might be caused by excited state contributions.

At large distances $y$, Lorentz symmetry is also intact for the $C_2$ graph, as can be seen in \fig{\ref{fig:tw2f_linv}} (c)-(e). However, once we go to smaller $y$, Lorentz invariance is clearly broken. The most extreme example for this is given by $A_{\delta u u}$, which is plotted in panel (d). Deviations start to show up for $y<5a$ and become large for $y<4a$.

The situation is even worse for the $S_2$ graph at $y<7a$, where in the most extreme cases the data for $P=0$ and $P\neq 0$ show different signs. As an example we show the corresponding data of $A_{qq}$ in \fig\ref{fig:tw2f_linvC2S2-S2-A_VV-smally}. For larger $y$, consistency with Lorentz invariance can be observed in all channels; an example is given in \fig\ref{fig:tw2f_linvC2S2-S2-A_VV-largey}.
\FloatBarrier

\subsection{Physical results for $py=0$}
\label{sec:twist2_py0}

\begin{figure}
\subfigure[graph comparison, $A_{qq^\prime}$, $\mvec{P} = \mvec{0}$, $C_1$, $C_2$, $S_2$ \label{fig:tw2f_graphs-A_VV}]{
\includegraphics[scale=0.25,trim={0.5cm 1.2cm 0.5cm 2.8cm},clip]{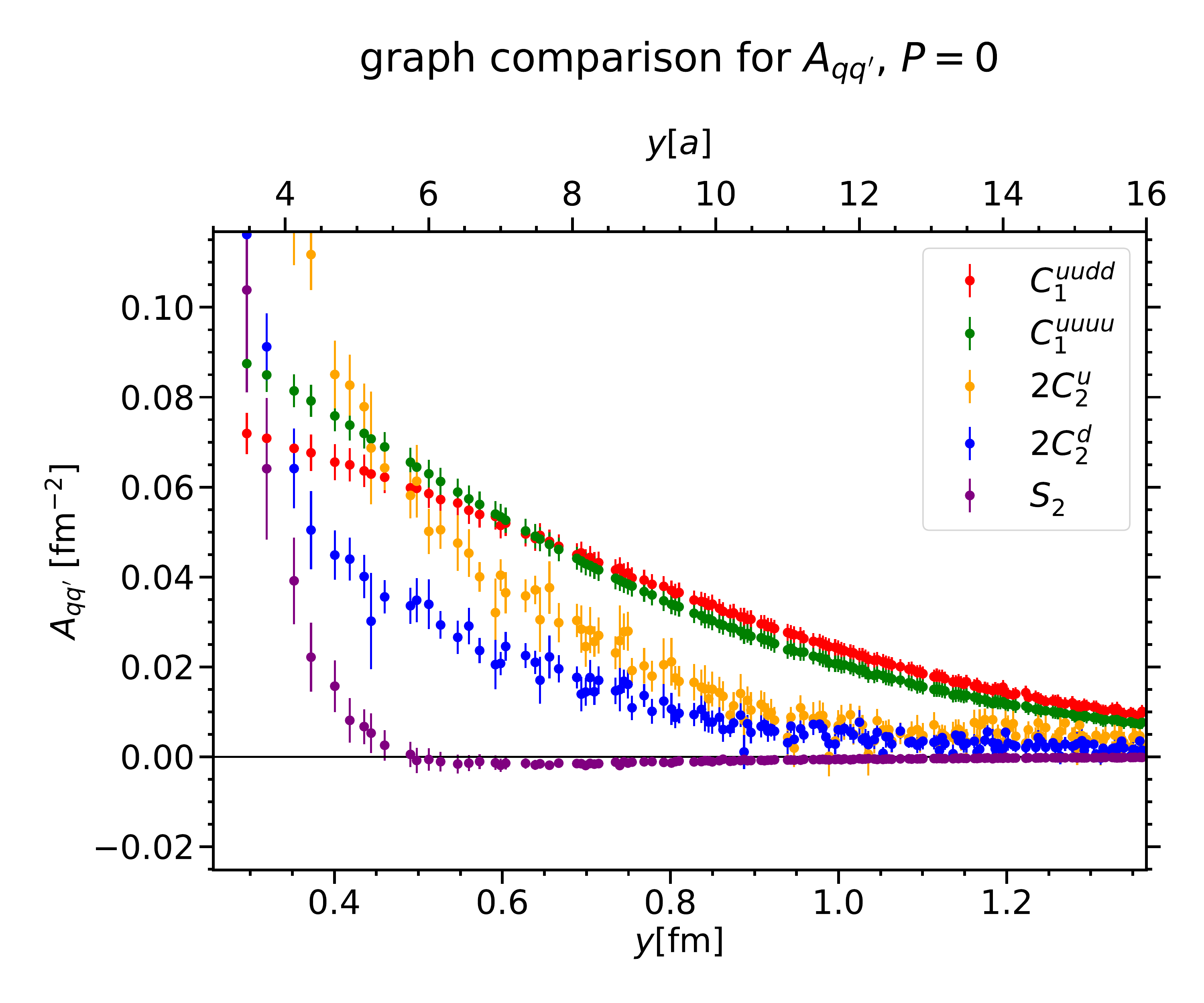}}\hfill 
\subfigure[graph comparison, $A_{\delta q q^\prime}$, $\mvec{P} = \mvec{0}$, $C_1$, $C_2$, $S_2$ \label{fig:tw2f_graphs-A_VT}]{
\includegraphics[scale=0.25,trim={0.5cm 1.2cm 0.5cm 2.8cm},clip]{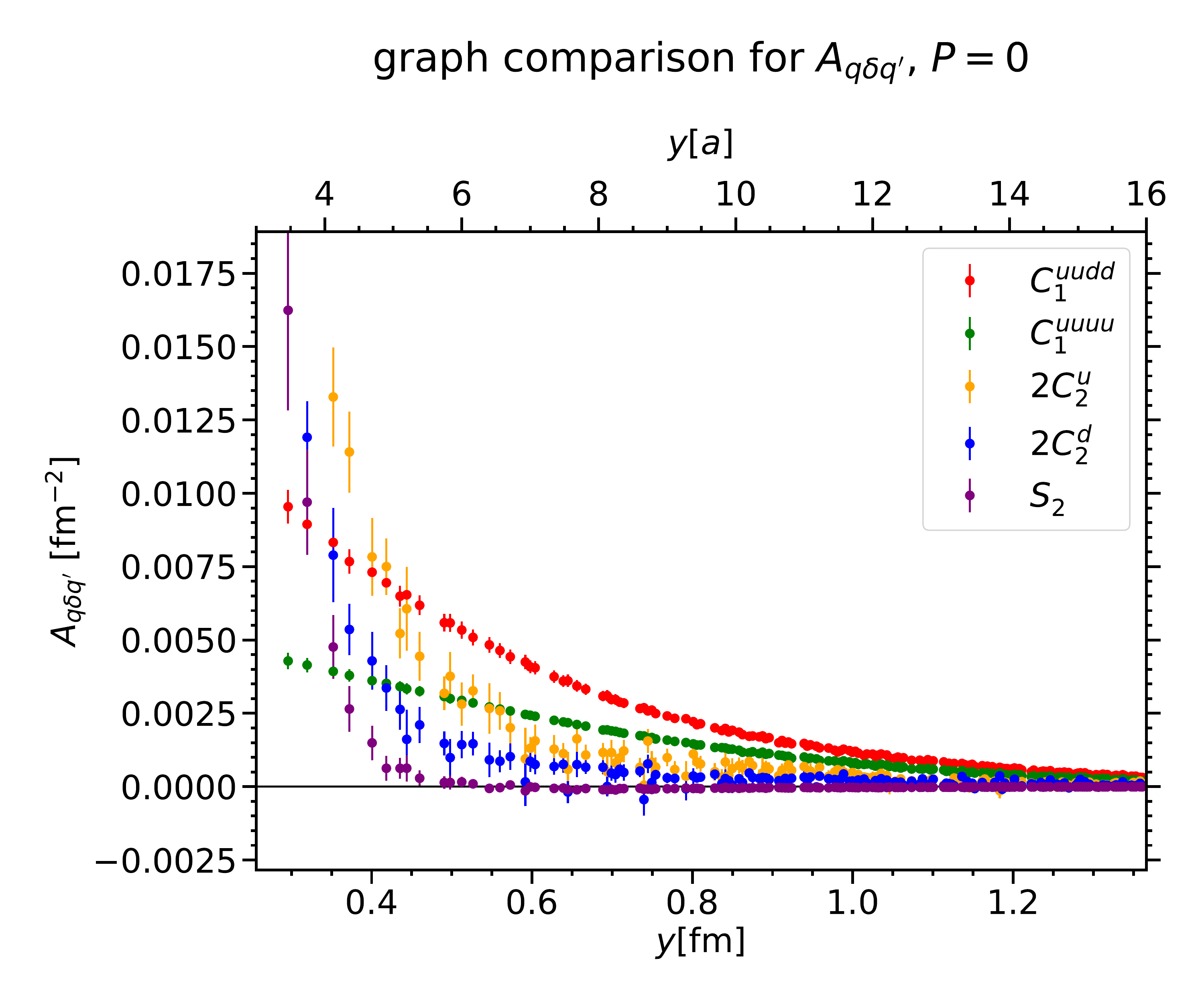}} \\ 
\subfigure[graph comparison, $A_{qq^\prime}$, $\mvec{P} = \mvec{0}$, $C_1$, $S_1$ \label{fig:tw2f_graphs-A_VV-CS1}]{
\includegraphics[scale=0.25,trim={0.5cm 1.2cm 0.5cm 2.8cm},clip]{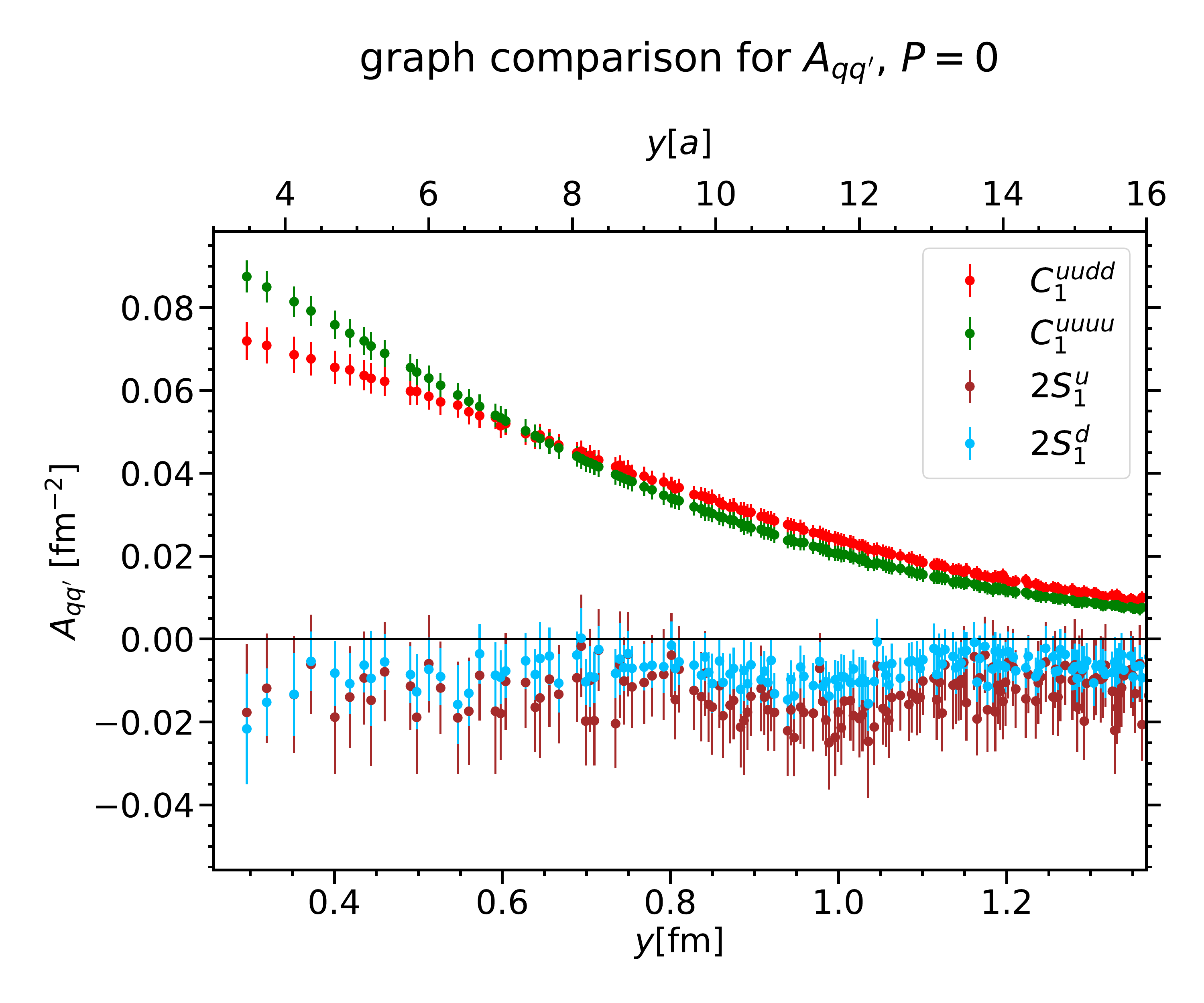}}\hfill 
\subfigure[graph comparison, $A_{\delta q q^\prime}$, $\mvec{P} = \mvec{0}$, $C_1$, $S_1$ \label{fig:tw2f_graphs-A_VT-CS1}]{
\includegraphics[scale=0.25,trim={0.5cm 1.2cm 0.5cm 2.8cm},clip]{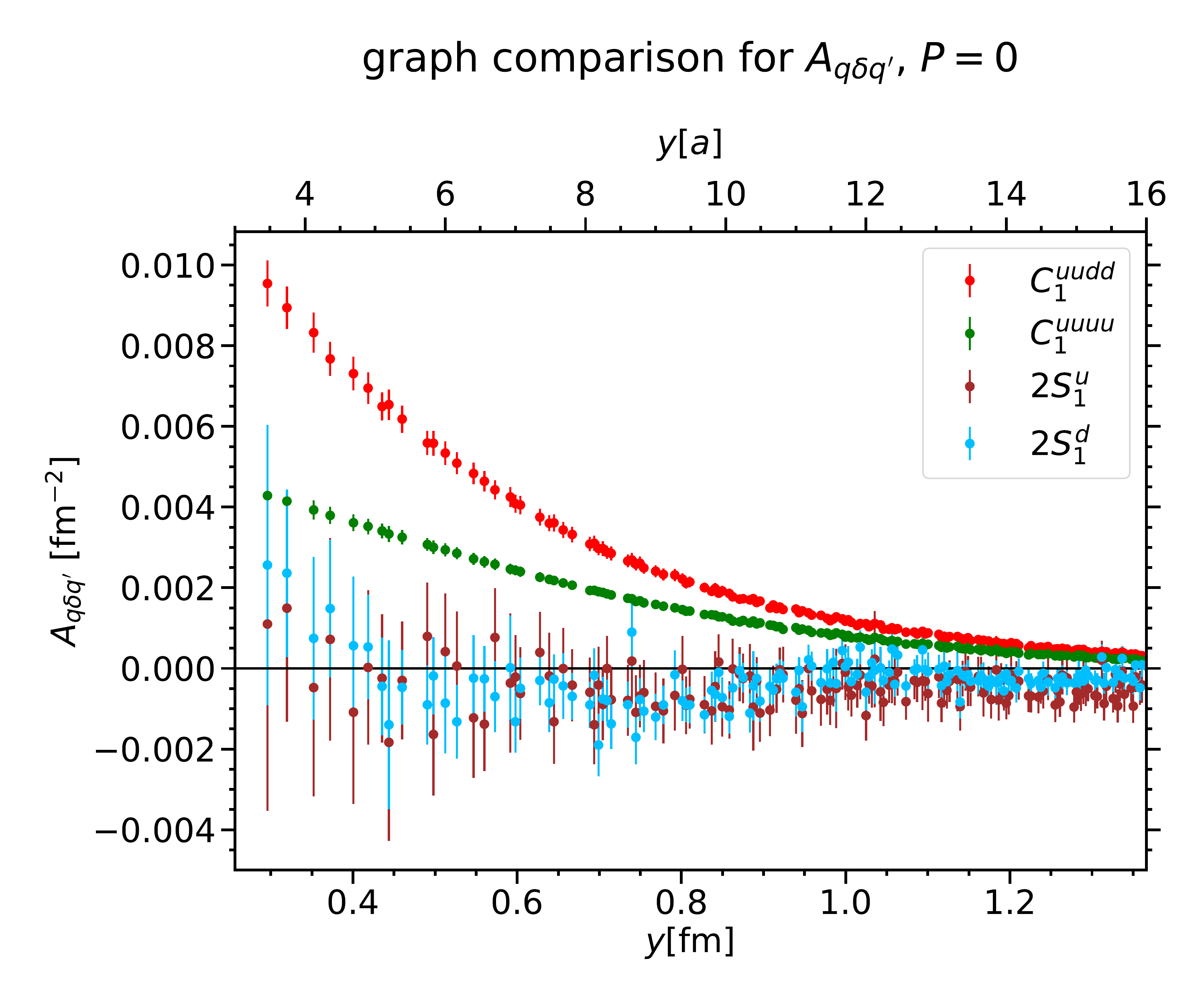}} \\ 
\caption{Comparison between the contributions of each Wick contraction to the twist-two functions for $\mvec{P} = \mvec{0}$ and $A_{qq^\prime}$ (left) and $A_{\delta qq^\prime}$ (right). Panels (a,b) show the data for $C_1$, $C_2$ and $S_2$, whereas in (c,d) we compare the data for $C_1$ and $S_1$.\label{fig:tw2f_graphs}}
\end{figure}

In the following, we consider the data of twist-two functions extracted for each single graph. For the moment we restrict ourselves to $\mvec{P} = \mvec{0}$. Again we take into account only the data points fulfilling \eqref{eq:cos-cut} and solve the system of equations \eqref{eq:tensor-decomp} for each value of $y^2$ and $py = 0$, \ie data points for equal $y = |\mvec{y}|$ are combined. In \fig{\ref{fig:tw2f_graphs}}(a) and (b) we show the results for $A_{qq^\prime}$ and $A_{\delta qq^\prime}$, where we compare the contributions of $C_1$, $C_2$ and $S_2$ for a specific flavor combination. Panels (c) and (d) show the same comparison for $C_1$ and $S_1$.

It is observed that the most dominant contributions are those of the two connected graphs $C_1$ and $C_2$. The $C_2$ data strongly increase towards small $y$, whereas $C_1$ is relatively large at all distances and shows a slow decay with increasing $y$. $S_2$ is smaller by orders of magnitude than the other contractions for $y>6a$ but very steeply increasing towards small $y$. Remember that in this region the $S_2$ graph strongly violates Lorentz invariance, as we have seen in the previous section. The $S_1$ contribution has rather large errors and is consistent with zero in all regions of $y$. For $A_{qq^\prime}$ we see a significant offset in the $S_1$ contribution. This offset is very small compared to the size of the connected contractions, except for very large distances, where the size of the offset and the decreasing signals of the connected contractions become comparable.

In the following discussion, we take into account only the $C_1$ and $C_2$ contributions, since all the other contractions are small compared to the connected graphs or, in the $S_2$ case, are not reliable due to violation of Lorentz symmetry. For our final result for the twist-two functions, we add up all considered contractions according to \eqref{eq:phys_me_decomp}, \emph{before} solving the system of equations \eqref{eq:tensor-decomp}. Furthermore, we include the data for all considered momenta, see \sect\ref{sec:latt_setup}.

\begin{figure}
\subfigure[flavor comparison, $A_{qq^\prime}$, $py=0$ \label{fig:tw2f_fcomp-A_VV}]{
\includegraphics[scale=0.25,trim={0.5cm 1.2cm 0.5cm 2.8cm},clip]{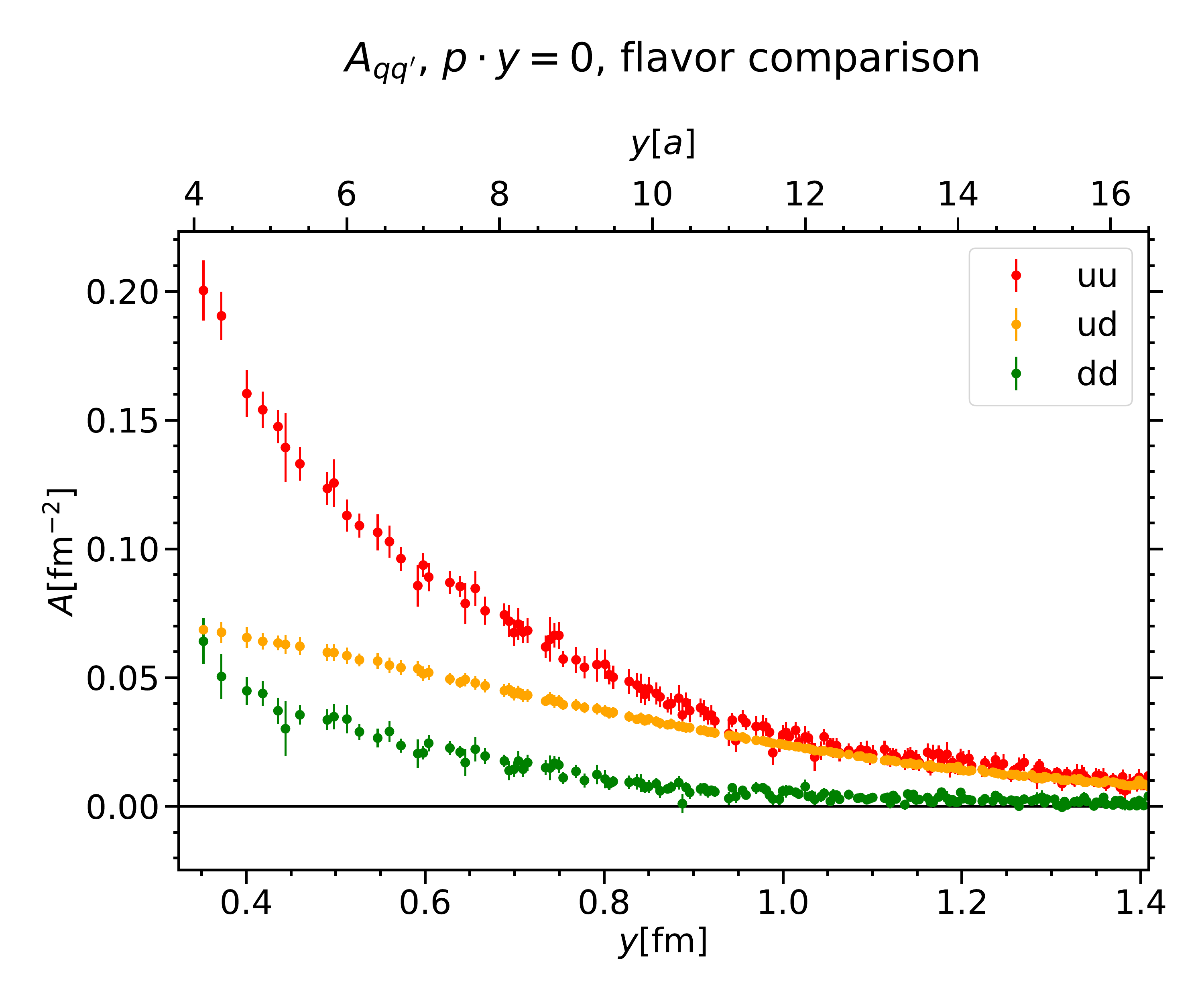}}\hfill 
\subfigure[flavor comparison, $A_{\delta q q^\prime}$, $py=0$ \label{fig:tw2f_fcomp-A_VT}]{
\includegraphics[scale=0.25,trim={0.5cm 1.2cm 0.5cm 2.8cm},clip]{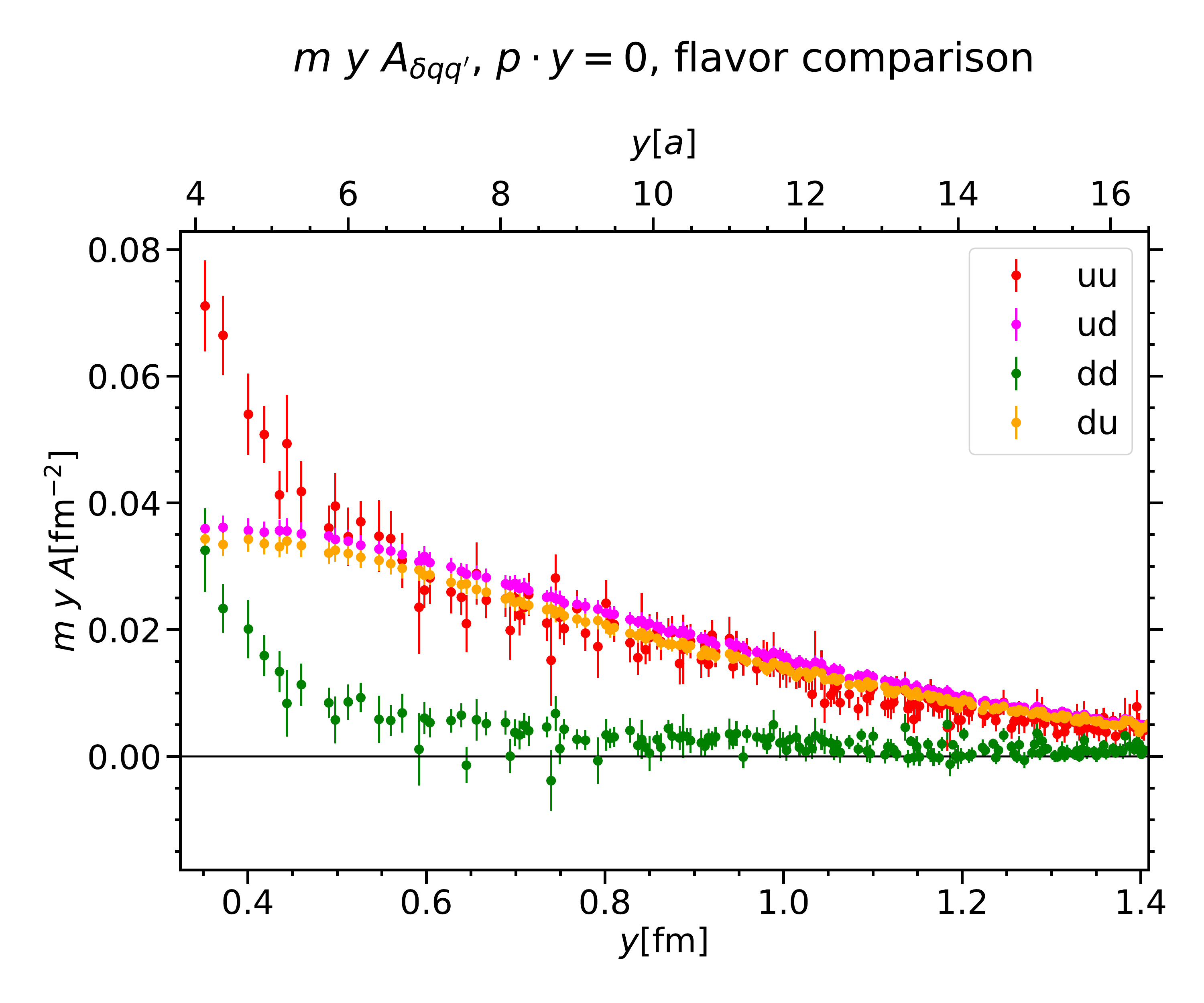}} \\ 
\caption{Twist-two functions $A_{qq^\prime}$ (a) and $A_{\delta q q^\prime}$ (b) for $py = 0$. In each panel we compare the results for all independent flavor combinations. For $A_{qq^\prime}$ these are $uu$, $ud$, and $dd$, whereas for $A_{\delta q q^\prime}$ we additionally have to consider the combination $du$. Here and in the following plots, only the contributions from the graphs $C_1$ and $C_2$ are included. \label{fig:tw2f_fcomp}}
\end{figure}

Let us first look at the flavor dependence of the twist-two functions at $py=0$. Since the spin-orbit correlations $A_{\delta qq^\prime}$ or $B_{\delta q \delta q^\prime}$ are multiplied by terms proportional to $m\,y$ or $m^2 |y^2|$ in the decomposition \eqref{eq:tensor-decomp}, we always consider $m\,y\,A_{\delta qq^\prime}$ and $m^2 |y^2| B_{\delta q \delta q^\prime}$ in the following discussion. The same applies to the corresponding DPD Mellin moments, see \eqref{eq:t2-mat-els}. In \fig{\ref{fig:tw2f_fcomp}} we show the results for the twist-two functions $A_{qq^\prime}$ (a) and $A_{\delta q q^\prime}$ (b) for the different flavor combinations. Notice that for $A_{\delta qq^\prime}$ we have the four combinations $uu$, $ud$, $du$, and $dd$, whereas in all other cases the functions for $ud$ and $du$ are equal by permutation symmetry between the two partons. At large distances we have comparably large signals for $uu$, $ud$ and $du$, while the ones for $dd$ are much smaller. This changes for smaller $y$, where both $uu$ and $dd$ strongly increase. The size of $dd$ becomes comparable to that of $ud$ and $du$ around $y = 4a = 0.342~\mathrm{fm}$.

\begin{figure}
\begin{center}
\subfigure[polarization dependence, $ud$, $py=0$ \label{fig:tw2f_polcomp-ud}]{
\includegraphics[scale=0.25,trim={0.5cm 1.2cm 0.5cm 2.8cm},clip]{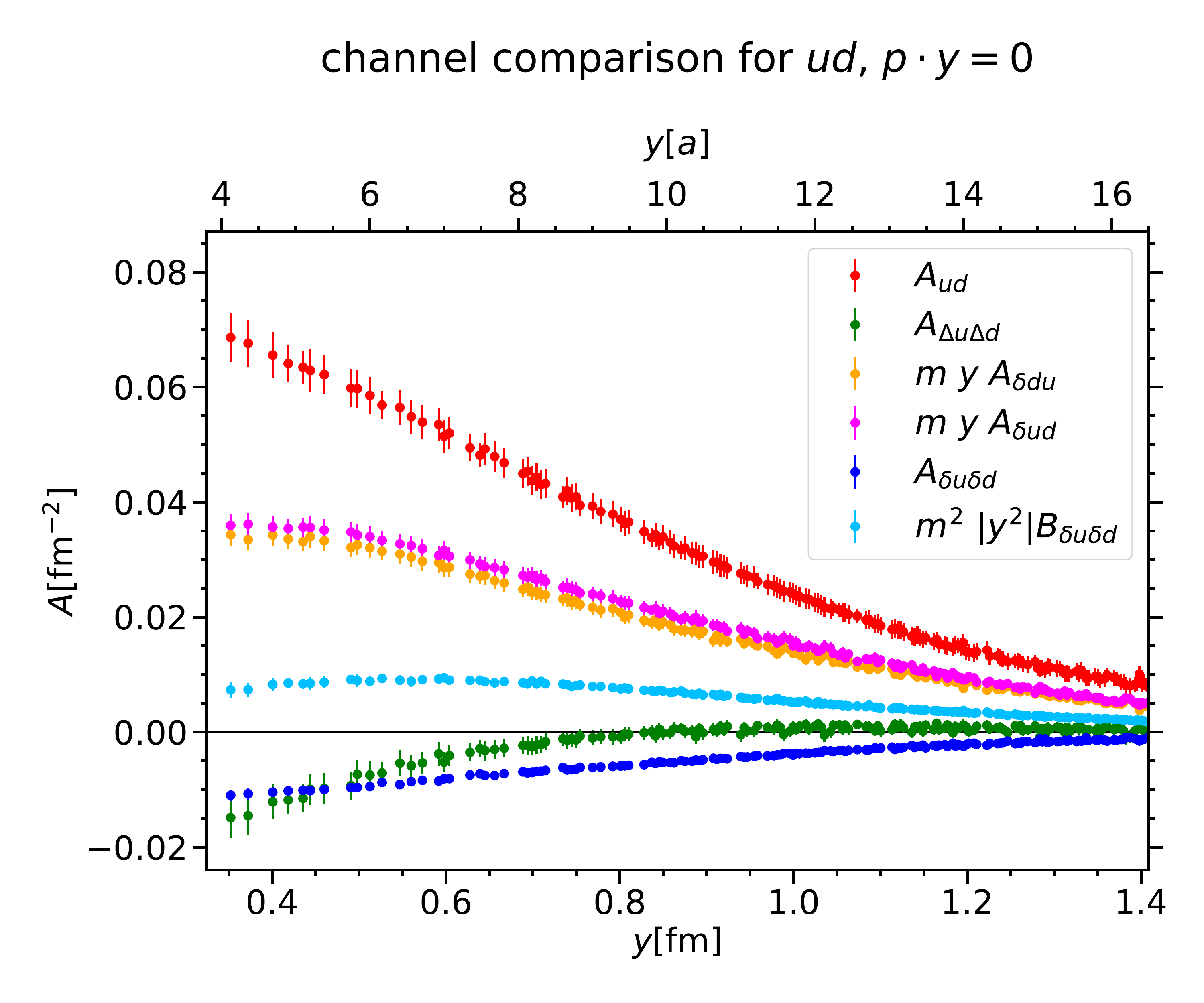}} \hfill 
\subfigure[polarization dependence, $uu$, $py=0$ \label{fig:tw2f_polcomp-uu}]{
\includegraphics[scale=0.25,trim={0.5cm 1.2cm 0.5cm 2.8cm},clip]{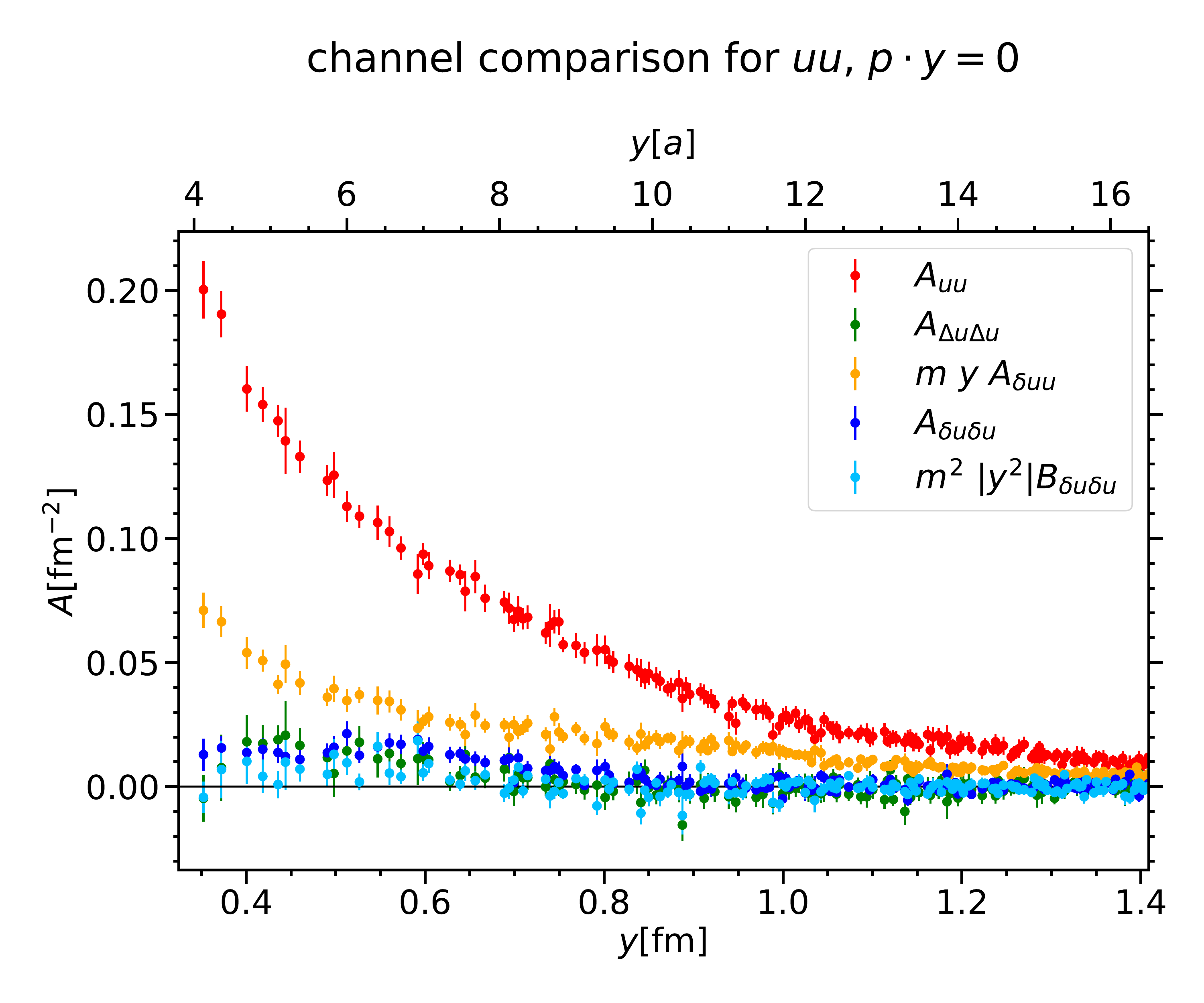}} \\ 
\subfigure[polarization dependence, $dd$, $py=0$ \label{fig:tw2f_polcomp-dd}]{
\includegraphics[scale=0.25,trim={0.5cm 1.2cm 0.5cm 2.8cm},clip]{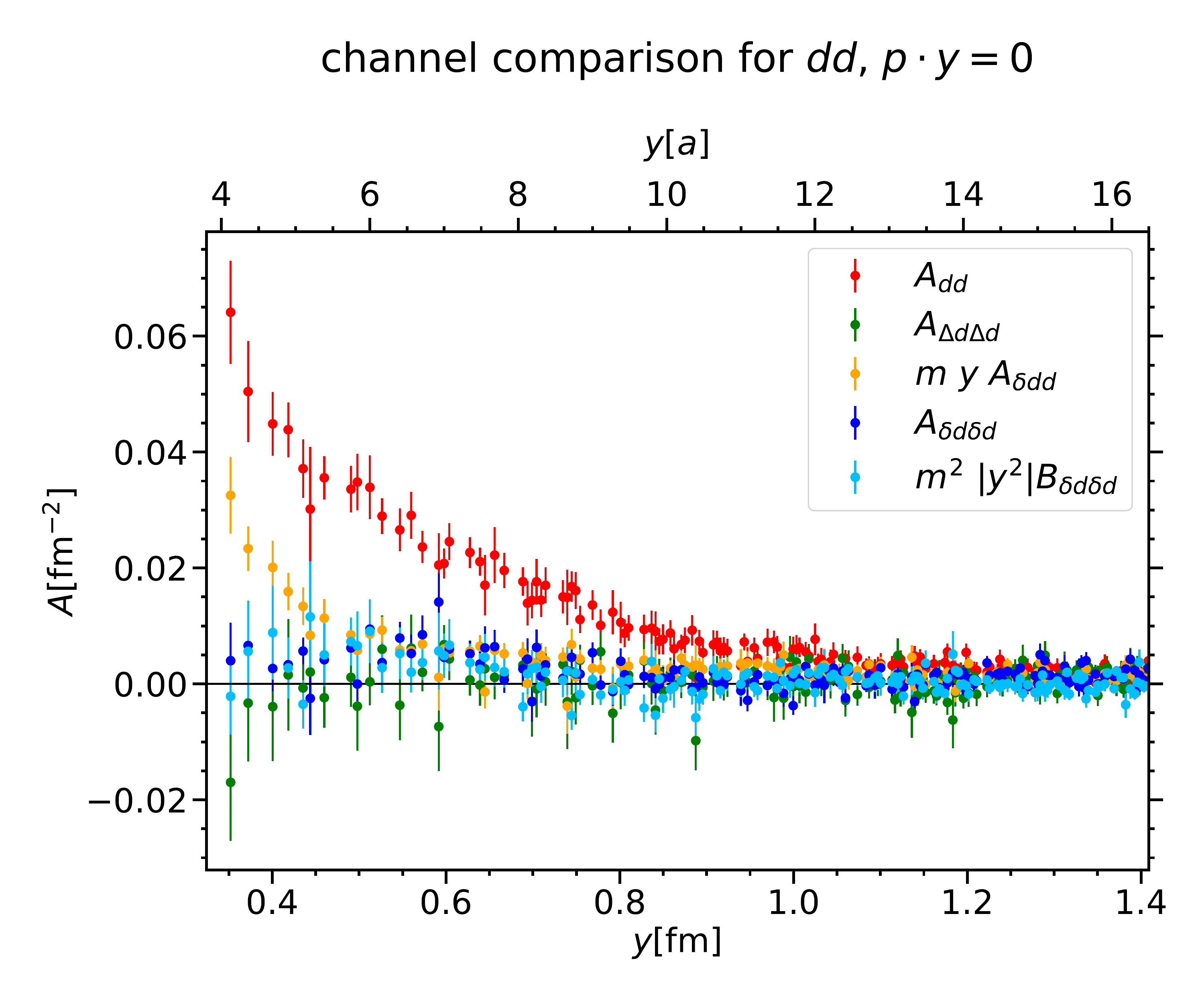}} \\  
\end{center}
\caption{Comparison of the results for different quark polarizations for the flavor combinations $ud$ (a), $uu$ (b) and $dd$ (c) at $py = 0$. \label{fig:tw2f_polcomp}}
\end{figure}
A very interesting aspect is the dependence on the quark polarization. We compare the corresponding channels in \fig{\ref{fig:tw2f_polcomp}} for $ud$ (a), $uu$ (b) and $dd$ (c). In all cases $A_{qq^\prime}$ is observed to be the channel with the largest signal. Polarization effects are significant in the case of $ud$, especially $A_{\delta u d}$ and $A_{\delta d u}$ are very large. The signal in the remaining channels is smaller but clearly different from zero. In the case of $uu$ and $dd$, polarization effects are suppressed. The largest polarized contribution is again $A_{\delta q q}$ in both cases. 
%\FloatBarrier

\subsection{Parameterization of the $y^2$ dependence}
\label{sec:twist2_y2_fit}

Further analysis steps require a parameterization of the results obtained for the twist-two functions. In the following, we adapt the approach we developed in \cite{Bali:2020mij}. For the description of the $y^2$-dependence at $py=0$ a sum of two exponentials is found to be suitable in most cases. For $A_{ud}$ and $A_{\delta u \delta d}$ it appears that this ansatz has to be slightly modified. As a general ansatz we write:

\begin{align}
A(py=0,y^2) = (\eta_1 y)^\delta A_1\ e^{-\eta_1(y-y_0)} + (\eta_2 y)^\delta A_2\ e^{-\eta_2(y-y_0)}\ ,
\label{eq:y2-ansatz}
\end{align}
where the fits are preformed for fixed $\delta$. In the cases of $A_{ud}$ and $A_{\delta u \delta d}$ it turns out that $\delta = 1.2$ is a suitable choice. In all other channels, a pure double exponential, \ie $\delta = 0$, is sufficient.

For most of the fits we take into account each point in the region $4a \le y \le 16a$. Thus, we ensure that the data points entering the fit are only mildly affected by the lattice artifacts that result in anisotropy effects or the breaking of boost invariance. For stability reasons the fit range is slightly modified in some channels. In all cases where the fit range is adjusted, we carefully checked that the data points within the modified fit range do not include such artifacts. An overview is given in \tab{\ref{tab:fit_ranges}}, where also the corresponding fixed value of $\delta$ is shown.
\begin{table}
\begin{center}
\begin{tabular}{c|c|c}
\hline \hline
channel & fit range & $\delta$ \\
\hline
$A_{ud}$ & $[1a,16a]$ & $1.2$ \\
$A_{dd}$ & $[3.5a,16a]$ & $0$ \\
$A_{\Delta d \Delta d}$ & $[3.5a,15a]$ & $0$ \\
$A_{\delta u \delta d}$ & $[3a,16a]$ & $1.2$ \\
$A_{\delta d \delta d}$ & $[3.5a,15a]$ & $0$ \\
$B_{\delta d \delta d}$ & $[4a,15a]$ & $0$ \\
\hline
else & $[4a,16a]$ & $0$ \\
\hline \hline
\end{tabular}
\end{center}
\caption{Fit ranges in $y$ used for the fit of each twist-two function for the double exponential \eqref{eq:y2-ansatz}. We also give the fixed parameter $\delta$. In all cases, $y_0 = 4a$.\label{tab:fit_ranges}}
\end{table}
In order to achieve that the parameters $A_i$ describe the relative weight of the two exponentials at the lower fit boundary, we introduce a shift $y_0 = 4a = 0.342~\mathrm{fm}$ in the exponent. In the fits we neglect correlations between the data points.

\begin{figure}
\subfigure[double-exponential fit for $A_{ud}(py=0,y^2)$ \label{fig:tw2f_y2fits-A_VV-ud}]{
\includegraphics[scale=0.25,trim={0.5cm 1.2cm 0.5cm 2.8cm},clip]{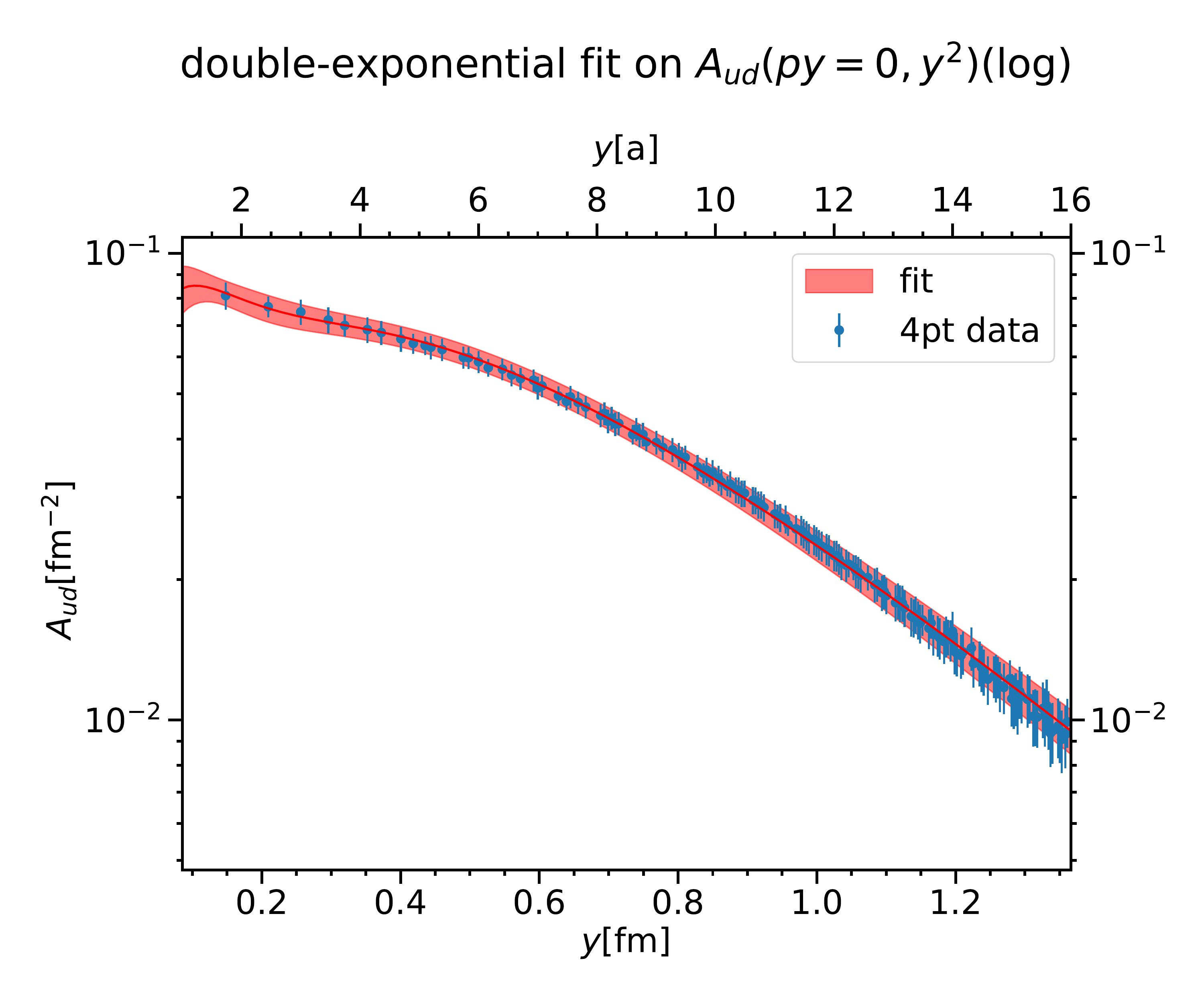}} \hfill 
\subfigure[double-exponential fit for $A_{uu}(py=0,y^2)$ \label{fig:tw2f_y2fits-A_VV-uu}]{
\includegraphics[scale=0.25,trim={0.5cm 1.2cm 0.5cm 2.8cm},clip]{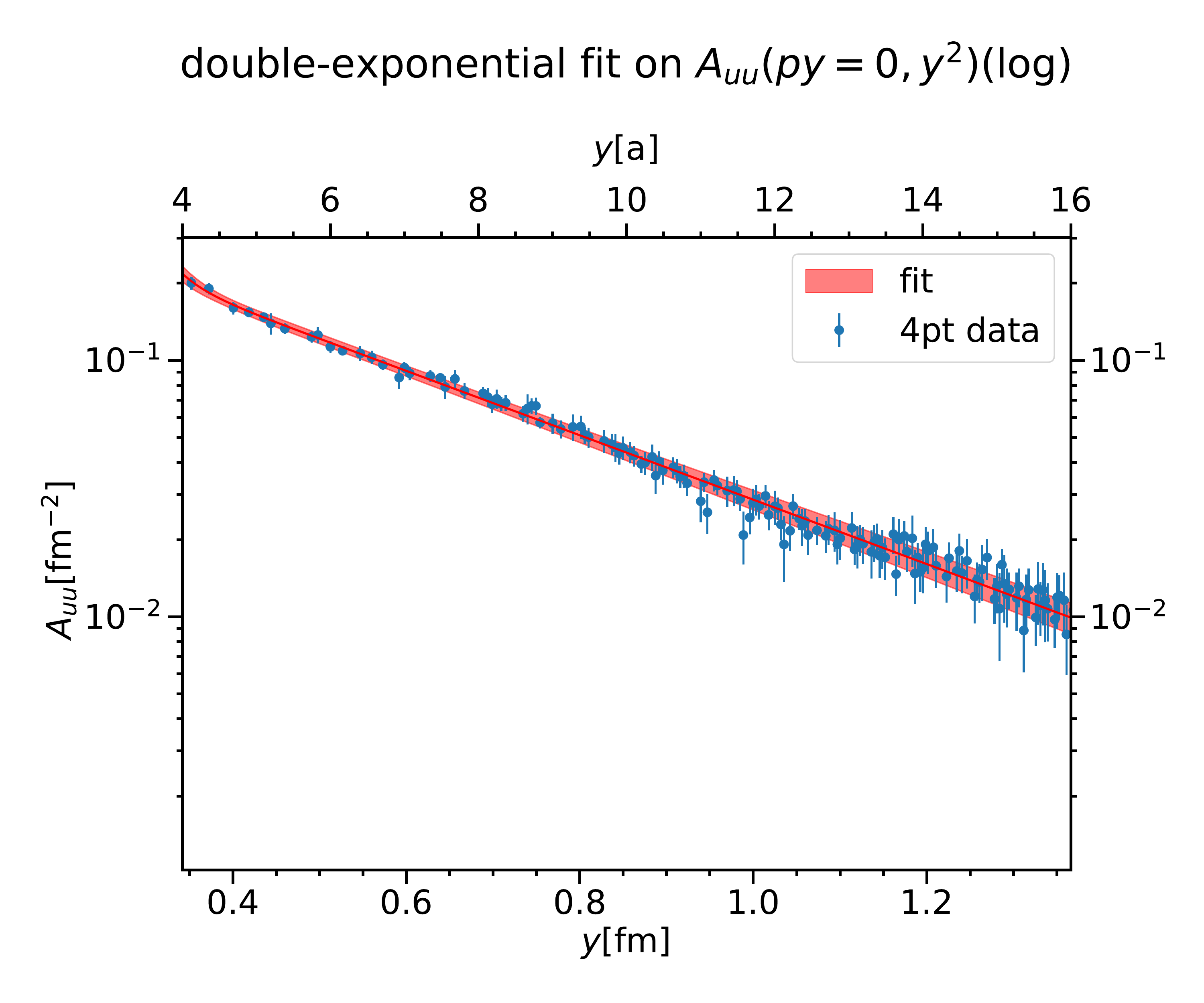}} \\ 
\subfigure[double-exponential fit for $A_{\delta d u}(py=0,y^2)$ \label{fig:tw2f_y2fits-A_VT-ud}]{
\includegraphics[scale=0.25,trim={0.5cm 1.2cm 0.5cm 2.8cm},clip]{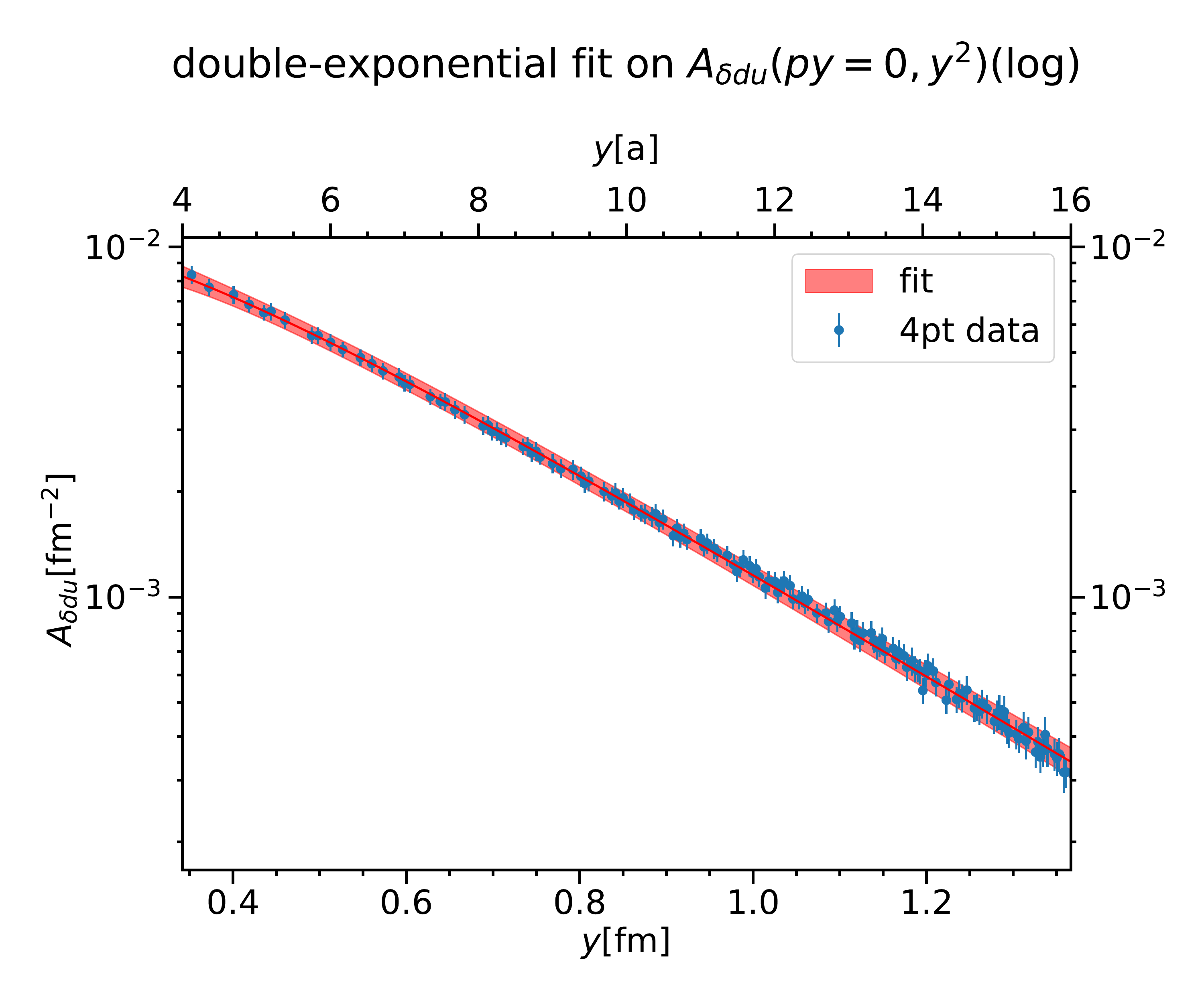}} \hfill 
\subfigure[double-exponential fit for $A_{\delta u u}(py=0,y^2)$ \label{fig:tw2f_y2fits-A_VT-uu}]{
\includegraphics[scale=0.25,trim={0.5cm 1.2cm 0.5cm 2.8cm},clip]{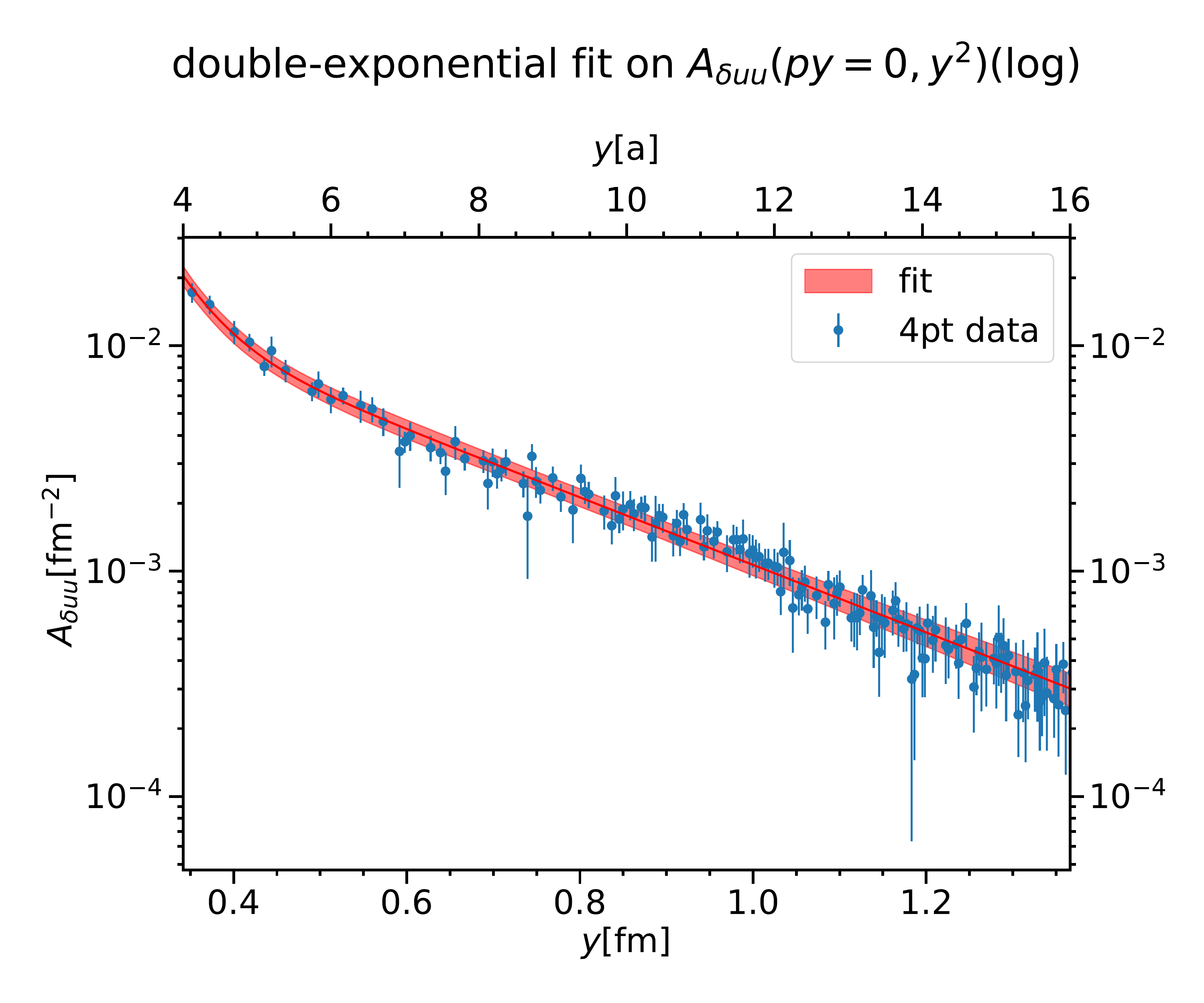}} \\ 
\caption{Data points for the twist-two functions compared to the corresponding curve resulting from a fit to the form \eqref{eq:y2-ansatz}. Each plot has a logarithmic scale for the vertical axis.\label{fig:tw2f_y2fits}}
\end{figure}

The data points of the twist-two functions at $py=0$ are plotted together with the curve resulting from the fit in \fig{\ref{fig:tw2f_y2fits}}. We take a logarithmic scale on the vertical axis to emphasize the double-exponential shape. As can be observed in the plots, the fitted curves describe the twist-two data reasonably well. The values obtained for the fit parameters $A_i$ and $\eta_i$ are listed in \tab{\ref{tab:tw2f_exp2fit}}, as well as the values of $\chi^2$ per degree of freedom. The corresponding errors are computed using the Jackknife procedure.
\clearpage
\begin{table}
\begin{center}
\begin{tabular}{c|cccc|c}
\hline
\hline
channel & $A_1[\mathrm{fm}^{-2}]$ & $\eta_1[\mathrm{fm}^{-1}]$ & $A_2[\mathrm{fm}^{-2}]$ & $\eta_2[\mathrm{fm}^{-1}]$ & $\chi^2 / \mathrm{dof}$\\ 
\hline
$A_{ u  u}$ & $0.026(17)$ & $39(20)$ & $0.1920(99)$ & $2.89(15)$ & $0.37$ \\
$A_{ u  d}$ & $0.00037(35)$ & $17.5(3.1)$ & $0.0530(28)$ & $3.52(12)$ & $0.07$ \\
$A_{ d  d}$ & $0.010(12)$ & $46(47)$ & $0.0573(64)$ & $3.66(36)$ & $0.48$ \\
\hline
$A_{\Delta u \Delta u}$ & $-0.62(57)$ & $13.7(3.8)$ & $0.61(58)$ & $12.3(2.3)$ & $0.63$ \\
$A_{\Delta u \Delta d}$ & $-0.0190(39)$ & $4.86(46)$ & $0.0026(24)$ & $1.30(73)$ & $0.30$ \\
$A_{\Delta d \Delta d}$ & $-0.029(61)$ & $14(15)$ & $0.010(61)$ & $4.6(8.2)$ & $0.61$ \\
\hline
$A_{\delta u  u}$ & $0.0208(46)$ & $21.8(6.9)$ & $0.0211(31)$ & $3.45(25)$ & $0.49$ \\
$A_{\delta d  u}$ & $-0.0059(27)$ & $6.80(37)$ & $0.0228(23)$ & $3.40(15)$ & $0.20$ \\
$A_{\delta u  d}$ & $-0.0085(27)$ & $6.85(23)$ & $0.0258(26)$ & $3.43(16)$ & $0.25$ \\
$A_{\delta d  d}$ & $0.0144(36)$ & $17.7(7.7)$ & $0.0036(26)$ & $3.6(1.1)$ & $0.64$ \\
\hline
$A_{\delta u \delta u}$ & $-0.193(99)$ & $9.5(1.3)$ & $0.196(98)$ & $7.5(1.3)$ & $0.74$ \\
$A_{\delta u \delta d}$ & $-0.000033(88)$ & $21(13)$ & $-0.00835(65)$ & $3.57(24)$ & $0.16$ \\
$A_{\delta d \delta d}$ & $-0.0027(82)$ & $18(35)$ & $0.0073(81)$ & $3.0(2.3)$ & $1.01$ \\
\hline
$B_{\delta u \delta u}$ & $-0.72(99)$ & $15.8(2.9)$ & $0.72(99)$ & $15.7(3.0)$ & $1.01$ \\
$B_{\delta u \delta d}$ & $-0.00074(71)$ & $7.9(2.1)$ & $0.00253(56)$ & $4.13(23)$ & $0.07$ \\
$B_{\delta d \delta d}$ & $0.73(41)$ & $16.9(1.5)$ & $-0.73(41)$ & $17.0(1.5)$ & $0.72$ \\
\hline
\hline
\end{tabular}

\end{center}
\caption{Results of the fit \eqref{eq:y2-ansatz} to the twist-two functions at $py=0$. The corresponding $\chi^2/\mathrm{dof}$ is listed in the rightmost column.\label{tab:tw2f_exp2fit}}
\end{table}
\FloatBarrier

\subsection{Parameterization of the $py$ dependence}
\label{sec:twist2_py_fit}

A parameterization of the twist-two functions is in particular mandatory for the evaluation of the $py$-integral in \eqref{eq:skewed-mellin-inv-fct}. The reason is that one has to extrapolate in $py$, since the accessible range is restricted by the largest proton momentum:

\begin{align}
|py| \underset{y^0 = 0}{\le}
	|\mvec{p}||\mvec{y}| 
\le 
	\frac{2\pi\ \sqrt{12}\ y}{La} 
\le 
	6\pi 
\approx 
	18.85\,.
\end{align}
In order to make an ansatz for the $py$-dependence, we consider the constraints on the $\zeta$-dependence of the skewed DPDs. These are the symmetry relation \eqref{eq:even-in-zeta} and the constraints \eqref{eq:support-region} restricting the support region in $\zeta$. Furthermore, we assume that the Mellin moment $I(\zeta,y^2)$ can be Taylor expanded around $\zeta=0$. Combining everything, we make the ansatz that the Mellin moment $I(\zeta,y^2)$ can be approximated by an even polynomial in $\zeta$ within the region $|\zeta| \le 1$:

\begin{align}
I(\zeta,y^2) = 
	\pi \sum_{n=0}^N a_n(y^2)\ \zeta^{2n}\ \Theta(1-\zeta^2)\,.
\label{eq:mellin-ansatz}
\end{align}
This implies for the twist-two functions, which are related to the Mellin moments by a Fourier transform:

\begin{align}
\label{eq:Apy-ansatz}
A(py,y^2) = \sum_{n=0}^N a_n(y^2)\ h_n(py)\,,
\end{align}
where the functions $h_n$ are defined as:

\begin{align}
h_n(x) := 
	\frac{1}{2} \int_{-1}^1 \dd \zeta\ e^{ix\zeta}\ \zeta^{2n} 
= 
	\sin(x)\ s_n(x) + \cos(x)\ c_n(x)\,
\label{eq:h_n-def}
\end{align}
with

\begin{align}
s_n(x) = 
	\sum_{m=0}^n \frac{(2n)!\, (-1)^m}{(2n-2m)!\, x^{1+2m}}\,, 
\qquad 
c_n(x) = 
	\sum_{m=0}^{n-1} \frac{(2n)!\, (-1)^m}{(2n-2m-1)!\, x^{2+2m}}\,.
\end{align}
%
%The functions $h_n$ fulfill the properties
It is easy to check that the functions $h_n(x)$ fulfill the following relations:

\begin{align}
h_n(0) = \frac{1}{1+2n}\,, 
\qquad 
\frac{\dd^2 h_n(x)}{\dd x^2} = -h_{n+1}(x)\,.
\label{eq:h_n-prop}
\end{align}
We recall that $A(py=0,y^2)$ is already completely described by the double exponential ansatz in \eqref{eq:y2-ansatz}. Therefore,
in the analysis of the $py$ dependence, we consider the normalized twist-two function

\begin{align}
\label{eq:Ahat}
\widehat{A}(py,y^2) := 
	\frac{A(py,y^2)}{A(0,y^2)} 
= 
	\sum_{n=0}^N \hat{a}_n(y^2)\ h_n(py)\,,
\end{align}
with the normalized coefficients

\begin{align}
\label{eq:ahat}
\hat{a}_n(y^2) = \frac{a_n(y^2)}{A(0,y^2)}\,.
\end{align}
A useful quantity to investigate in the context of the $py$-analysis is the $2m$-th moment in $\zeta$ of the DPD Mellin moment, which can be written as:

\begin{align}
\langle \zeta^{2m} \rangle(y^2) := 
	\frac{
		\int_{-1}^1 \dd \zeta\ \zeta^{2m} I(\zeta,y^2)
	}{
		\int_{-1}^1 \dd \zeta\ I(\zeta,y^2)
	} 
=
	\left. 
		(-1)^m \frac{
			\partial^{2m} \widehat{A}(py,y^2)
		}{
			\partial (py)^{2m}
		}
	\right|_{py=0} \,.
\label{eq:mellin-zeta-mom}
\end{align}
If we insert our ansatz \eqref{eq:Apy-ansatz} combined with \eqref{eq:Ahat} and replace the $2m$-th derivative of $\widehat{A}$ according to \eqref{eq:h_n-prop}, we find that $\langle \zeta^{2m} \rangle$ can be expressed as: 

\begin{align}
\langle \zeta^{2m} \rangle(y^2) = \sum_{n=0}^N T_{mn} \hat{a}_n(y^2)\,,
\label{eq:zeta-T-a}
\end{align}
where we defined the $(N+1)\times(N+1)$-matrix $T$

\begin{align}
\label{eq:Tmn}
T_{mn} = (1+2n+2m)^{-1}\,.
\end{align}
Equation \eqref{eq:zeta-T-a} can be inverted, so that we are able to express the coefficients $\hat{a}_n$ in terms of the $\zeta$-moments:

\begin{align}
\hat{a}_n(y^2) = 
	\sum_{m=0}^N \left(T^{-1}\right)_{nm} 
	\langle \zeta^{2m} \rangle(y^2)\,,
\end{align}
and hence

\begin{align}
\widehat{A}(py,y^2) = 
	\sum_{n,m=0}^N \left(T^{-1}\right)_{nm} 
	\langle \zeta^{2m} \rangle(y^2)\ h_n(py)\,.
\label{eq:A-local-fit}
\end{align}
One has $\langle\zeta^0\rangle(y^2) \equiv 1$ by definition. Thus, the first non-trivial term in \eqref{eq:A-local-fit} is the one with $m=1$. For each value of $y^2$ we can perform a fit with the functional form \eqref{eq:A-local-fit} with $N$ fit parameters. These kind of fits are referred to as "local" fits in the following. Furthermore, we parameterize the moments of $\zeta$ in terms of powers of the distance $y = \sqrt{-y^2}$, \ie we write

\begin{align}
\langle\zeta^{2m}\rangle(y^2) = \sum_{k=0}^K c_{mk} \sqrt{-y^2}^k\,,
\end{align}
such that we obtain a global parameterization describing both the $y^2$ and $py$-dependence:

\begin{align}
\widehat{A}(py,y^2) = 
	\sum_{n,m=0}^N \sum_{k=0}^K \left(T^{-1}\right)_{nm} c_{mk} \sqrt{-y^2}^k h_n(py)\,.
\label{eq:A-global-fit}
\end{align}
Since $c_{0k} = \delta_{0k}$ by definition, there are $N(K+1)$ parameters to be determined in a "global" fit to the parameterization \eqref{eq:A-global-fit}.
%\paragraph{$y^2$-fit.} 
%We first discuss the double-exponential fit on the data for $A(py=0,y^2)$. 

\paragraph{Local $py$-fits:} 

The results obtained for the $y^2$-fit are used to calculate the normalized function $\widehat{A}(py,y^2)$, which is then fitted to the functional form \eqref{eq:A-local-fit} for certain values of $y^2$. We perform two sets of fits using $N=2$ or $N=3$, \ie there are two or three free fit parameters, respectively. The free fit parameters are the moments in $\zeta$, \ie $\langle \zeta^{2m} \rangle$ with $m=1,\dots ,N$. For each accessible value of $y^2$, there is a number of available data points that can be used to fit the $py$-dependence. This number strongly varies with $y^2$. In order to avoid fluctuations caused by this circumstance we do not only consider the data points with $y=y_{\mathrm{fit}}$, but take into account all data points in a band $y_{\mathrm{fit}}-0.5a \le y \le y_{\mathrm{fit}}+0.5a$. The fit is carried out for $y_{\mathrm{fit}} = \nu a$, where $\nu \in [4,16]$ is an integer.
\begin{figure}
\subfigure[local fit on $\widehat{A}_{ud}$ at $y_{\mathrm{fit}} = 10a = 0.856~\mathrm{fm}$ \label{fig:local_py_fit-A_VV-ud-y100}]{
\includegraphics[scale=0.25,trim={0.5cm 1.2cm 0.5cm 2.8cm},clip]{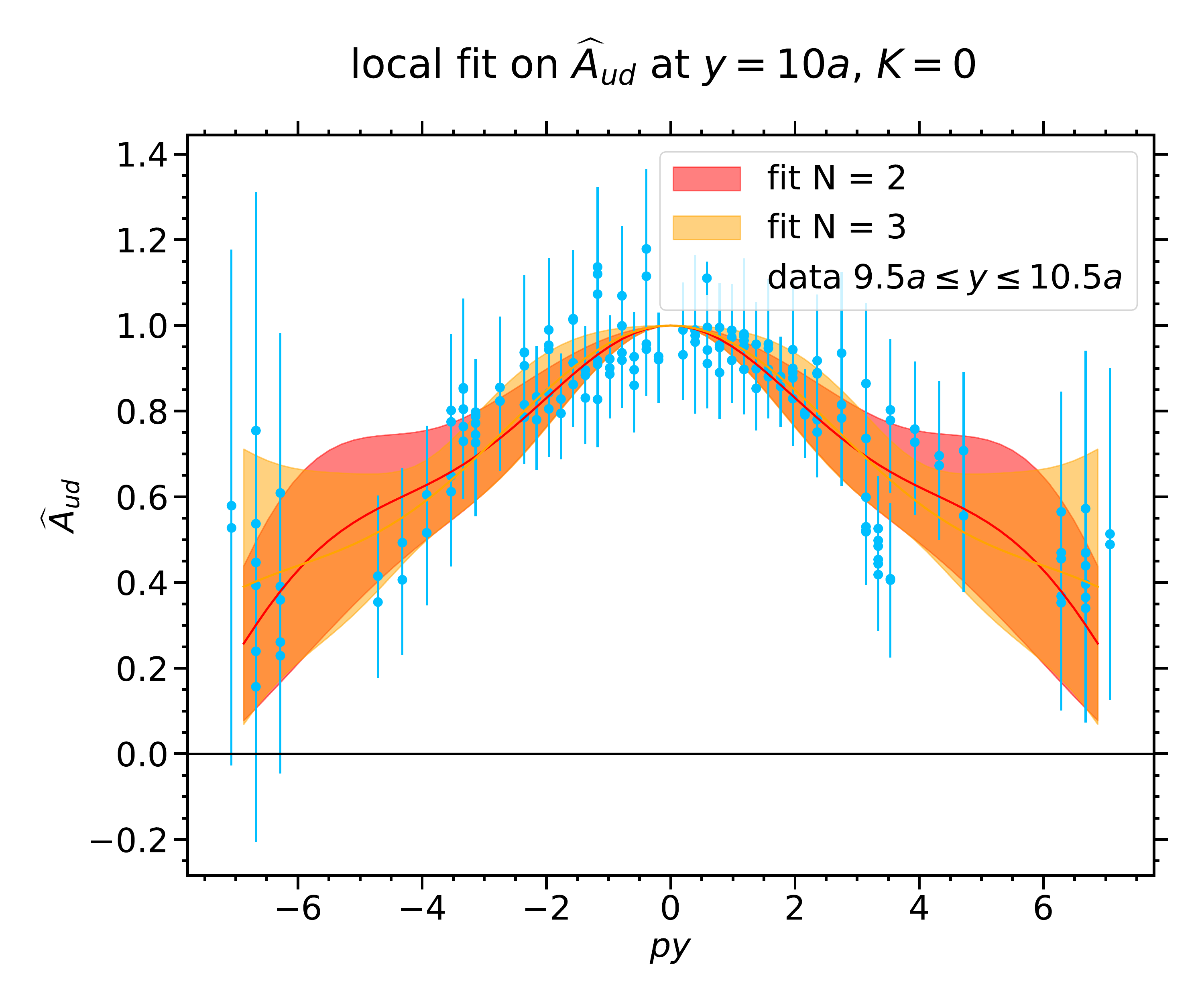}} \hfill 
\subfigure[local fit on $\widehat{A}_{ud}$ at $y_{\mathrm{fit}} = 12a = 1.027~\mathrm{fm}$ \label{fig:local_py_fit-A_VV-ud-y144}]{
\includegraphics[scale=0.25,trim={0.5cm 1.2cm 0.5cm 2.8cm},clip]{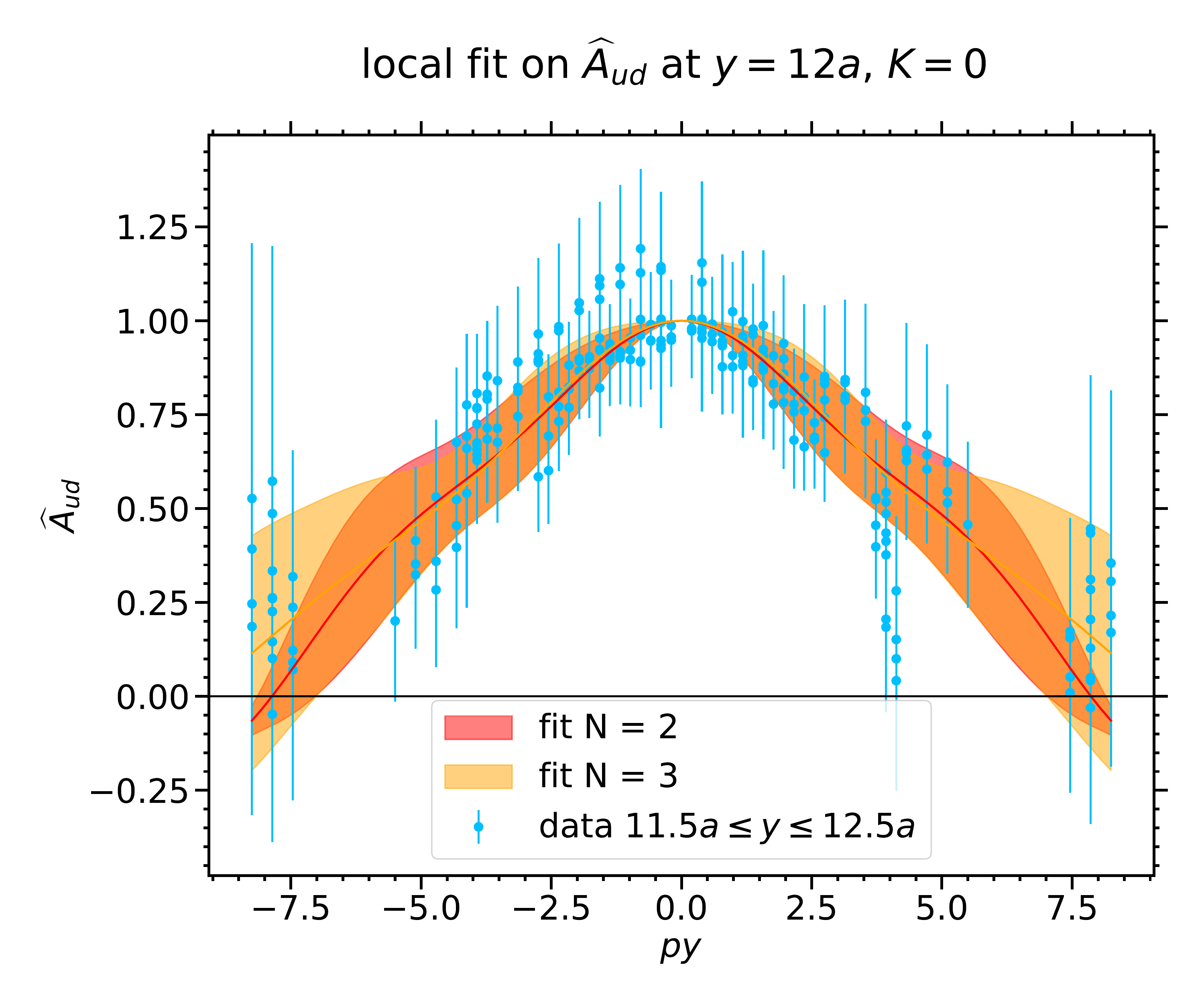}} \\ 
\subfigure[local fit on $\widehat{A}_{uu}$ at $y_{\mathrm{fit}} = 10a = 0.856~\mathrm{fm}$ \label{fig:local_py_fit-A_VV-uu}]{
\includegraphics[scale=0.25,trim={0.5cm 1.2cm 0.5cm 2.8cm},clip]{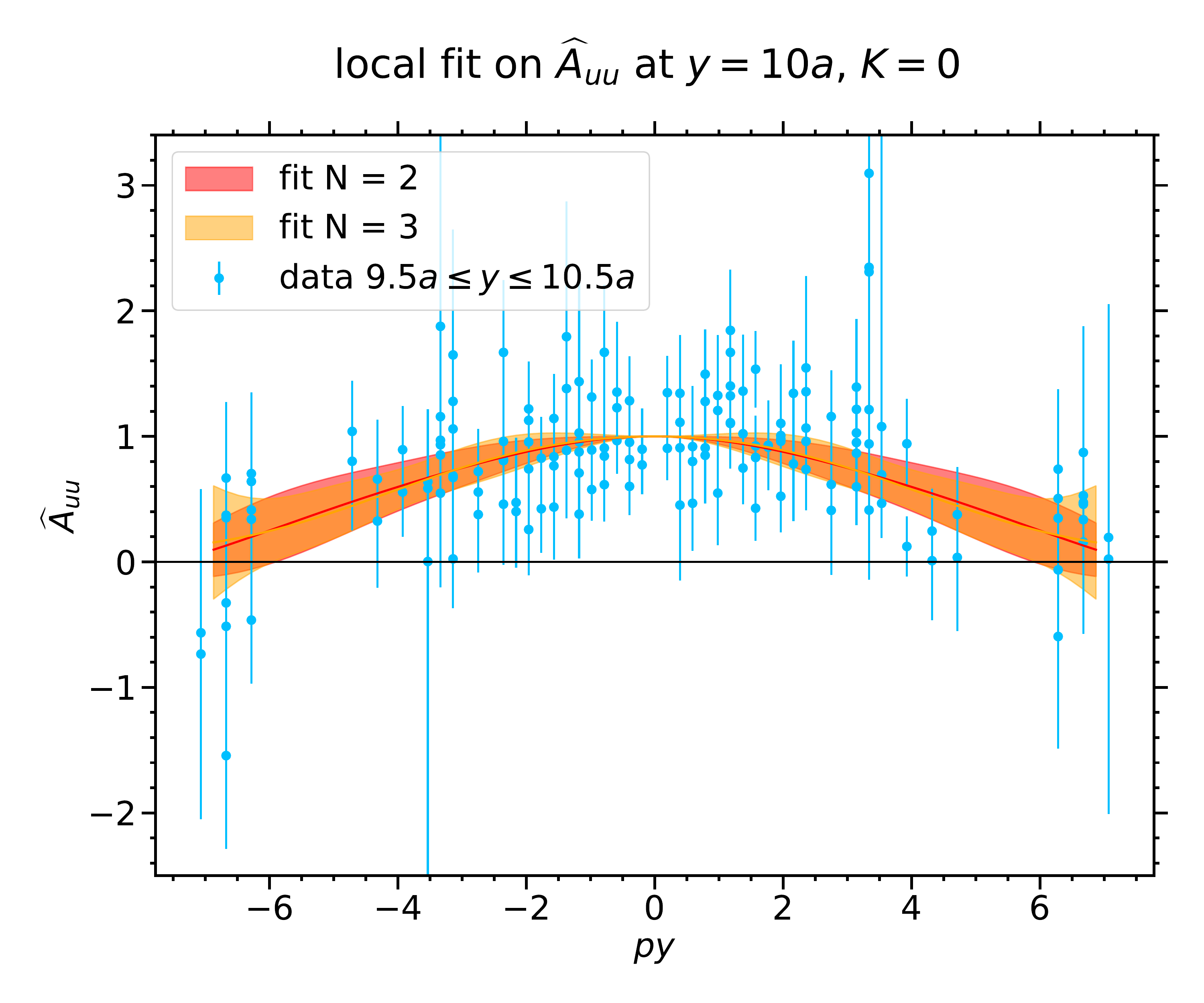}} \hfill 
\subfigure[local fit on $\widehat{A}_{\delta d u}$ at $y_{\mathrm{fit}} = 10a = 0.856~\mathrm{fm}$ \label{fig:local_py_fit-A_VT-ud}]{
\includegraphics[scale=0.25,trim={0.5cm 1.2cm 0.5cm 2.8cm},clip]{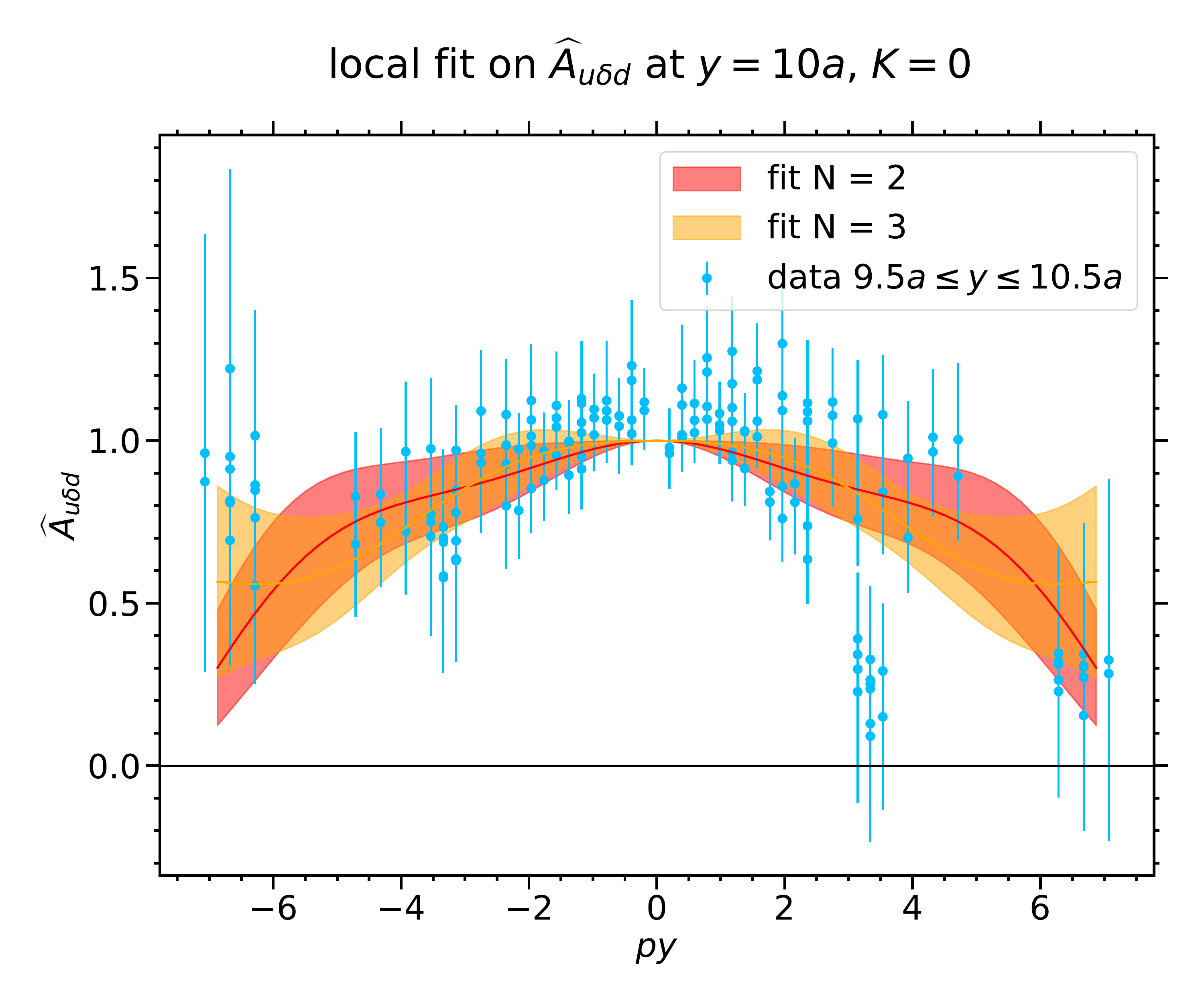}} \\ 
\caption{$py$ dependence of the twist-two function data and the corresponding local fits with $N=2,3$. This is shown for the functions $\widehat{A}_{ud}$ at $y_{\mathrm{fit}} = 10a$~(a) and $y_{\mathrm{fit}}=12a$~(b), as well as for $\widehat{A}_{uu}$~(c) and $\widehat{A}_{\delta d u}$~(d) both at $y_{\mathrm{fit}}=10a$. We plot all data points included by the fits for a given $y_{\mathrm{fit}}$, \ie all data points in the range $y_{\mathrm{fit}}\pm 0.5a$ (see the text). \label{fig:local_py_fit}}
\end{figure}

In \fig{\ref{fig:local_py_fit}} we show for selected channels the data points of $\widehat{A}(py,y^2)$ entering the fit for a given $y^2$ in comparison to the resulting fit bands for $N=2$ and $N=3$. We observe that the $\widehat{A}$ data are reasonably described and the two fits are consistent within the statistical error. For $N=3$ the fit tends to be sensitive to the data points at large $py$, which causes visible deviations relative to the fit with $N=2$.

There are channels where the data of $\widehat{A}$ are compatible with zero, which leads to a dominance of fluctuations. In these cases a reliable fit of the $py$-dependence is not feasible. We refer to these channels as the "bad" channels. Explicitly, they are given by the functions $\widehat{A}_{\Delta q \Delta q^\prime}$ and $\widehat{B}_{\delta u \delta u}$, as well as all polarized channels for the flavor combination $dd$. These channels will not be considered in the subsequent physics discussions.
\begin{figure}
\subfigure[$\langle \zeta^2 \rangle$ for $I_{ud}$, $N=2$ \label{fig:zeta_mom_VVud-N2}]{
\includegraphics[scale=0.25,trim={0.5cm 1.2cm 0.5cm 2.8cm},clip]{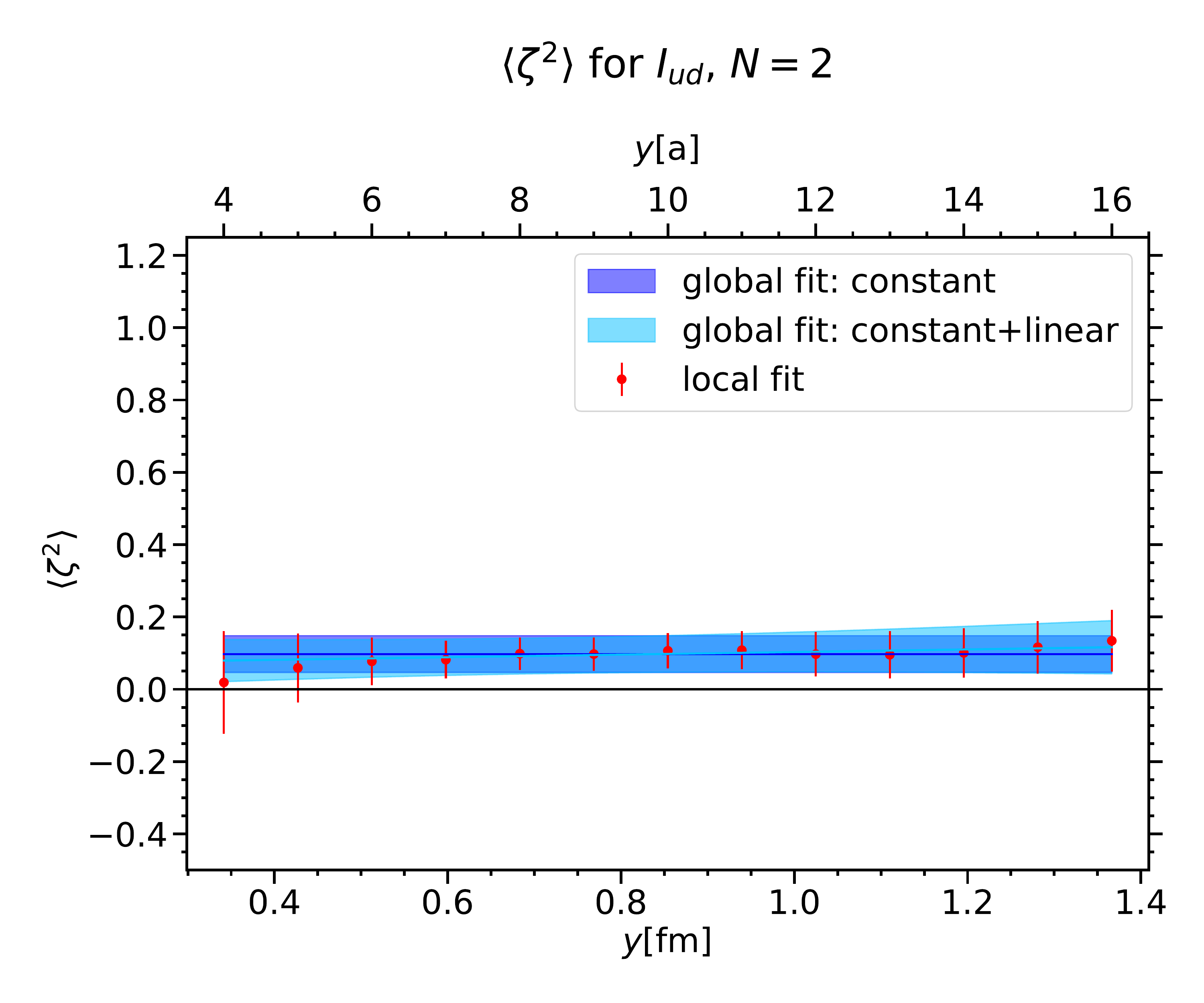}}\hfill 
\subfigure[$\langle \zeta^4 \rangle$ for $I_{ud}$, $N=2$ \label{fig:zeta_mom_VVud-N2-j2}]{
\includegraphics[scale=0.25,trim={0.5cm 1.2cm 0.5cm 2.8cm},clip]{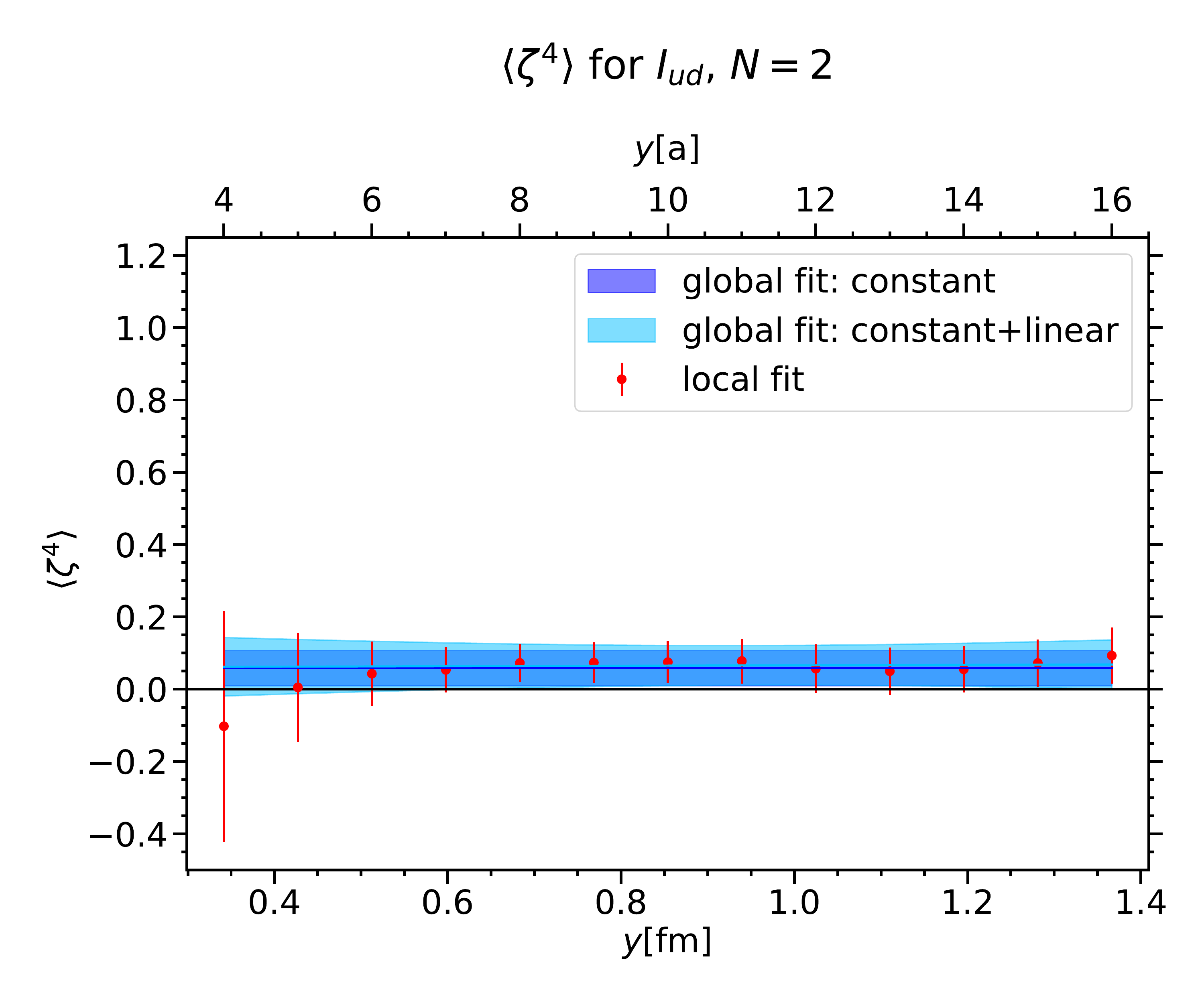}} \\
\subfigure[$\langle \zeta^2 \rangle$ for $I_{ud}$, $N=3$ \label{fig:zeta_mom_VVud-N3}]{
\includegraphics[scale=0.25,trim={0.5cm 1.2cm 0.5cm 2.8cm},clip]{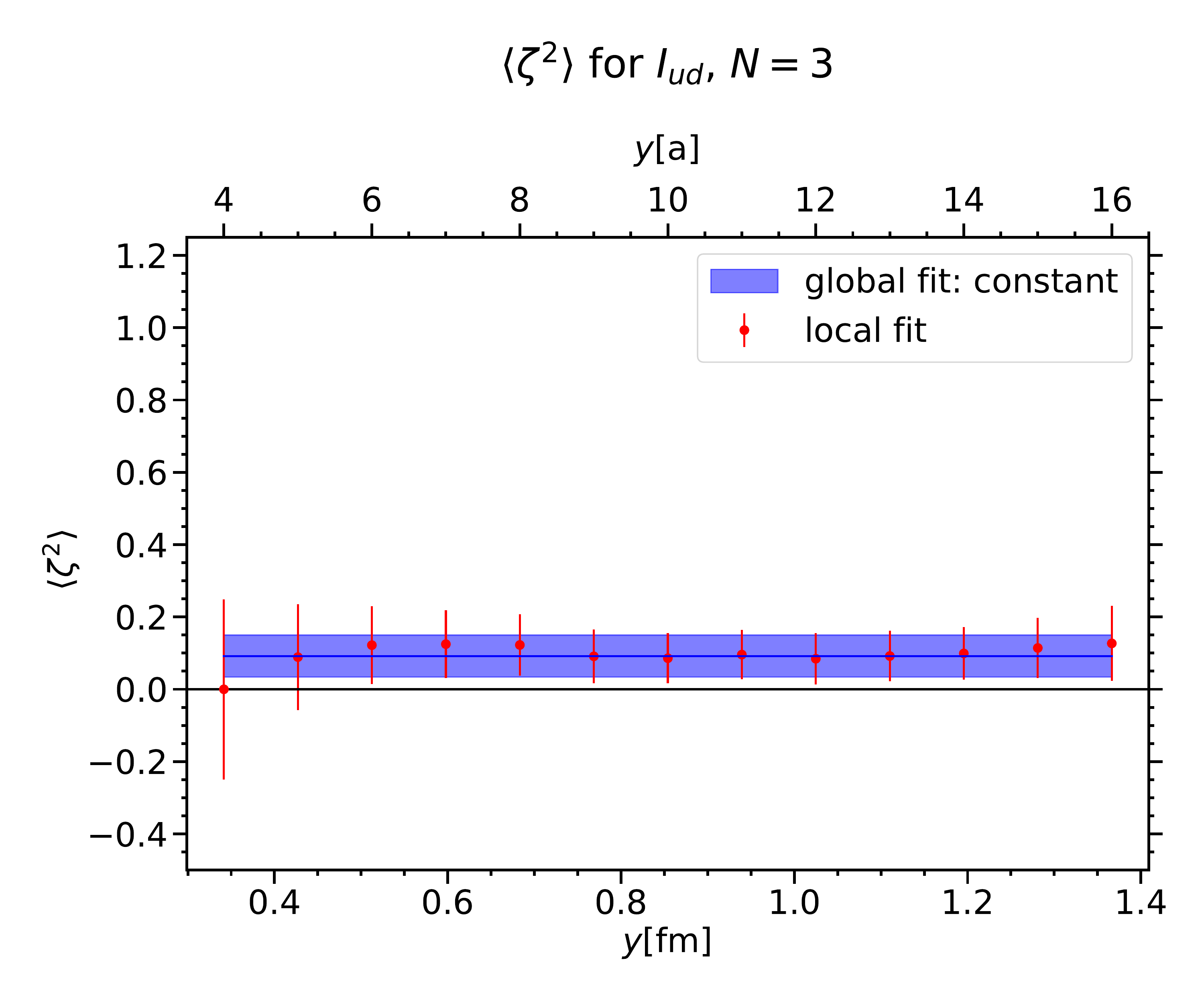}} \hfill 
\subfigure[$\langle \zeta^4 \rangle$ for $I_{ud}$, $N=3$ \label{fig:zeta_mom_VVud-N3-j2}]{
\includegraphics[scale=0.25,trim={0.5cm 1.2cm 0.5cm 2.8cm},clip]{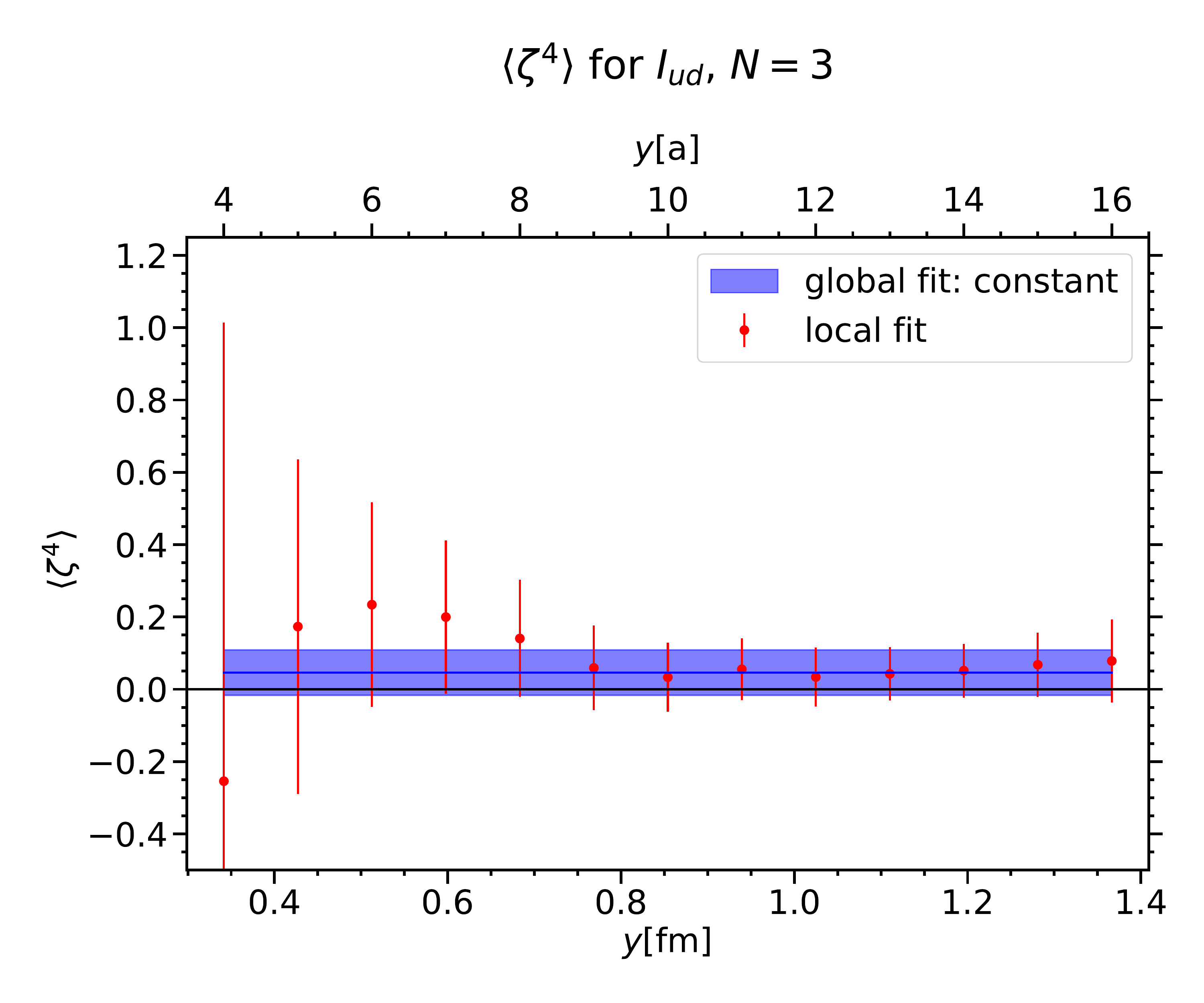}} \\ 
\caption{Results for $\langle \zeta^{2m} \rangle$ obtained from the local fit (red points) compared to the global fits (bands). For $N=2$ (top) we performed fits for $K=0$ (dark blue), as well as $K=1$ (light blue), whereas for $N=3$ (bottom) we fixed $K=0$. The results are shown for the second (a,c) and the fourth moment in $\zeta$ (b,d) of the Mellin moment $I_{ud}$. \label{fig:zeta_mom_VVud}}
\end{figure}
\begin{figure}
\subfigure[$\langle \zeta^2 \rangle$ for $I_{uu}$, $N=2$ \label{fig:zeta_mom_VVuuVTud-I_VV}]{
\includegraphics[scale=0.25,trim={0.5cm 1.2cm 0.5cm 2.8cm},clip]{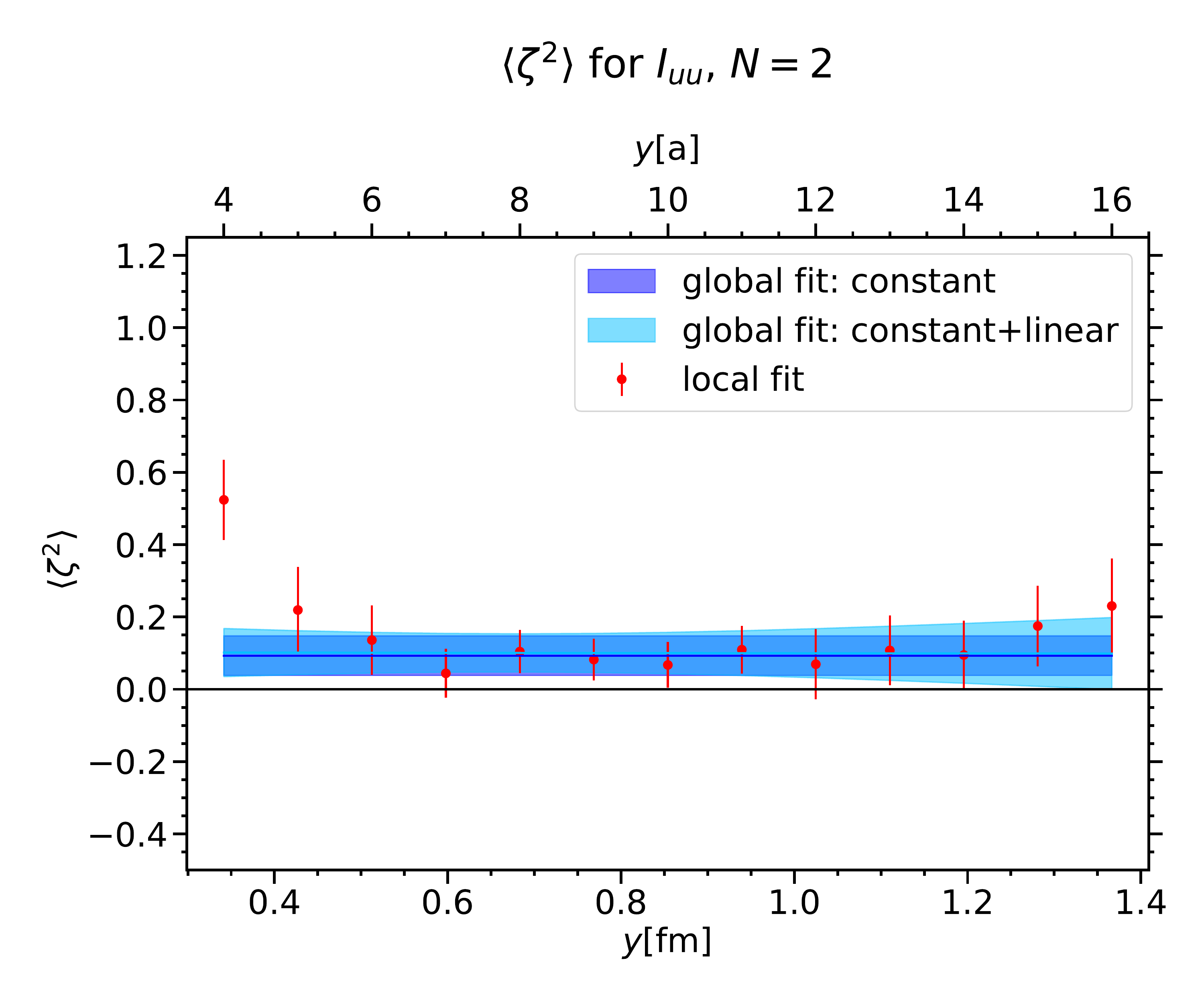}}\hfill 
\subfigure[$\langle \zeta^4 \rangle$ for $I_{uu}$, $N=2$ \label{fig:zeta_mom_VVuuVTud-I_VV-j2}]{
\includegraphics[scale=0.25,trim={0.5cm 1.2cm 0.5cm 2.8cm},clip]{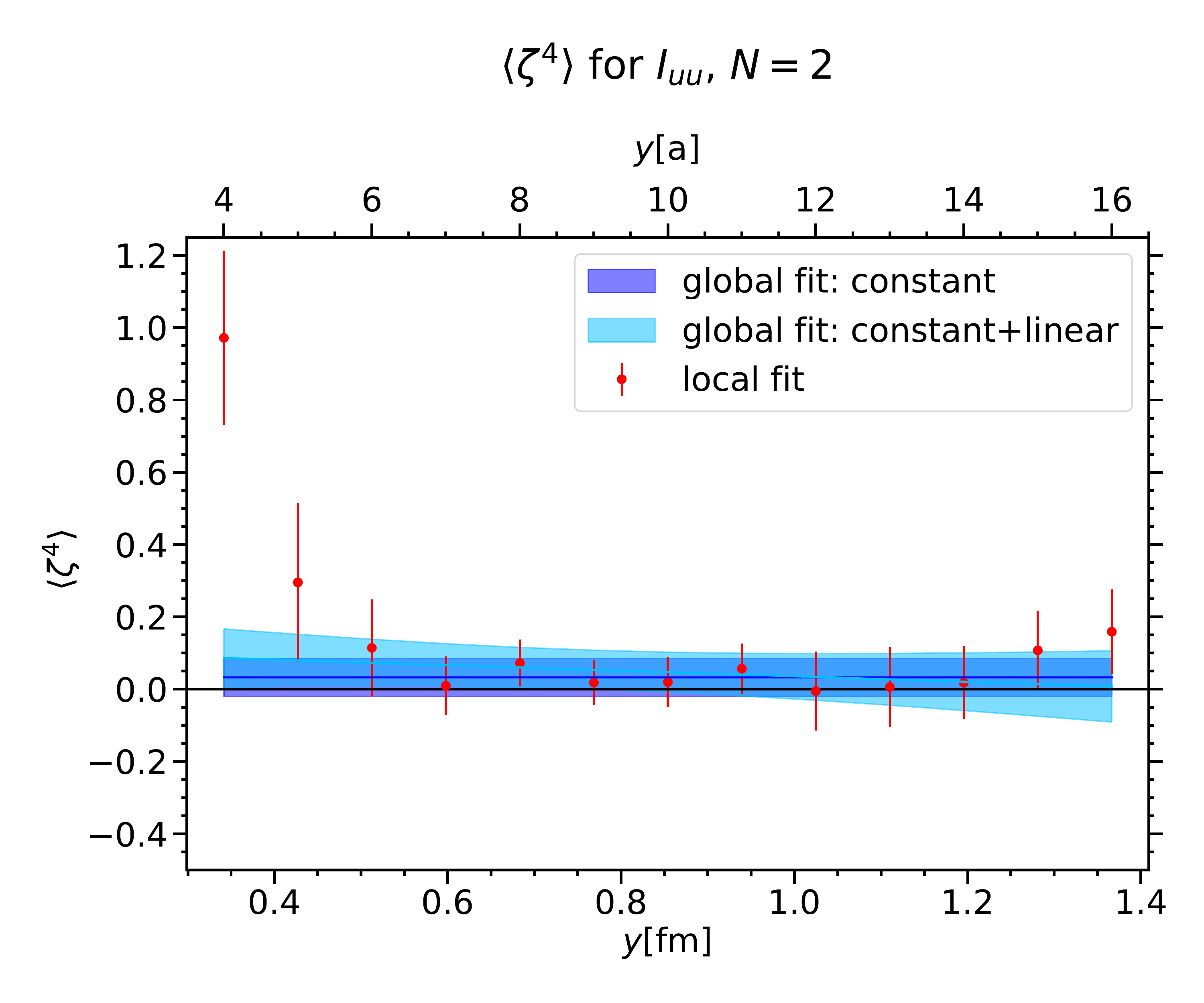}} \\
\subfigure[$\langle \zeta^2 \rangle$ for $I_{\delta d u}$, $N=2$ \label{fig:zeta_mom_VVuuVTud-I_VT}]{
\includegraphics[scale=0.25,trim={0.5cm 1.2cm 0.5cm 2.8cm},clip]{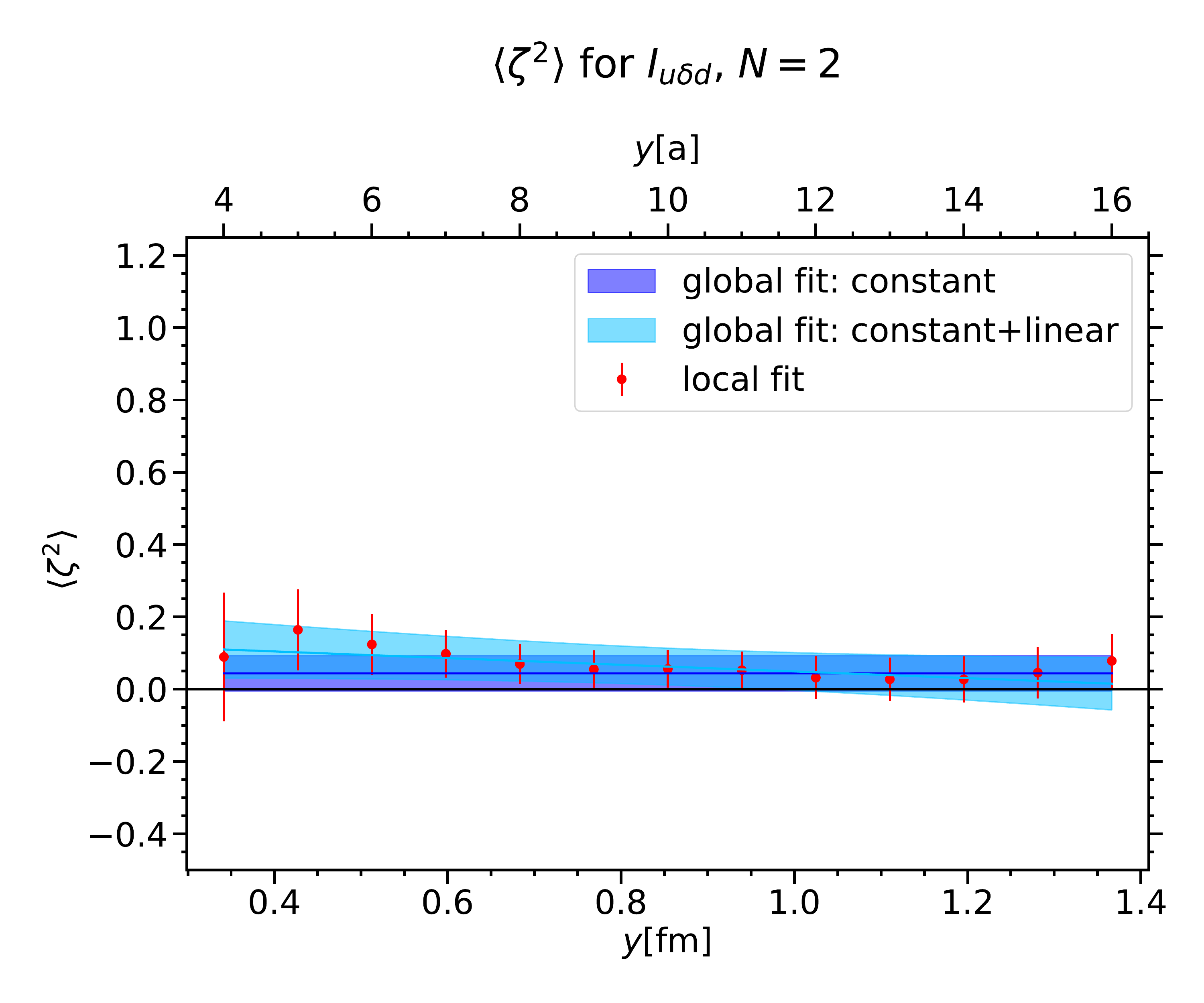}}\hfill 
\subfigure[$\langle \zeta^4 \rangle$ for $I_{\delta d u}$, $N=2$ \label{fig:zeta_mom_VVuuVTud-I_VT-j2}]{
\includegraphics[scale=0.25,trim={0.5cm 1.2cm 0.5cm 2.8cm},clip]{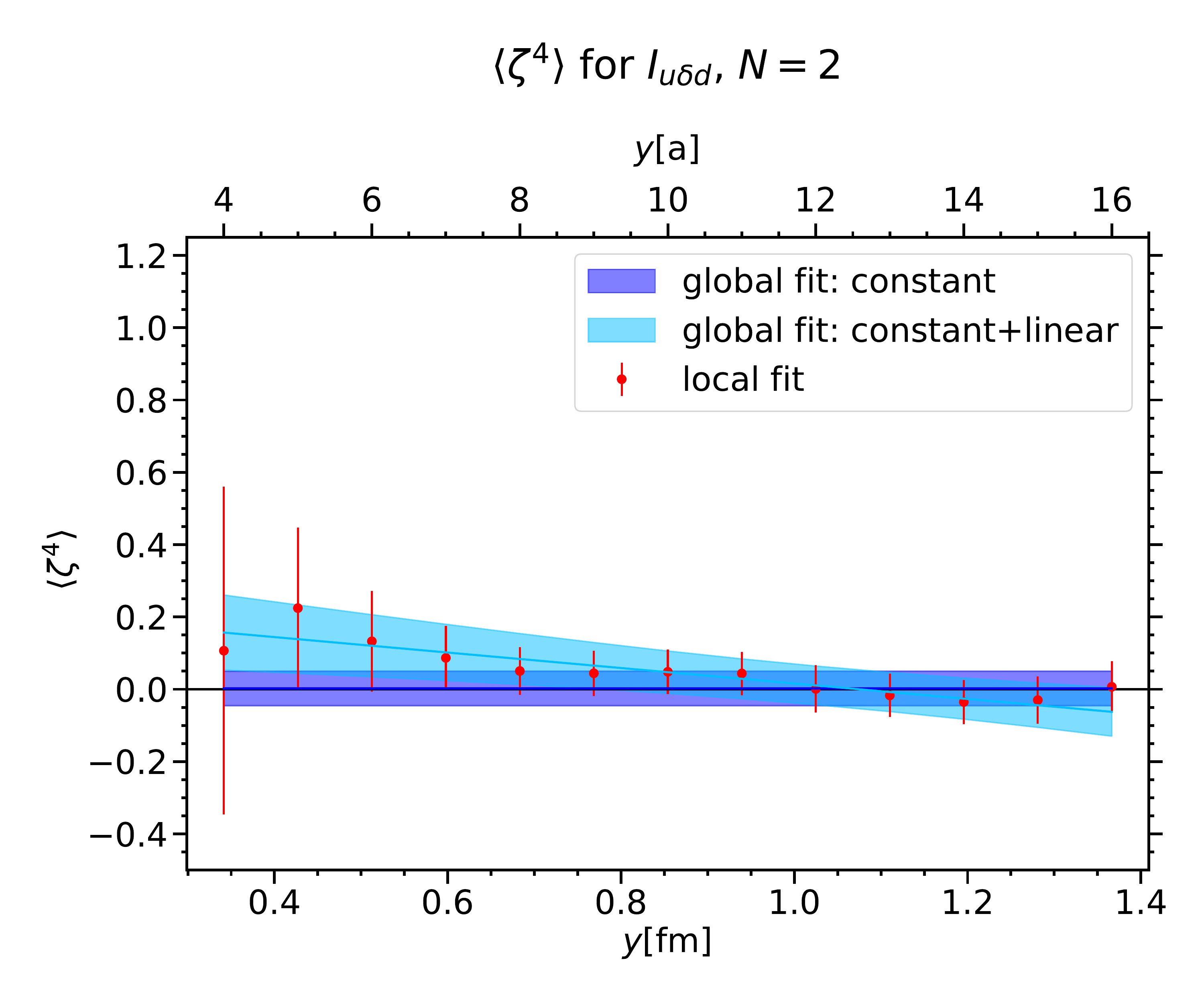}} \\
\subfigure[$\langle \zeta^2 \rangle$ for $I^t_{\delta u \delta d}$, $N=2$ \label{fig:zeta_mom_BTTud-N2}]{
\includegraphics[scale=0.25,trim={0.5cm 1.2cm 0.5cm 2.8cm},clip]{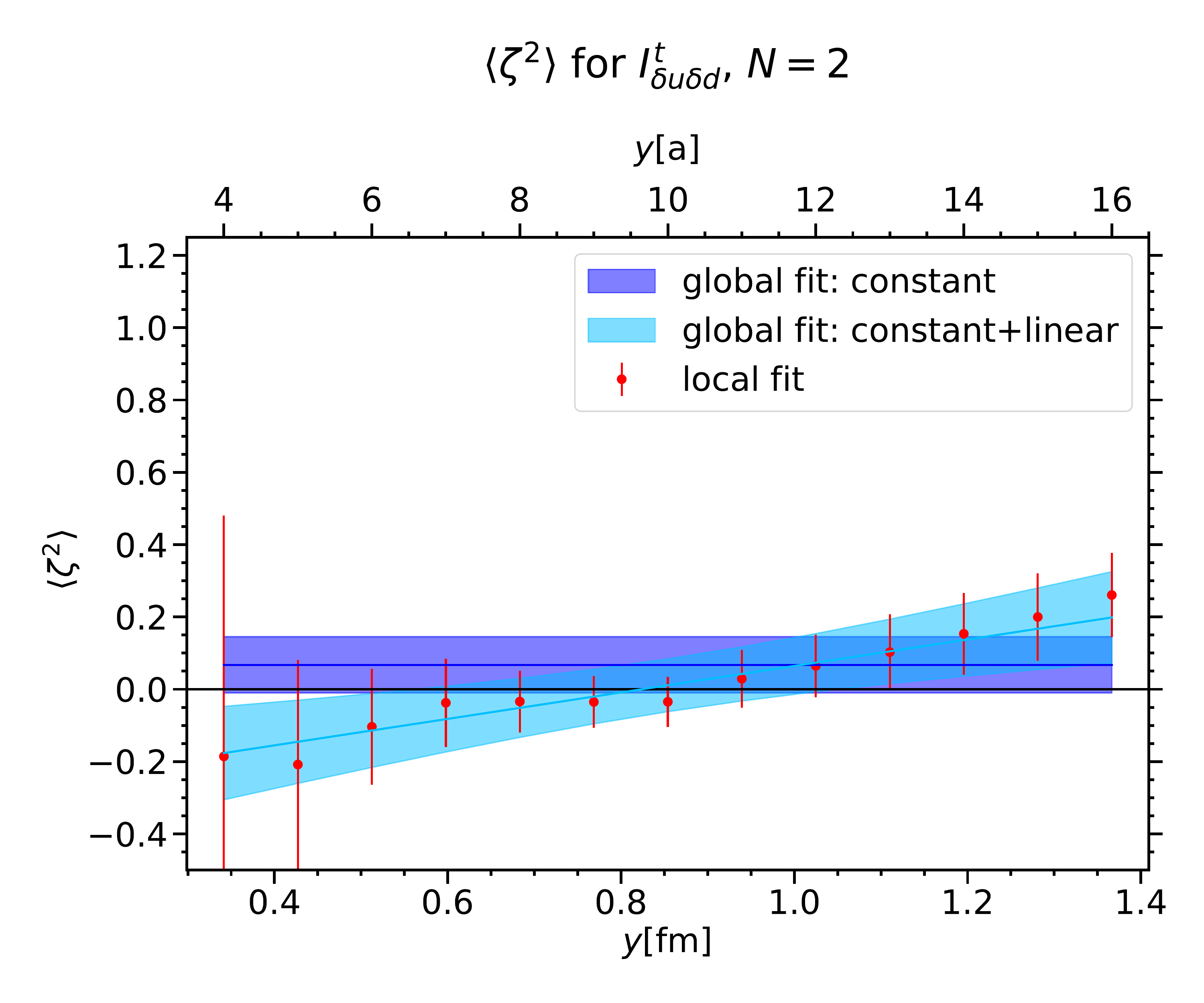}}\hfill 
\subfigure[$\langle \zeta^4 \rangle$ for $I^t_{\delta u \delta d}$, $N=2$ \label{fig:zeta_mom_BTTud-N2-j2}]{
\includegraphics[scale=0.25,trim={0.5cm 1.2cm 0.5cm 2.8cm},clip]{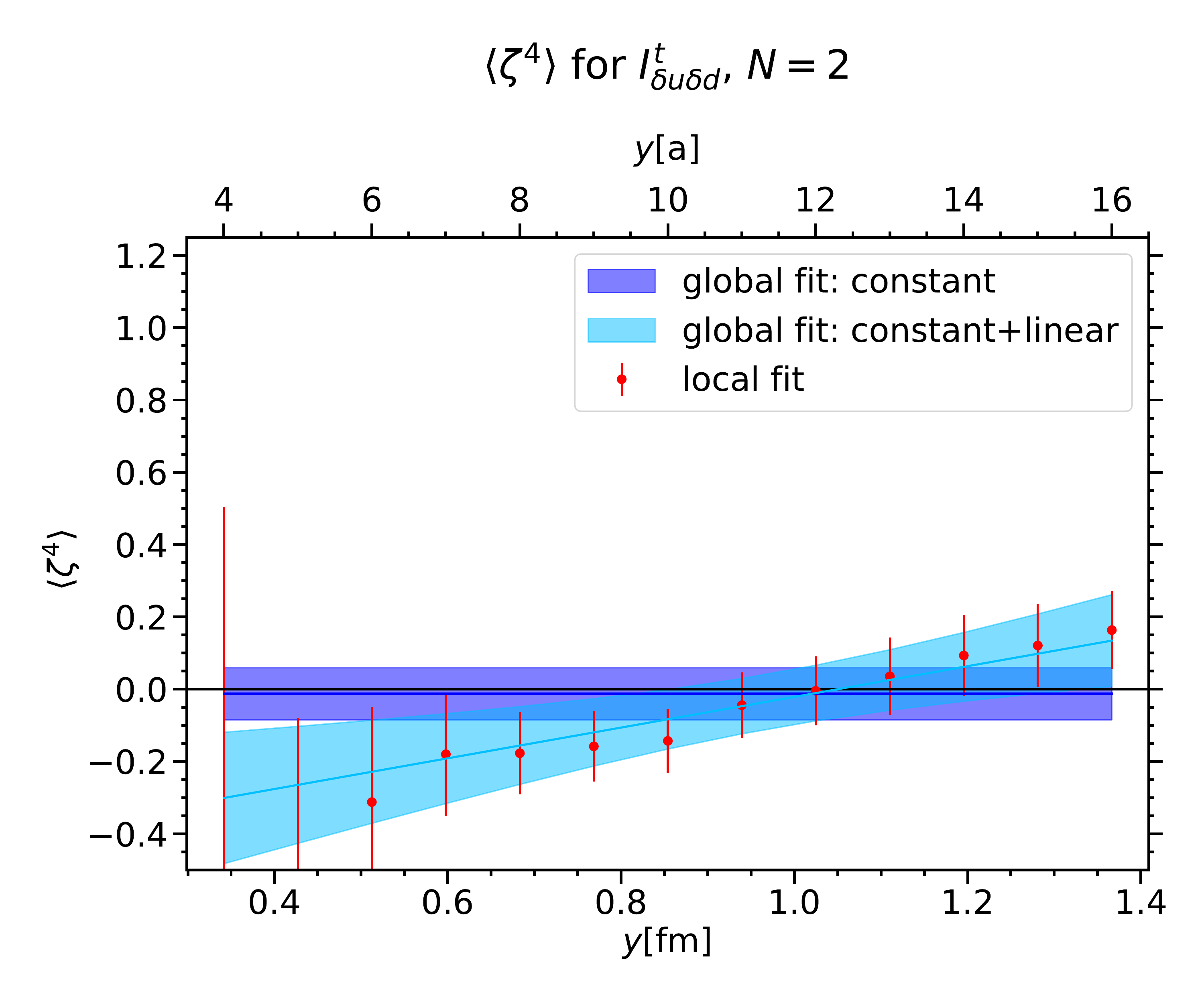}} \\
\caption{The same as \fig\ref{fig:zeta_mom_VVud} for the second (left) and fourth (right) moment of $\zeta$ in $I_{uu}$, $I_{\delta d u}$, and $I^t_{\delta u \delta d}$, where only results for $N=2$ are shown. \label{fig:zeta_mom_VVuuVTudBTTud}}
\end{figure}

The resulting values of $\langle \zeta^{2m} \rangle$ are plotted in \figs{\ref{fig:zeta_mom_VVud}} and \ref{fig:zeta_mom_VVuuVTudBTTud} (red data points). It appears that the moments are rather small ($\langle \zeta^{2m} \rangle < 0.25$) and in almost all cases these show a linear dependence on the distance $y$. In most cases they are nearly constant. Deviations from that behavior are seen for $uu$ at small $y$, where the data tend to increase. However, this is the region where the violation of Lorentz invariance starts to show up in the corresponding channels, as we have discussed earlier. This might skew the $py$-dependence. The results that are not shown in the plots look very similar. An exception to this are the data for $\widehat{A}_{\delta u \delta u}$, which carry large statistical errors.

The results for the $\zeta$-moments are quite different from those we obtained for the pion \cite{Bali:2020mij}, where we found a clear linear rise with increasing $y$. In that case, for $y>1~\mathrm{fm}$, values of $\langle \zeta^2 \rangle > 0.5$ were observed. 

\begin{figure}
\subfigure[global fit on $\widehat{A}_{ud}$ at $y = 10a \approx 0.86~\mathrm{fm}$, $K=0$ \label{fig:global_py_fit_M0-A_VV-ud-y100}]{
\includegraphics[scale=0.25,trim={0.5cm 1.2cm 0.5cm 2.8cm},clip]{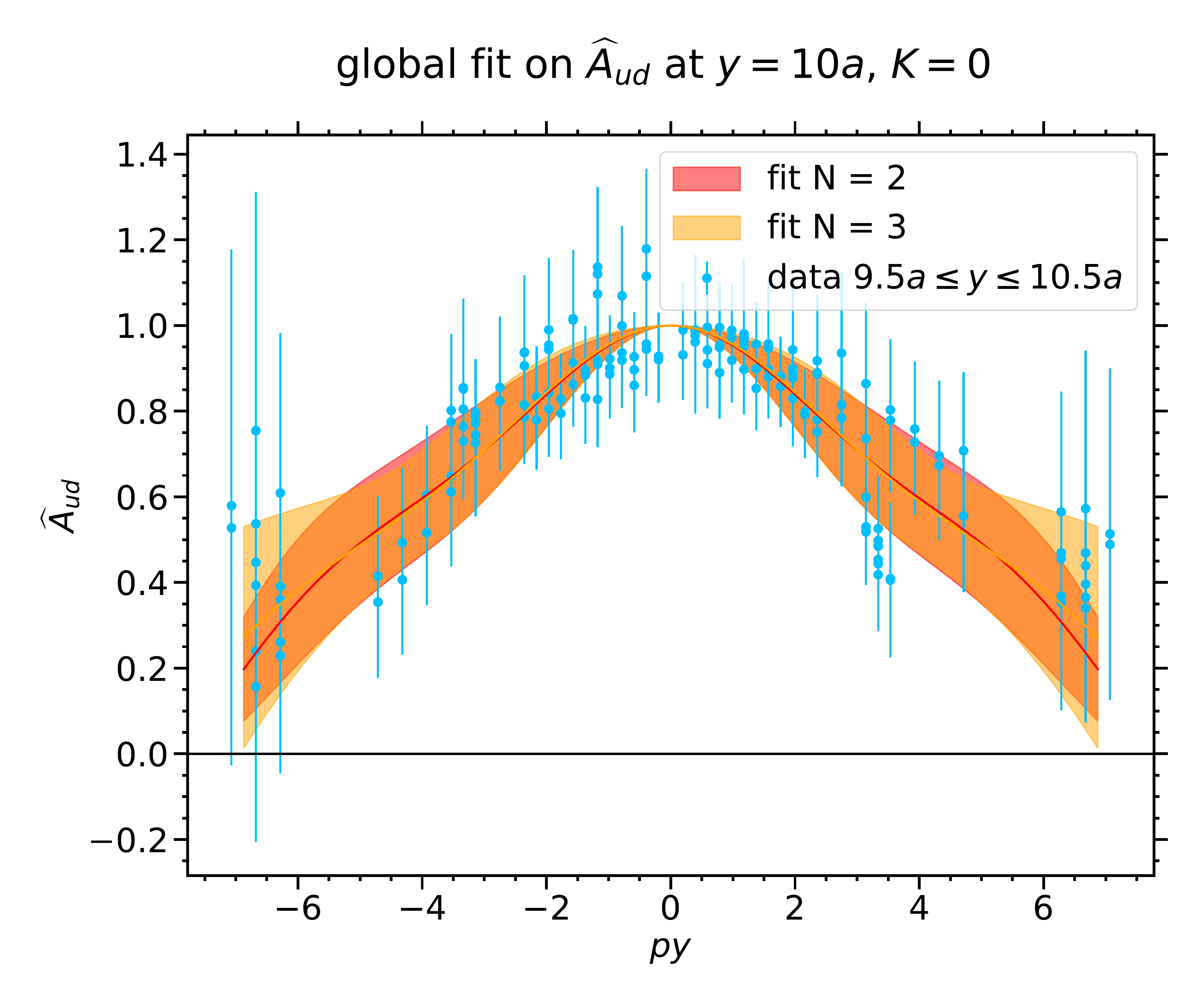}}\hfill 
\subfigure[global fit on $\widehat{A}_{ud}$ at $y = 12a \approx 1.03~\mathrm{fm}$, $K=0$ \label{fig:global_py_fit_M0-A_VV-ud-y144}]{
\includegraphics[scale=0.25,trim={0.5cm 1.2cm 0.5cm 2.8cm},clip]{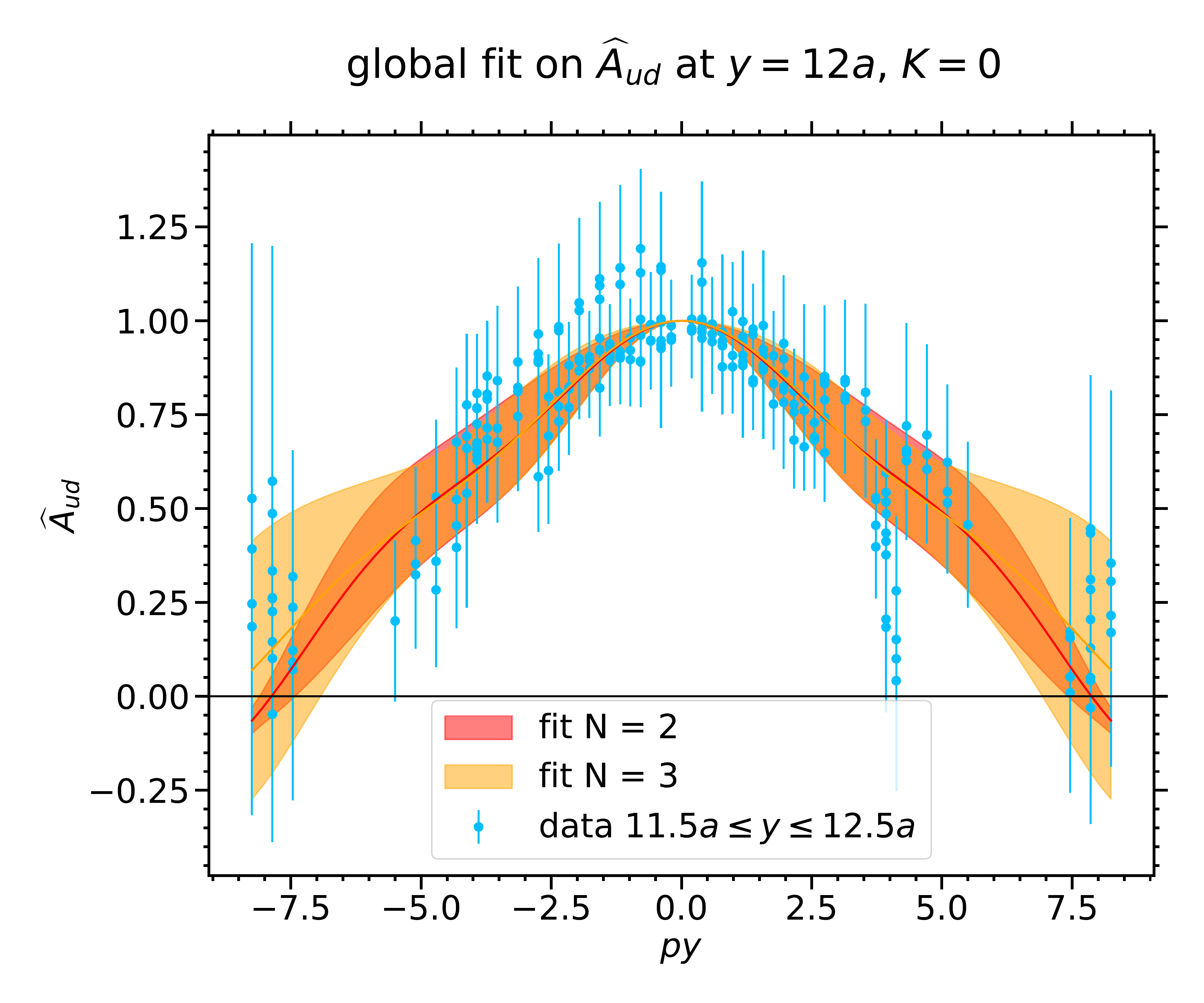}} \\
\subfigure[global fit on $\widehat{A}_{uu}$ at $y = 10a \approx 0.86~\mathrm{fm}$, $K=0$ \label{fig:global_py_fit_M0-A_VV-uu}]{
\includegraphics[scale=0.25,trim={0.5cm 1.2cm 0.5cm 2.8cm},clip]{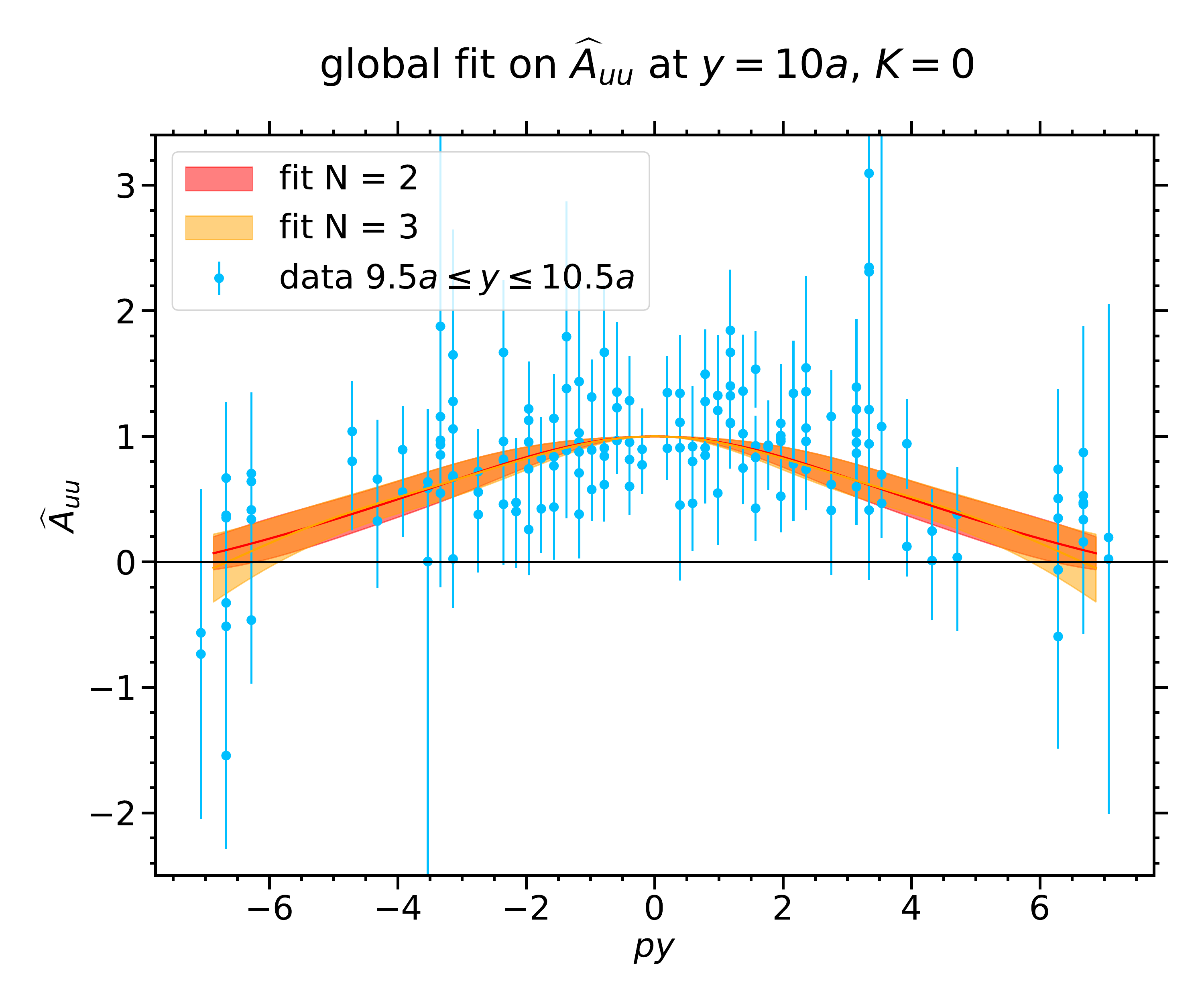}}\hfill 
\subfigure[global fit on $\widehat{A}_{\delta d u}$ at $y = 10a \approx 0.86~\mathrm{fm}$, $K=0$ \label{fig:global_py_fit_M0-A_VT-ud}]{
\includegraphics[scale=0.25,trim={0.5cm 1.2cm 0.5cm 2.8cm},clip]{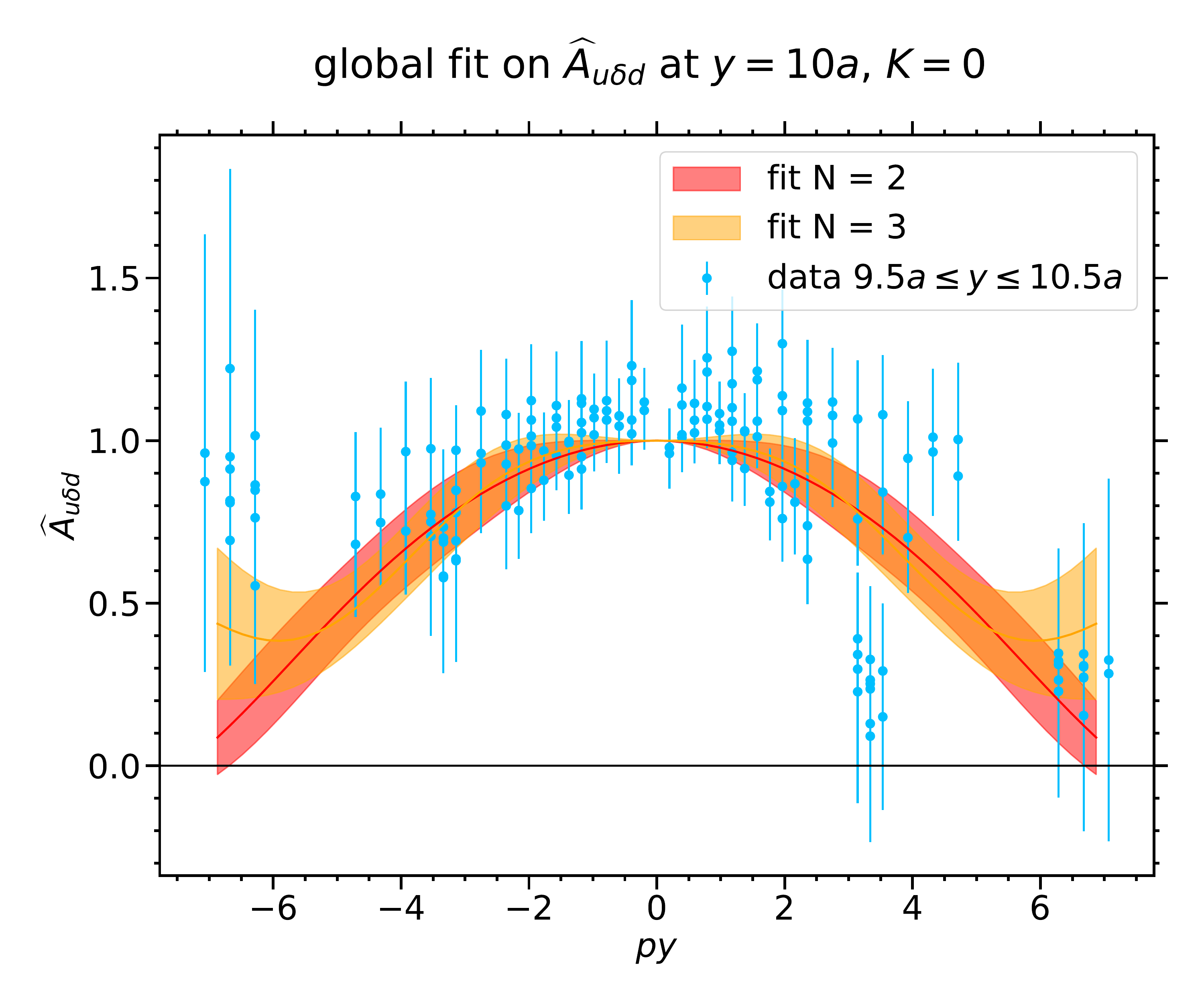}} \\
\caption{The same as \fig\ref{fig:local_py_fit} for slices of the global $py$-fit with $K=0$. \label{fig:global_py_fit_M0}}
\end{figure}

\paragraph{Global $py$-fits:} 
In order to reduce the number of parameters entering our analysis, we perform a global fit on the $\widehat{A}$ data using the functional form given in \eqref{eq:A-global-fit}. This is again carried out for $N=2,3$. We have seen in the previous discussion that a linear dependence on $y$ is sufficient to describe the $\langle \zeta^{2m} \rangle$ behavior. Therefore, we take $K=0,1$ for the global fits. For $N=3$ we restrict ourselves to $K=0$, \ie a constant, since for $K=1$ we find that the data are overfitted. In total we have three fits, where we use $(N,K) = (2,0), (2,1), (3,0)$ with $2$, $4$ or $3$ free fit parameters, respectively. In each fit we take into account all data points for which $4a < y <16a$. The resulting curves for $K=0$ are plotted in \fig{\ref{fig:global_py_fit_M0}}, where again we show the $py$-dependence for fixed values of $y^2$. As for the local fits, the two possibilities $N=2$ and $N=3$ yield comparable results; small deviations are found for large $py$.

In general, the value of $\chi^2/\mathrm{dof}$ differs only weakly between different fits of the same channel. In most channels, the differences are marginal ($\lesssim 0.01$). Hence, we consider the fit with $(N,M) = (2,0)$ as reliable; the other two fits might already overfit the data. Exceptions are given by $B_{\delta u \delta d}$ (see the discussion below), and $A_{\delta d u}$, $A_{\delta u d}$, where discrepancies up to $0.11$ in $\chi^2/\mathrm{dof}$ are found. This can also be observed in the slightly different behavior of the fit bands for large $py$, see \fig\ref{fig:global_py_fit_M0-A_VT-ud}. In the last two cases, fits with $N=3$ yield the smallest value for $\chi^2/\mathrm{dof}$.

The $\langle \zeta^{2m}\rangle$ curves resulting from the global fits are also shown in \figs{\ref{fig:zeta_mom_VVud}} and \ref{fig:zeta_mom_VVuuVTudBTTud} (blue and light blue bands). The results for the fit parameters $c_{mk}$ are listed in \tab{\ref{tab:cjl_A_VV}} to \ref{tab:cjl_B_TT}, where for completeness also the results of the "bad" channels (see the discussion above) are shown. In most cases, the linear fit barely differs from the fit to a constant. For a few exceptions, there is a better overlap with the data if the linear term is included. The most extreme example is given by $I^t_{\delta u \delta d}$, which is shown in \fig{\ref{fig:zeta_mom_BTTud-N2}} and \ref{fig:zeta_mom_BTTud-N2-j2}. The corresponding $\chi^2$, see \tab{\ref{tab:cjl_B_TT}}, is slightly smaller. However, the linear fit must be considered with some caution, since there is a wide region in $y$ where the moments $\langle \zeta^{2m}\rangle$ become negative. For even moments this is mathematically inconsistent. The constant fit still covers the data points sufficiently well.
\clearpage
\begin{table}
\begin{center}
\begin{tabular}{c|cc|ccccc|c}
\hline
\hline
 & $N$ & $K$ & $c_{10}$ & $c_{11}[\mathrm{fm}^{-1}]$ & $c_{20}$ & $c_{21}[\mathrm{fm}^{-1}]$ & $c_{30}$ & $\chi^2 / \mathrm{dof}$\\ 
\hline
$A_{ u  u}$ & 2 & 0 & $0.093(55)$ & - & $0.032(53)$ & - & - & $0.96$ \\
 & & 1 & $0.102(98)$ & $-0.00(12)$ & $0.11(12)$ & $-0.08(14)$ & - & $0.95$ \\
 & 3 & 0 & $0.104(66)$ & - & $0.056(76)$ & - & $0.059(88)$ & $0.96$ \\
\hline
$A_{ u  d}$ & 2 & 0 & $0.097(51)$ & - & $0.058(49)$ & - & - & $0.47$ \\
 & & 1 & $0.067(77)$ & $0.036(84)$ & $0.06(11)$ & $0.006(97)$ & - & $0.46$ \\
 & 3 & 0 & $0.092(58)$ & - & $0.046(63)$ & - & $0.038(69)$ & $0.46$ \\
\hline
$A_{ d  d}$ & 2 & 0 & $-0.029(99)$ & - & $-0.13(12)$ & - & - & $0.93$ \\
 & & 1 & $-0.03(27)$ & $0.02(34)$ & $-0.03(34)$ & $-0.10(42)$ & - & $0.93$ \\
 & 3 & 0 & $0.05(10)$ & - & $0.03(13)$ & - & $0.10(17)$ & $0.92$ \\
\hline
\hline
\end{tabular}

\end{center}
\caption{Fit results for the parameters $c_{mk}$ of our global fit ansatz \eqref{eq:A-global-fit} obtained for the unpolarized channels $A_{uu}$, $A_{ud}$ and $A_{dd}$. We take into account $(N,K) = (2,0),(2,1),(3,0)$. \label{tab:cjl_A_VV}}
\end{table}
\begin{table}
\begin{center}
\begin{tabular}{c|cc|ccccc|c}
\hline
\hline
 & $N$ & $K$ & $c_{10}$ & $c_{11}[\mathrm{fm}^{-1}]$ & $c_{20}$ & $c_{21}[\mathrm{fm}^{-1}]$ & $c_{30}$ & $\chi^2 / \mathrm{dof}$\\ 
\hline
$A_{\Delta u \Delta u}$ & 2 & 0 & $-0.36(87)$ & - & $-0.8(1.3)$ & - & - & $0.24$ \\
 & & 1 & $-0.3(3.2)$ & $0.4(6.8)$ & $1.6(4.6)$ & $-3.6(9.2)$ & - & $0.24$ \\
 & 3 & 0 & $0.21(90)$ & - & $0.9(1.7)$ & - & $2.4(3.3)$ & $0.24$ \\
\hline
$A_{\Delta u \Delta d}$ & 2 & 0 & $0.15(42)$ & - & $0.07(62)$ & - & - & $0.17$ \\
 & & 1 & $0.26(96)$ & $-0.2(1.5)$ & $0.2(1.2)$ & $-0.2(1.5)$ & - & $0.17$ \\
 & 3 & 0 & $0.09(49)$ & - & $-0.10(84)$ & - & $-0.2(1.2)$ & $0.17$ \\
\hline
$A_{\Delta d \Delta d}$ & 2 & 0 & $0.4(1.3)$ & - & $0.2(1.5)$ & - & - & $0.12$ \\
 & & 1 & $1.3(2.6)$ & $-1.4(5.1)$ & $1.7(4.0)$ & $-2.0(6.4)$ & - & $0.12$ \\
 & 3 & 0 & $0.6(1.2)$ & - & $0.7(2.0)$ & - & $0.8(3.0)$ & $0.12$ \\
\hline
\hline
\end{tabular}

\end{center}
\caption{The same as \tab\ref{tab:cjl_A_VV}, but for the twist-two function $A_{\Delta q \Delta q^\prime}$. \label{tab:cjl_A_AA}}
\end{table}
\begin{table}
\begin{center}
\begin{tabular}{c|cc|ccccc|c}
\hline
\hline
 & $N$ & $K$ & $c_{10}$ & $c_{11}[\mathrm{fm}^{-1}]$ & $c_{20}$ & $c_{21}[\mathrm{fm}^{-1}]$ & $c_{30}$ & $\chi^2 / \mathrm{dof}$\\ 
\hline
$A_{\delta u  u}$ & 2 & 0 & $0.126(82)$ & - & $0.080(84)$ & - & - & $0.98$ \\
 & & 1 & $-0.14(24)$ & $0.29(24)$ & $-0.25(31)$ & $0.36(31)$ & - & $0.97$ \\
 & 3 & 0 & $0.137(85)$ & - & $0.102(97)$ & - & $0.11(11)$ & $0.98$ \\
\hline
$A_{\delta d  u}$ & 2 & 0 & $0.044(49)$ & - & $0.002(48)$ & - & - & $1.02$ \\
 & & 1 & $0.14(11)$ & $-0.09(11)$ & $0.23(14)$ & $-0.21(13)$ & - & $0.95$ \\
 & 3 & 0 & $0.017(54)$ & - & $-0.056(60)$ & - & $-0.086(67)$ & $0.91$ \\
\hline
$A_{\delta u  d}$ & 2 & 0 & $0.106(49)$ & - & $0.048(49)$ & - & - & $1.06$ \\
 & & 1 & $0.013(99)$ & $0.099(96)$ & $-0.02(11)$ & $0.077(99)$ & - & $1.03$ \\
 & 3 & 0 & $0.123(54)$ & - & $0.085(60)$ & - & $0.095(66)$ & $1.01$ \\
\hline
$A_{\delta d  d}$ & 2 & 0 & $-0.36(26)$ & - & $-0.51(28)$ & - & - & $0.76$ \\
 & & 1 & $-0.70(61)$ & $0.49(67)$ & $-0.52(83)$ & $0.15(85)$ & - & $0.75$ \\
 & 3 & 0 & $-0.31(31)$ & - & $-0.42(42)$ & - & $-0.38(53)$ & $0.76$ \\
\hline
\hline
\end{tabular}

\end{center}
\caption{The same as \tab\ref{tab:cjl_A_VV}, but for the twist-two function $A_{\delta q q^\prime}$. \label{tab:cjl_A_VT}}
\end{table}
\begin{table}
\begin{center}
\begin{tabular}{c|cc|ccccc|c}
\hline
\hline
 & $N$ & $K$ & $c_{10}$ & $c_{11}[\mathrm{fm}^{-1}]$ & $c_{20}$ & $c_{21}[\mathrm{fm}^{-1}]$ & $c_{30}$ & $\chi^2 / \mathrm{dof}$\\ 
\hline
$A_{\delta u \delta u}$ & 2 & 0 & $0.18(29)$ & - & $0.37(33)$ & - & - & $0.58$ \\
 & & 1 & $-0.2(1.1)$ & $0.6(1.7)$ & $0.2(1.5)$ & $0.4(2.2)$ & - & $0.58$ \\
 & 3 & 0 & $0.45(38)$ & - & $1.01(68)$ & - & $1.5(1.0)$ & $0.58$ \\
\hline
$A_{\delta u \delta d}$ & 2 & 0 & $0.057(87)$ & - & $0.024(95)$ & - & - & $0.80$ \\
 & & 1 & $0.04(19)$ & $0.01(21)$ & $-0.15(22)$ & $0.15(22)$ & - & $0.78$ \\
 & 3 & 0 & $0.085(93)$ & - & $0.08(11)$ & - & $0.12(13)$ & $0.78$ \\
\hline
$A_{\delta d \delta d}$ & 2 & 0 & $0.38(49)$ & - & $0.35(56)$ & - & - & $0.47$ \\
 & & 1 & $-0.1(1.5)$ & $0.7(1.8)$ & $0.1(1.9)$ & $0.4(2.2)$ & - & $0.47$ \\
 & 3 & 0 & $0.69(62)$ & - & $0.96(95)$ & - & $1.2(1.3)$ & $0.46$ \\
\hline
\hline
\end{tabular}

\end{center}
\caption{The same as \tab\ref{tab:cjl_A_VV}, but for the twist-two function $A_{\delta q \delta q^\prime}$. \label{tab:cjl_A_TT}}
\end{table}
\clearpage
\begin{table}
\begin{center}
\begin{tabular}{c|cc|ccccc|c}
\hline
\hline
 & $N$ & $K$ & $c_{10}$ & $c_{11}[\mathrm{fm}^{-1}]$ & $c_{20}$ & $c_{21}[\mathrm{fm}^{-1}]$ & $c_{30}$ & $\chi^2 / \mathrm{dof}$\\ 
\hline
$B_{\delta u \delta u}$ & 2 & 0 & $0.5(1.0)$ & - & $0.5(1.5)$ & - & - & $0.25$ \\
 & & 1 & $5.2(4.4)$ & $-7.7(7.8)$ & $11.4(8.1)$ & $-18(14)$ & - & $0.25$ \\
 & 3 & 0 & $1.6(1.3)$ & - & $3.9(2.6)$ & - & $6.9(4.6)$ & $0.25$ \\
\hline
$B_{\delta u \delta d}$ & 2 & 0 & $0.068(78)$ & - & $-0.012(72)$ & - & - & $0.71$ \\
 & & 1 & $-0.30(19)$ & $0.37(20)$ & $-0.45(27)$ & $0.42(26)$ & - & $0.66$ \\
 & 3 & 0 & $0.080(92)$ & - & $0.01(10)$ & - & $0.01(12)$ & $0.70$ \\
\hline
$B_{\delta d \delta d}$ & 2 & 0 & $-0.4(1.3)$ & - & $-0.7(2.0)$ & - & - & $0.20$ \\
 & & 1 & $6(12)$ & $-11(23)$ & $13(20)$ & $-23(35)$ & - & $0.20$ \\
 & 3 & 0 & $1.3(1.6)$ & - & $4.3(5.4)$ & - & $9(11)$ & $0.20$ \\
\hline
\hline
\end{tabular}

\end{center}
\caption{The same as \tab\ref{tab:cjl_A_VV}, but for the twist-two function $B_{\delta q \delta q^\prime}$. \label{tab:cjl_B_TT}}
\end{table}

\FloatBarrier

\subsection{Results for Mellin moments}
\label{sec:mellin_res}

From the fits described in the previous section, we are able to reconstruct the Mellin moments $I(\zeta,y^2)$. Combining \eqref{eq:A-global-fit}, \eqref{eq:y2-ansatz}, \eqref{eq:Ahat} and executing the Fourier transform \eqref{eq:skewed-mellin-inv-fct} we arrive at:

\begin{align}
\label{eq:I-from-ansatz}
I_{qq^\prime}(\zeta,y^2) = 
	\pi \sum_{i=1,2} A_i e^{-\eta_i (y-y_0)}
	\sum_{n,m=0}^N \sum_{k=0}^K
	\zeta^{2n} \left( T^{-1} \right)_{nm} c_{mk} 
	\sqrt{-y^2}^{k+\delta} \eta_i^\delta\ \Theta(1-\zeta^2)\,.
\end{align}
In the following we discuss the corresponding results and physics implications. We take into account every channel except for those we characterized as "bad" channels in \sect\ref{sec:twist2_py_fit}.

\begin{figure}
\subfigure[fit comparison $I_{ud}(\zeta=0,y^2)$ \label{fig:mellin_fit_comp-I_VV-ud}]{
\includegraphics[scale=0.25,trim={0.5cm 1.2cm 0.5cm 2.8cm},clip]{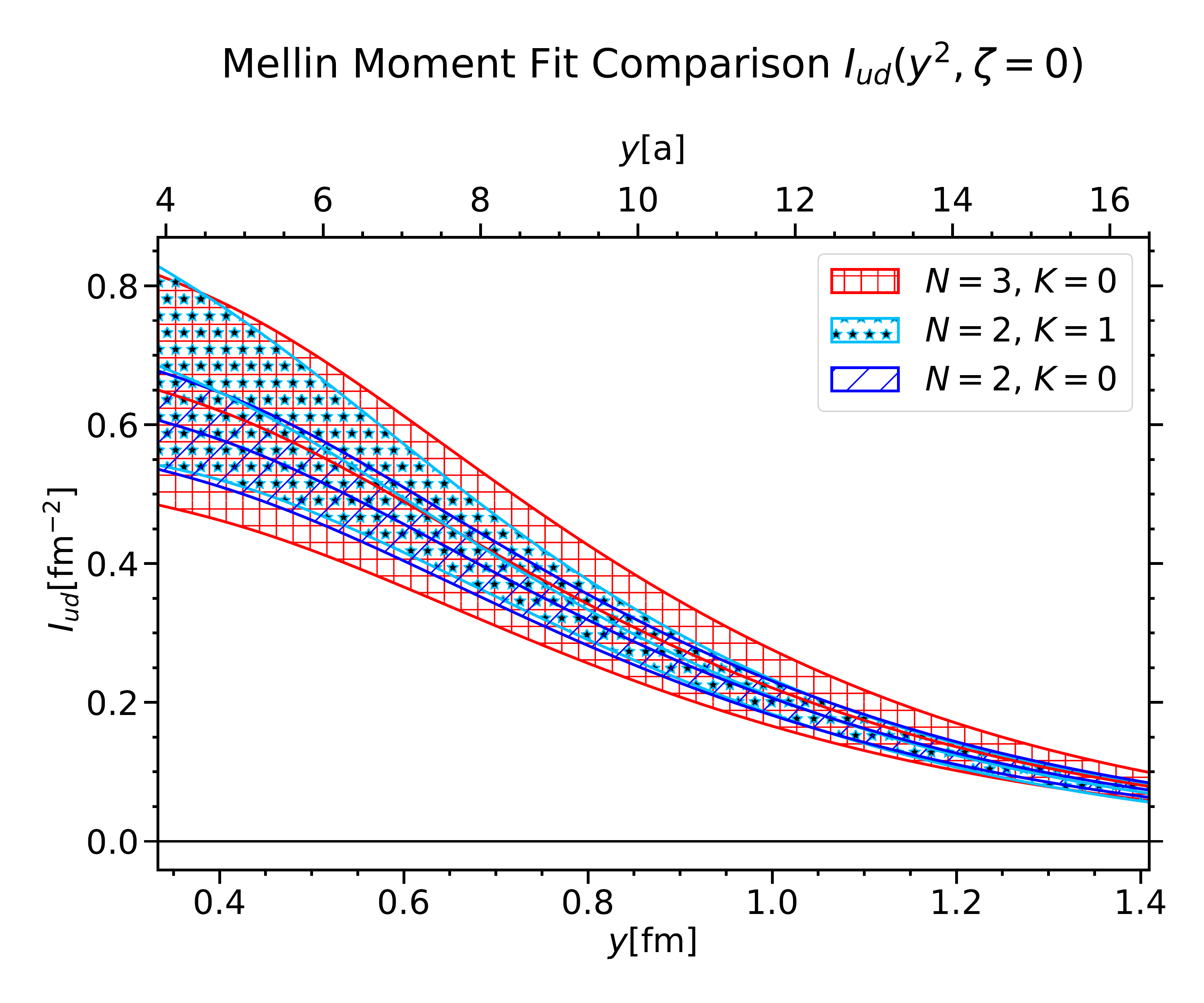}} \hfill 
\subfigure[fit comparison $I^t_{\delta u \delta d}(\zeta=0,y^2)$ \label{fig:mellin_fit_comp-It_TT-ud}]{
\includegraphics[scale=0.25,trim={0.5cm 1.2cm 0.5cm 2.8cm},clip]{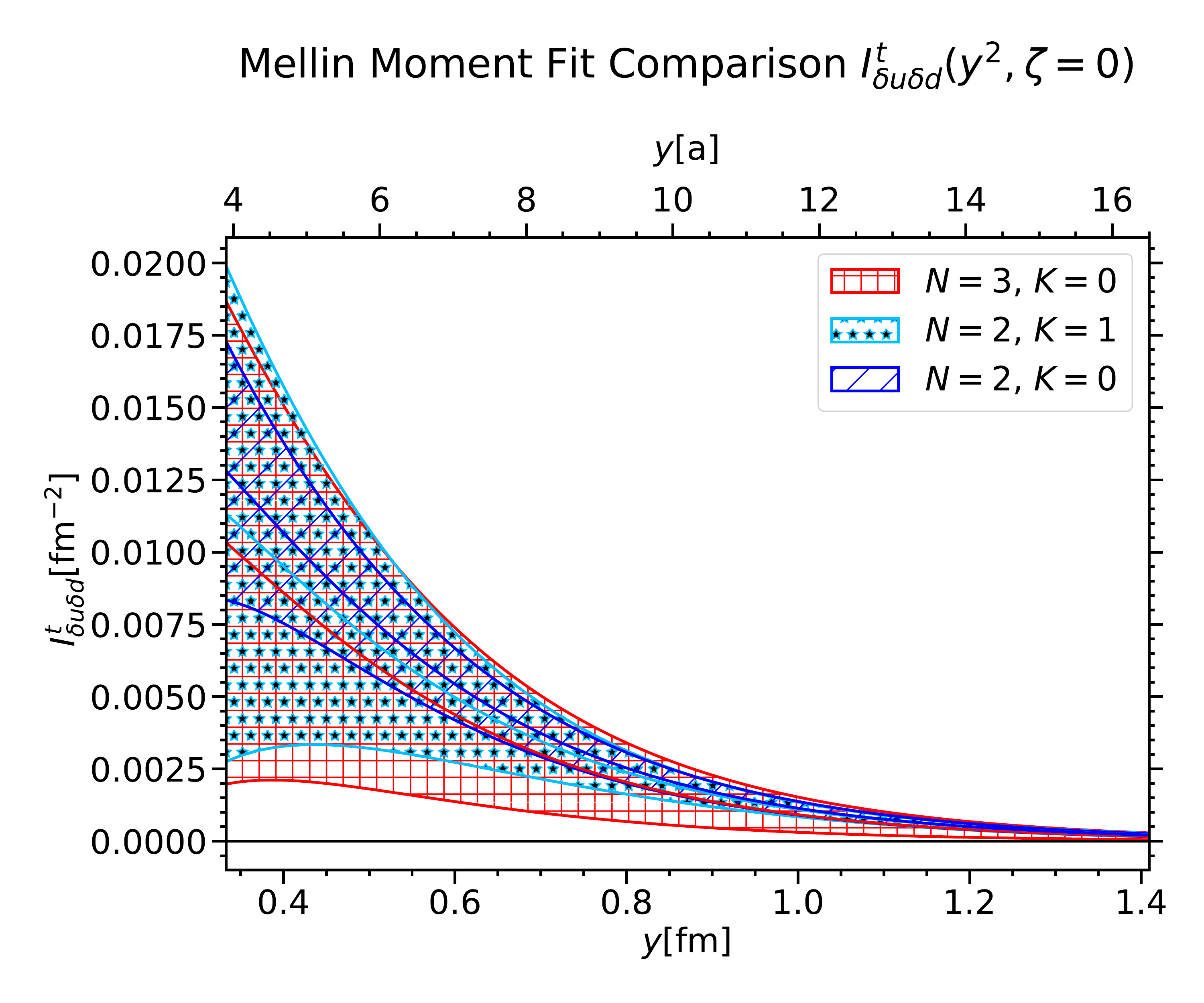}} \\
\subfigure[fit comparison $I_{uu}(\zeta=0,y^2)$ \label{fig:mellin_fit_comp-I_VV-uu}]{
\includegraphics[scale=0.25,trim={0.5cm 1.2cm 0.5cm 2.8cm},clip]{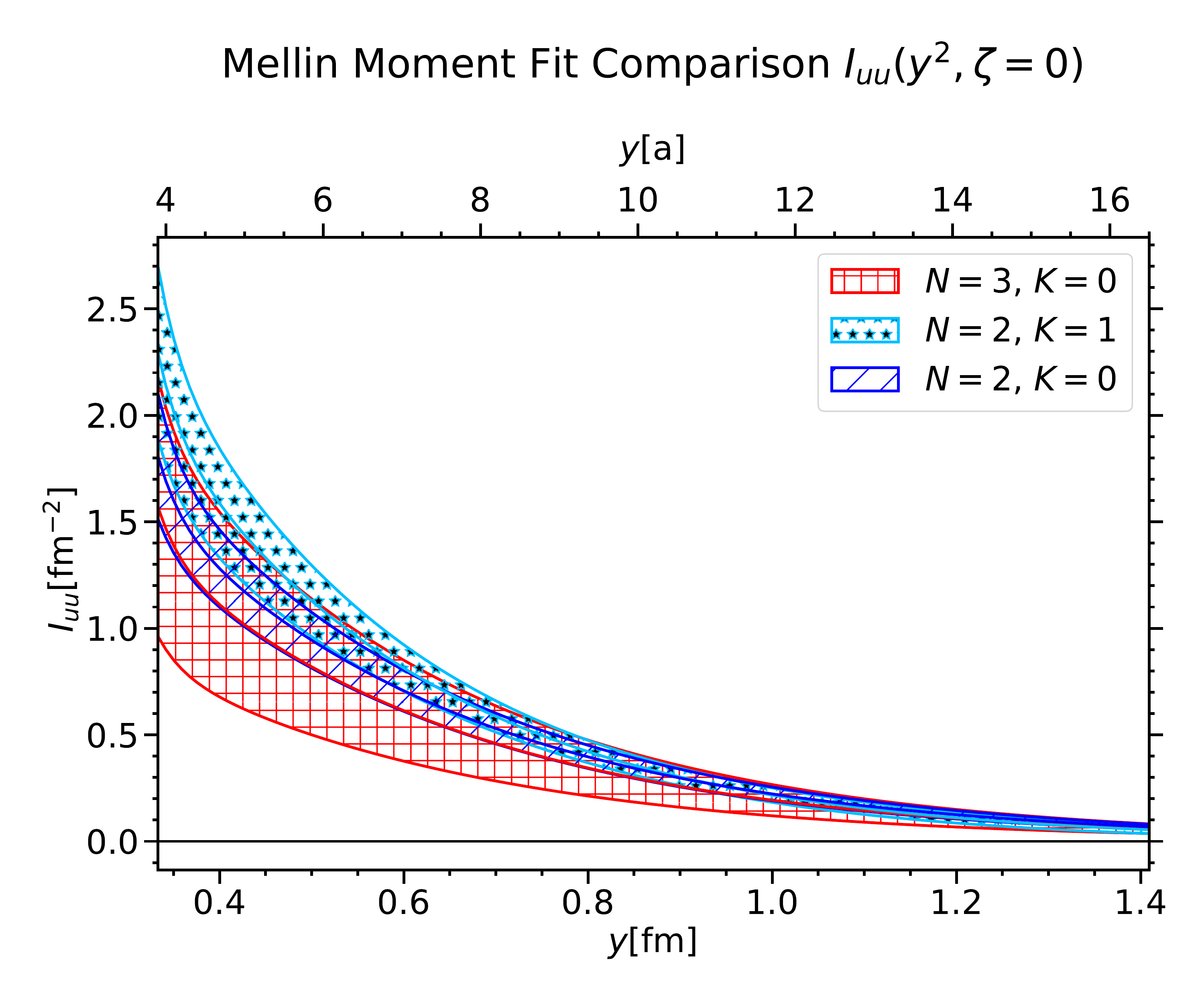}} \hfill 
\subfigure[fit comparison $I_{\delta u u}(\zeta=0,y^2)$ \label{fig:mellin_fit_comp-I_VT-uu}]{
\includegraphics[scale=0.25,trim={0.5cm 1.2cm 0.5cm 2.8cm},clip]{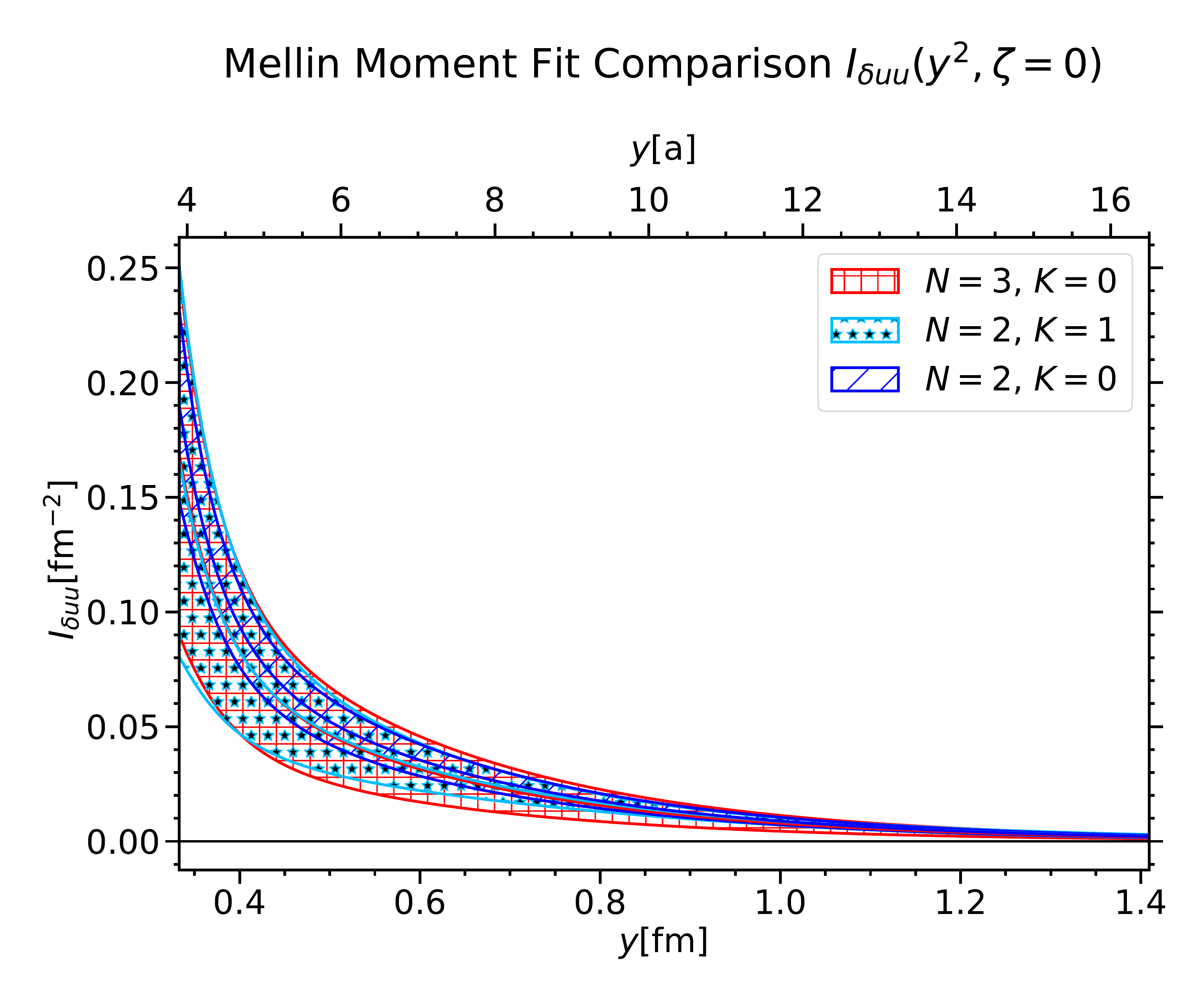}} \\
\caption{Results for selected DPD Mellin moments, where we compare the curves obtained from the fits with $(N,K)=(2,0),(2,1),(3,0)$. \label{fig:mellin_fit_comp}}
\end{figure}

\paragraph*{Fit dependence:}
\Fig\ref{fig:mellin_fit_comp} shows the results for the Mellin moments $I(\zeta = 0,y^2)$ for selected channels. We compare the bands obtained from the three different fits in order to estimate the systematic error introduced by the extrapolation in $py$. In each channel we observe consistency between the different fits, \ie the three curves coincide within the error bands. The situation is the same for the channels which are not shown in the plots. Notice that also the bands for $I^t_{\delta u \delta d}$ match within the statistical error, despite the fact that a linear dependence of the moments $\langle \zeta^{2m} \rangle$ on $y$ seemed to give a better description.

\begin{figure}
%\captionsetup[subfigure]{justification=centering}
\subfigure[fit comparison $I_{ud}(\zeta,y=10a)$ \label{fig:mellin_zeta_fit_comp-I_VV-ud}]{
\includegraphics[scale=0.25,trim={0.5cm 1.2cm 0.5cm 2.8cm},clip]{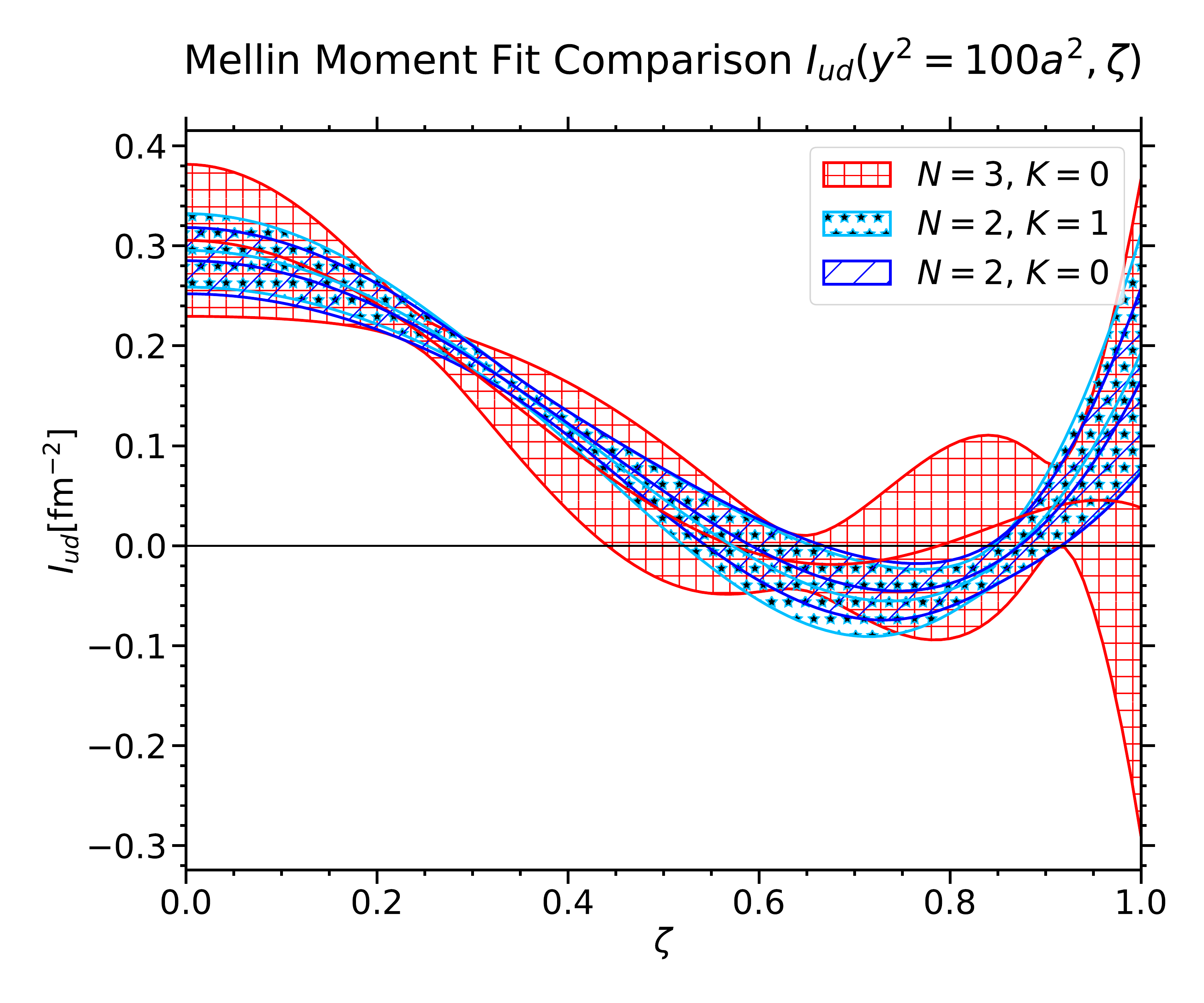}} \hfill 
\subfigure[fit comparison $I^t_{\delta u \delta d}(\zeta,y=10a)$ \label{fig:mellin_zeta_fit_comp-It_TT-ud}]{
\includegraphics[scale=0.25,trim={0.5cm 1.2cm 0.5cm 2.8cm},clip]{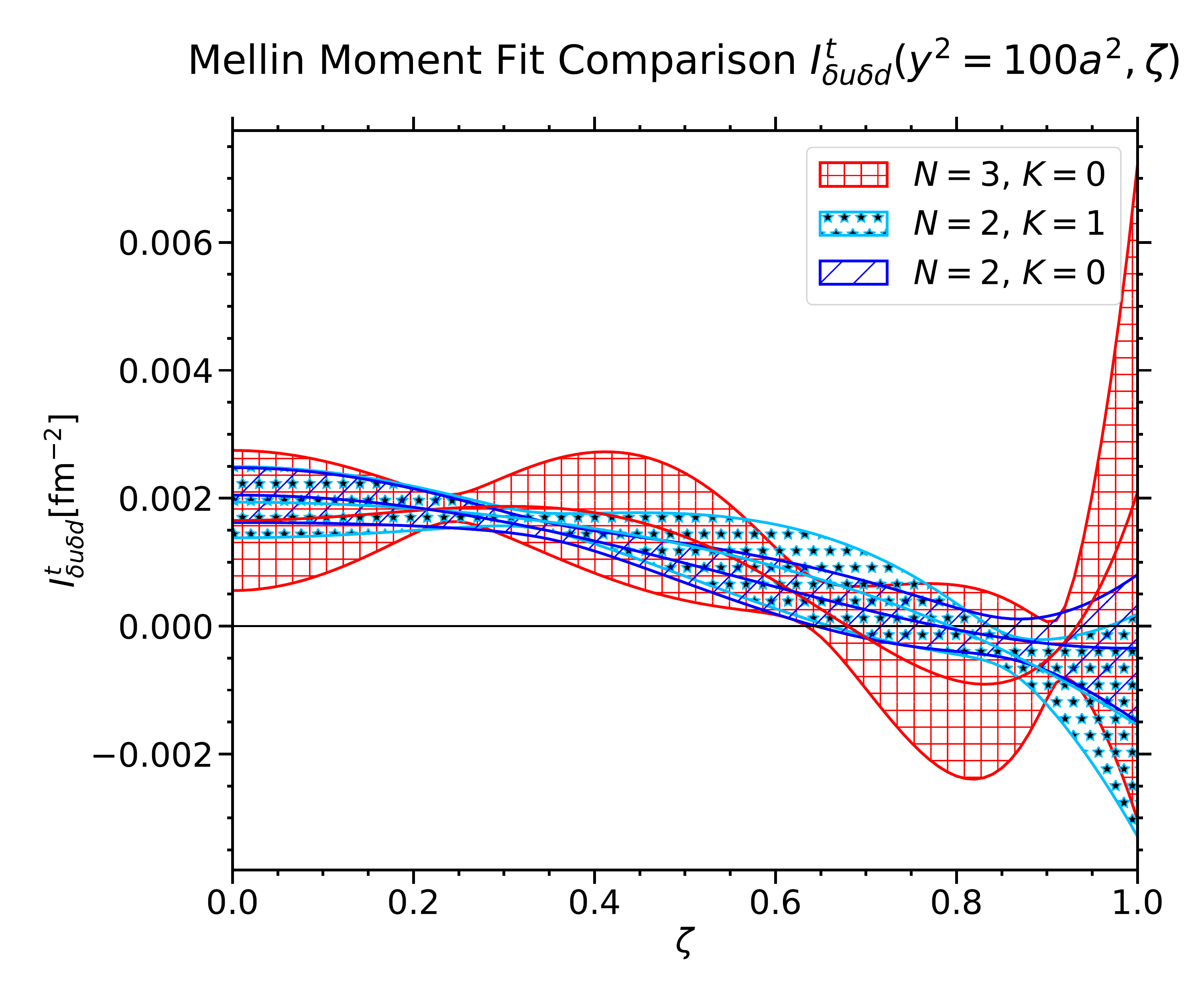}} \\
\subfigure[fit comparison $I_{uu}(\zeta,y=10a)$ \label{fig:mellin_zeta_fit_comp-I_VV-uu}]{
\includegraphics[scale=0.25,trim={0.5cm 1.2cm 0.5cm 2.8cm},clip]{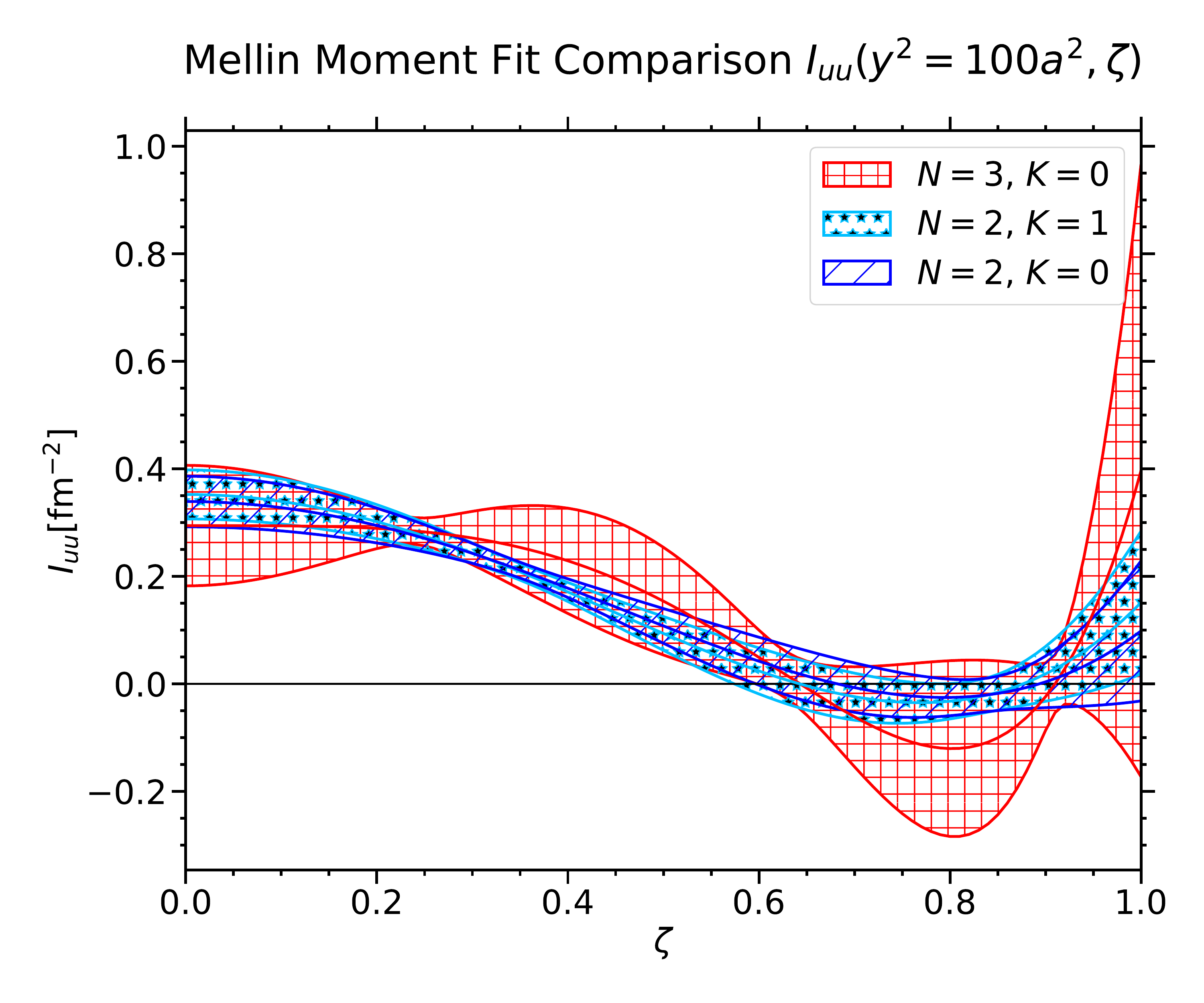}} \hfill 
\subfigure[fit comparison $I_{\delta d u}(\zeta,y=10a)$ \label{fig:mellin_zeta_fit_comp-I_VT-ud}]{
\includegraphics[scale=0.25,trim={0.5cm 1.2cm 0.5cm 2.8cm},clip]{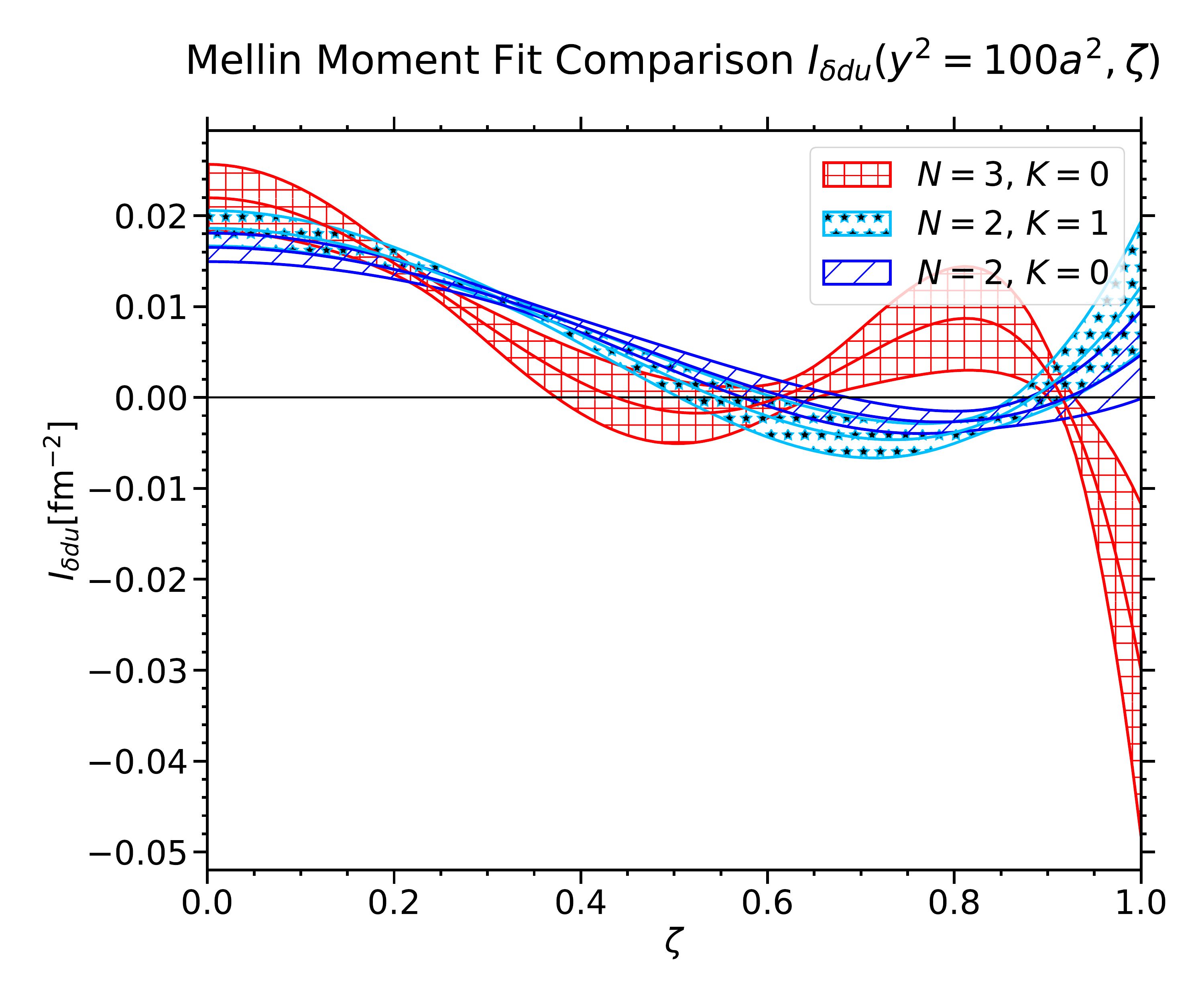}} \\
\caption{$\zeta$ dependence of selected DPD Mellin moments, where we compare the curves obtained from the fits with $(N,K)=(2,0),(2,1),(3,0)$. This is shown for $y=10a$. \label{fig:mellin_zeta_fit_comp}}
%\end{figure}
%
%\begin{figure}
\subfigure[{\parbox[t]{5cm}{flavor comparison, $I_{qq^\prime}(\zeta=0,y^2)$, $N=2$, $K=0$\label{fig:mellin_fcomp-A_VV}}} ]{
\includegraphics[scale=0.25,trim={0.5cm 1.2cm 0.5cm 2.8cm},clip]{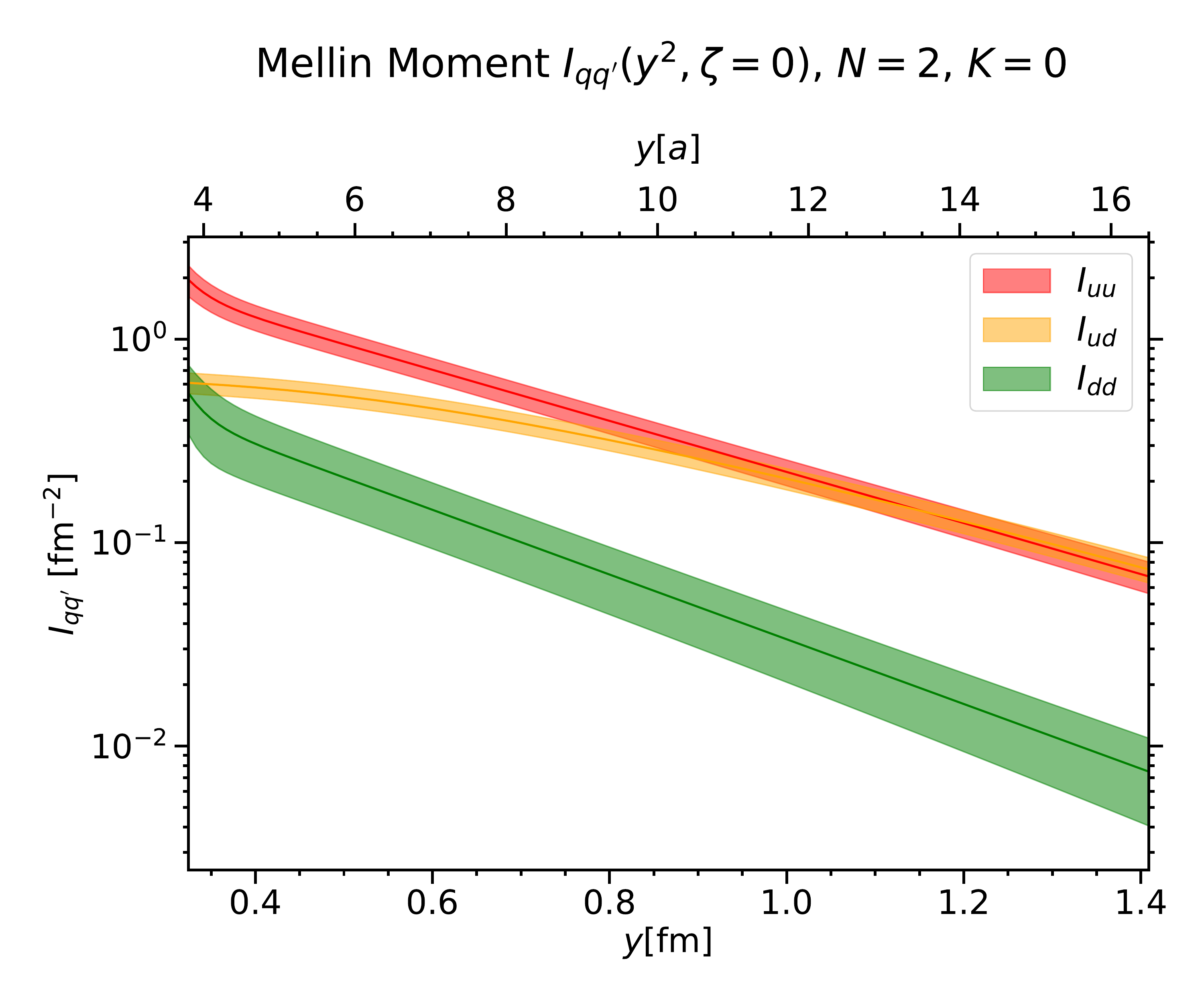}} \hfill 
\subfigure[{\parbox[t]{5cm}{flavor comparison, $I_{\delta q q^\prime}(\zeta=0,y^2)$, $N=2$, $K=0$\label{fig:mellin_fcomp-A_VT}}} ]{
\includegraphics[scale=0.25,trim={0.5cm 1.2cm 0.5cm 2.8cm},clip]{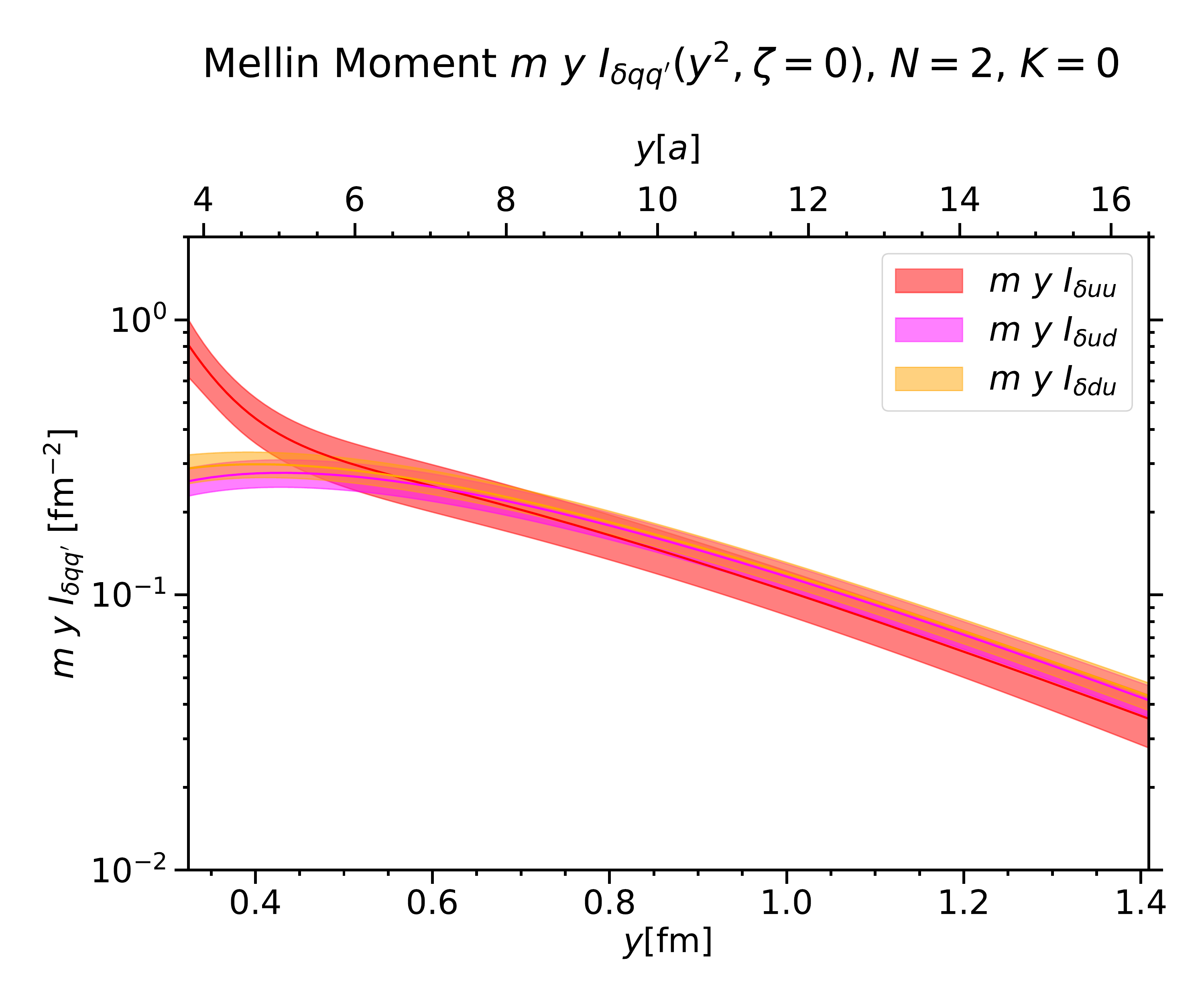}} \\
\caption{Results for the Mellin moments $I_{qq^\prime}$ (a) and $I_{\delta qq^\prime}$ (b) obtained from fits with $(N,K) = (2,0)$. In each panel we compare contributions for different flavor combinations using a logarithmic scale on the vertical axis. \label{fig:mellin_fcomp}}
\end{figure}

The agreement of the results for different fits also holds for $\zeta\lesssim 0.6$ in most of the channels that we have not excluded. As an example we show the results for $I_{ud}$, $I^t_{\delta u \delta d}$ and $I_{uu}$ in \fig\ref{fig:mellin_zeta_fit_comp} (a-c). An exception is found for $I_{\delta d u}$ plotted in \fig\ref{fig:mellin_zeta_fit_comp} (d), where clear deviations between the fits with $N=3$ and $N=2$ are found for $\zeta > 0.2$. Notice that in this channel we found the largest variations between the values of $\chi^2/\mathrm{dof}$ of the different fits. At this point, we emphasize again that the ansatz \eqref{eq:mellin-ansatz} for the functional form of the DPD Mellin moments represents an expansion around $\zeta = 0$. Consequently, the more terms of this expansion are taken into account, the more sensitive the results for large $\zeta$ become to fluctuations of the corresponding coefficients.

Since the fit for $(N,K) = (2,0)$ yields already a consistent description of the data, we will base our physics discussion on the corresponding results.

\paragraph*{Flavor comparison:}

We compare the results for the DPD Mellin moments \wrt the quark flavor in \fig\ref{fig:mellin_fcomp}, using a logarithmic scale on the vertical axes. The results for $I_{\delta q q^\prime}$ are multiplied by $m\,y$, which follows from the decomposition \eqref{eq:t2-mat-els}. Like for the twist-two functions, we observe that in the case of two unpolarized quarks (see panel (a)) the $dd$ signal is much smaller than that of $ud$ and $uu$ for large distances. At small $y$, the Mellin moments for $uu$ and $dd$ show a steeper slope than $I_{ud}$. The same behavior is observed for $I_{\delta q q^\prime}$ in panel (b), where we compare only $uu$, $ud$ and $du$, since we have classified $dd$ as a "bad" channel.

A very interesting result is the different behavior of the Mellin moments $I_{ud}$ and $I_{uu}$. In factorization assumptions as they are made in the pocket formula (see \sect\ref{sec:dps}) it is required that the dependence of DPDs on the transverse quark distance is independent of the quark flavor, see \eqref{eq:dpd-pocket}. Our results clearly exclude this. %This clearly disagrees with our results.

\paragraph*{Polarization effects:}

\begin{figure}
\subfigure[{\parbox[t]{6cm}{polarization dependence, $uu$, $\zeta = 0$,\\ fit with $N=2$, $K=0$ \label{fig:mellin_polcomp-uu}}}]{
\includegraphics[scale=0.248,trim={0.5cm 1.1cm 0.5cm 3.3cm},clip]{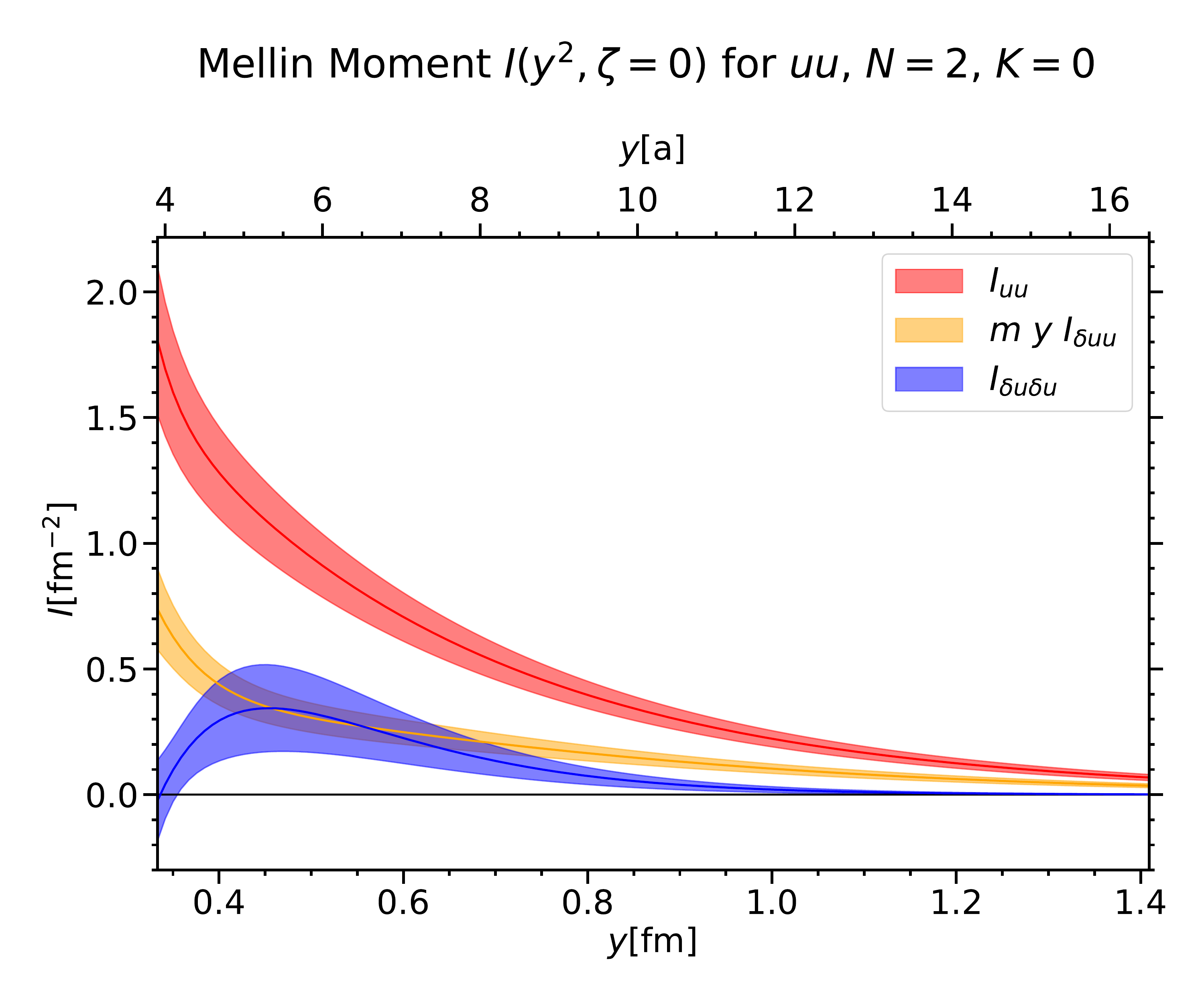}}\hfill 
\subfigure[{\parbox[t]{6cm}{polarization dependence, $uu$, \\ twist-two function at $py=0$ \label{fig:mellin_polcomp-uu-tw2}}}]{
\includegraphics[scale=0.248,trim={0.5cm 1.1cm 0.5cm 3.3cm},clip]{plots/chan_comp_phys-uu.pdf}} \\
\subfigure[{\parbox[t]{6cm}{polarization dependence, $ud$, $\zeta = 0$,\\ fit with $N=2$, $K=0$ \label{fig:mellin_polcomp-ud}}}]{
\includegraphics[scale=0.248,trim={0.5cm 1.1cm 0.5cm 3.3cm},clip]{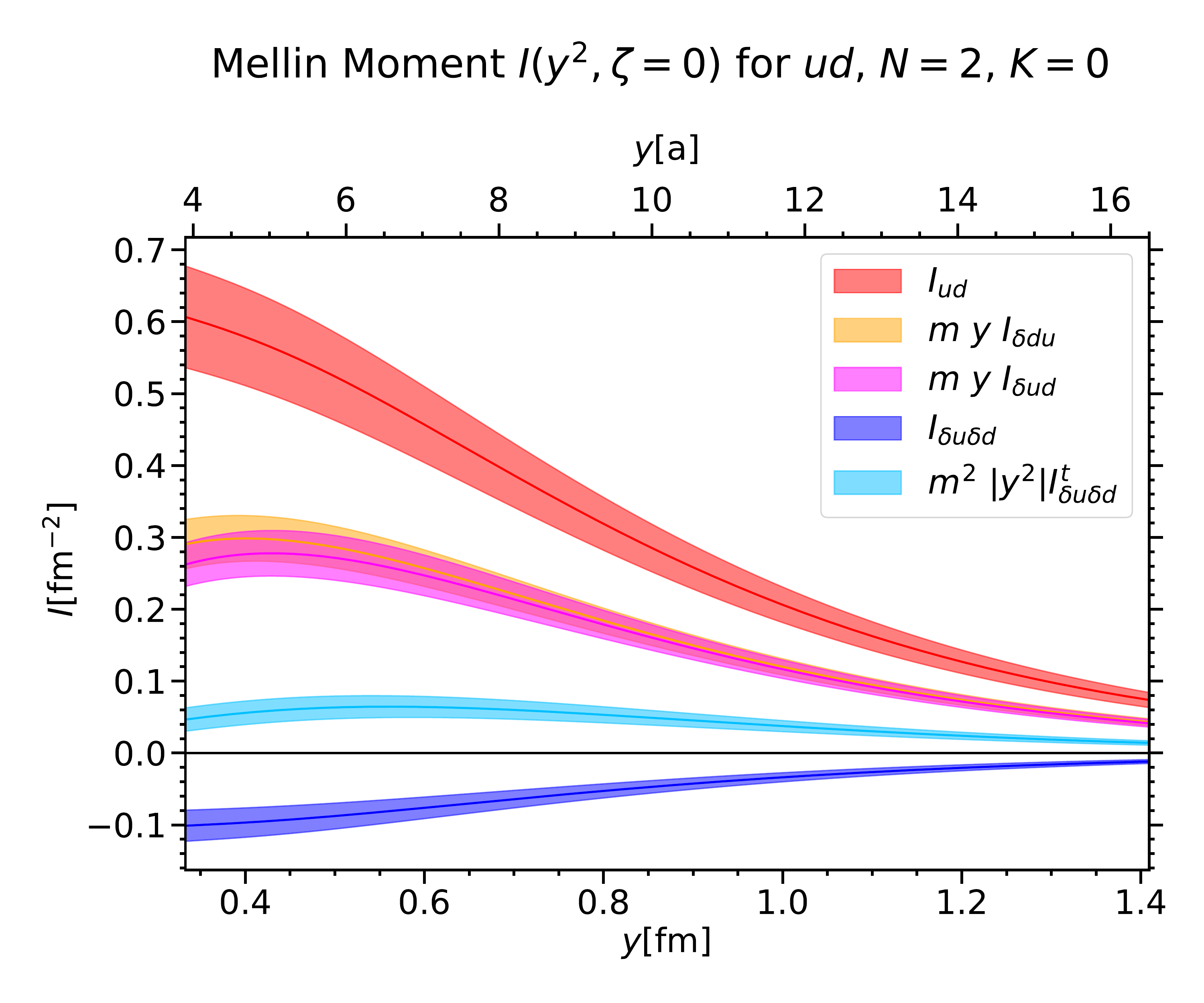}}\hfill 
\subfigure[{\parbox[t]{6cm}{polarization dependence, $ud$, \\ twist-two function at $py=0$ \label{fig:mellin_polcomp-ud-tw2}}}]{
\includegraphics[scale=0.248,trim={0.5cm 1.1cm 0.5cm 3.3cm},clip]{plots/chan_comp_phys-ud.pdf}} \\
\subfigure[{\parbox[t]{6cm}{polarization dependence, $u\bar{d}$ in $\pi^+$, $\zeta = 0$,\\ fit with $N=2$, $K=0$ \label{fig:mellin_polcomp-pion}}}]{
\includegraphics[scale=0.248,trim={0.5cm 1.1cm 0.5cm 3.3cm},clip]{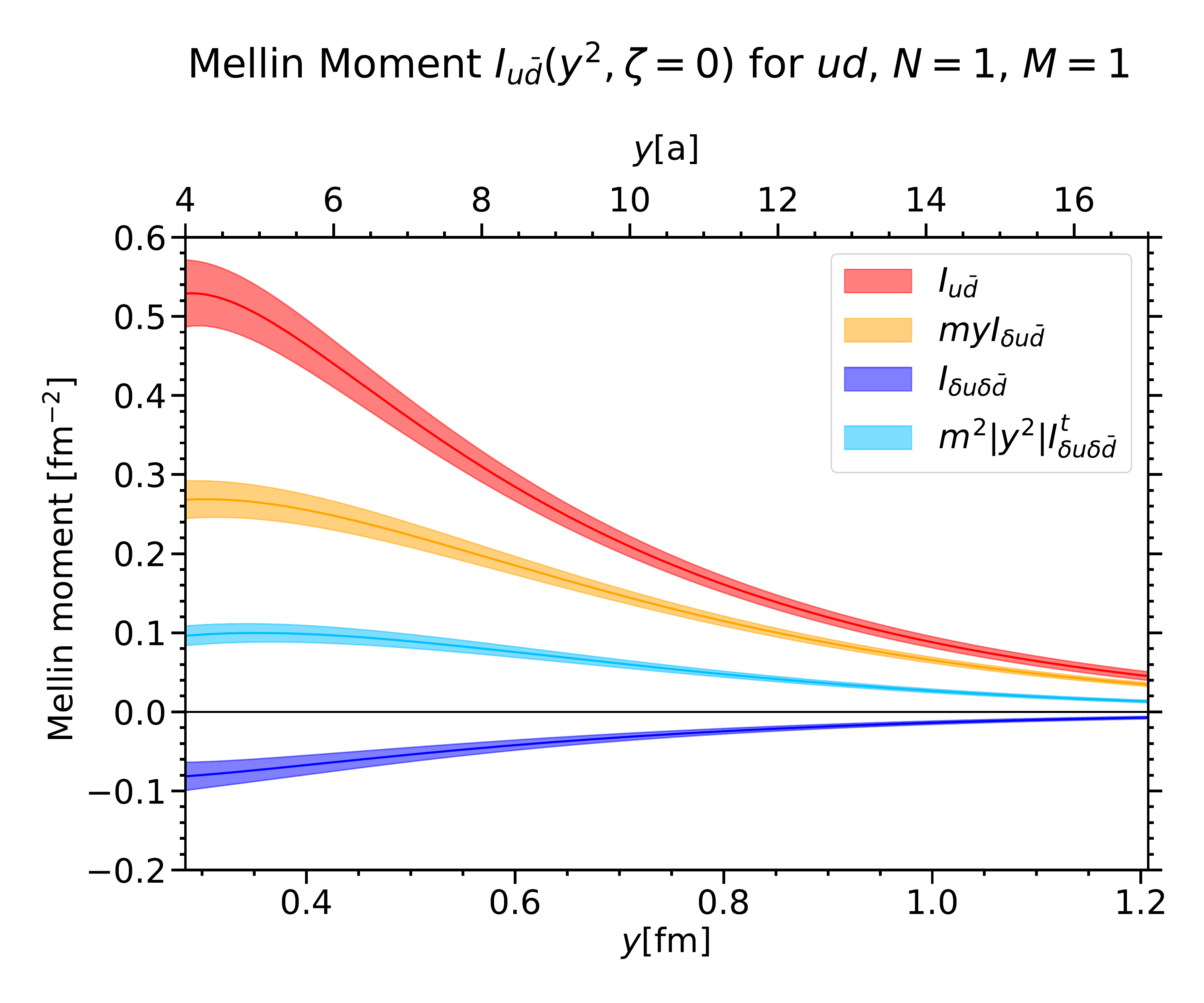}}\hfill 
\subfigure[{\parbox[t]{6cm}{polarization dependence, $u\bar{d}$ in $\pi^+$,\\ twist-two function at $py=0$ \label{fig:mellin_polcomp-pion-tw2}}}]{
\includegraphics[scale=0.6,trim={0 0 0 0.8cm},clip]{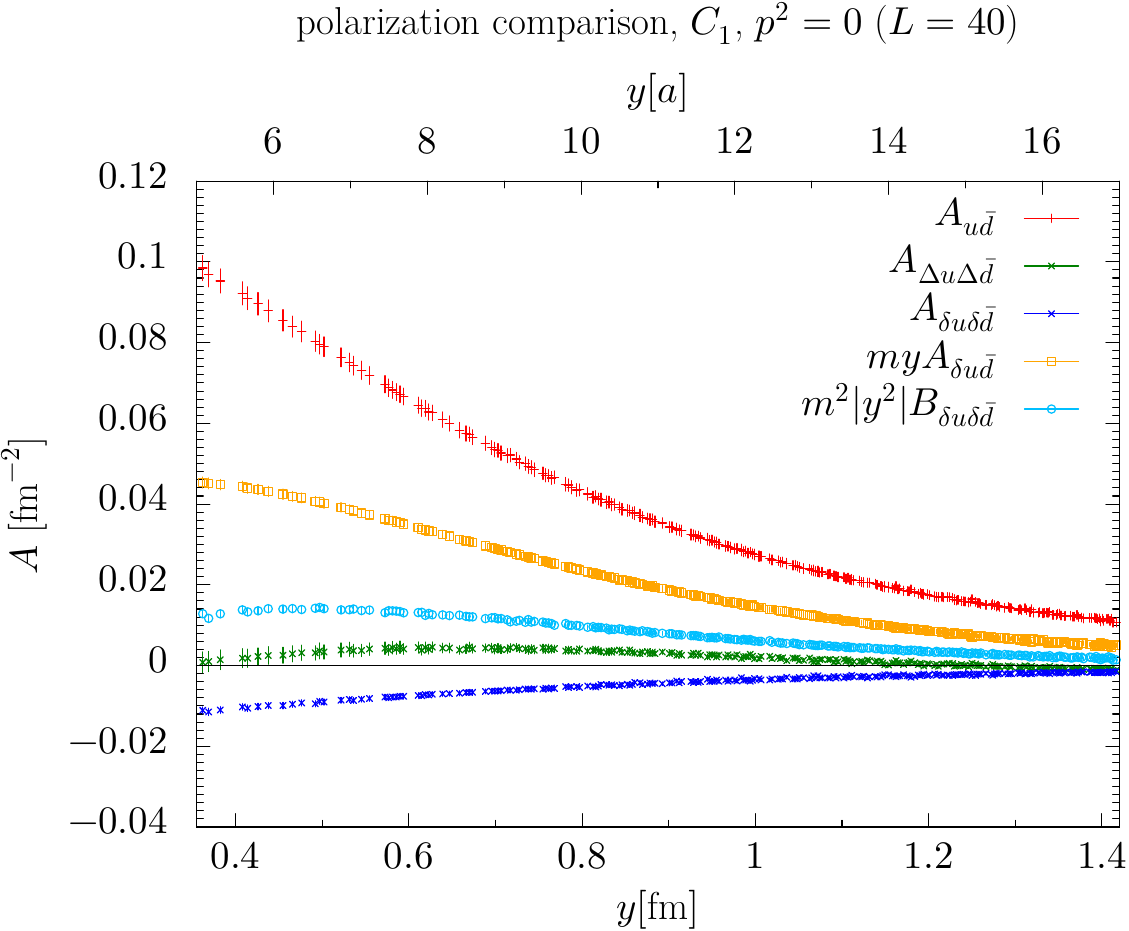}} \\
\caption{Comparison between different quark polarizations for the flavor combinations $uu$ (a,b) and $ud$ (c,d). The left panels show the results for the Mellin moments obtained from the fit with $(N,K)=(2,0)$. In the right panels we again show the data for the corresponding twist-two functions, which was already plotted in \fig\ref{fig:tw2f_polcomp}. Panels (e) and (f) show the results for $u\bar{d}$ in the $\pi^+$, which were calculated in \cite{Bali:2020mij}. \label{fig:mellin_polcomp}}
\end{figure}

In \fig\ref{fig:mellin_polcomp} we show the dependence of the Mellin moments on the quark polarization for $ud$ (c) and $uu$ (a). Again we only show the results for $N=2$ and $K=0$. As in the discussion of the twist-2 functions, we multiply the DPD Mellin moments $I_{\delta qq^\prime}$ or $I^t_{\delta q \delta q^\prime}$ by $m\,y$ or $m^2 |y^2|$, respectively, which follows from the decomposition \eqref{eq:t2-mat-els}. The polarization dependence of the Mellin moments is very similar to that of the twist-two functions, which we already gave in \fig\ref{fig:tw2f_polcomp}. These are again shown in panel (d) and (b). We see that the unpolarized channels are clearly dominant for both flavor combinations. However, in the case of $ud$, there are visible polarization effects. They are especially large for $I_{\delta u d}$ and $I_{\delta d u}$, whereas Mellin moments $I_{\delta u \delta d}$ and $I^t_{\delta u \delta d}$ are smaller but still significantly different from zero.
At this point, we want to compare with the situation for $u\bar{d}$ in the $\pi^+$, which was calculated in \cite{Bali:2020mij}. The corresponding results are also plotted in \fig\ref{fig:mellin_polcomp}. Remarkably, the behavior of the Mellin moments (e), as well as the twist-two functions (f), for $u\bar{d}$ in a $\pi^+$ is comparable to the one for $ud$ in a nucleon.

In the case of $uu$ in the proton, polarization effects appear to be less important. Notice that in the corresponding plots we only show the results for $I_{\delta u u}$ and $I_{\delta u \delta u}$, since the remaining functions belong to "bad" channels, as we have discussed before. The largest polarized Mellin moment is again $I_{\delta u u}$. $I_{\delta u \delta u}$ is clearly non-zero for small distances, but the corresponding statistical error is quite large ($>50\%$). The sign of $I_{\delta u \delta u}$ indicates that the quark spins are more aligned than anti-aligned, which agrees with expectations from $SU(6)$ symmetric valence quark wave functions \cite{Diehl:2011yj}. However, the ratios $I_{\Delta u \Delta d}/I_{ud} = -2/3$ or $I_{\Delta u \Delta u}/I_{uu} = +1/3$ predicted by this model are clearly not observed in our results. The same conclusion can be drawn from the corresponding data of the twist-two functions.

\subsection{The number sum rule}
\label{sec:sumrule}

We consider the DPD number sum rule, which we have already stated in \eqref{eq:dpd-sr} in position space. We look at the flavor combination $ud$. The remaining two flavor combinations $uu$ and $dd$ cannot be investigated, since the corresponding expressions include sea quark contributions that would lead to diverging integrals over $x_1$. In the considered case of one $u$ and one $d$ quark, splitting contributions are at least of second order in $\alpha_s$. Inserting the sum rule for ordinary PDFs in \eqref{eq:dpd-sr} we can write: 

\begin{align}
\int_{-1}^1 \dd x_1 \int_{-1}^1 \dd x_2 
\int_{b_0/\mu} \dd^2 \tvec{y}\  F_{ud}(x_1,x_2,\tvec{y};\mu) = 
	2 + \Op(\alpha_s^2(\mu)) + \Op((\Lambda / \mu)^2)\,.
\label{eq:sr-proton}
\end{align}
By executing the integrals over $x_1$ and $x_2$, we can identify the DPD Mellin moments for $\zeta = 0$. The Fourier transform in $py$ \eqref{eq:skewed-mellin-inv-fct} then yields up to corrections of order $\Lambda^2 / \mu^2$ and $\alpha_s^2$:

\begin{align}
2\pi \int^\infty_{b_0/\mu} \dd y\ y 
\int_{-\infty}^{\infty} \dd (py)\ A_{ud}(py,y^2) = 2\,.
\label{eq:sr-check}
\end{align}
The verification that this equations holds for the results we presented in the previous sections can be seen as a consistency check of our lattice calculations and our fitting ansatz. We evaluate the expression on the \lhs of \eqref{eq:sr-check} by inserting the parameters obtained from the $y^2$ fit and each of the three global $py$ fits. The corresponding values are summarized in \tab{\ref{tab:sr_res}}.

\begin{table}[ht]
\begin{center}
\begin{tabular}{cc|c|c}
\hline
\hline
$N$ & $K$ & $\chi^2/\mathrm{dof}$ & integral \\ 
\hline
2 & 0 & 0.47 & 1.93(23) \\
3 & 0 & 0.46 & 2.07(51) \\
2 & 1 & 0.46 & 1.98(24) \\
\hline
\hline
\end{tabular}

\end{center}
\caption{Results for the integral on the \lhs of \eqref{eq:sr-check} obtained for the fits with $(N,K)=(2,0),(3,0),(2,1)$. In the center column we again list the values of $\chi^2/\mathrm{dof}$ for the fit. \label{tab:sr_res}}
\end{table}

Each of the obtained results is very close to the value predicted by the sum rule with a largest absolute deviation of the mean of $0.07$. The statistical error varies between $12\%$ and $25\%$, \ie it is larger than the systematic error which is introduced by the extrapolation in $py$. Evaluating the integral \eqref{eq:sr-check} implicitly includes an extrapolation for $y>16a$. In order to estimate the corresponding systematic error, we decrease the upper integration boundary of the $y$ integral to $16a = 1.37~\mathrm{fm}$. We obtain values which are at most $16\%$ smaller. Thus, the systematic error from the extrapolation in $y$ is at most of the size of the statistical error. Notice that there is no extrapolation to the lower boundary $b_0/\mu \approx 1.29a$, since the lower boundary of the fit range is $1a$ in the unpolarized $ud$ case.

\FloatBarrier

\section{Factorization Tests}
\label{sec:factorization}

A crucial aspect to be studied in the context of DPDs is the strength of parton-parton correlations. These are neglected in factorization assumptions like \eqref{eq:dpd-fact}. In the following we want to check to what extent this factorization ansatz is valid.

\subsection{Derivation}
\label{sec:fact_deriv}

Equation \eqref{eq:dpd-fact} can be derived by inserting a complete set of states in the two-current matrix element appearing in \eqref{eq:dpd-def} or \eqref{eq:dpd-skew-def} and then assuming that the intermediate nucleon states dominate, \ie omitting all remaining contributions:

\begin{align}
&\sum_{\lambda} \bra{p,\lambda} 
	\Op_{a_1}(y,z_1)\ \Op_{a_2}(0,z_2) 
\ket{p,\lambda} 
\stackrel{?}{=} \nonumber \\
	&\stackrel{?}{=} \sum_{\lambda,\lambda^\prime} 
	\int \frac{
		\dd p^{\prime +}\dd^2 \tvec{p}^\prime
	}{
		2 p^{\prime +} (2\pi)^3} e^{-iy(p^\prime - p)
	} 
	\bra{p,\lambda} 
		\Op_{a_1}(0,z_1) 
	\ket{p^\prime,\lambda^\prime} 
	\bra{p^\prime,\lambda^\prime} 
		\Op_{a_2}(0,z_2) 
	\ket{p,\lambda}\ .
\label{eq:ans_f1}
\end{align}
By writing $\stackrel{?}{=}$, we emphasize that \eqref{eq:ans_f1} is an assumption; its validity is investigated in this section. For the remaining derivation steps, we substitute the intermediate momentum $p^\prime$ by:

\begin{align}
p^{\prime +} = (1-\zeta)p^+\,, \qquad 
\tvec{p}^\prime = \tvec{p} - \tvec{r}\,.
\end{align}
Furthermore, we set $\tvec{p} = \tvec{0}$ and identify:

\begin{align}
\bar{x}_i = \frac{x_i}{1-\frac{\zeta}{2}}\,, 
\qquad
\xi = \frac{\zeta}{2-\zeta}\,, 
\qquad
t(\zeta,\tvec{r}^2) = -\frac{\zeta^2 m^2 + \tvec{r}^2}{1-\zeta}\,.
\label{eq:gpdvar}
\end{align}
This enables us to write a factorized expression of the skewed DPD defined in \eqref{eq:dpd-skew-def} in terms of GPD matrix elements $f^{\lambda^\prime\lambda}(\bar{x},\xi,\tvec{p}^\prime,\tvec{p})$:

\begin{align}
&F_{a_1 a_2}(x_1,x_2,\zeta,\tvec{y}) 
\stackrel{?}{=} 
	\frac{1}{2(1-\zeta)} 
	\int \frac{\dd^2 \tvec{r}}{(2\pi)^2} e^{-i\tvec{r}\tvec{y}} 
	\left[ 
		\prod_{i=1,2} \int \frac{\dd z_i^-}{2\pi} e^{i x_i p^+ z_i^-} 
	\right] \nonumber \\
&\quad\times
	\sum_{\lambda \lambda^\prime} \bra{p,\lambda} 
		\Op_{a_1}(0,z_1) 
	\ket{(1-\zeta)p^+,-\tvec{r},\lambda^\prime} 
	\bra{(1-\zeta)p^+,-\tvec{r},\lambda^\prime} 
		\Op_{a_2}(0,z_2) 
	\ket{p,\lambda} \nonumber \\
&= 
	\frac{1}{2(1-\zeta)} 
	\int \frac{\dd^2 \tvec{r}}{(2\pi)^2} e^{-i\tvec{r}\tvec{y}}
	\sum_{\lambda \lambda^\prime}
	f_{a_1}^{\lambda\lambda^\prime}(\bar{x}_1,-\xi,\tvec{0},-\tvec{r})\ 
	f_{a_2}^{\lambda^\prime \lambda}(\bar{x}_2,\xi,-\tvec{r},\tvec{0})\,
\label{eq:F_fact}
\end{align}
with

\begin{align}
f_{a}^{\lambda^\prime\lambda}(\bar{x},\xi,\tvec{p}^\prime,\tvec{p}) := 
	\int \frac{\dd z^-}{2\pi} e^{i\bar{x} (p^\prime+p)^+ z^-/2} 
	\bra{p^\prime,\lambda^\prime} \Op_a(0,z) \ket{p,\lambda} \,,
\end{align}
where $\xi = (p-p^\prime)^+/(p+p^\prime)^+$.
\begin{figure}
\subfigure[\label{fig:dpd-approx-ud}]{
\includegraphics[width=0.48\textwidth]{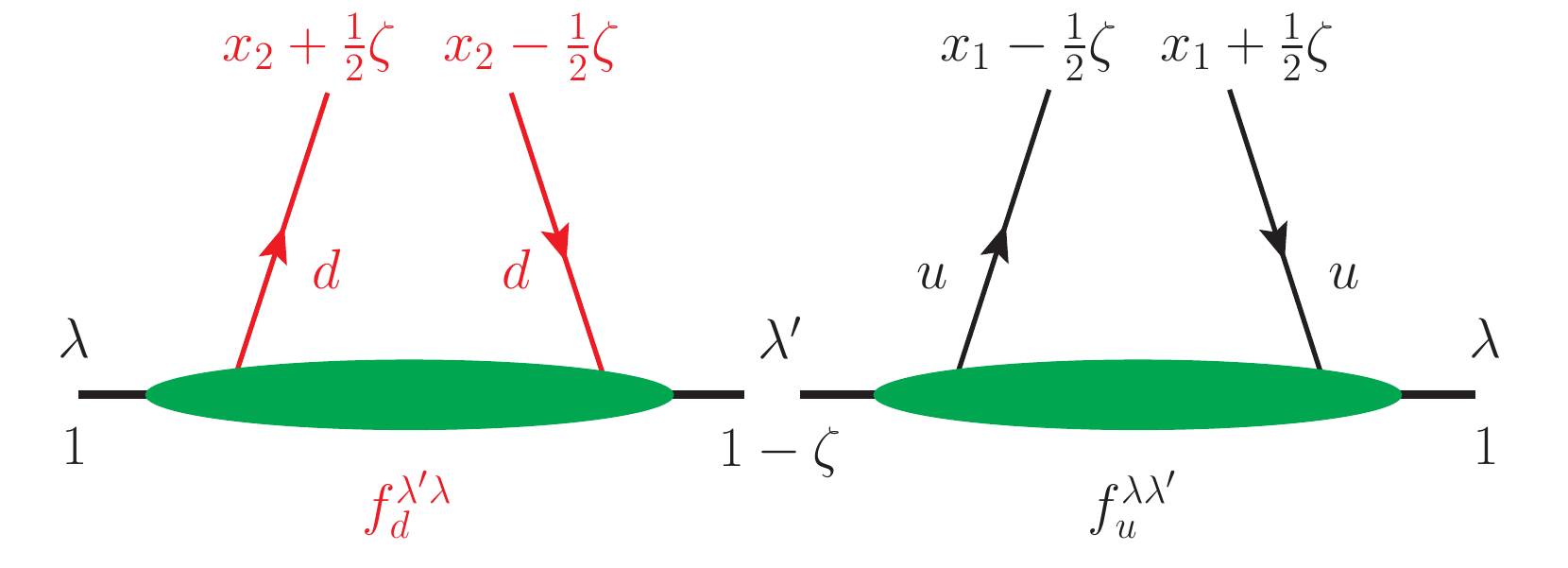}
}
\subfigure[\label{fig:dpd-approx-du}]{
\includegraphics[width=0.48\textwidth]{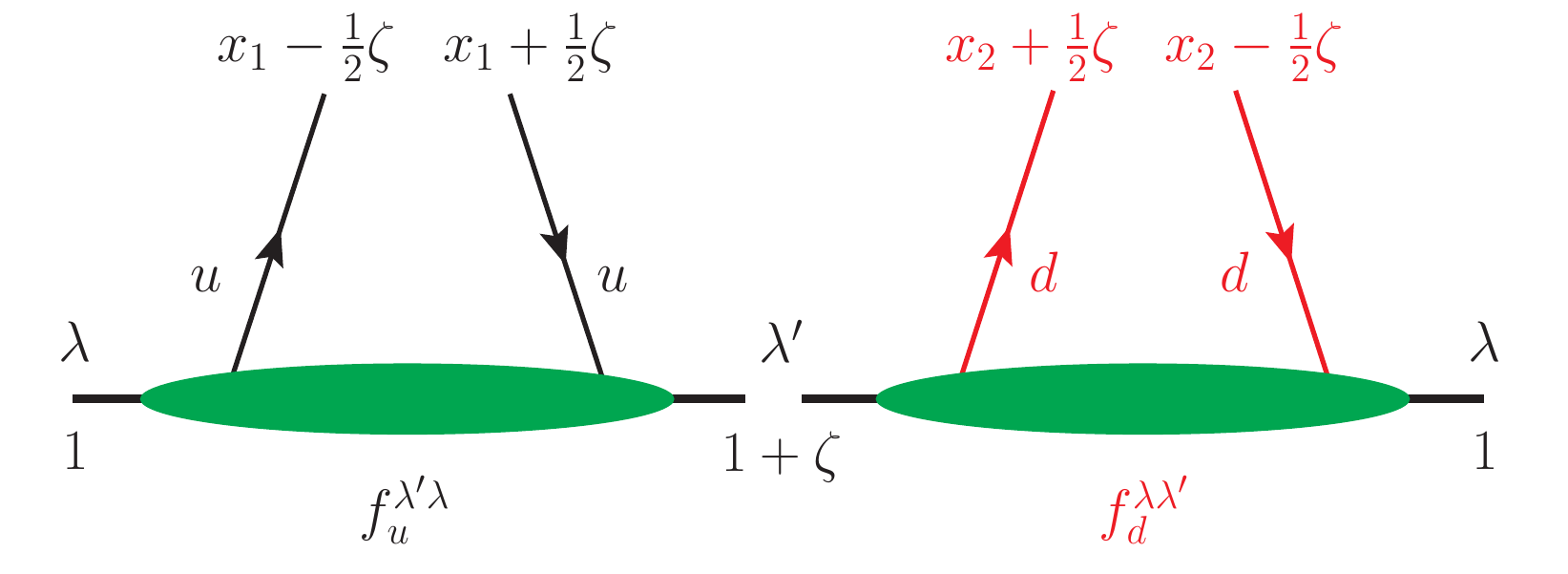}
}
\caption{\label{fig:dpd-approx} Illustration of the approximation of a DPD in terms of GPD matrix elements $f^{\lambda\lambda^\prime}$ for the flavor combination $ud$. Panel (a) shows the factorization ansatz to be used for the case $\zeta > 0$, whereas (b) depicts the variant that we employ if $\zeta < 0$.}
\end{figure}
This factorization is shown pictorially in \fig\ref{fig:dpd-approx-ud} for the flavor combination $ud$. In the following we concentrate on the case of two unpolarized quarks or two longitudinally polarized quarks. In these cases, the GPD matrix elements can be decomposed in terms of the GPDs $H$ and $E$ or $\tilde{H}$ and $\tilde{E}$, respectively. For details we refer to equation (14) in \cite{Diehl:2003ny}. The polarization sum in \eqref{eq:F_fact} can be replaced by:

\begin{align}
\label{eq:gpd_prod_spin_sum_VV}
&\frac{1}{2}\sum_{\lambda \lambda^\prime} 
	f_q^{\lambda \lambda^\prime}(\bar{x}_1,-\xi,\tvec{0},-\tvec{r})\ 
	f_{q^\prime}^{\lambda^\prime \lambda}
		(\bar{x}_2,\xi,-\tvec{r},\tvec{0}) 
= 
	(1-\xi^2)\ H_q(\bar{x}_1,-\xi,t)\ H_{q^\prime}(\bar{x}_2,\xi,t) 
	\nonumber \\
&\qquad 
	- \xi^2 H_q(\bar{x}_1,-\xi,t)\ E_{q^\prime}(\bar{x}_2,\xi,t) 
	- \xi^2 E_q(\bar{x}_1,-\xi,t)\ H_{q^\prime}(\bar{x}_2,\xi,t) 
	\nonumber \\
&\qquad 
	+ \left( 
		\frac{\xi^4}{1-\xi^2} + 
		\frac{1+\xi}{1-\xi} \frac{\tvec{r}^2}{4m^2} 
	\right)  
	E_q(\bar{x}_1,-\xi,t)\ E_{q^\prime}(\bar{x}_2,\xi,t)\,, \\
\label{eq:gpd_prod_spin_sum_AA}
&\frac{1}{2}\sum_{\lambda \lambda^\prime} 
	f_{\Delta q}^{\lambda \lambda^\prime}
		(\bar{x}_1,-\xi,\tvec{0},-\tvec{r})\ 
	f_{\Delta q^\prime}^{\lambda^\prime \lambda}
		(\bar{x}_2,\xi,-\tvec{r},\tvec{0}) 
= 
	(1-\xi^2)\ 
	\tilde{H}_q(\bar{x}_1,-\xi,t)\ 
	\tilde{H}_{q^\prime}(\bar{x}_2,\xi,t) \nonumber \\
&\qquad 
	- \xi^2
		\tilde{H}_q(\bar{x}_1,-\xi,t)\ 
		\tilde{E}_{q^\prime}(\bar{x}_2,\xi,t) 
	- \xi^2
		\tilde{E}_q(\bar{x}_1,-\xi,t)\ 
		\tilde{H}_{q^\prime}(\bar{x}_2,\xi,t) \nonumber \\
&\qquad 
	+ \left( 
		\frac{\xi^4}{1-\xi^2} + 
		\frac{1+\xi}{1-\xi}\ \xi^2 \frac{\tvec{r}^2}{4m^2} 
	\right) 
	\tilde{E}_q(\bar{x}_1,-\xi,t)\ 
	\tilde{E}_{q^\prime}(\bar{x}_2,\xi,t)\,,
\end{align}
with $t = t(\zeta,\tvec{r}^2)$ from \eqref{eq:gpdvar}. Notice that for $\xi = 0$ the cross terms between $H$ and $E$ in \eqref{eq:gpd_prod_spin_sum_VV}, as well as the last three terms in \eqref{eq:gpd_prod_spin_sum_AA} vanish. This is the case if the skewness parameter $\zeta$ is zero. For that case, the expressions in \eqref{eq:gpd_prod_spin_sum_VV} and \eqref{eq:gpd_prod_spin_sum_AA} have already been derived in \cite{Diehl:2011yj}, see equations (4.48) and (4.49) therein. 

Before we continue, we have to discuss an issue regarding the support region \wrt $x_i$ and $\zeta$, which is different on the two sides of \eqref{eq:F_fact}. On the \rhs the support region is constrained by $-1+\zeta/2 < x_i < 1-\zeta/2$, whereas on the \lhs it is given by \eqref{eq:support-region}. Except for the case where $\zeta = 1$, the two regions are distinct. Their mismatch is even more pronounced if $\zeta < 0$. For this reason, we derive an alternative factorization formula by commuting the two operators in the two-current matrix element. Following the same steps as in the derivation of \eqref{eq:F_fact}, we obtain:

\begin{align}
F_{a_1 a_2}(x_1,x_2,\zeta,\tvec{y}) &\stackrel{?}{=} 
	\frac{1}{2(1+\zeta)} 
	\int \frac{\dd^2 \tvec{r}}{(2\pi)^2}
	e^{-i\tvec{r}\tvec{y}} \sum_{\lambda \lambda^\prime} 
	f_{a_2}^{\lambda\lambda^\prime}
		(\bar{x}_2^\prime,-\xi^\prime,\tvec{0},-\tvec{r})\ 
	f_{a_1}^{\lambda^\prime \lambda}
		(\bar{x}_1^\prime,\xi^\prime,-\tvec{r},\tvec{0})\,, \nonumber \\
\bar{x}^\prime_i &= 
	\frac{x_i}{1+\frac{\zeta}{2}}\,,
\qquad 
\xi^\prime = 
	-\frac{\zeta}{2+\zeta}\,.
\label{eq:F_fact2}
\end{align}
The corresponding support regions show the same relative behavior as for \eqref{eq:F_fact} and $\zeta > 0$. Hence, we shall use \eqref{eq:F_fact} for $\zeta > 0$ and \eqref{eq:F_fact2} if $\zeta < 0$ for the following calculations. A graphical representation of \eqref{eq:F_fact2} can be found in \fig\ref{fig:dpd-approx-du}.
Taking the first Mellin moments on both sides in \eqref{eq:F_fact}, we find

\begin{align}
I_{a_1 a_2}(\zeta,-\tvec{y}^2) &\stackrel{?}{=} 
	\frac{(1-\frac{\zeta}{2})^2}{2(1-\zeta)} 
	\int \frac{\dd^2 \tvec{r}}{(2\pi)^2}
	e^{-i\tvec{r}\tvec{y}} 
	\int \dd x_1 \int \dd x_2 
	\nonumber \\
	&\qquad\times 
	\sum_{\lambda \lambda^\prime}\ 
	f_{a_1}^{\lambda \lambda^\prime}(x_1,-\xi,\tvec{0},-\tvec{r} )\ 
	f_{a_2}^{\lambda^\prime \lambda}(x_2,\xi,-\tvec{r},\tvec{0})\,.
\label{eq:I_fact}
\end{align}
and an analogous expression for \eqref{eq:F_fact2}. The integrals over $x_i$ of the corresponding GPD matrix elements can be expressed in terms of the Pauli and Dirac form factors $F_1$ and $F_2$ (for $f_q$) or the axial and pseudoscalar\footnote{This is also called the \emph{induced} pseudoscalar form factor.} form factors $g_A$ and $g_P$ (for $f_{\Delta q}$), which are the lowest Mellin moments of the GPDs $H$ and $E$ or $\tilde{H}$ and $\tilde{E}$, respectively. Since the GPDs are invariant under rotations in the transverse plane, we can evaluate the angular part of the $\tvec{r}$-integral. Considering $I_{qq^\prime}$ or $I_{\Delta q \Delta q^\prime}$ and inserting $\zeta=0$ we can write:

\begin{align}
\label{eq:facttest-IVV}
I_{qq^\prime}(\zeta=0,-\tvec{y}^2) &\stackrel{?}{=} 
	\int \frac{\dd r}{2\pi}\ r J_0(ry) 
	\left[ 
		F_1^q(-\tvec{r}^2)\ F_1^{q^\prime}(-\tvec{r}^2) +
		\frac{\tvec{r}^2}{4m^2}
		F_2^q(-\tvec{r}^2)\ F_2^{q^\prime}(-\tvec{r}^2)  
	\right]\,, \\
\label{eq:facttest-IAA}
I_{\Delta q \Delta q^\prime}(\zeta=0,-\tvec{y}^2) &\stackrel{?}{=} 
	\int \frac{\dd r}{2\pi}\ r J_0(ry)\ 
	g_A^q(-\tvec{r}^2)\ g_A^{q^\prime}(-\tvec{r}^2) \,
\end{align}
with the Bessel function $J_0$. The validity of the equations \eqref{eq:facttest-IVV} and \eqref{eq:facttest-IAA} is one subject to be investigated in this section. Another relation can be derived by using \eqref{eq:py-zero-fct} and performing the angular part of the $\tvec{r}$-integral in \eqref{eq:I_fact}. This yields:

\begin{align}
&A_{a_1 a_2}(py=0,-\tvec{y}^2) \stackrel{?}{=} 
	\frac{1}{2\pi^2} \int_0^1 \dd \zeta\ 
	\frac{(1-\frac{\zeta}{2})^2}{2(1-\zeta)} 
	\int \dd r\ r J_0 (y r) 
	\nonumber \\
&\quad 
	\times 
	\int \dd x_1\ \int \dd x_2\ 
	\sum_{\lambda \lambda^\prime} 
	f_{a_1}^{\lambda \lambda^\prime} (x_1,-\xi,\tvec{0},-\tvec{r})\ 
	f_{a_2}^{\lambda^\prime \lambda} (x_2,\xi,-\tvec{r},\tvec{0})\,.
\label{eq:fact-test-A}
\end{align}
Considering $A_{qq}$ and $A_{\Delta q \Delta q^\prime}$ and replacing the integrals over $x_i$ of the GPD matrix elements by $F_{1}$, $F_{F_2}$, $g_{A}$, or $g_{P}$, we arrive at:

\begin{align}
\label{eq:facttest-AVV}
&A_{qq^\prime}(py=0,-\tvec{y}^2) \stackrel{?}{=} 
	\frac{1}{2\pi^2} \int_0^1 \dd \zeta\ 
	\frac{(1-\frac{\zeta}{2})^2}{1-\zeta} 
	\int \dd r\ r J_0 (y r) 
	\left[ 
		K_1(\zeta)\ F_1^q(t)\ F_1^{q^\prime}(t) 
		\vphantom{\frac{\tvec{r}^2}{4m^2}} 
	\right. \nonumber \\
&\qquad\left. 
	- K_2(\zeta) 
		\left( 
			F_1^{q}(t)\ F_2^{q^\prime}(t) + 
			F_1^{q^\prime}(t)\ F_2^{q}(t) 
		\right)
	+ \left( 
		K_3(\zeta) + K_4(\zeta) \frac{\tvec{r}^2}{4m^2} 
	\right) F_2^q(t)\ F_2^{q^\prime}(t)  
\right]\,, \\
\label{eq:facttest-AAA}
&A_{\Delta q \Delta q^\prime}(py=0,-\tvec{y}^2) \stackrel{?}{=} 
	\frac{1}{2\pi^2} \int_0^1 \dd \zeta\ 
	\frac{(1-\frac{\zeta}{2})^2}{1-\zeta} 
	\int \dd r\ r J_0 (y r) 
	\left[ 
		K_1(\zeta)\ g_A^q(t)\ g_A^{q^\prime}(t)
		\vphantom{\frac{\tvec{r}^2}{4m^2}}
	\right. \nonumber \\ 
&\qquad\left. 
	- K_2(\zeta) 
		\left( 
			g_A^{q}(t)\ g_P^{q^\prime}(t) + 
			g_A^{q^\prime}(t)\ g_P^{q}(t) 
		\right)
	+ \left( 
		K_3(\zeta) + K_5(\zeta) \frac{\tvec{r}^2}{4m^2} 
	\right) 
	g_P^q(t)\ g_P^{q^\prime}(t) 
\right]\,,
\end{align}
where $t$ is a function of $\zeta$ and $\tvec{r}^2$ as defined in \eqref{eq:gpdvar}, and

\begin{align}
\label{eq:factZ}
K_1(\zeta) &:= 1 - K_2(\zeta)\,, 
&\ 
&K_2(\zeta)~:=~\frac{\zeta^2}{(2-\zeta)^2}\,, 
&\ 
&K_3(\zeta)~:=~\frac{(K_2(\zeta))^2}{K_1(\zeta)}\,, 
\nonumber \\
K_4(\zeta) &:= \frac{1}{1-\zeta}\,, 
&\ &K_5(\zeta)~:=~K_2(\zeta)\ K_4(\zeta)\,.
\end{align}
In our lattice study we obtained data for the \lhs of \eqref{eq:facttest-AVV}, \eqref{eq:facttest-AAA}, \eqref{eq:facttest-IVV}, and \eqref{eq:facttest-IAA}. In the remainder of this section we investigate differences relative to the corresponding factorized expressions given on the r.h.s.. These can be calculated form the nucleon form factors, which can be evaluated in lattice studies.

\subsection{The nucleon form factor}
\label{sec:nucl_ff}

As already mentioned, the nucleon form factors as functions of the virtuality $t$ can be obtained from lattice calculations. In this study we use the form factor data \cite{Wurm:privcom} which has been generated in the simulation described in \cite{Bali:2019yiy}. In that work various gauge ensembles have been investigated; we take the form factor data for gauge ensemble H102, which is the same ensemble that is used in our DPD study. The form factor analysis carefully takes account of excited state contributions. The absolute value of the largest initial proton momentum that has been used is $|\mvec{p}| = \sqrt{6}\cdot 2\pi/(La) \approx 1.11~\mathrm{GeV}$. Notice that the final momentum is set to $\mvec{p}^\prime = 0$. In this setup, the largest available virtuality is $t = -\Delta^2 \approx 1.02~\mathrm{GeV}^2$.

In order to evaluate the integrals \eqref{eq:facttest-AVV}, \eqref{eq:facttest-AAA}, \eqref{eq:facttest-IVV}, and \eqref{eq:facttest-IAA}, we need to extrapolate the lattice results in $t$. To this end, we fit the form factor data to a power law of the form

\begin{align}
\label{eq:ff-fit-ansatz}
F(t) = \frac{F(0)}{\left(1-\frac{t}{M^2}\right)^n}\,,
\end{align}
\begin{table}
\begin{center}
\begin{tabular}{c|ccc|c}
\hline
\hline
form factor & $F(0)$ & $M^2[\mathrm{GeV}^2]$ & $n$(fixed) & $\chi^2/\mathrm{dof}$\\ 
\hline
$F_1^u$ & $1.977(12)$ & $1.063(19)$ & $2$ & $1.09$ \\
 & $1.936(11)$ & $1.747(29)$ & $3$ & $1.79$ \\
$F_2^u$ & $1.764(38)$ & $0.982(44)$ & $2$ & $1.63$ \\
 & $1.711(34)$ & $1.674(68)$ & $3$ & $0.52$ \\
\hline
$F_1^d$ & $1.0421(70)$ & $0.766(13)$ & $2$ & $7.15$ \\
 & $1.0035(60)$ & $1.300(19)$ & $3$ & $2.06$ \\
 & $0.9860(57)$ & $1.837(26)$ & $4$ & $0.94$ \\
$F_2^d$ & $-1.744(23)$ & $0.834(19)$ & $2$ & $2.51$ \\
 & $-1.658(20)$ & $1.456(29)$ & $3$ & $1.30$ \\
\hline
\hline
\end{tabular}

\end{center}
\caption{Results for the fit parameters $F(0)$ and $M^2$ obtained from a fit on the data of the Pauli and Dirac form factors using the ansatz \eqref{eq:ff-fit-ansatz} with fixed $n$. The corresponding $\chi^2/\mathrm{dof}$, which takes into account the complete covariance matrix, is also listed. \label{tab:fffits_V}}
%\end{table}
%
%\begin{table}
\begin{center}
\begin{tabular}{c|ccc|c}
\hline
\hline
form factor & $F(0)$ & $M^2[\mathrm{GeV}^2]$ & $n$(fixed) & $\chi^2/\mathrm{dof}$\\ 
\hline
$g_A^u$ & $0.8999(82)$ & $1.971(64)$ & $2$ & $1.61$ \\
 & $0.8920(78)$ & $3.161(97)$ & $3$ & $0.82$ \\
$g_P^u$ & $29.84(94)$ & $0.327(11)$ & $2$ & $0.30$ \\
 & $24.73(62)$ & $0.688(17)$ & $3$ & $1.03$ \\
\hline
$g_A^d$ & $-0.2930(41)$ & $1.800(81)$ & $2$ & $1.05$ \\
 & $-0.2896(39)$ & $2.90(12)$ & $3$ & $0.93$ \\
$g_P^d$ & $-9.62(77)$ & $0.305(27)$ & $2$ & $0.13$ \\
 & $-7.88(49)$ & $0.638(44)$ & $3$ & $0.60$ \\
\hline
\hline
\end{tabular}

\end{center}
\caption{The same as \tab\ref{tab:fffits_V} for the axial and pseudoscalar form factors.\label{tab:fffits_A}}
\end{table}
\begin{figure}
\subfigure[form factor fit, vector current, $n=3$]{
\includegraphics[scale=0.25,trim={0.5cm 1.2cm 0.5cm 2.8cm},clip]{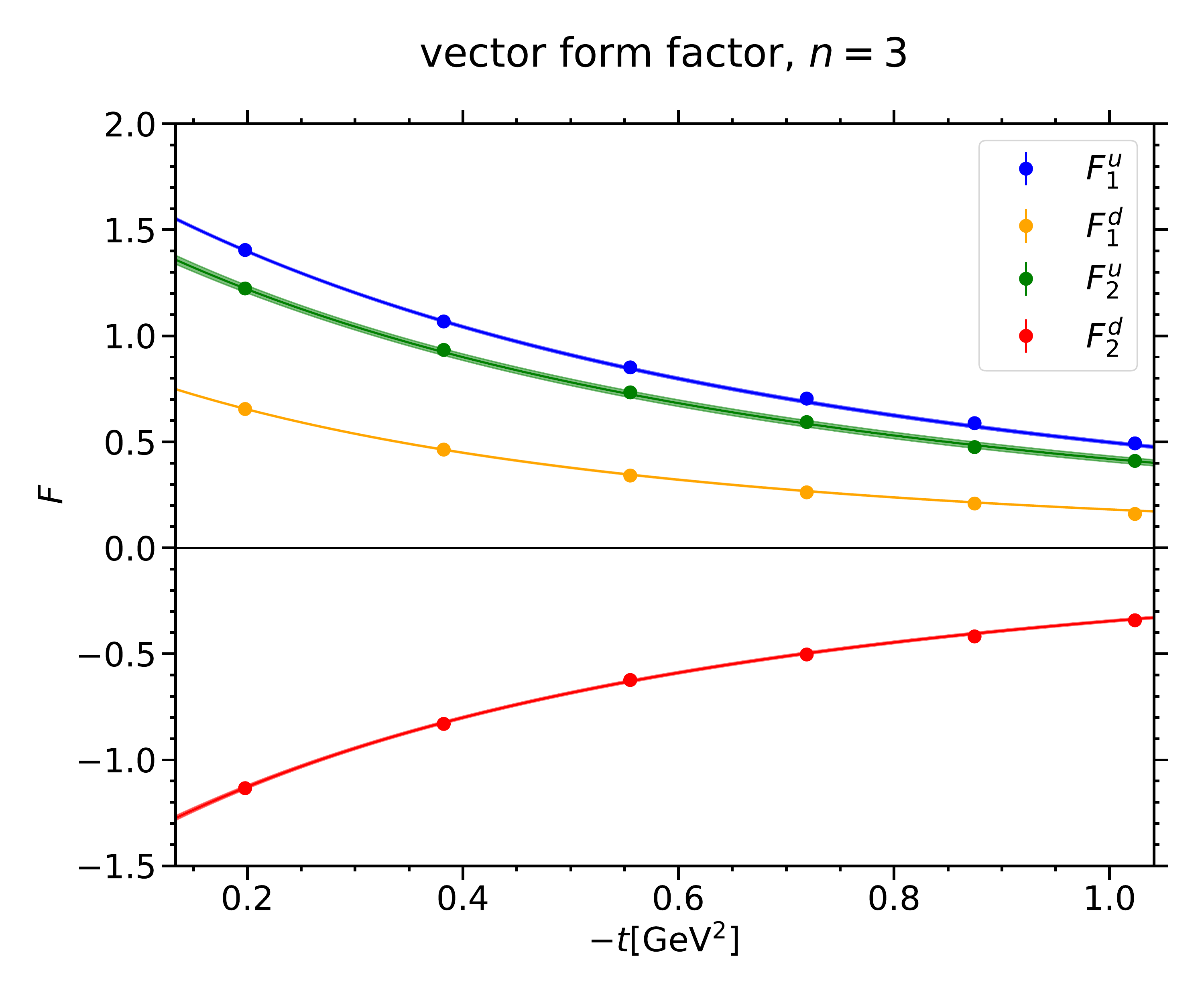}\label{fig:fffits-V}} \hfill
\subfigure[form factor fit, axial vector current, $n=3$]{
\includegraphics[scale=0.25,trim={0.5cm 1.2cm 0.5cm 2.8cm},clip]{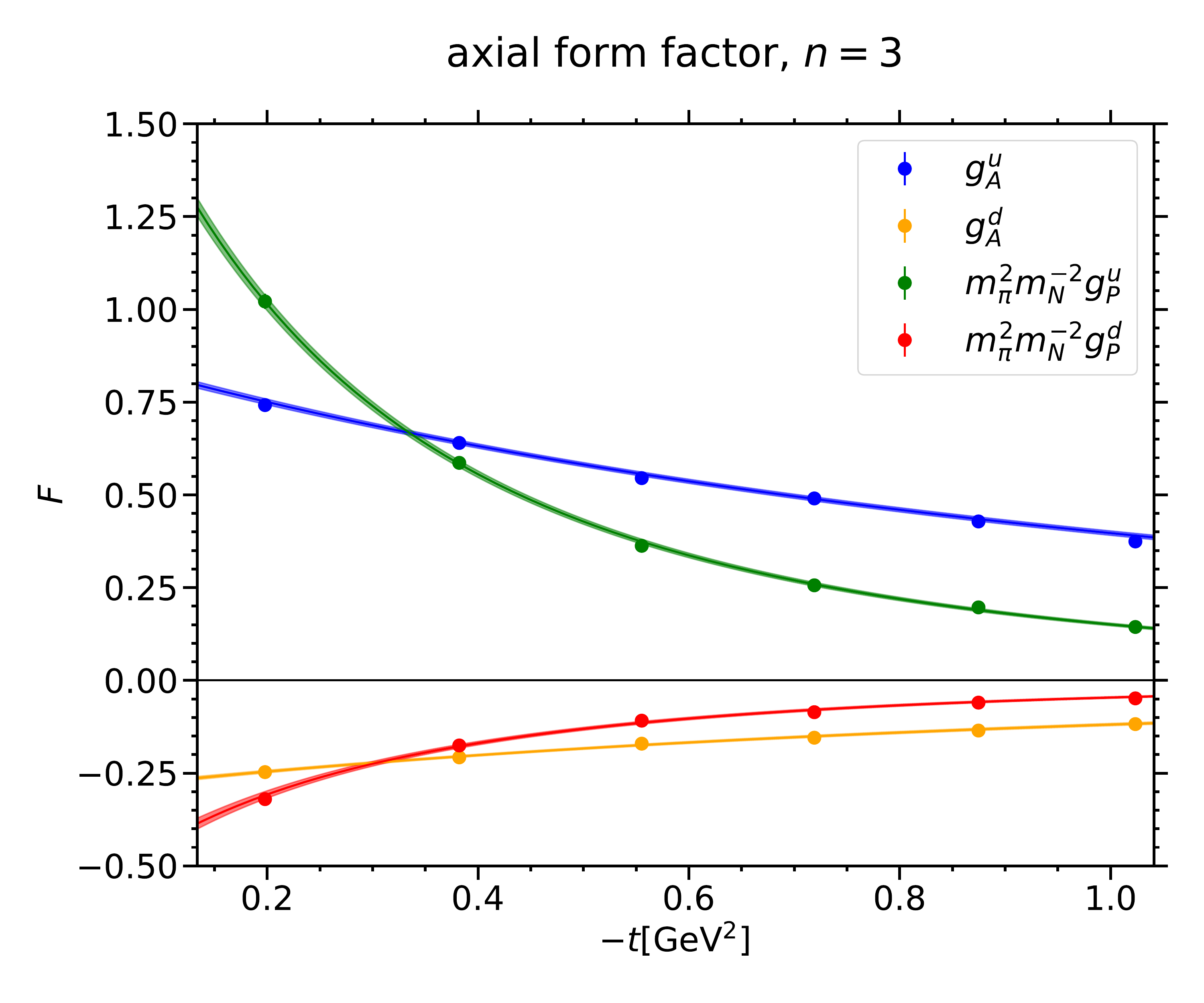}\label{fig:fffits-A}} \\
\caption{$t$-dependence of the form factor data points and the corresponding curves obtained from a fit to the ansatz \eqref{eq:ff-fit-ansatz} with $n=3$. This is shown for the Pauli and Dirac form factors in panel (a), as well as for the axial and pseudoscalar form factors (b). \label{fig:fffits}}
\vspace*{1.5\baselineskip}
%\end{figure}
%
%\begin{figure}
\subfigure[data/fit ratio for $F_1$]{
\includegraphics[scale=0.24,trim={0.5cm 1.2cm 0.5cm 2.8cm},clip]{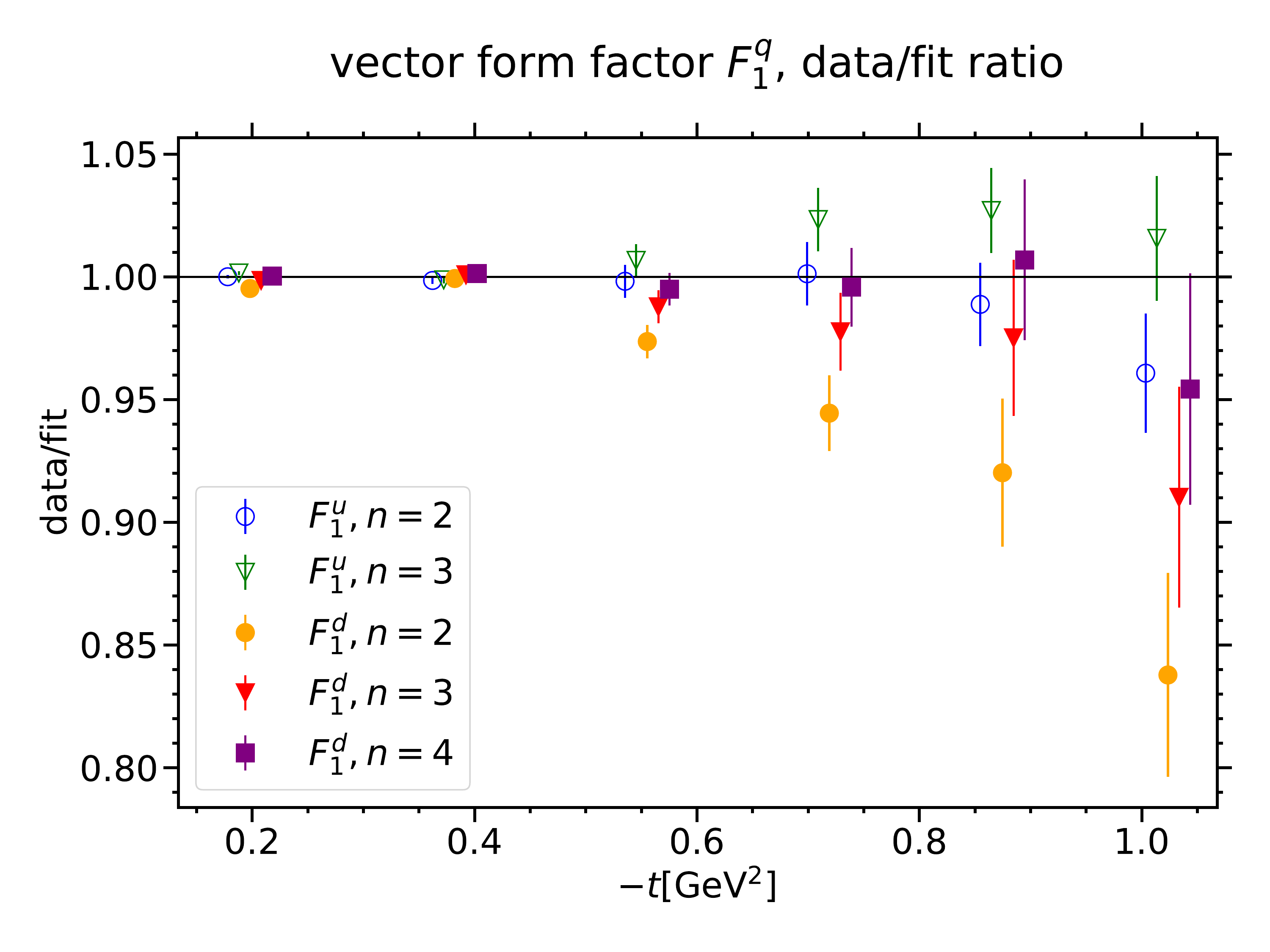}\label{fig:fffits_quality-V-1}} \hfill
\subfigure[data/fit ratio for $F_2$]{
\includegraphics[scale=0.24,trim={0.5cm 1.2cm 0.5cm 2.8cm},clip]{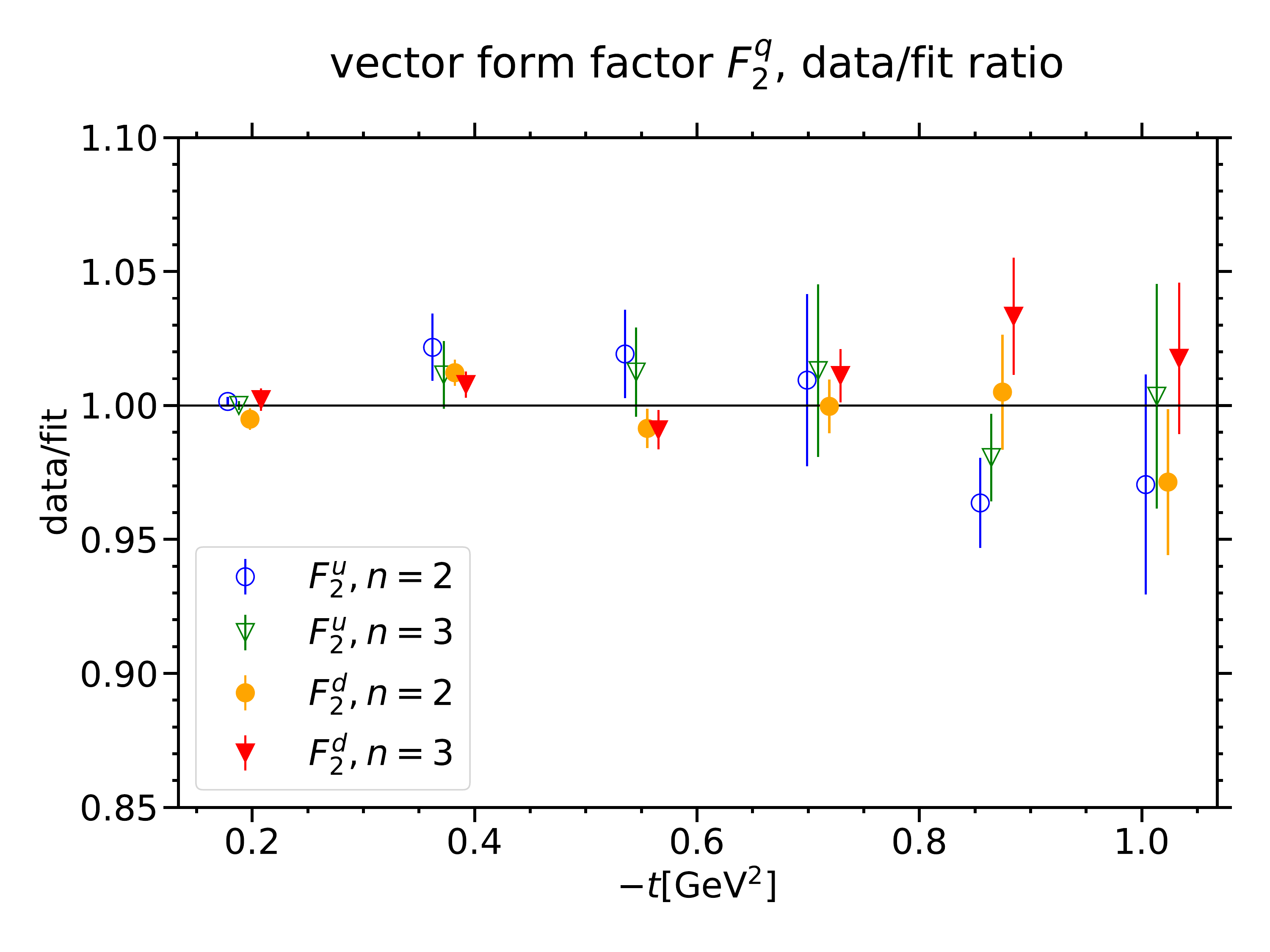}\label{fig:fffits_quality-V-2}} \\
\subfigure[data/fit ratio for $g_A$]{
\includegraphics[scale=0.24,trim={0.5cm 1.2cm 0.5cm 2.8cm},clip]{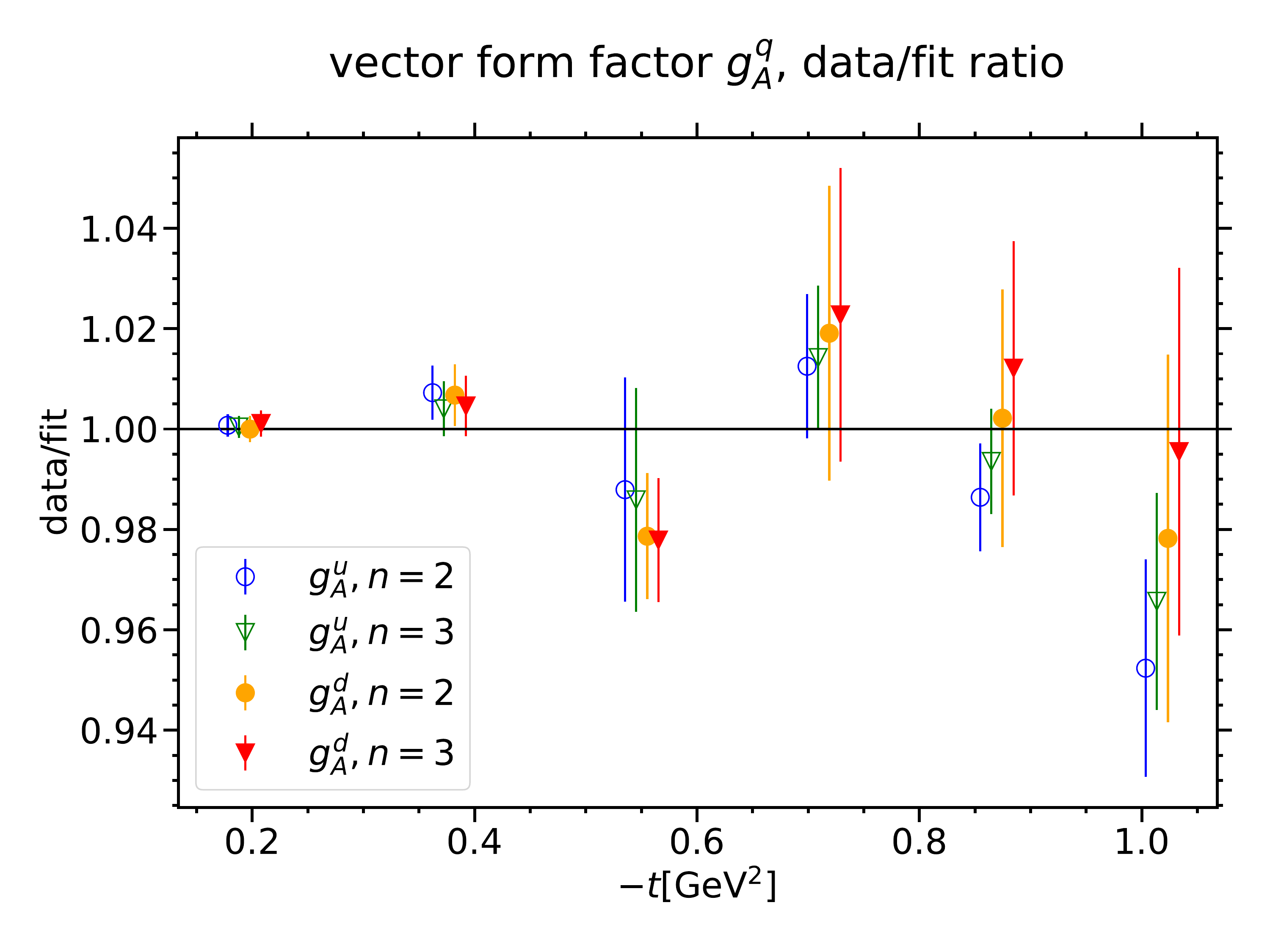}\label{fig:fffits_quality-A-1}} \hfill
\subfigure[data/fit ratio for $g_P$]{
\includegraphics[scale=0.24,trim={0.5cm 1.2cm 0.5cm 2.8cm},clip]{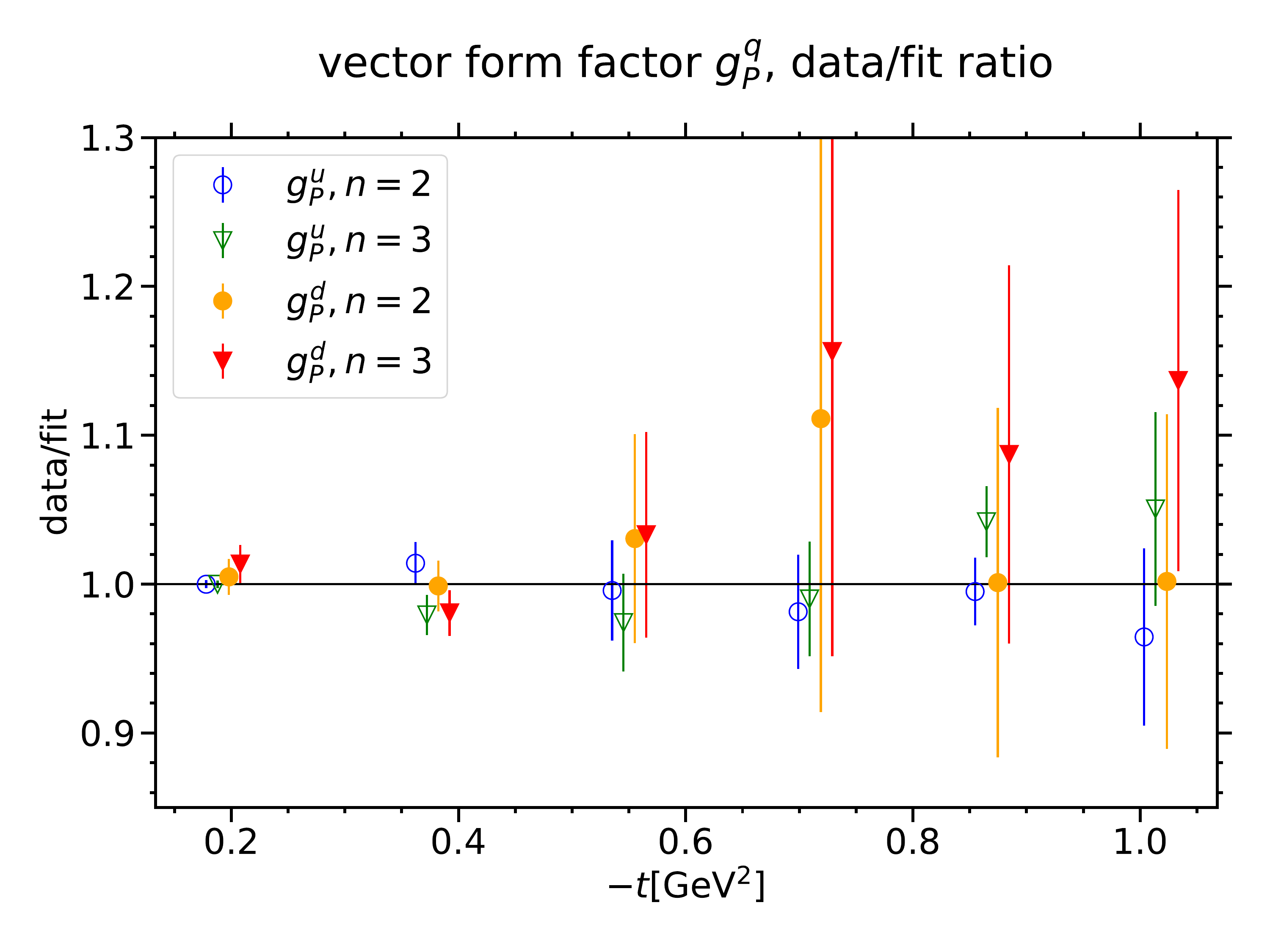}\label{fig:fffits_quality-A-2}} \\
\caption{$t$-dependence of the ratio of the form factor data and the corresponding fit, where we compare results for different values of the exponent $n$. For better distinguishability data points for different fits are shifted with different offsets. The results are shown for $F_1$ (a), $F_2$ (b), $g_A$ (c) and $g_P$ (d) for each flavor. \label{fig:fffits_quality}}
\end{figure}
which is frequently used for parameterization of form factors. For each channel we perform two different fits with fixed values for the exponent, $n=2$ and $n=3$, whereas $F(0)$ and $M$ enter the fit as free fit parameters. The fits are performed employing the complete covariance matrix, \ie taking into account correlations between the data points. The resulting curves are shown together with the form factor data in \fig\ref{fig:fffits} for $n=3$. The corresponding values of the fit parameters and of the $\chi^2/\mathrm{dof}$ are summarized in \tab\ref{tab:fffits_V} (vector current) and \tab\ref{tab:fffits_A} (axial current), respectively. In order to analyze the quality of the fit, we plot for each fit the ratio of the data and the fit value. This is shown in \fig\ref{fig:fffits_quality}.

From most of the fits we obtain a sufficiently good description of the form factor data. The only exception is found for $F_1^d$ and $n=2$, where we observe a relatively large discrepancy between the data and the resulting curve, see \fig\ref{fig:fffits_quality-V-1}. Consequently, the corresponding $\chi^2/\mathrm{dof}$ has the very large value of $7.15$. Hence, we perform an alternative fit using $n=4$, which again yields a reasonable result. For the remainder of this section we discard the fit for $F_1^d$ with $n=2$ and instead use the fit for $n=4$ in this channel.

\subsection{Results}
\label{sec:fact_res}

\begin{figure}
\subfigure[$A_{ud}$]{
\includegraphics[scale=0.25,trim={0.5cm 1.2cm 0.5cm 2.8cm},clip]{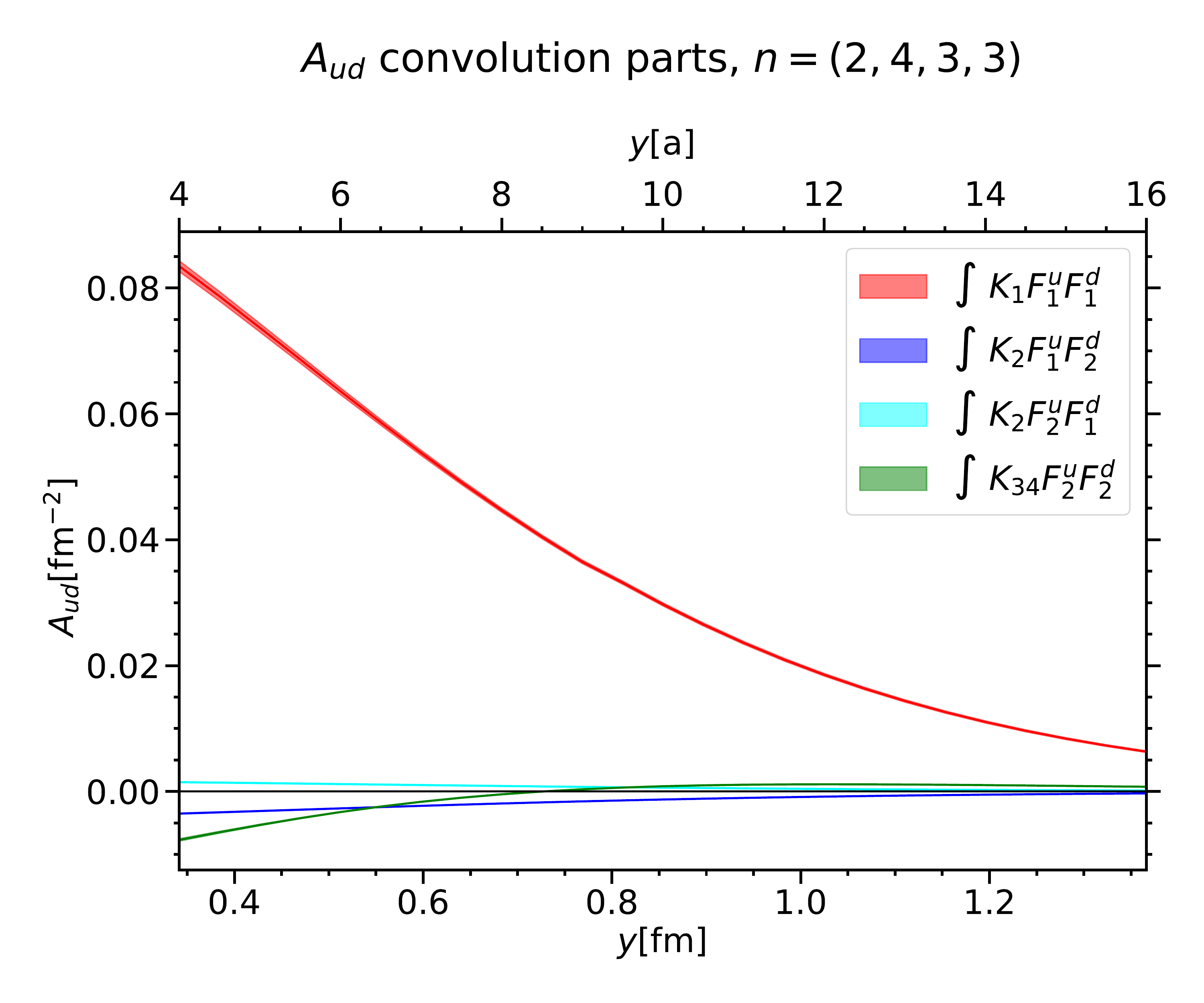}\label{fig:fact_parts-A_VVud}} \hfill
\subfigure[$A_{\Delta u \Delta d}$]{
\includegraphics[scale=0.25,trim={0.5cm 1.2cm 0.5cm 2.8cm},clip]{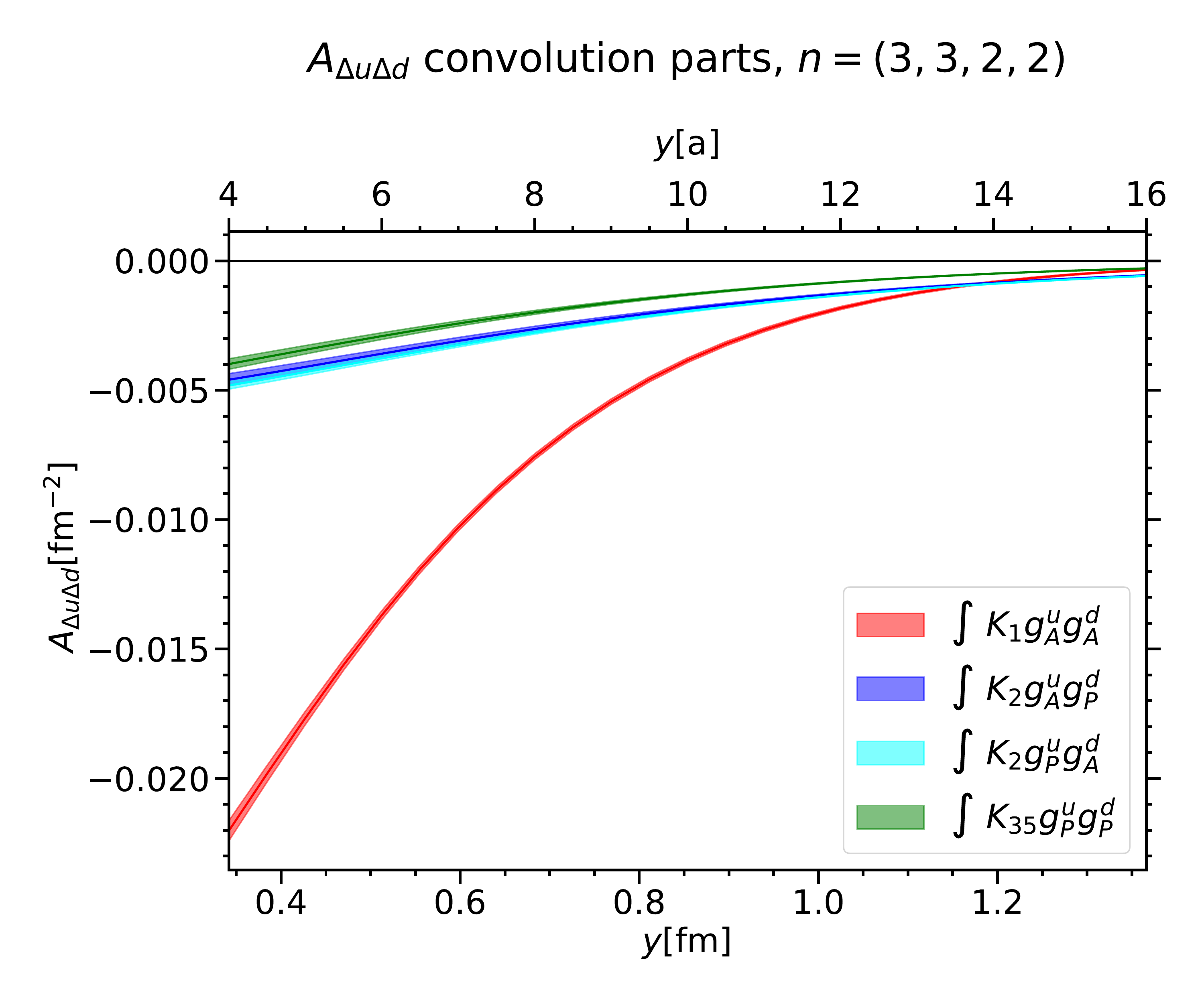}\label{fig:fact_parts-A_AAud}} \\
\subfigure[$A_{uu}$]{
\includegraphics[scale=0.25,trim={0.5cm 1.2cm 0.5cm 2.8cm},clip]{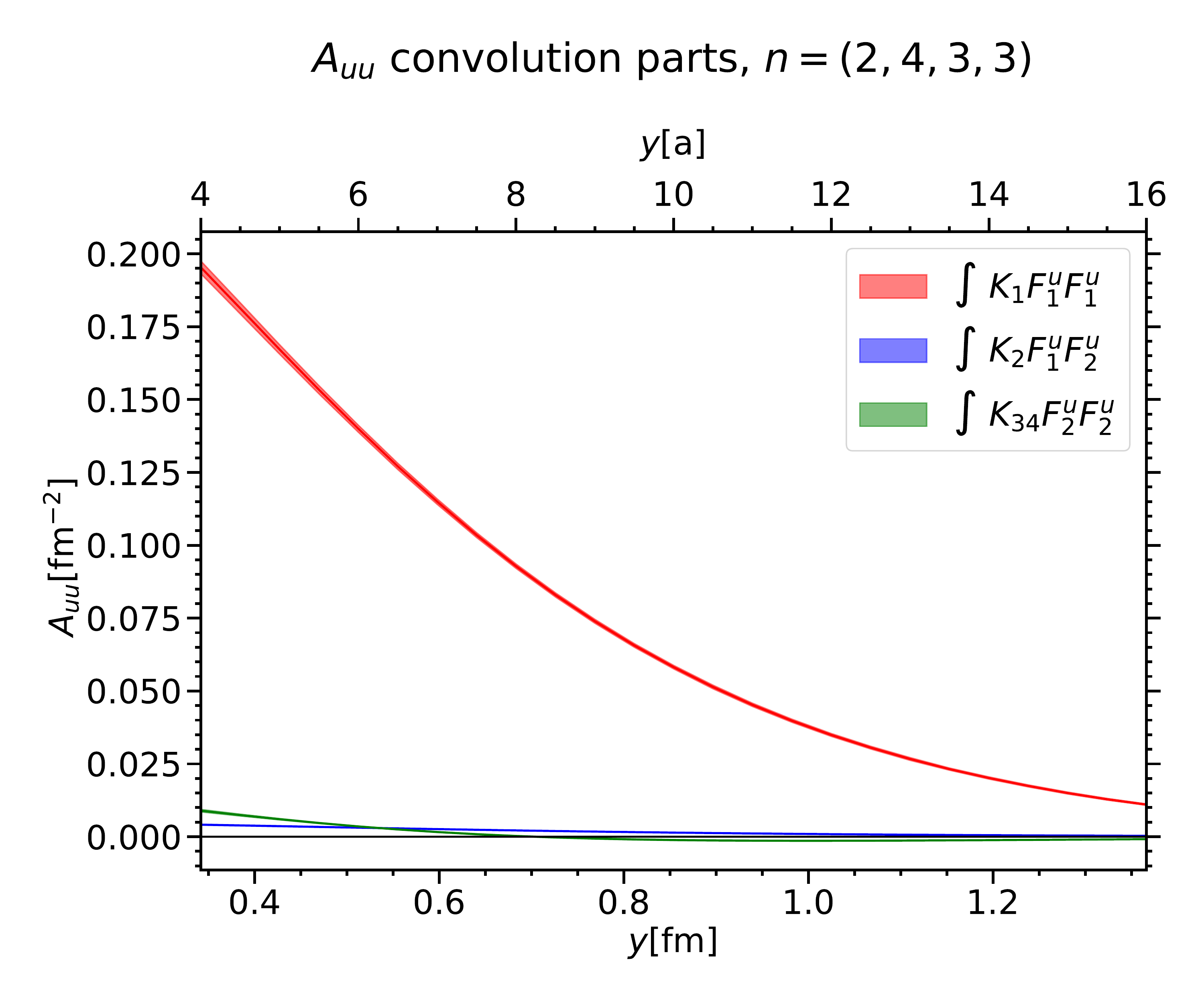}\label{fig:fact_parts-A_VVuu}} \hfill
\subfigure[$I_{ud}$, $\zeta = 0$]{
\includegraphics[scale=0.25,trim={0.5cm 1.2cm 0.5cm 2.8cm},clip]{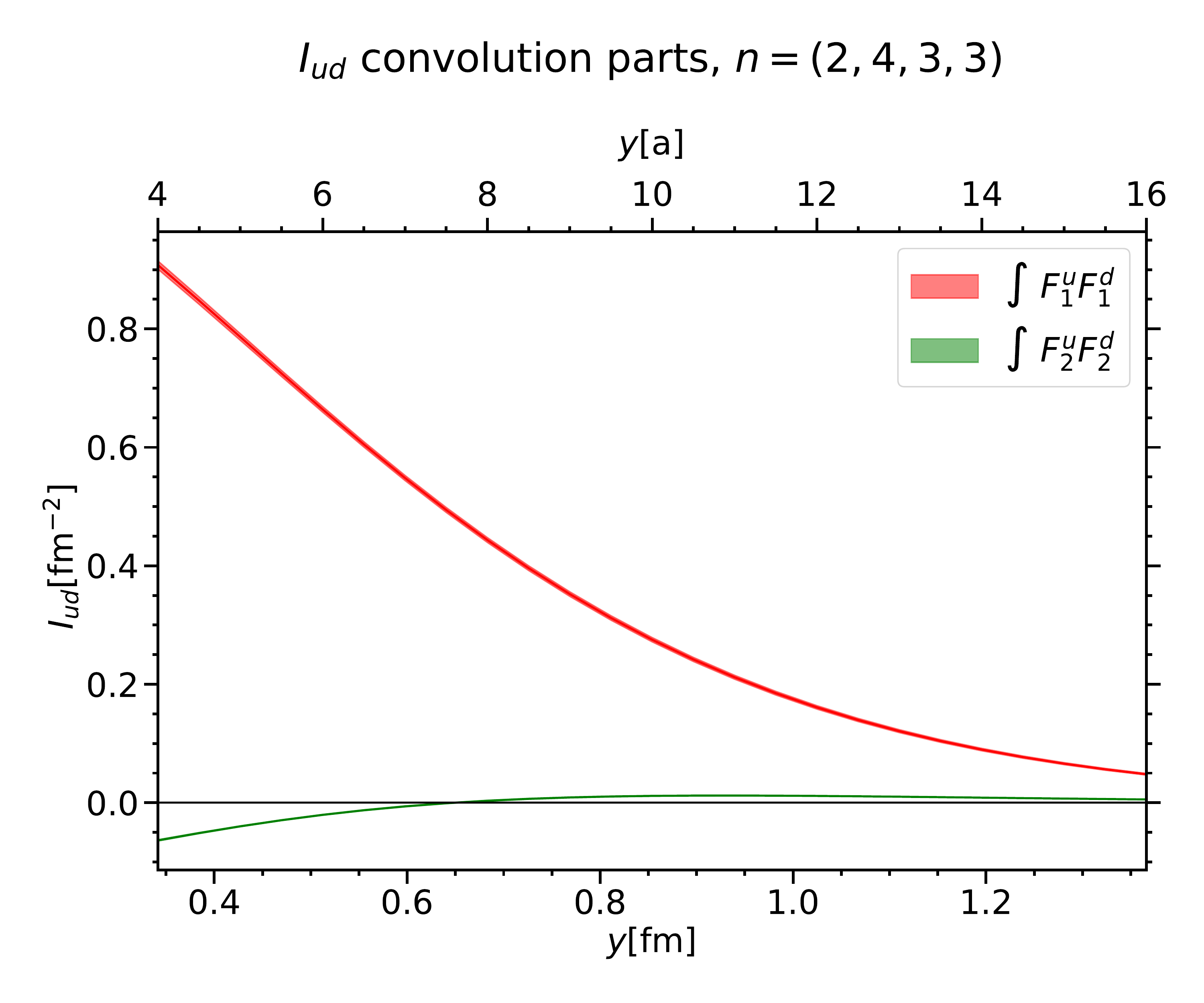}\label{fig:fact_parts-I_VVud}} \\
\caption{Comparison between the different terms contributing to the factorized expressions for the twist-two functions $A_{ud}$, $A_{\Delta u \Delta d}$, $A_{uu}$, and for the Mellin moment $I_{ud}$ at zero skewness $\zeta$. In the keys we use the short notation $K_{34} := K_3(\zeta) + K_4(\zeta)\, \tvec{r}^2 / (4m^2)$ and $K_{35} := K_3(\zeta) + K_5(\zeta)\, \tvec{r}^2 / (4m^2)$.\label{fig:fact_parts}}
\end{figure}

Before comparing the two sides of the factorization formulae \eqref{eq:facttest-AVV}, \eqref{eq:facttest-AAA}, \eqref{eq:facttest-IVV}, and \eqref{eq:facttest-IAA}, let us investigate the different terms on their r.h.s.. In \fig\ref{fig:fact_parts} we compare the size of the integrals over these terms. Notice that the shown results are based on the form factor fits with the smallest $\chi^2$. In the unpolarized channels the $F_1 F_1$-term is found to be dominant, whereas the remaining contributions are very small. As an example we show $A_{ud}$ (a) and $A_{dd}$ (c), as well as $I_{ud}$ (d). In the longitudinally polarized case, the $g_A g_A$-term is also the most relevant one, but the relative size of the other contributions is larger than in the unpolarized cases. This can be observed \eg in the result for $A_{\Delta u \Delta d}$, which is plotted in \fig\ref{fig:fact_parts-A_AAud}. A similar behavior is found in the other channels that are not shown in the plots.

\begin{figure}
\subfigure[$A_{uu}$ vs $\int F_u F_u$, $py = 0$]{
\includegraphics[scale=0.25,trim={0.5cm 1.2cm 0.5cm 2.8cm},clip]{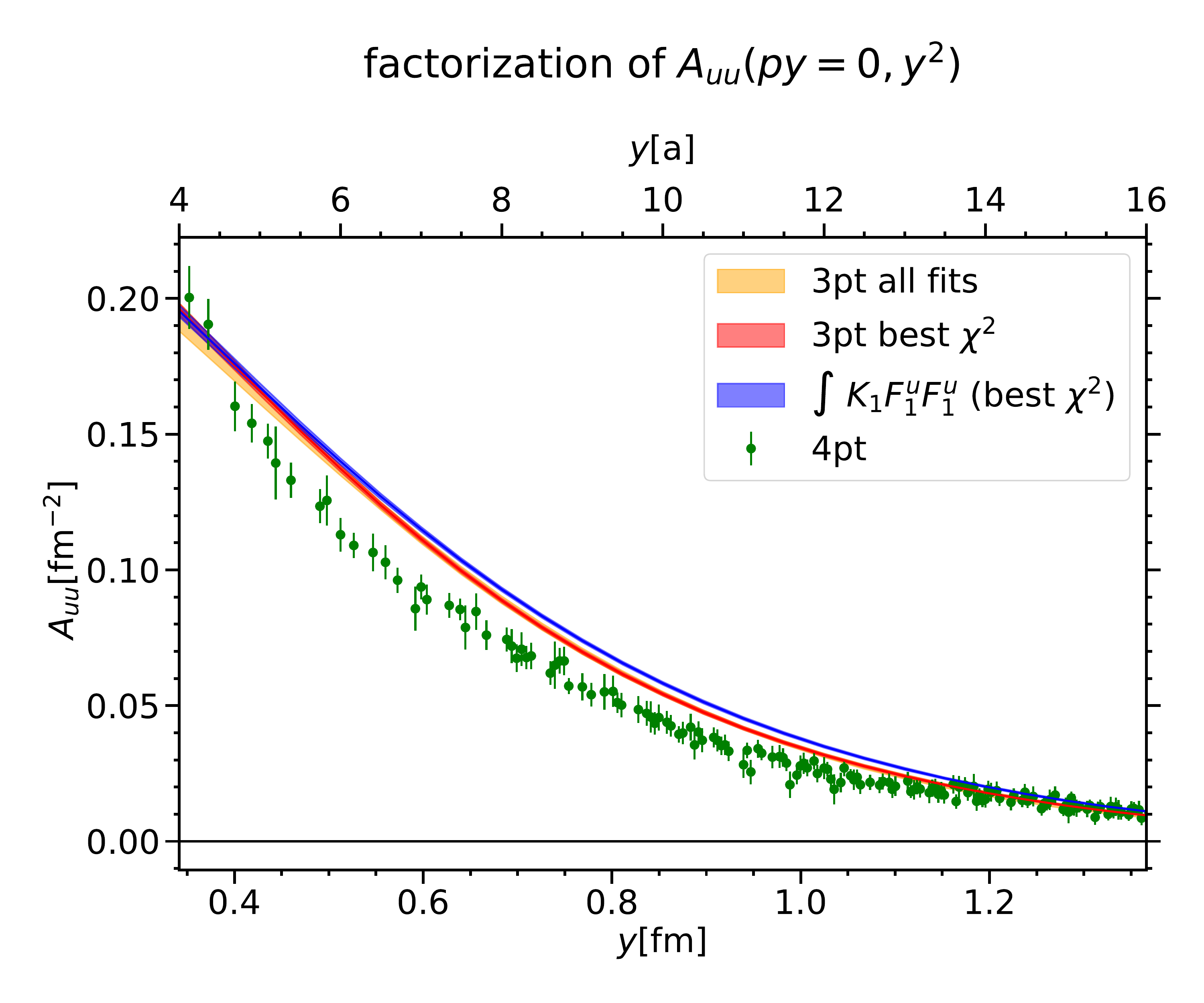}\label{fig:factAVVuduu-A_VVuu}} \hfill
\subfigure[Ratio $\int F_u F_u/A_{uu}$, $py = 0$]{
\includegraphics[scale=0.25,trim={0.5cm 1.2cm 0.5cm 2.8cm},clip]{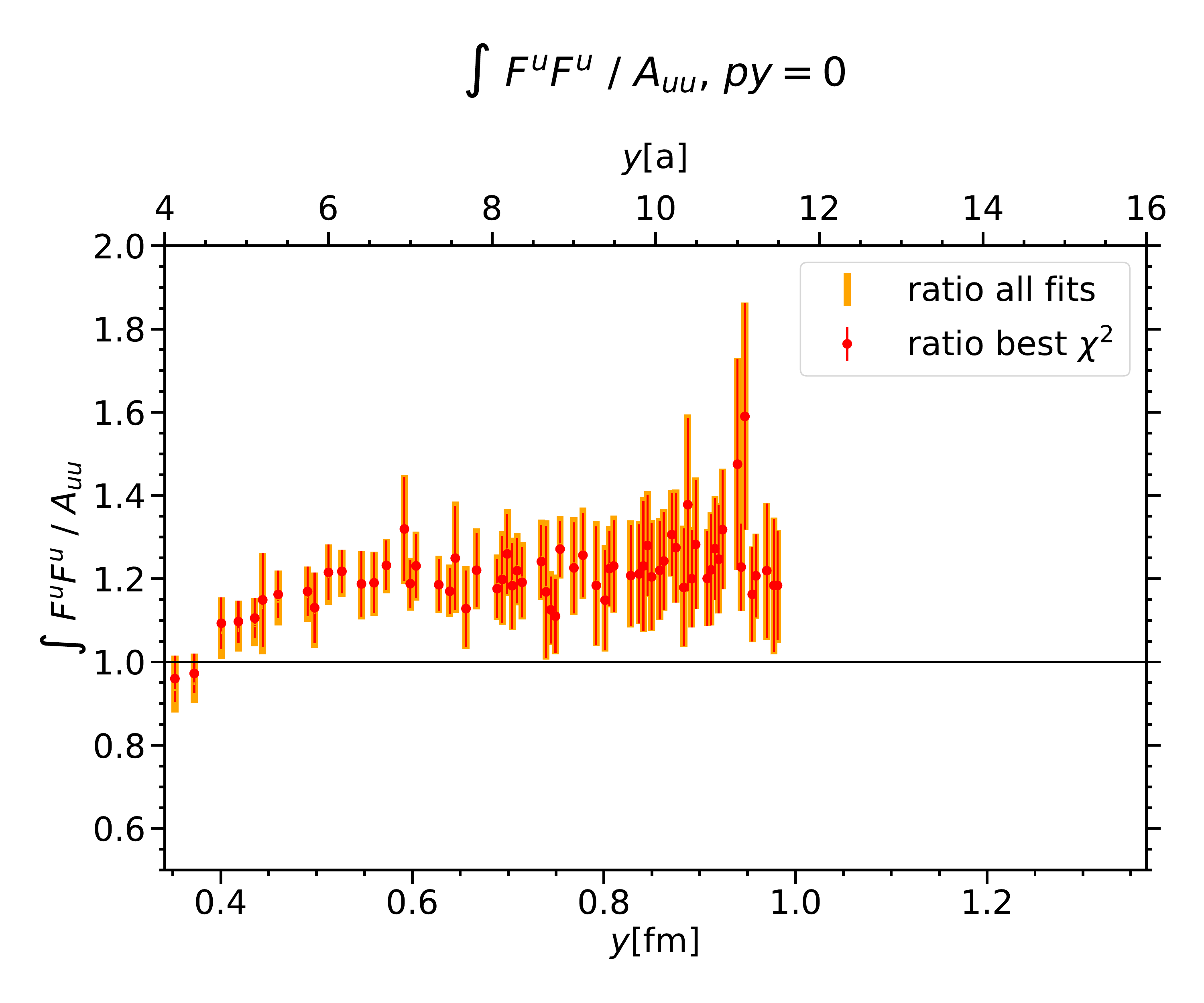}\label{fig:factAVVuduu-A_VVuu-rat}} \\
\subfigure[$A_{ud}$ vs $\int F_u F_d$, $py = 0$]{
\includegraphics[scale=0.25,trim={0.5cm 1.2cm 0.5cm 2.8cm},clip]{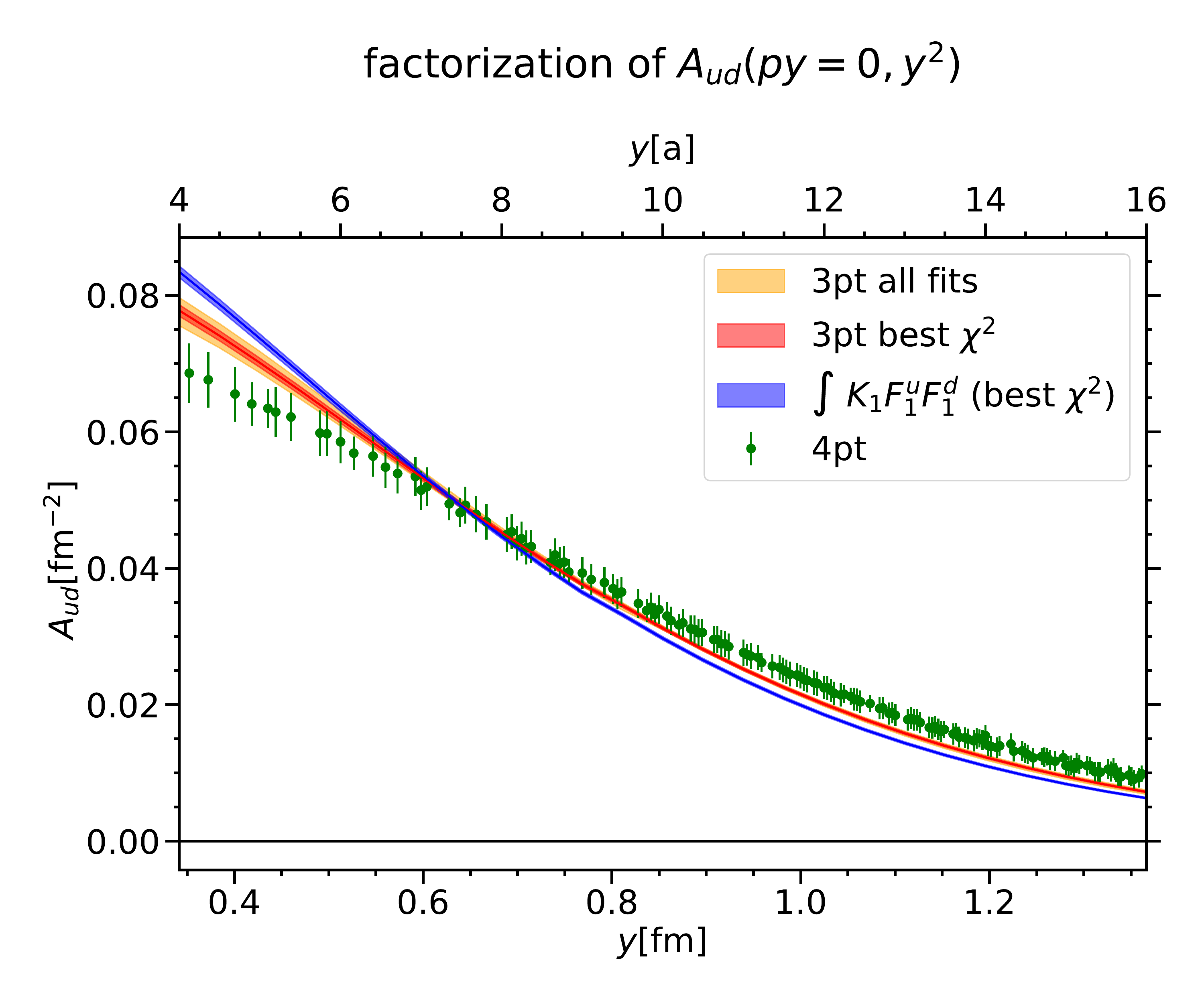}\label{fig:factAVVuduu-A_VVud}} \hfill
\subfigure[Ratio $\int F_u F_d/A_{ud}$, $py = 0$]{
\includegraphics[scale=0.25,trim={0.5cm 1.2cm 0.5cm 2.8cm},clip]{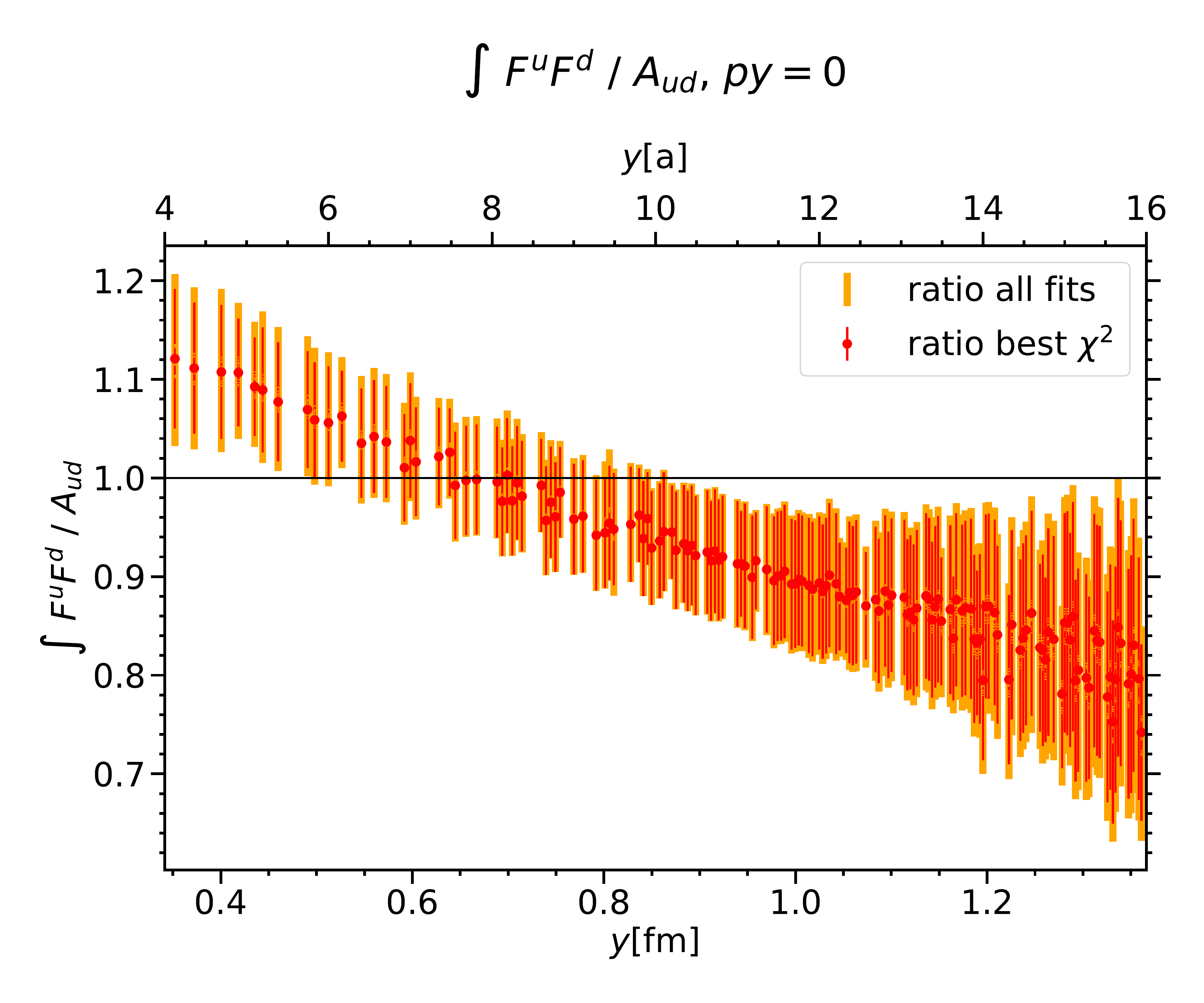}\label{fig:factAVVuduu-A_VVud-rat}} \\
\subfigure[$A_{u\bar{d}}$ vs $\int F_u F_{\bar{d}}$ for $\pi^+$, $py = 0$]{
\includegraphics[scale=0.25,trim={0.5cm 1.2cm 0.5cm 2.8cm},clip]{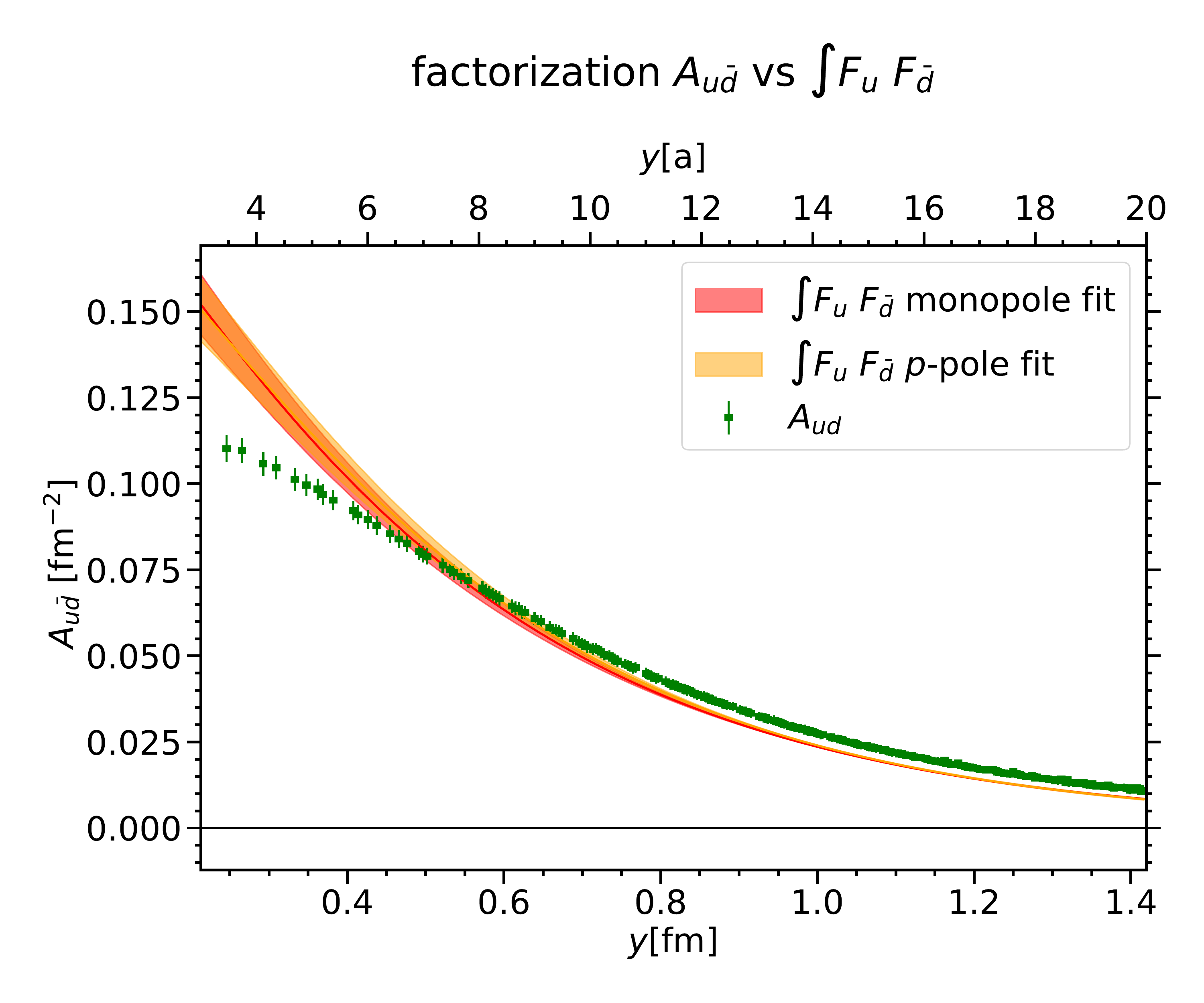}\label{fig:factAVVuduu-A_VVud-pi}} \hfill
\subfigure[Ratio $\int F_u F_{\bar{d}}/A_{u\bar{d}}$ for $\pi^+$, $py = 0$]{
\includegraphics[scale=0.25,trim={0.5cm 1.2cm 0.5cm 2.8cm},clip]{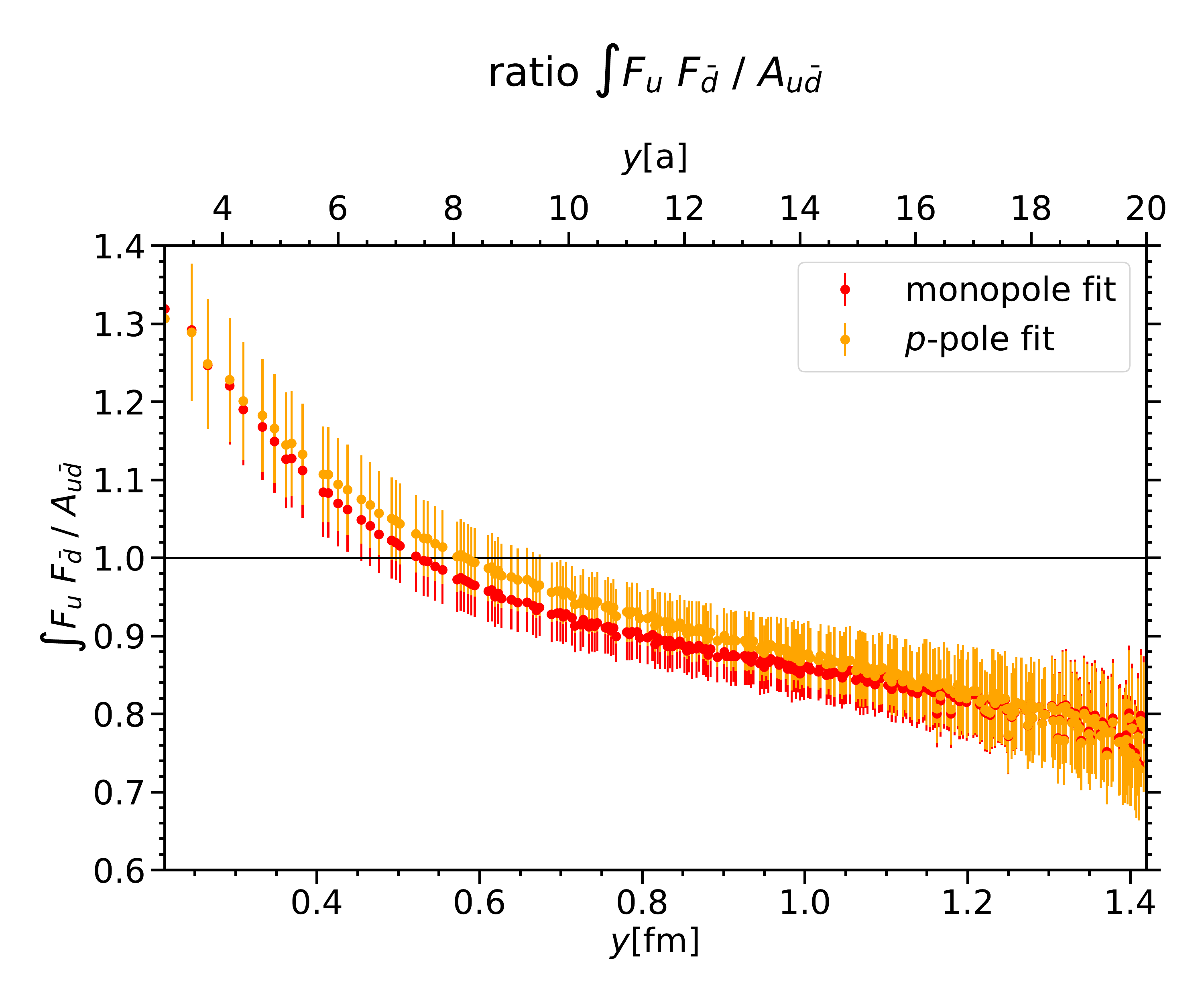}\label{fig:factAVVuduu-A_VVud-rat-pi}} \\
\caption{Left: Comparison of the twist-two functions $A_{uu}$ (a) and $A_{ud}$ (c) (green points) and the factorization results obtained by the integral \eqref{eq:facttest-AVV}. The red curve is obtained from the form factor fits with best $\chi^2/\mathrm{dof}$. The orange band represents the envelope of the error bands of the different fits. Right: Ratio of the form factor integral and the corresponding twist-two functions, again shown for $A_{uu}$ (b) and $A_{ud}$ (d). In the panels (a) and (c) we also present the integration result taking into account only the first term \eqref{eq:facttest-AVV} (blue curve). In panel (e) and (f), we show the corresponding results for the $\pi^+$ obtained in \cite{Bali:2020mij} for two different fits. \label{fig:factAVVuduu}}
\end{figure}

\begin{figure}
\subfigure[$A_{dd}$ vs $\int F_d F_d$, $py = 0$]{
\includegraphics[scale=0.25,trim={0.5cm 1.2cm 0.5cm 2.8cm},clip]{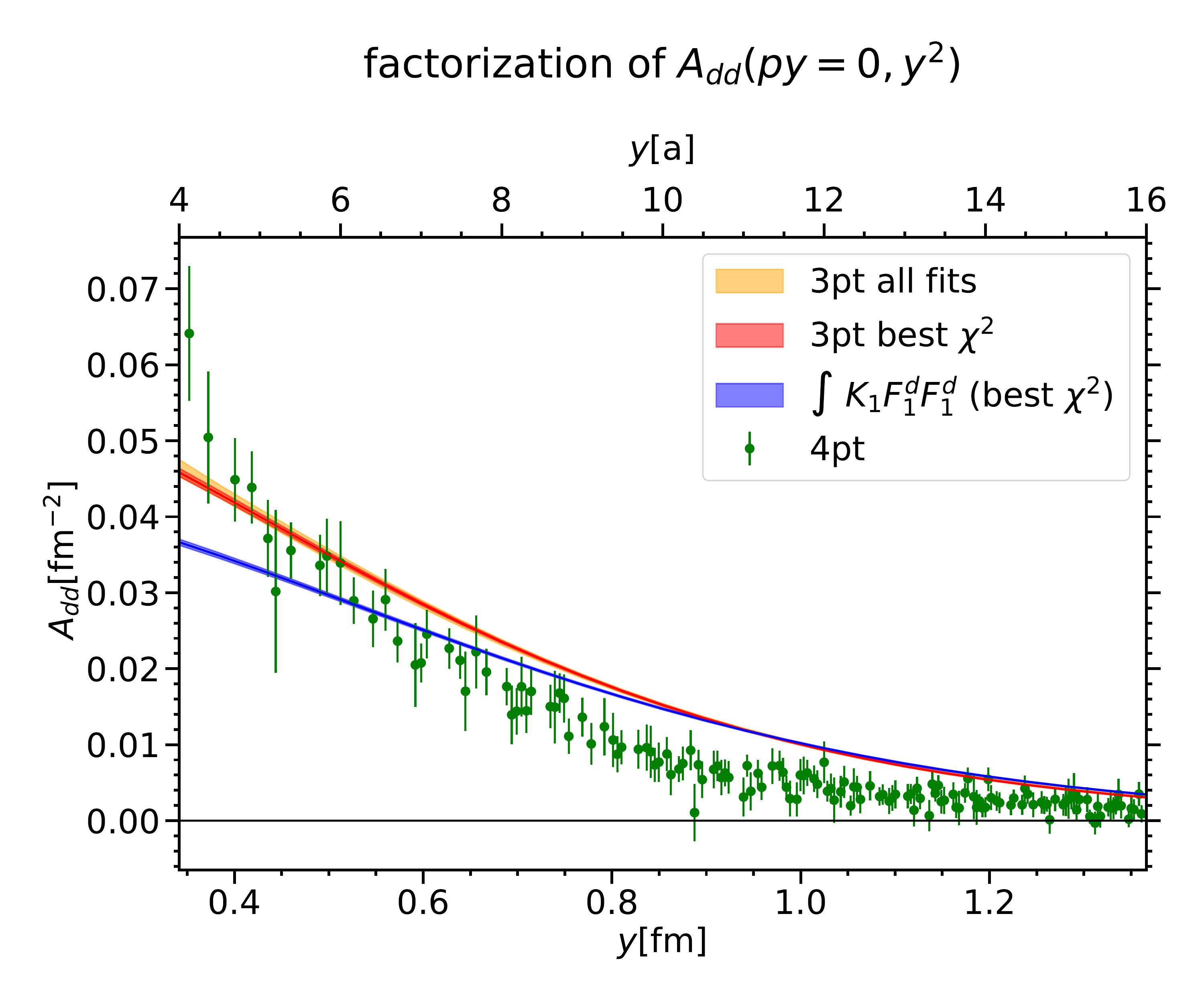}\label{fig:factAVVddAAA-A_VVdd}} \hfill
\subfigure[$A_{\Delta u \Delta d}$ vs $\int g_{u} g_{d}$, $py = 0$]{
\includegraphics[scale=0.25,trim={0.5cm 1.2cm 0.5cm 2.8cm},clip]{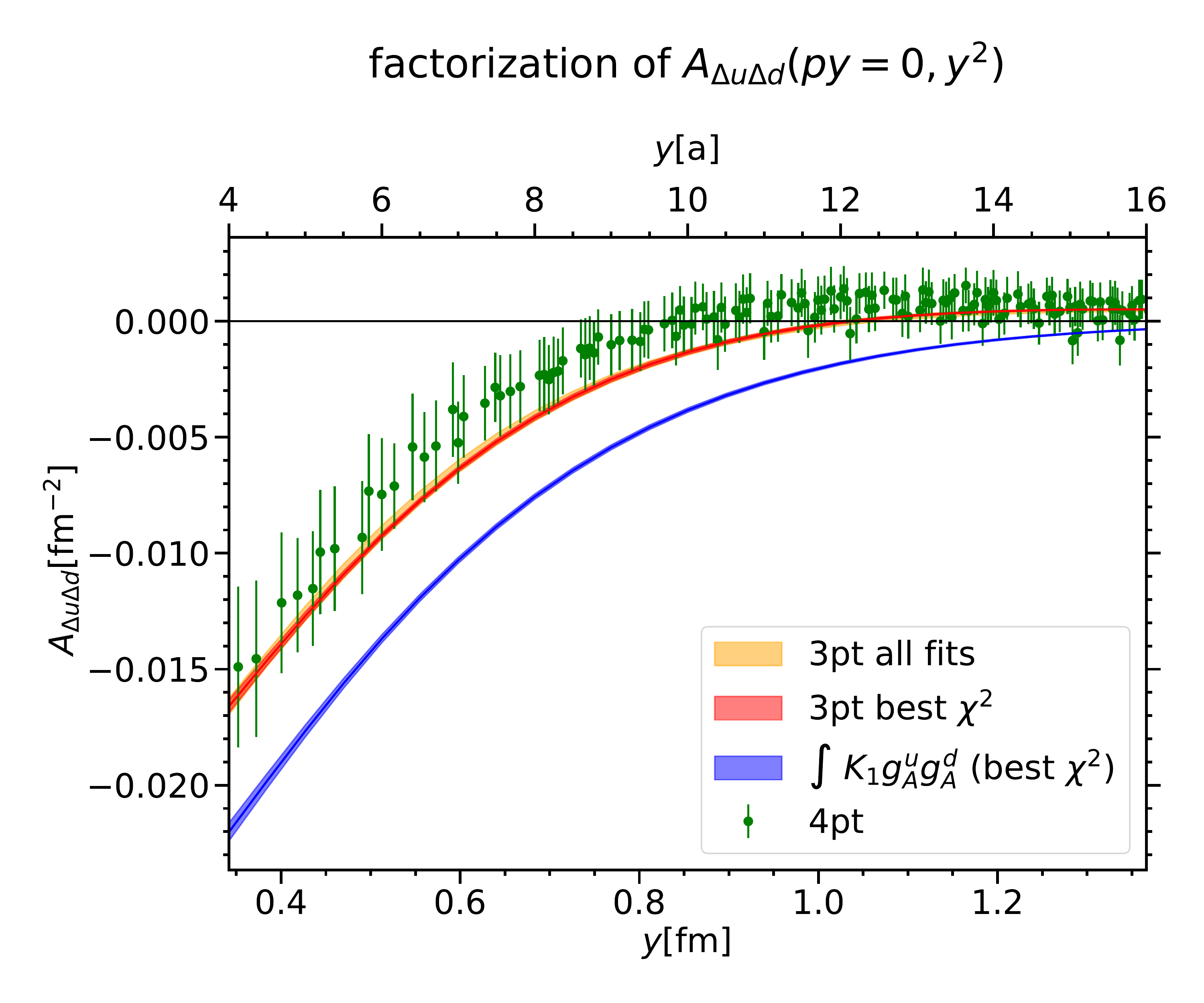}\label{fig:factAVVddAAA-A_AAud}} \\
\subfigure[$A_{\Delta d \Delta d}$ vs $\int g_{d} g_{d}$, $py = 0$]{
\includegraphics[scale=0.25,trim={0.5cm 1.2cm 0.5cm 2.8cm},clip]{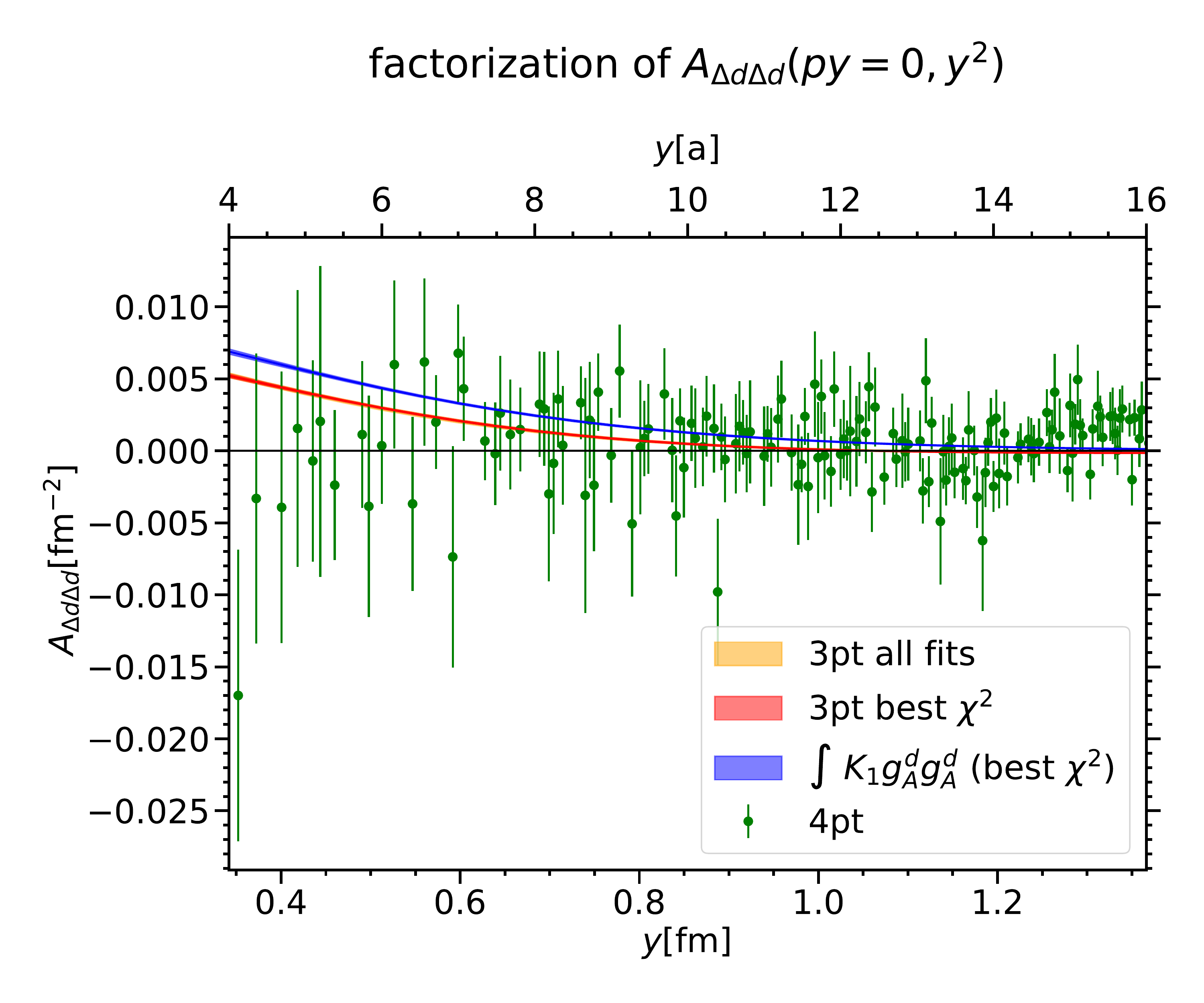}\label{fig:factAVVddAAA-A_AAdd}} \hfill
\subfigure[$A_{\Delta u \Delta u}$ vs $\int g_{u} g_{u}$, $py = 0$]{
\includegraphics[scale=0.25,trim={0.5cm 1.2cm 0.5cm 2.8cm},clip]{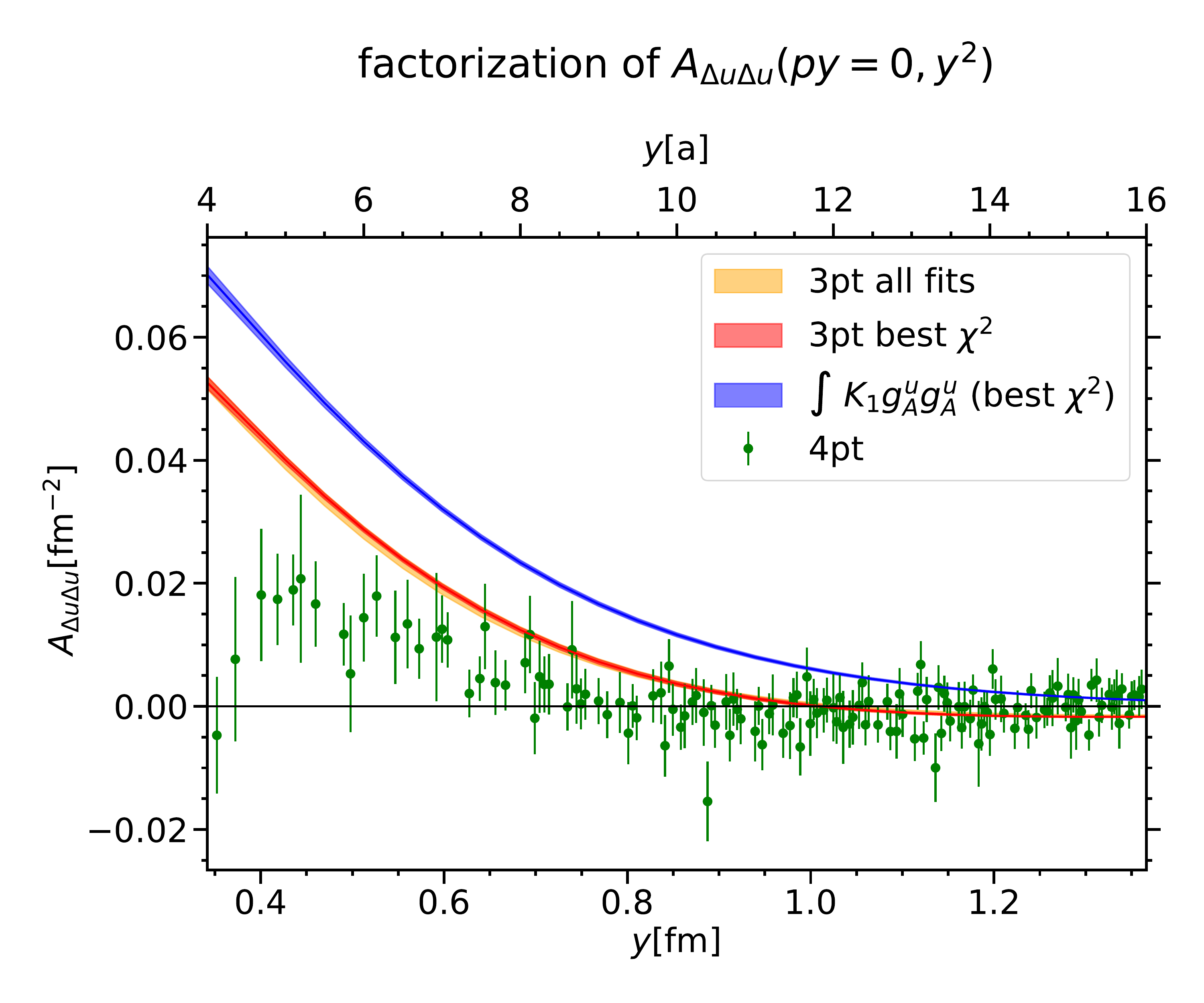}\label{fig:factAVVddAAA-A_AAuu}} \\
\caption{The twist-two functions $A_{dd}$ (a), $A_{\Delta u \Delta d}$ (b), $A_{\Delta d \Delta d}$ (c) and $A_{\Delta u \Delta u}$ (d) compared to the corresponding form factor integral \eqref{eq:facttest-AVV} or \eqref{eq:facttest-AAA}. The orange band again represents the envelope of the error bands for the different fits. The blue curve shows again the integration result of the first term in \eqref{eq:facttest-AVV} or \eqref{eq:facttest-AAA}.\label{fig:factAVVddAAA}}
\end{figure}

In the following, we consider the complete results of the \rhs of \eqref{eq:facttest-AVV}, \eqref{eq:facttest-AAA}, \eqref{eq:facttest-IVV}, and \eqref{eq:facttest-IAA} obtained from the corresponding integrals over the form factors and compare them to the l.h.s.. The observed difference can be interpreted as a measure of the strength of the quark-quark correlations. If the values of the involved data points are large enough compared to the statistical error, we also compute the ratio of both sides, in order to better see similarities and differences. We start with \eqref{eq:facttest-AVV}, where the two sides, as well as the ratio of both sides is shown in \fig\ref{fig:factAVVuduu} for $A_{ud}$ and $A_{uu}$. The result for $A_{dd}$ (without the ratio, since the signal is not sufficiently clean) is plotted in \fig\ref{fig:factAVVddAAA-A_VVdd}. For all flavor combinations, the form factor result correctly reproduces the size of the two-current data. Deviations are observed to be very small. From the ratio, we can read off the relative deviation, which is at most $\sim 20\%$ for $ud$. For $uu$, deviations are seen to be typically around $\sim 20\%$. Notice that the $F_2 F_2$-term and the mixed term play only a minor role in the integral formula, \ie the $F_1 F_1$-term (blue curve) is almost equal to the complete result.

The size of the two results also matches in the longitudinally polarized channels, as can be seen in \fig\ref{fig:factAVVddAAA} for $A_{\Delta u \Delta d}$ (b), $A_{\Delta d \Delta d}$ (c), and $A_{\Delta u \Delta u}$ (d). A remarkable observation is the nearly perfect agreement within statistical errors in the case of $A_{\Delta u \Delta d}$. Notice that the two-current signal of $A_{\Delta d \Delta d}$ is consistent with zero. Hence, the agreement of the corresponding curves and data points should be interpreted with some caution. In contrast to the unpolarized case, taking the complete integral instead of only the $g_A g_A$-term is crucial. Evaluating the integral over the $g_A g_A$-term only (the corresponding result is again shown by the blue curve) yields a significant difference between the two sides of \eqref{eq:facttest-AAA}. In \fig\ref{fig:factAVVuduu-A_VVud-pi} and \ref{fig:factAVVuduu-A_VVud-rat-pi} we show again the factorization results for $A_{u\bar{d}}$ for the $\pi^+$, which has been investigated in \cite{Bali:2020mij}. The results obtained there are comparable with those of $A_{ud}$ in the nucleon that we have described above.

\begin{figure}
\begin{center}
\subfigure[$I_{ud}$ vs $\int F_{u} F_{d}$, $\zeta = 0$]{
\includegraphics[scale=0.25,trim={0.5cm 1.2cm 0.5cm 2.8cm},clip]{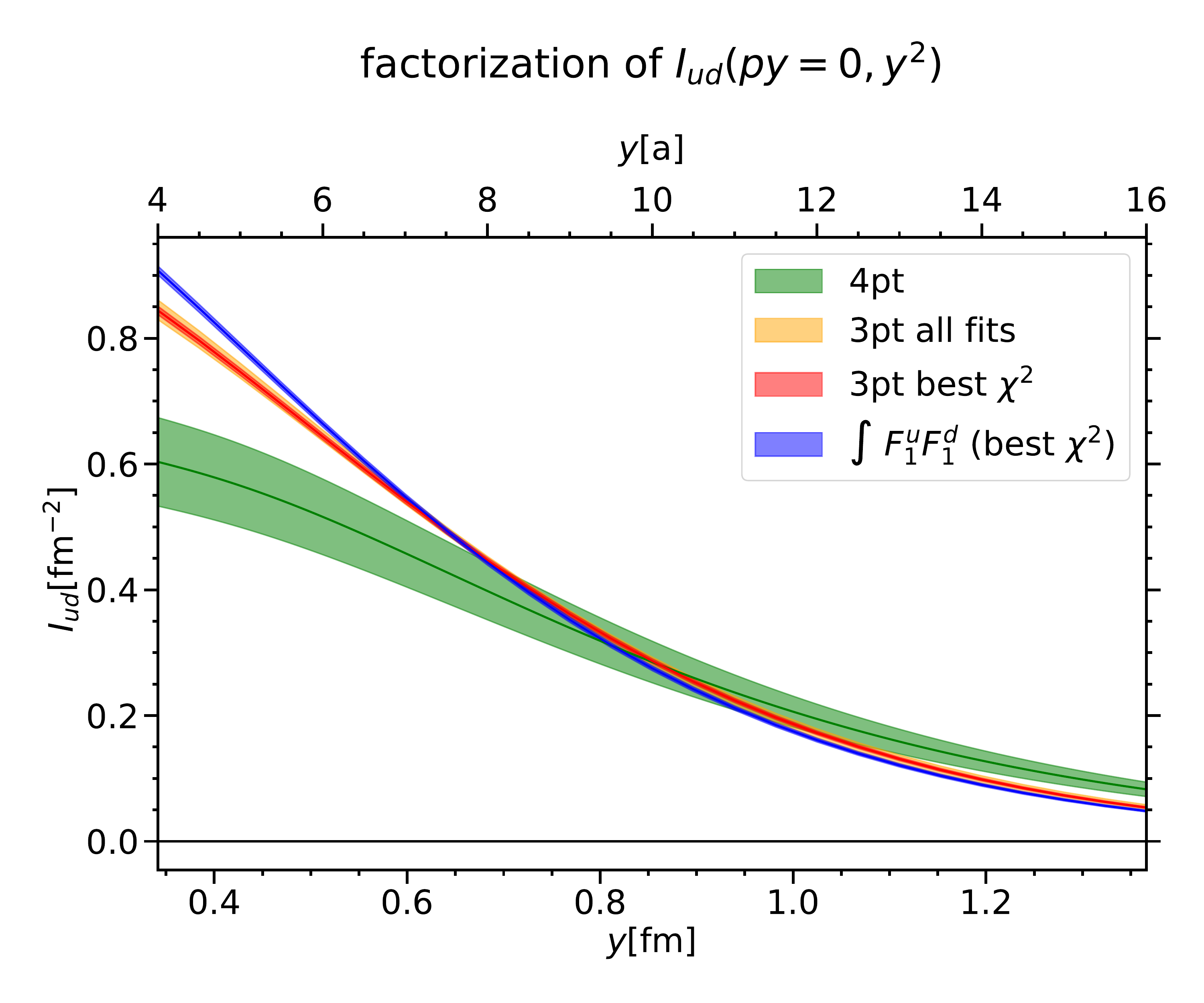}\label{fig:factIVV-I_VVud}} \hfill
\subfigure[Ratio $\int F_{u} F_{d}/I_{ud}$, $\zeta = 0$]{
\includegraphics[scale=0.25,trim={0.5cm 1.2cm 0.5cm 2.8cm},clip]{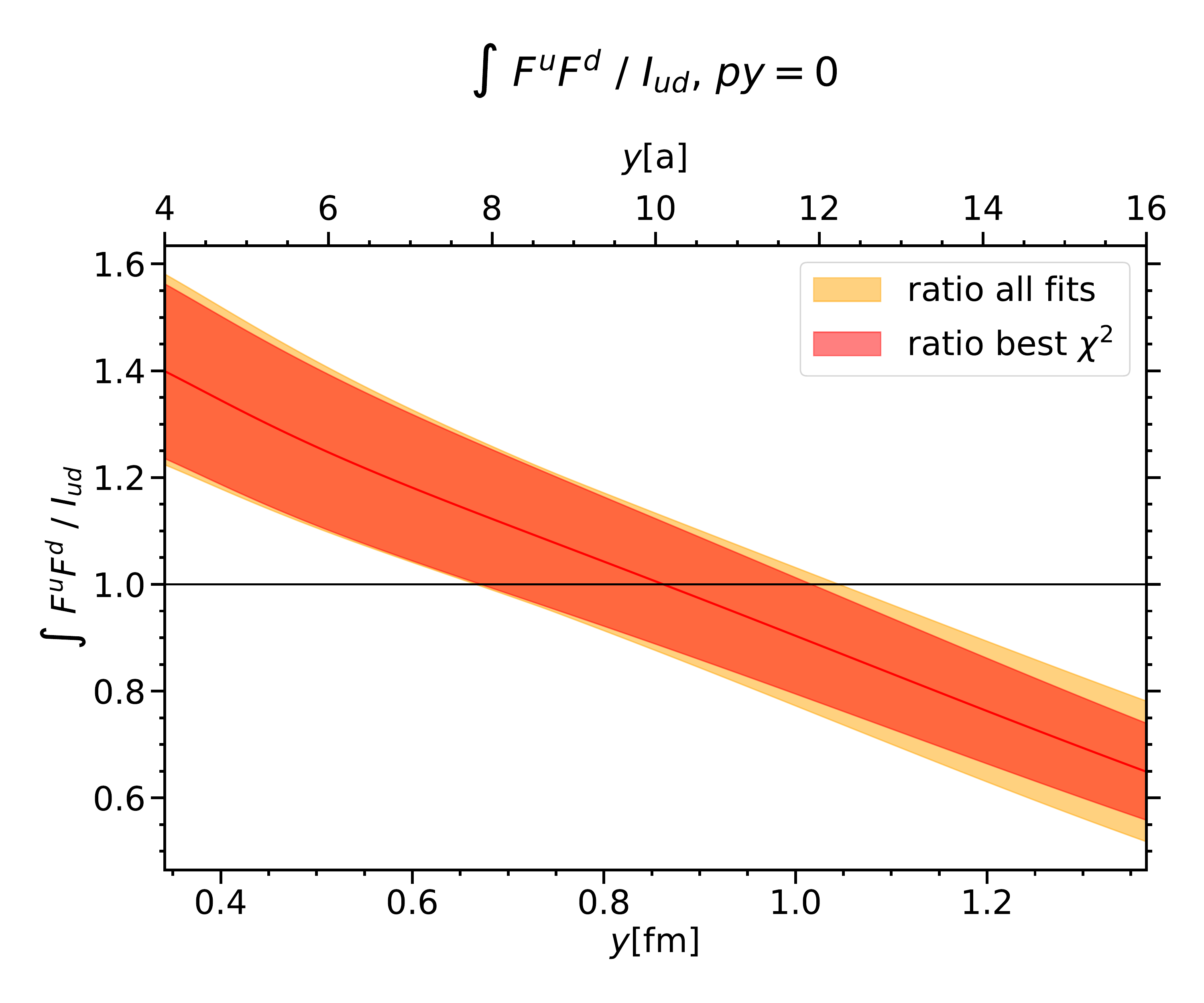}\label{fig:factIVV-I_VVud-rat}} \\
\subfigure[$I_{uu}$ vs $\int F_{u} F_{u}$, $\zeta = 0$]{
\includegraphics[scale=0.25,trim={0.5cm 1.2cm 0.5cm 2.8cm},clip]{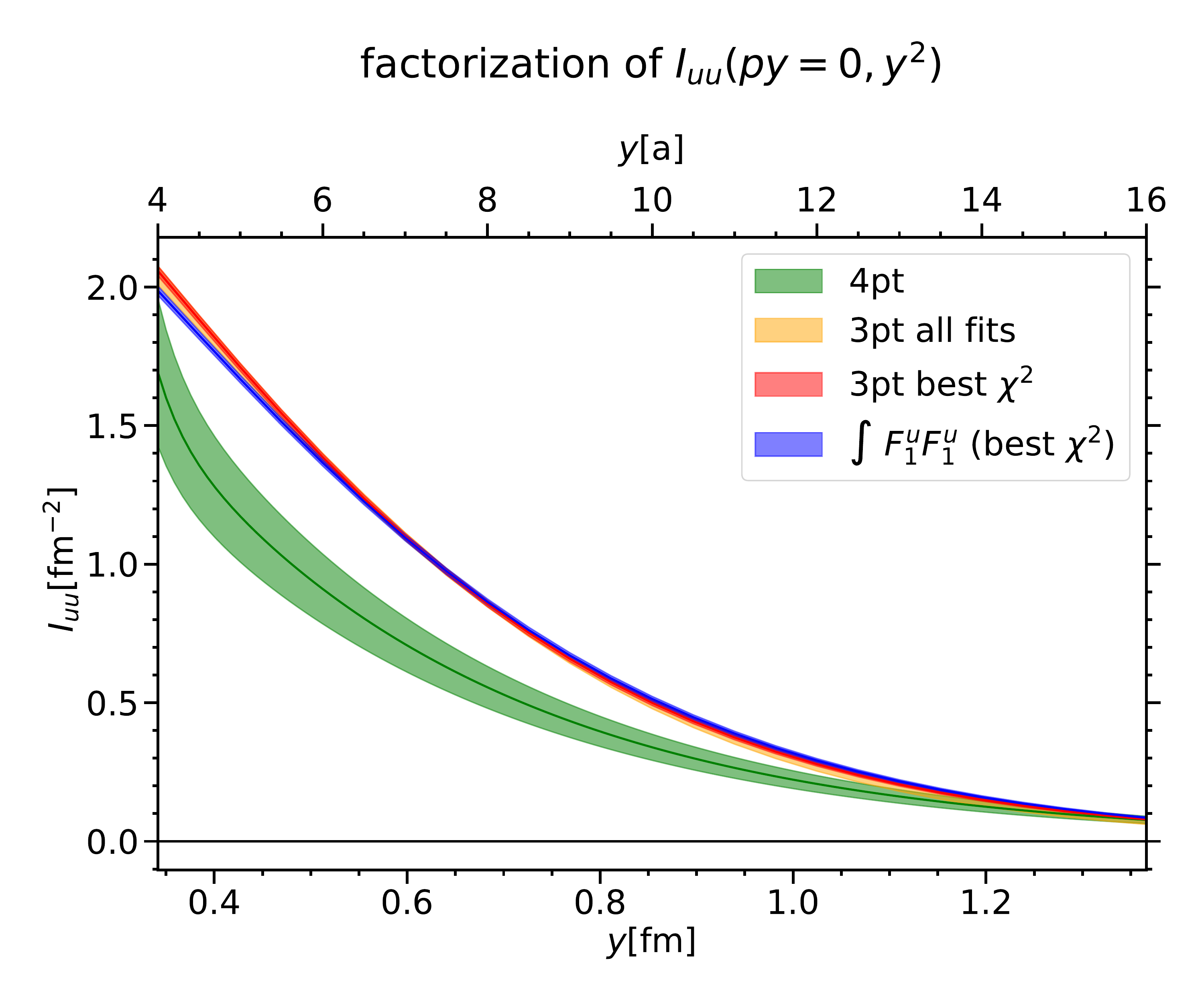}\label{fig:factIVV-I_VVuu}} \hfill
\subfigure[Ratio $\int F_{u} F_{u}/I_{uu}$, $\zeta = 0$]{
\includegraphics[scale=0.25,trim={0.5cm 1.2cm 0.5cm 2.8cm},clip]{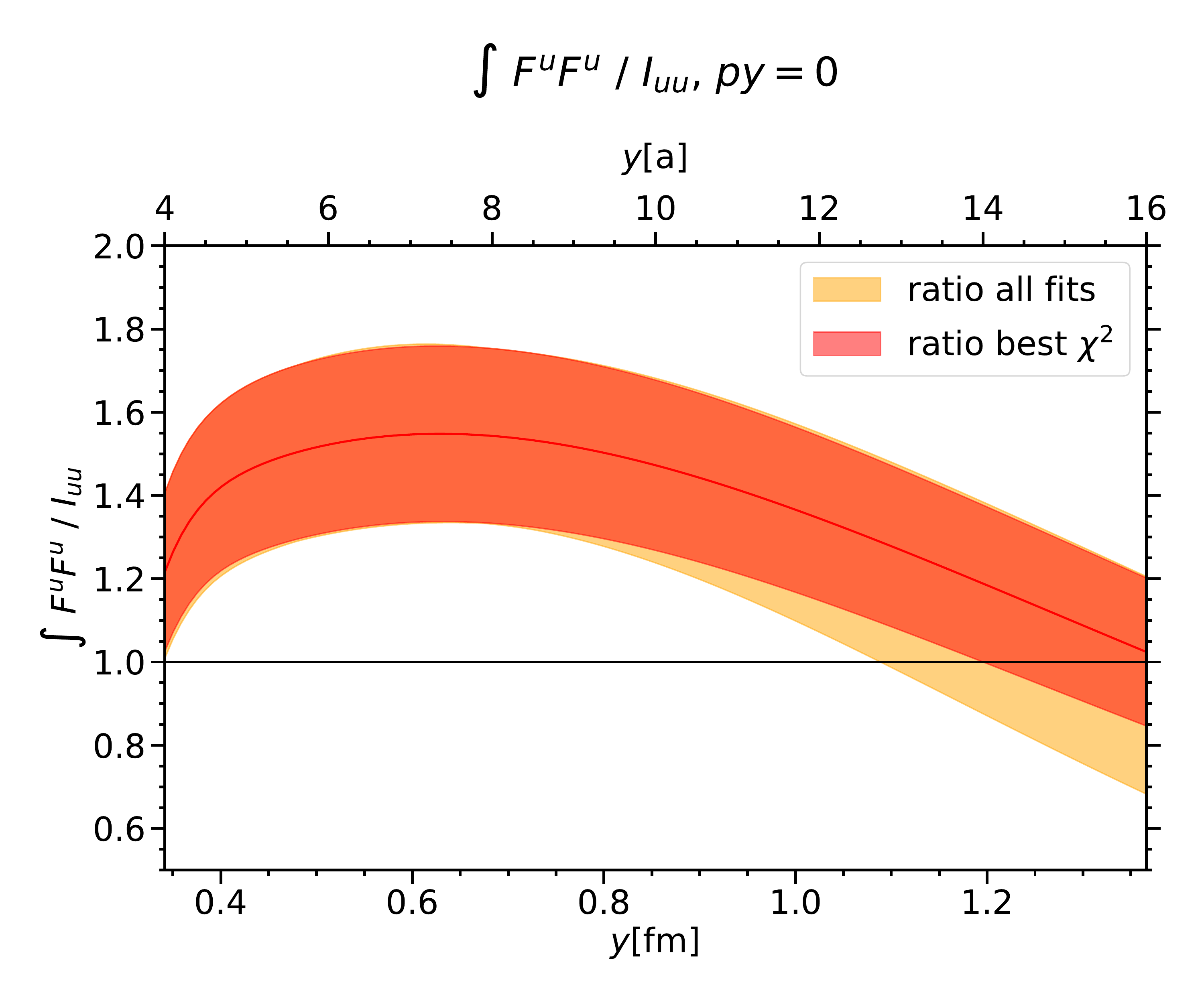}\label{fig:factIVV-I_VVuu-rat}} \\
\subfigure[$I_{dd}$ vs $\int F_{d} F_{d}$, $\zeta = 0$]{
\includegraphics[scale=0.25,trim={0.5cm 1.2cm 0.5cm 2.8cm},clip]{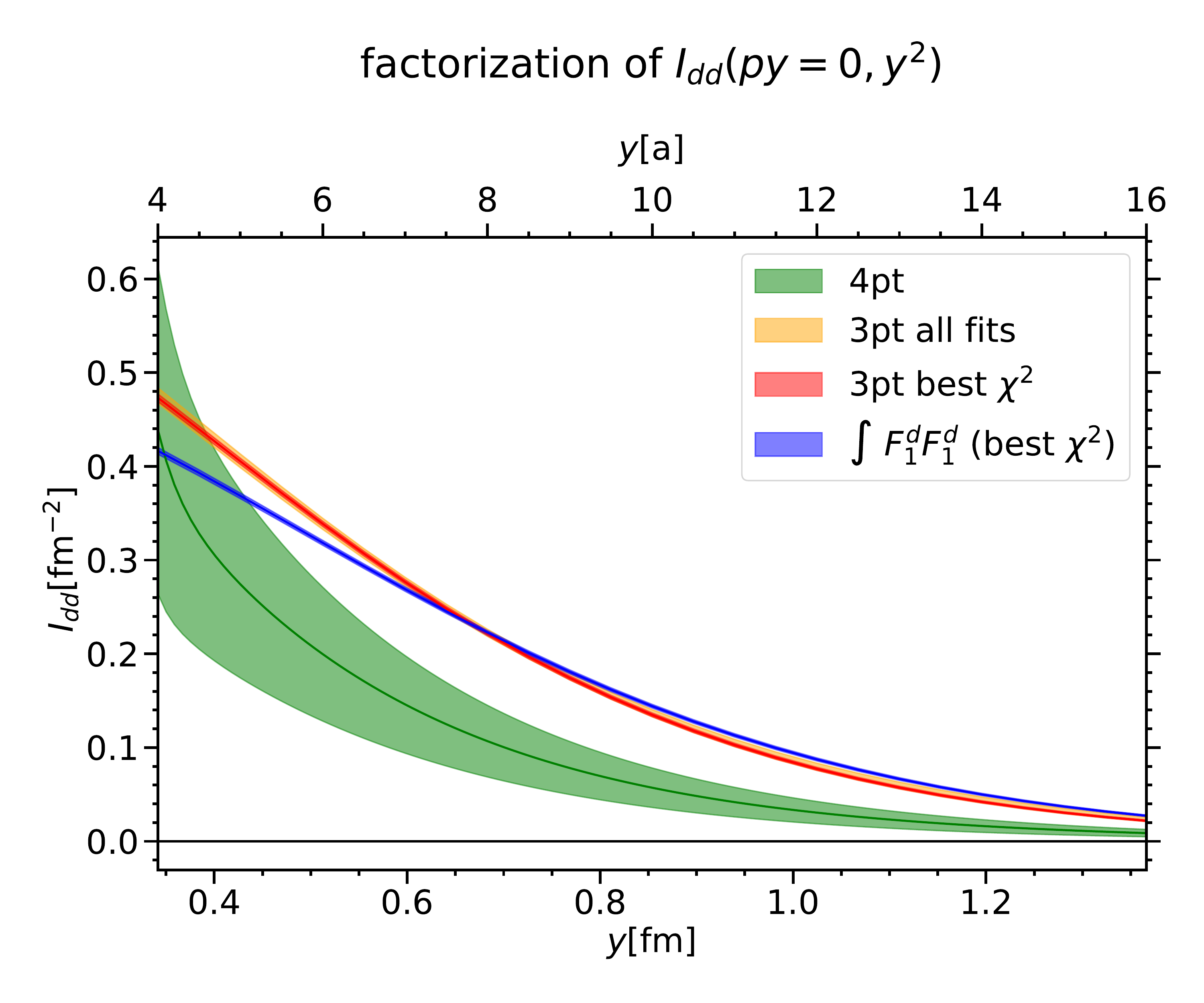}\label{fig:factIVV-I_VVdd}} \hfill
\subfigure[$I_{u\bar{d}}$ vs $\int F_{u} F_{\bar{d}}$ for $\pi^+$, $\zeta = 0$]{
\includegraphics[scale=0.25,trim={0.5cm 1.2cm 0.5cm 2.8cm},clip]{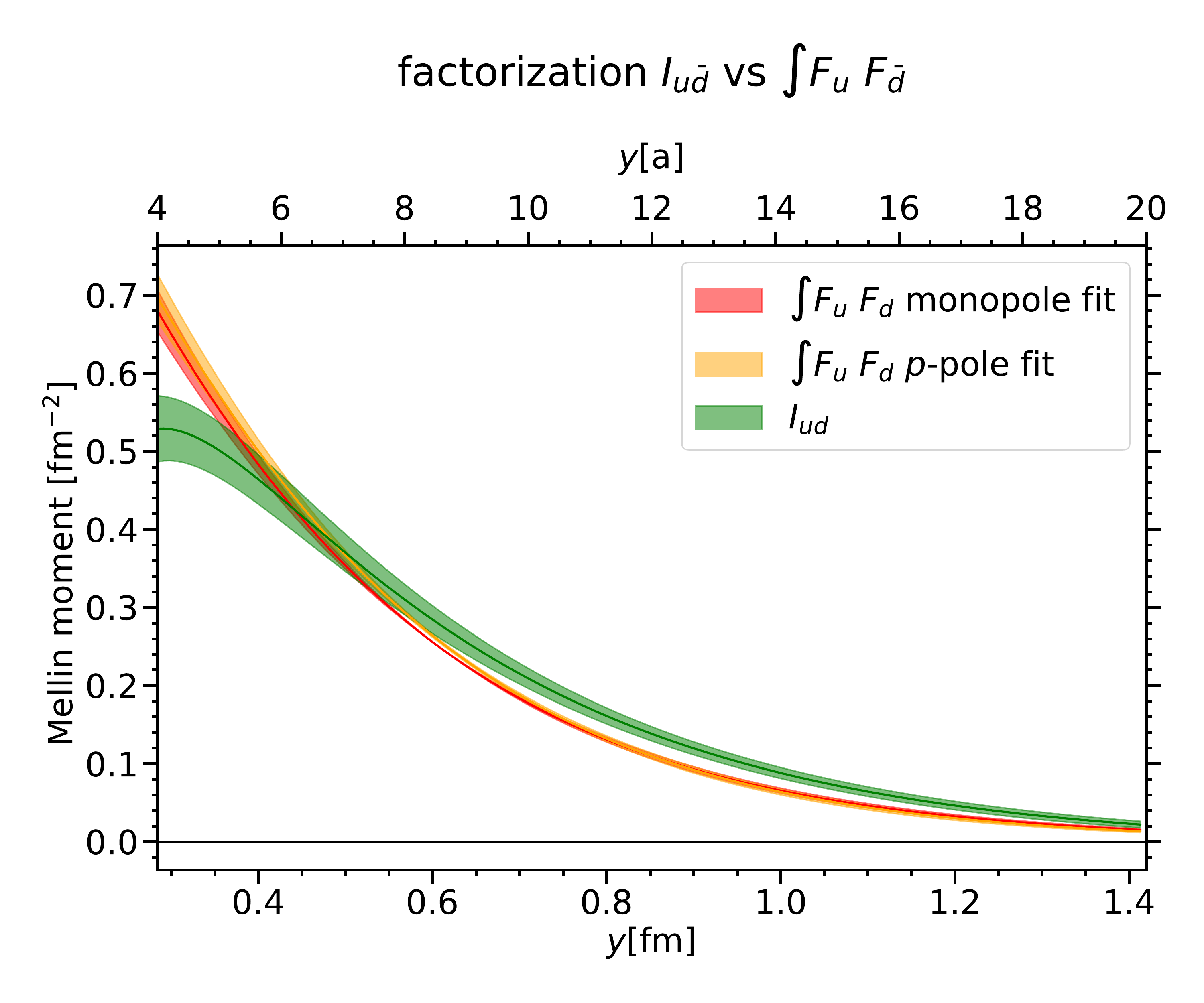}\label{fig:factIVV-I_VVud-pi}} \\
\end{center}
\caption{Mellin moment $I_{ud}$ at $\zeta = 0$ compared to its factorized result obtained from the corresponding integral \eqref{eq:facttest-IVV} (a) and the ratio of the integral and the Mellin moment (b). The same is plotted for $I_{uu}$ (c,d) and $I_{dd}$ (e). For the latter the ratio is not shown. The orange curve shows the envelope of the error bands for every fit. The result of a integral where only the first term in \eqref{eq:facttest-IVV} is taken into account is represented by the blue curve. Panel (f) shows the factorization result for $I_{u\bar{d}}$ in the $\pi^+$ obtained in \cite{Bali:2020mij} for the two form factor fits considered in that work. \label{fig:factIVV}}
\end{figure}

Finally, we want to consider the factorization for the Mellin moments $I_{qq^\prime}$ at $\zeta = 0$ according to \eqref{eq:facttest-IVV}. We shall not discuss \eqref{eq:facttest-IAA}, since we do not have results of sufficient quality for $I_{\Delta q \Delta q^\prime}$, as we have concluded in \sect\ref{sec:twist2_py_fit}. \Fig\ref{fig:factIVV} shows the two sides of \eqref{eq:facttest-IVV} (a), as well as the ratio (b) for quark flavor $ud$, while the analogous results for $uu$ and $dd$ (the latter again without the ratio) are shown in (c), (d) and (e). The integral again yields a consistent order of magnitude. However, the deviations of the two curves are found to be larger than for the factorization ansatz of the twist-two functions. The relative deviations are at most $\sim 40\%$  for $ud$ and $\sim 60\%$ for $uu$. Again we compare with the situation for the $\pi^+$, which is shown in \fig\ref{fig:factIVV-I_VVud-pi}. Especially for small distances $y$, the factorization result of $I_{u\bar{d}}$ is closer to the two-current result for the Mellin moment in the pion case than it is observed for $ud$ in the proton.

Notice that regions where the integral gives a higher value than the two-current data, or vice versa, are consistently the same for the twist-two functions and the Mellin moments. For $ud$ we observe the integral to be larger for $y<8a$, while it is smaller if $y>8a$. This means that in a joint observation of an $u$ and a $d$ quark, we find the two quarks farther apart than we would if they were uncorrelated. This is similar to $u\bar{d}$ in the $\pi^+$ described in \cite{Bali:2020mij}. For two quarks of the same flavor, the integration results are generally larger than the two-current data. An exception is given by the region $y<5a$, where at least the twist-two function results indicate a sign change in the absolute difference.

\FloatBarrier

\section{Conclusions}
\label{sec:summary}

This paper presents the first lattice calculation that provides information about double parton distributions in the proton.  The distributions in the neutron are readily obtained from isospin symmetry.  Our simulations are done on a $32^3 \times 96$ lattice with spacing $a \approx 0.086 \fm$ and a pion mass of $m_\pi = 355 \mev$.  We compute the correlation functions \eqref{eq:mat-els} of two spatially separated currents in the proton and project out their twist-two parts.  Our primary observables are the invariant functions $A$ and $B$ associated with that projection, see \eqref{eq:tensor-decomp}.  They depend on the distance $y^\mu$ between the two currents and on proton four-momentum $p^\mu$ via the scalar products $y^2$ and $py$.  We consider the vector, axial, and tensor current, whose twist-two components respectively correspond to unpolarized, longitudinally polarized, and transversely polarized quarks.

\paragraph{Lattice aspects.} We evaluate all Wick contractions that contribute to the two-current correlation functions, making heavy use of stochastic sources, sequential sources, and the hopping parameter expansion.  The statistical signal we obtain is in general very good for the connected graphs $C_1$ and $C_2$ and the disconnected graph $S_2$, and fair for the disconnected graph $S_1$ (see \fig\ref{fig:graphs}).  Only for the doubly disconnected graph $D$ are the errors so large that we must exclude it from our analysis.  Lattice artifacts manifest themselves in the invariant functions as a breaking of rotation invariance (i.e.\ a dependence on direction of $\mvec{y}$) and a breaking of boost invariance (at given $y = |\mvec{y}|$ and $py$ the functions must be independent of $p^\mu$).  We find a significant amount of anisotropy in the $C_1$ data at large $y$ and in the $C_2$ and $S_2$ data at small $y$.  These can be interpreted as a finite size effect in the first case and as due to the anisotropy of the lattice propagator in the second case.  We can largely remove these effects by selecting points $\mvec{y}$ close to the lattice diagonals and by imposing a lower cutoff on $y$, which depending on the polarization channel is taken of order $4 a \approx 0.34 \fm$.  After this selection, the violation of boost invariance is at an acceptable level, except for graph $S_2$, where a momentum dependence is seen up to about $y \sim 7 a$.  For larger $y$, the contribution of $S_2$ to physical matrix elements is small compared with the one from $C_1$ and $C_2$.  The contribution of $S_1$ is found to be small at the scale of $C_1$ and $C_2$, except for larger $y$, where the errors on $S_1$ prevent us from drawing strong conclusions.  For our final physics analysis, we restrict ourselves to the contributions of the connected graphs $C_1$ and $C_2$, where $C_2$ is absent for the parton combination $u d$ and $C_1$ is absent for $d d$.

\paragraph{Results.} In a first stage, we analyze the invariant twist-two functions $A$ and $B$ at $py = 0$, where the statistical signal is best and the data can be plotted as a function of the single variable $y$.  To connect these functions with DPDs, we slightly deform them by a skewness in the parton momentum fractions that is parameterized by $\zeta$ (see \fig\ref{fig:zeta-distrib}).  Twist-two functions at $py = 0$ are then equal to the Mellin moments of skewed DPDs integrated over $\zeta$.  The size of these functions is seen to be largest for $A_{q q'}$ and $A_{\delta q\ms q'}$, with the former corresponding to unpolarized partons and the latter to the correlation between the transverse polarization of one parton and the parton separation.  Our results exhibit a clear flavor dependence, with $A_{u d}$ and $A_{\delta u\ms d} \approx A_{\delta d\ms u}$ decreasing more slowly with $y$ than their counterparts for two $u$ or two $d$ quarks (see \fig\ref{fig:tw2f_fcomp}).  For unpolarized quarks, this finding is of particular importance, because one of the assumptions made for deriving the pocket formula \eqref{eq:dpd-pocket-formula} for DPS cross sections is a universal $y$ dependence of DPDs for all flavor combinations.  Interestingly, $A_{u u}$ and $A_{d d}$ have a rather similar $y$ dependence, although the former receives a contribution from $C_1$ but the latter does not.

The signal for spin dependent functions other than $A_{\delta q\ms q'}$ is best for the $u d$ combination, whereas for $q q^\prime = u u$ and $d d$ it is mostly consistent with zero (see \fig\ref{fig:tw2f_polcomp}).  In the $u d$ channel, the invariant functions for two polarized quarks are significantly smaller than $A_{\delta u\ms d}$.  We see a clear difference between the spin-spin correlations $A_{\Delta u \Delta d}$ and $A_{\delta u \delta d}$ for longitudinal and transverse polarization, which shows the inadequacy of simple non-relativistic pictures that predict them to be equal.  Moreover, we find that the longitudinal polarization ratios $A_{\Delta u \Delta d} / A_{u d}$ and $A_{\Delta u \Delta u} / A_{u u}$ are significantly smaller in size than the values $-2/3$ and $+1/3$ obtained with a static SU(6) invariant wave function for the three valence quarks in the proton \cite{Diehl:2011yj}.  Interestingly, the pattern of polarization dependence for $u d$ in the proton is quite similar to the one we found for $u \bar{d}$ in a $\pi^+$ in our previous work \cite{Bali:2020mij}.

In the second stage of our analysis, we assume a parametric form for the $y$ and $py$ dependence of the twist-two functions (see \eqref{eq:y2-ansatz} and \eqref{eq:A-global-fit}).  We use this to fit our data and to extrapolate it to the full range of $py$, which is needed to compute the Mellin moments $I_{qq'}$, $I_{\delta q q'}$, \ldots of DPDs at given skewness $\zeta$.  For flavor and polarization combinations with sufficiently small statistical errors, the results of fits with different numbers of parameters are consistent with each other for small to moderate $\zeta$ (see \figs\ref{fig:mellin_fit_comp} and \ref{fig:mellin_zeta_fit_comp}).  This gives us confidence in analyzing the corresponding Mellin moments at $\zeta = 0$ and thus to make closer contact with the physics of double parton scattering.

The flavor and polarization dependence of Mellin moments at $\zeta=0$ is very similar to the one of the associated twist-two functions at $py = 0$, which corroborates the physics conclusions discussed above (see \figs\ref{fig:mellin_fcomp} and \ref{fig:mellin_polcomp}).  From the moment $I_{u d}$, we can also evaluate the $x$ integral of the number sum rule for DPDs \cite{Gaunt:2009re,Diehl:2018kgr}.  We find excellent agreement with the predicted value of the sum rule (see \tab\ref{tab:sr_res}) and regard this as a strong check of our fitting ansatz and analysis procedure.

\paragraph{Correlation effects.} Many models for DPDs rest on the assumption that the two partons are independent of each other.  This assumption can be formalized and leads to factorization formulae for the twist-two functions $A_{q q'}$ and $A_{\Delta q \Delta q'}$ (\eqref{eq:facttest-AVV} and \eqref{eq:facttest-AAA}), and for the associated Mellin moments (\eqref{eq:facttest-IVV} and \eqref{eq:facttest-IAA}).  These functions are then expressed in terms of the nucleon Dirac and Pauli form factors $F_1$ and $F_2$ for unpolarized quarks, and of the axial and pseudoscalar form factors $g_A$ and $g_P$ for longitudinal quark polarization.  We fit these form factors to lattice data from the same ensemble used for computing the two-current correlators, and then extrapolate the form factors in the momentum transfer.  We find that the factorization formula for unpolarized quarks is to a good approximation saturated by the contribution from $F_1$, whilst for longitudinal polarization it is important to include the contributions from both $g_A$ and $g_P$ (see \fig\ref{fig:fact_parts}).

We find that the factorization assumption for $A_{u d}$ and $A_{u u}$ at $py = 0$ works remarkably well, with deviations not larger than $20 \%$ in the $y$ range considered (see \fig\ref{fig:factAVVuduu}).  It works rather well also for $A_{\Delta u \Delta d}$, whereas for $A_{d d}$ and $A_{\Delta u \Delta u}$ larger deviations from factorization are observed (see \fig\ref{fig:factAVVddAAA}).  The factorization for the Mellin moments $I_{u d}$ and $I_{u u}$ at $\zeta=0$ works rather well, albeit with deviations up to almost $60\%$, whereas for $I_{d d}$ the discrepancies are again larger (see \fig\ref{fig:factIVV}).  In other channels, the errors in our data or fits are too large for drawing solid conclusions.

\paragraph{Summary and outlook.} In summary, we find that the calculation of two-current correlators on the lattice can provide valuable physics insight into two-quark correlations inside the proton, which are essential for understanding double parton scattering. Our main results are as follows. (i) The dependence of two-parton distributions on the distance $y$ is \emph{not} the same for different flavors. (ii) Spin-spin correlations between two quarks are remarkably small, in contrast to spin-orbit correlations. (iii) The functions we studied approximately factorize into separate functions for the individual partons.

Important challenges for future work are to perform simulations at smaller lattice spacings, so as to extend the $y$ range where lattice artifacts can be controlled, and to move closer to the physical pion mass.  Improvements that will allow the inclusion of disconnected graphs in the physics analysis are also highly desired.  The results obtained in the present study strongly motivate us to make efforts in these directions.

%%%%%%%%%%%%%%%%%%%%%%%%%%%%%%%%%%%%%%%%%%%%%%%%%%%%%%%%%%%%%%%%%%%%%%%%%%

\section*{Acknowledgments}

We thank Thomas Wurm for providing the nucleon form factor data generated in the context of the simulation in \cite{Bali:2019yiy}. Furthermore, we gratefully acknowledge the CLS effort (\href{http://wiki-zeuthen.desy.de/CLS/CLS}{http://wiki-zeuthen.desy.de/CLS/CLS}) for generating the $n_f = 2+1$ ensembles \cite{Bruno:2014jqa}, one of which we employed for the present study. This work was supported in parts by the German Research Foundation (DFG, SFB/TRR-55), the German Federal Ministry of Education and Research (BMBF, grant 05P18WRCA1), and the European Union’s Horizon 2020 research and innovation programme under grant agreement no. 824093 (STRONG-2020). Our simulations were performed on the QPACE 3 systems of the SFB/TRR-55. Here we used an extended version of the Chroma software stack \cite{Edwards:2004sx} together with a KNL adaption of the Multi-Grid solver \cite{Babich:2010qb,Frommer:2013fsa,Heybrock:2015kpy,Richtmann:2016kcq,Georg:2017zua}. The graphs in this paper were generated using Jaxodraw \cite{Binosi:2003yf,Binosi:2008ig} and the tikz library. All plots were created using the Matplotlib library \cite{Hunter:2007}.

%%%%%%%%%%%%%%%%%%%%%%%%%%%%%%%%%%%%%%%%%%%%%%%%%%%%%%%%%%%%%%%%%%%%%%%%%%

\begin{appendices}
\section{Notation and lattice technicalities}
\label{sec:appendix}

In the following, we list expressions that are useful for the calculation of the baryon four-point contractions introduced in \sect\ref{sec:lattice}. This includes symmetry relations, as well as ingredients that are used to evaluate the four-point contractions on the lattice. Furthermore, we give details on the notation used in this paper.

\subsection{Notation}
\label{sec:notation}

In this work we use the following notation conventions:
\begin{itemize}
\item Indices: Lorentz indices are denoted by Greek letters $\mu,\nu,\dots$, spinor indices by $\alpha,\beta,\dots$, and color indices (fundamental) by Latin letters $a,b,c,\dots$.
\item Spacetime dependencies are indicated by an argument if it represents a degree of freedom. If the corresponding variable is fixed (\eg the source position of a point-to-all propagator) an index  is used instead.
\item Unless stated otherwise, traces and transpositions are taken \wrt spinor \emph{and} color indices.
\item For a given 4-vector $y^\mu$ we denote the spatial components by $\mvec{y}$ (identical in Minkowski and Euclidean spacetime). The spatial distance is denoted by $y:=|\mvec{y}|$. If $y^0 = 0$, we have $|\mvec{y}|^2 = -y^2 := -y^\mu y_\mu$. In order to avoid confusion with the usual Minkowski scalar product $y^2$, we explicitly write $\sqrt{-y^2}^n$ for the $n$-th power of $y=|\mvec{y}|$.
\end{itemize}
For better readability, spinor and color indices, as well as spacetime arguments are not always explicitly written in \sect\ref{sec:tcme_wick}. This applies if the considered objects have matrix or vector character \wrt these indices or arguments. We list some of the objects that are considered in this work and display their explicit notation in \tab\ref{tab:notation}.
\begin{table}
\begin{center}
\begin{tabular}{c|ccc}
\hline\hline
Object & Symbol & Degrees of freedom & Explicit \\
\hline
Gauge link & $U_\mu$ & $\mathrm{space} \times \mathrm{color}^2$ & $\left(U_\mu\right)^{ab}(x)$ \\
Generic source & $S$ & $\mathrm{space} \times \mathrm{spinor}^2 \times \mathrm{color}^2$ & $S_{\alpha\beta}^{ab}(x)$ \\
Smearing function & $\Phi$ & $\mathrm{space}^2 \times \mathrm{color}^2$ & $\Phi^{ab}(x|y)$ \\
Dirac operator & $\mathcal{D}$ & $\mathrm{space}^2 \times \mathrm{spin}^2 \times \mathrm{color}^2$ & $\mathcal{D}_{\alpha\beta}^{ab}(x|y)$ \\
Propagator ($x\rightarrow y$) & $M$ & $\mathrm{space}^2 \times \mathrm{spin}^2 \times \mathrm{color}^2$ & $M_{\alpha\beta}^{ab}(y|x)$ \\
Point($x$)-to-all($y$) propagator & $M_x$ & $\mathrm{space} \times \mathrm{spinor}^2 \times \mathrm{color}^2$ & $\left(M_x\right)_{\alpha\beta}^{ab}(y)$ \\
Stochastic source/propagator & $\eta$, $\psi$ & $\mathrm{space} \times \mathrm{spinor} \times \mathrm{color}$ & $\eta_{\alpha a}(x)$, $\psi_{\alpha a}(x)$ \\
Sequential propagator (at time $t$) & $X_t$ & $\mathrm{space} \times \mathrm{spinor}^2 \times \mathrm{color}^2$ & $\left( X_t \right)_{\alpha\beta}^{ab}(x)$ \\
Gamma matrices & $\Gamma$ & $\mathrm{spinor}^2$ & $\Gamma_{\alpha\beta}$ \\ 
\hline\hline
\end{tabular}
\hspace*{0.5cm}\\
\end{center}
\caption{Lattice objects and their degrees of freedom regarding spacetime (for brevity, we write "space" in the table), spinor and color indices. Notice that Lorentz indices, \eg of the gauge link $U_\mu$, are always written explicitly.\label{tab:notation}}
\end{table}
Notice that each of the mentioned expressions may have further dependencies which are not stated above. A product of these quantities is considered to be a matrix-matrix or matrix-vector product. As an example we rewrite \eqref{eq:seq_3pt_inv} using the compact and the explicit notation, respectively:

\begin{align}
\D X^{\Phi,\mvec{p}}_{t,3\mathrm{pt}} &= 
	\Phi^{\mvec{p}} \gamma_5 S^{\dagger,\mvec{p}}_{t,3\mathrm{pt}} \nonumber \\
\Leftrightarrow \quad
\sum_{y,\beta, b} \mathcal{D}_{\alpha\beta}^{ab}(x|y) 
\left( 
	X_{t,3\mathrm{pt}}^{\Phi,\mvec{p}} (y) 
\right)_{\beta\gamma}^{bc} &= 
	\sum_{y,\beta, b}
	\left( \Phi^{\mvec{p}}(x|y) \right)^{ab}\ 
	\left( \gamma_5 \right)_{\alpha\beta}\ 
	\left[
		\left( 
			S_{t,3\mathrm{pt}}^{\mvec{p}} (y)
		\right)_{\gamma\beta}^{cb}
	\right]^{*}\,.
\end{align}
Each of the two sides carries the (implicit) indices or arguments $x$, $\alpha$, $\gamma$, $a$, and $c$, \ie :

\begin{align}
\left[ 
	\D X^{\Phi,\mvec{p}}_{t,3\mathrm{pt}} (x)
\right]_{\alpha\gamma}^{ac} \,,
\qquad
\left[ 
	\Phi^{\mvec{p}} \gamma_5 S^{\dagger,\mvec{p}}_{t,3\mathrm{pt}} (x)
\right]_{\alpha\gamma}^{ac} \,.
\end{align}
In some cases where spinor and color indices are written explicitly, we make use of the Einstein summation convention, \ie indices that appear twice are to be summed over. 

\subsection{Explicit expressions for four-point Wick contractions}
\label{sec:fptcontr}

\begin{figure}[ht]
\includegraphics[scale=1]{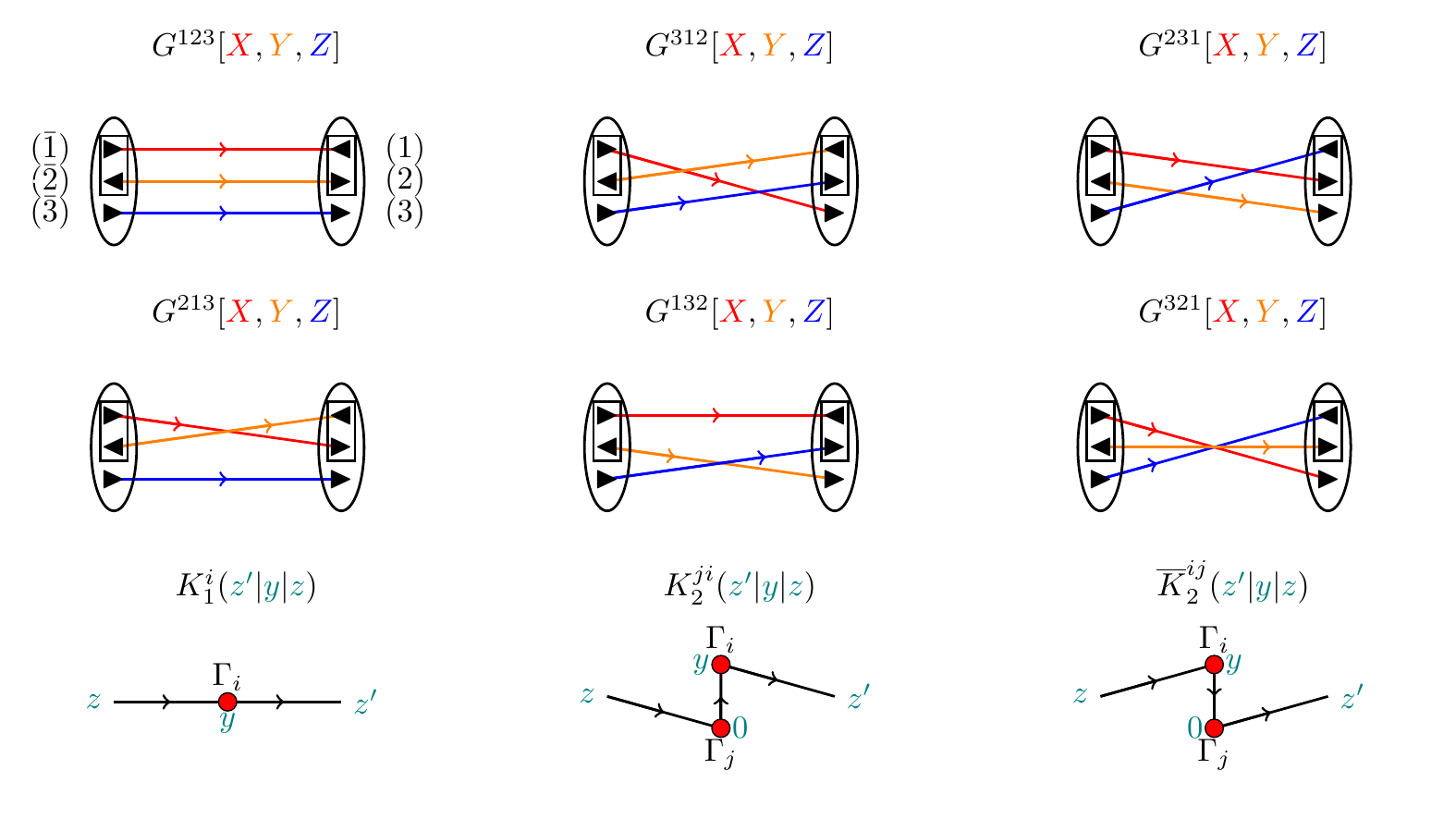}
\caption{Depiction of expressions \eqref{eq:gen_bar_contr} and \eqref{eq:4pt_parts} used for the construction of each baryonic four-point function graph. The blobs at the left (right) of each graph in the two first lines denote the baryon source (sink). Each symbol $\blacktriangleright$ and $\blacktriangleleft$ represents a quark field, where $\blacktriangleleft$ means that the corresponding spinor has to be transposed. The boxes denote the di-quark. For the three quark line types (bottom) we also indicate the positions of the propagator end points and the indices of the current insertions. \label{fig:baryon_wick}}
\end{figure}

The baryons are created and annihilated by the interpolators \eqref{eq:interpdef}. Referring to this equation, we assign the following integer numbers to the quark fields: 
\begin{align}
\label{eq:quark_numbers}
\bar{u}_a \rightarrow \bar{1}\,, \quad
\bar{d}_b^{\,T} \rightarrow \bar{2}\,, \quad
\bar{u}_c \rightarrow \bar{3}\,, \nonumber \\
u_b^T \rightarrow 1\,, \quad
d_c \rightarrow 2\,, \quad
u_a \rightarrow 3\,.
\end{align}
These numbers are also shown in the upper left panel of \fig\ref{fig:baryon_wick} and are used in the following to indicate the permutation of the annihilator fields \wrt the creator fields. The connected part of a generic baryon Wick contraction can be written in terms of the expressions (traces and transpositions are taken \wrt to spinor indices only):

\begin{align}
G^{123}[X,Y,Z] &:= 
	\epsilon^{abc}\ \epsilon^{a^\prime b^\prime c^\prime} 
	\tr \left\{ 
		\left(\Gamma^B\right)^T X_{a^\prime a}\ 
		\Gamma^B\ Y_{b^\prime b}^T 
	\right\} \tr\left\{ 
		Z_{c^\prime c}\ \Gamma^A 
	\right\}\ , \nonumber\\
G^{213}[X,Y,Z] &:= 
	-\epsilon^{abc}\ \epsilon^{a^\prime b^\prime c^\prime} 
	\tr \left\{ 
		\Gamma^B\ X_{b^\prime a}\ \Gamma^B\ Y_{a^\prime b}^T 
	\right\} \tr\left\{ 
		Z_{c^\prime c}\ \Gamma^A 
	\right\}\ , \nonumber\\
G^{321}[X,Y,Z] &:= 
	-\epsilon^{abc}\ \epsilon^{a^\prime b^\prime c^\prime} 
	\tr \left\{ 
		\Gamma^A\ X_{c^\prime a}\ \Gamma^B\ Y_{b^\prime b}^T\ 
		\left(\Gamma^B\right)^T Z_{a^\prime c} 
	\right\}\ , \nonumber\\
G^{132}[X,Y,Z] &:= 
	-\epsilon^{abc}\ \epsilon^{a^\prime b^\prime c^\prime} 
	\tr \left\{ 
		\left(\Gamma^B\right)^T X_{a^\prime a}^T\ 
		\Gamma^B\ Z_{b^\prime c}\ \Gamma^A\ Y_{c^\prime b} 
	\right\}\ , \nonumber\\
G^{231}[X,Y,Z] &:= 
	\epsilon^{abc}\ \epsilon^{a^\prime b^\prime c^\prime} 
	\tr \left\{ 
		\left(\Gamma^B\ X_{b^\prime a}\ 
		\Gamma^B\right)^T Z_{a^\prime c}\ \Gamma^A\ Y_{c^\prime b} 
	\right\}\ , \nonumber\\
G^{312}[X,Y,Z] &:= 
	\epsilon^{abc}\ \epsilon^{a^\prime b^\prime c^\prime} 
	\tr \left\{ 
		\Gamma^A\ X_{c^\prime a}\ 
		\Gamma^B\ Y_{a^\prime b}^T\ \Gamma^B\ Z_{b^\prime c} 
	\right\}\,.
\label{eq:gen_bar_contr}
\end{align}
For the nucleon we have $\Gamma^B = C \gamma_5$ and $\Gamma^A = P_+$, where $C$ is the charge conjugation matrix and $P_+$ selects positive parity. As a consequence, we can relate:

\begin{align}
G^{321}[X,Y,Z] = G^{132}[Y,X,Z]\,, 
\qquad 
G^{312}[X,Y,Z] = G^{231}[Y,X,Z]\,.
\end{align}
$X$, $Y$ and $Z$ can be either a propagator $M(z^\prime|z)$ connecting the source at $z$ and the sink at $z^\prime$ or one of the following terms:

\begin{align}
K_{1}^i(z^\prime|y|z) &:= 
	M(z^\prime|y)\ \Gamma_i\ M(y|z)\ ,\nonumber \\
K_{2}^{ji}(z^\prime|y|z) &:= 
	M(z^\prime|y)\ \Gamma_i\ M(y|0)\ \Gamma_j\ M(0|z)\ ,\nonumber \\
\overline{K}_{2}^{ij}(z^\prime|y|z) &:= 
	M(z^\prime|0)\ \Gamma_j\ M(0|y)\ \Gamma_i\ M(y|z) = 
	K_{2}^{ij}(z^\prime|-y|z)\,.
\label{eq:4pt_parts}
\end{align}
Each of the expressions $G^{ijk}$, $K_1$, $K_2$ and $\overline{K}_2$ is pictorially represented in \fig\ref{fig:baryon_wick}. The second identity in the last line of \eqref{eq:4pt_parts} is a consequence of translational invariance. We now consider the effects of $PT$ transformations and the combination of complex conjugation and $CP$ transformation on the previously defined expressions. The following relations are understood to be valid after integrating over the gauge fields:

\begin{align}
\label{eq:K_sym}
X \xrightarrow{PT} S^{-1} \left[U_{PT}(X)\right]^T S 
\qquad 
X^{*} \xrightarrow{CP} A^{-1}\ U_{CP}(X)\ A \,,
\end{align}
where

\begin{align}
\label{eq:K_sym_2}
U_{PT}(M(z^\prime|z)) &:= M(-z|-z^\prime) \nonumber \,, \\
U_{CP}(M(z^\prime|z)) &:= M(\tilde{z}^\prime|\tilde{z}) \nonumber \,, \\
U_{PT}(K_1^i(z^\prime|y|z)) 
	&:= \etapt^i\ \etaf^i\ K_1^i(-z|-y|-z^\prime) \nonumber \,, \\
U_{CP}(K_1^i(z^\prime|y|z)) 
	&:= \etapt^i\ K_1^i(\tilde{z}^\prime|\tilde{y}|\tilde{z}) 
	\nonumber \,, \\
U_{PT}(K_2^{ji}(z^\prime|y|z)) 
	&:= \etapt^{ij}\ \etaf^{ij}\ K_2^{ij}(-z|y|-z^\prime) 
	\nonumber \,, \\
U_{CP}(K_2^{ji}(z^\prime|y|z)) 
	&:= \etapt^{ij}\ K_2^{ji}(\tilde{z}^\prime|\tilde{y}|\tilde{z}) \,,
\end{align}
with $\tilde{z} := (-\mvec{z},z^4)$, and

\begin{align}
S := \gamma_4 T \,,
\qquad 
A := \gamma_4 C \gamma_5\,.
\end{align}
$T$ is the time reflection matrix. We use a
chiral basis for the Dirac matrices, where $C = \gamma_2 \gamma_4$ and $T = \gamma_1 \gamma_3 \gamma_4$. The sign factors $\etapt$, $\etaf$ are defined in \eqref{eq:dpd-eta-pt} and \eqref{eq:eta_4}, respectively. Considering the generic connected baryon contractions \eqref{eq:gen_bar_contr} we find for the nucleon:

\begin{align}
\left( G^{ijk}[X,Y,Z] \right)^{*} &\xrightarrow{CP} 
	G^{ijk}
		[U_{CP}(X),U_{CP}(Y),U_{CP}(Z)]
	\,,
\label{eq:Gijk_ccp}
\end{align}
and moreover

\begin{align}
G^{ijk}[X,Y,Z] &\xrightarrow{PT} 
	G^{ijk}
		[U_{PT}(X),U_{PT}(Y),U_{PT}(Z)] 
	&\ \mathrm{for}\ (ijk) = (123),(213)\,, \nonumber \\
G^{ijk}[X,Y,Z] &\xrightarrow{\mathcal{PT}} 
	G^{ijk}
		[U_{PT}(Z),U_{PT}(Y),U_{PT}(X)] 
	&\ \mathrm{for}\ (ijk) = (321),(312)\,, \nonumber \\
G^{ijk}[X,Y,Z] &\xrightarrow{\mathcal{PT}} 
	G^{ijk}
		[U_{PT}(X),U_{PT}(Z),U_{PT}(Y)] 
	&\ \mathrm{for}\ (ijk) = (132),(231)\,.
\label{eq:Gijk_pt}
\end{align}
Notice the different orderings of $X$, $Y$, $Z$ on the \rhs. Furthermore, we define the loops:

\begin{align}
L_{1}^i(y) &:= 
	\tr\left\{ \Gamma_i\ M(y|y) \right\}\,, \nonumber\\
L_{2}^{ij}(y) &:= 
	\tr\left\{ \Gamma_i\ M(y|0)\ \Gamma_j\ M(0|y) \right\}\,.
\label{eq:loop_def}
\end{align}
As discussed in \sect\ref{sec:fourpt_def}, there are five types of Wick contractions, which can be represented by the graphs depicted in \fig\ref{fig:graphs}. The explicit contributions depend on the quark flavors of the inserted operators and the baryon, which in our case is always a proton. Since the end points are always connected to the source at $z$ and the sink at $z^\prime$, we shall not write the corresponding arguments of $K_{1,2}$ and $\overline{K}_2$ in the following for brevity. For $C_1$-type graphs we define:

\begin{align}
C_{1,uudd}^{ij}(z,z^\prime,y) &:= 
	\left\langle 
		G^{123}[K_{1}^i(y),K_{1}^j(0),M] + 
		G^{321}[K_{1}^i(y),K_{1}^j(0),M] 
	\right. \nonumber\\
&\quad
	\left.+
		G^{321}[M,K_{1}^j(0),K_{1}^i(y)] +
		G^{123}[M,K_{1}^j(0),K_{1}^i(y)] 
	\right\rangle\,, \nonumber\\
C_{1,uuuu}^{ij}(z,z^\prime,y) &:= 
	\left\langle  
		G^{123}[K_{1}^i(y),M,K_{1}^j(0)] + 
		G^{321}[K_{1}^i(y),M,K_{1}^j(0)] 
	\right. \nonumber\\
&\quad
	\left. + 
		G^{321}[K_{1}^j(0),M,K_{1}^i(y)] + 
		G^{123}[K_{1}^j(0),M,K_{1}^i(y)] 
	\right\rangle\,, \nonumber\\
C_{1,uddu}^{ij}(z,z^\prime,y) &:= 
	\left\langle  
		G^{213}[K_{1}^i(y),K_{1}^j(0),M] + 
		G^{231}[K_{1}^i(y),K_{1}^j(0),M] 
	\right. \nonumber\\
&\quad
	\left. + 
		G^{312}[M,K_{1}^j(0),K_{1}^i(y)] + 
		G^{132}[M,K_{1}^j(0),K_{1}^i(y)] 
	\right\rangle \nonumber\\
&= C_{1,duud}^{ij}(z,z^\prime,-y)\,. 
\label{eq:graph_C1}
\end{align}
The contribution for a certain proton momentum is obtained by a discrete Fourier transform: 

\begin{align}
C_{1,uudd}^{ij,\mvec{p}}(\mvec{y},t,\tau) := 
	a^6 \sum_{\mvec{z}\mvec{z}^\prime} 
	e^{-i\mvec{p}(\mvec{z}^\prime-\mvec{z})} 
	C_{1,uudd}^{ij}(z,z^\prime,y) 
	|_{y^4 = \tau, z^4 = 0, z^{\prime 4} = t}\,,
\end{align}
with analogous expressions for the remaining contractions, which shall be defined in the following. The contributions for $C_2$ and $S_1$ can be written as:

\begin{align}
C_{2,u}^{ij}(z,z^\prime,y) &:= 
	\left\langle 
		G^{123}[K_{2}^{ji}(y),M,M] + 
		G^{321}[K_{2}^{ji}(y),M,M] 
	\right.\nonumber \\
&\quad
	\left. +
		G^{321}[M,M,K_{2}^{ji}(y)] + 
		G^{123}[M,M,K_{2}^{ji}(y)] 
	\right\rangle\,,\nonumber \\
C_{2,d}^{ij}(z,z^\prime,y) &:= 
	\left\langle 
		G^{123}[M,K_{2}^{ji}(y),M] + 
		G^{321}[M,K_{2}^{ji}(y),M] 
	\right\rangle\,,\nonumber \\
S_{1,u}^{ij}(z,z^\prime,y) &:= 
	-\left\langle 
		\left[ 
			G^{123}[K_{1}^i(y),M,M] + 
			G^{321}[K^i_{1}(y),M,M] 
		\right. 
	\right.\nonumber \\
&\quad
	\left.  
		\left. + 
			G^{321}[M,M,K^i_{1}(y)] + 
			G^{123}[M,M,K_{1}^i(y)] 
		\right] 
		L_{1}^j(0)
	\right\rangle\,,\nonumber \\
S_{1,d}^{ij}(z,z^\prime,y) &:= 
	-\left\langle 
		\left[ 
			G^{123}[M,K_{1}^i(y),M] + 
			G^{321}[M,K^i_{1}(y),M] 
		\right] 
		L_{1}^j(0) 
	\right\rangle\,.
\label{eq:graph_C2S1}
\end{align}
The last two contractions we consider are purely disconnected and are defined as:

\begin{align}
S_{2}^{ij}(z,z^\prime,y) &:= 
	-\left\langle
		\left[ 
			G^{123}[M,M,M] + 
			G^{321}[M,M,M] 
		\right] 
		L_{2}^{ij}(y)
	\right\rangle\ \,, \nonumber\\
D^{ij}(z,z^\prime,y) &:= 
	\left\langle
		\left[ 
			G^{123}[M,M,M] + 
			G^{321}[M,M,M] 
		\right] 
		L_{1}^{i}(y) L_{1}^{j}(0)
	\right\rangle\ \,.
\label{eq:graph_S2D}
\end{align}
For completeness, we also give the expression for the two-point function:

\begin{align}
C_{2\mathrm{pt}}(z,z^\prime) &:= 
	\left\langle G^{123}[M,M,M] + G^{321}[M,M,M] \right\rangle\,.
\label{eq:graph_twopt}
\end{align}

\subsection{Baryon sources and sinks}
\label{sec:seqsources}

In the following, we list the terms that are used to construct the sequential sources and contractions needed for the evaluation of baryonic four-point graphs. Notice that each quantity given in the following is based on a point-to-all propagator with point source at $z$. After correcting the momentum phase by multiplying with $\pha^{\mvec{p}}(z)$ and averaging over all gauge fields, the complete contraction is independent of the source position. This is why on the \lhs in \eqref{eq:c1_src} $z$ is not written as argument or index.

\paragraph*{$C_1$-sink (sequential source):} For the sequential source required by the $C_1$ graph at the baryon sink, we have six possibilities to attach the quark lines to the sink kernel \eqref{eq:bar-ann-kern}. Since each of the quark lines is evaluated in a technically different manner, there are also six possible expressions that can appear in the construction of the sequential source from which the sequential propagator of quark line $\mathfrak{c}$ \eqref{eq:c1_seq_inv} is calculated. In terms of the forward propagator $M$ (quark line $\mathfrak{a}$) and the stochastic source $\eta$ (quark line $\mathfrak{b}$), the sequential sources are given by:

\begin{align}
{\srcph}_{\sigma}\left( 
	S_{123}^{\mvec{p},(\ell)} 
\right)_{\bar{\beta}\alpha^\prime }^{\bar{b} a^\prime }
	(z^\prime) 
&:= 
	\pha^{\mvec{p}}(z^\prime) 
	\left( P_+ \gamma_5 \right)_{\sigma\alpha^\prime} 
	\left[ 
		\left(
			\Phi^{\mvec{p}} M_z^{\Phi,\mvec{p}} (z^\prime)
		\right)^T\ 
		E^{a^\prime} \Phi^{\mvec{p}} \gamma_5 \eta^{(\ell)}(z^\prime) 
	\right]_{\bar{\beta}}^{\bar{b}}\,,\nonumber \\
{\srcph}_{\sigma}\left( 
	S_{213}^{\mvec{p},(\ell)} 
\right)_{\bar{\gamma}\alpha^\prime }^{\bar{c}a^\prime }
	(z^\prime) 
&:= 
	\pha^{\mvec{p}}(z^\prime) 
	\left( P_+ \gamma_5 \right)_{\sigma\alpha^\prime} 
	\left[ 
		\left( \Phi^{\mvec{p}} \eta^{(\ell)} (z^\prime) \right)^T 
		\gamma_5^T E^{a^\prime} 
		\Phi^{\mvec{p}} M_z^{\Phi,\mvec{p}}(z^\prime) 
	\right]_{\bar{\gamma}}^{\bar{c}}\,,\nonumber \\
{\srcph}_{\sigma}\left( 
	S_{231}^{\mvec{p},(\ell)} 
\right)_{\bar{\gamma}\beta^\prime}^{\bar{c}b^\prime}(z^\prime) 
&:= 
	\pha^{\mvec{p}}(z^\prime) 
	\left( 
		P_+ \Phi^{\mvec{p}} \gamma_5 \eta^{(\ell)}(z^\prime) 
	\right)_{\sigma}^a 
	\left[ 
		\gamma_5^T E^{a} \Phi^{\mvec{p}} M_z^{\Phi,\mvec{p}}(z^\prime) 
	\right]_{\beta^\prime\bar{\gamma}}^{b^\prime\bar{c}}\,,  
\nonumber \\
{\srcph}_{\sigma}\left( 
	S_{132}^{\mvec{p},(\ell)} 
\right)_{\bar{\beta}\gamma^\prime}^{\bar{b}c^\prime}(z^\prime) 
&:= 
	\pha^{\mvec{p}}(z^\prime) 
	\left( 
		P_+ \Phi^{\mvec{p}} \gamma_5 \eta^{(\ell)}(z^\prime) 
	\right)_{\sigma}^a 
	\left[ 
		\left(\Phi^{\mvec{p}}M_z^{\Phi,\mvec{p}} (z^\prime) \right)^T\ 
		E^{a} \gamma_5 
	\right]_{\bar{\beta}\gamma^\prime}^{\bar{b}c^\prime}\,,
\nonumber \\
{\srcph}_{\sigma}\left( 
	S_{312}^{\mvec{p},(\ell)} 
\right)_{\bar{\alpha}\gamma^\prime}^{\bar{a}c^\prime}(z^\prime) 
&:= 
	\pha^{\mvec{p}}(z^\prime) 
	\left( 
		P_+ \Phi^{\mvec{p}} M^{\Phi,\mvec{p}}_z(z^\prime) 
	\right)^{a\bar{a}}_{\sigma\bar{\alpha}} 
	\left[ 
		\left( \Phi^{\mvec{p}} \eta^{(\ell)} (z^\prime) \right)^T 
		\gamma_5^T E^{a} \gamma_5 
	\right]_{\gamma^\prime}^{c^\prime}\,,\nonumber \\
{\srcph}_{\sigma}\left( 
	S_{321}^{\mvec{p},(\ell)} 
\right)_{\bar{\alpha}\beta^\prime}^{\bar{a}b^\prime}(z^\prime) 
&:= 
	\pha^{\mvec{p}}(z^\prime) 
	\left( 
		P_+ \Phi^{\mvec{p}} M^{\Phi,\mvec{p}}_z(z^\prime) 
	\right)^{a\bar{a}}_{\sigma\bar{\alpha}} 
	\left[ 
		\gamma_5^T E^{a} \Phi^{\mvec{p}} 
		\gamma_5 \eta^{(\ell)}(z^\prime) 
	\right]_{\beta^\prime}^{b^\prime}\,.
\label{eq:seq_c1_src}
\end{align}
The integer indices denote which quark line is connected to which part according to the pattern $n(\mathfrak{a})n(\mathfrak{b})n(\mathfrak{c})$, where $n(\mathfrak{a})$ indicates the number of the quark field (see \eqref{eq:quark_numbers}) to which the quark line $\mathfrak{a}$ is connected, and similar for $n(\mathfrak{b})$, $n(\mathfrak{c})$. For instance, in case of the expression $S_{312}$ we have the forward quark line $\mathfrak{a}$ attached to quark field (3), the stochastic quark line $\mathfrak{b}$ to quark field (1), and the sequential quark line $\mathfrak{c}$ to quark field (2). A sequential source $\mathcal{S}$ for a specific flavor combination is represented by a sum of a certain subset of terms given in \eqref{eq:seq_c1_src}.

\paragraph*{$C_1$-source:} 

Analogous combinatorics lead to the six possible expressions used to construct the quantity $q_1$ from the forward propagator $M$ and the sequential propagator $X$. In terms of the quantity $Y$ defined in \eqref{eq:c1_Y}, the contractions read:

\begin{align}
\left( 
	\overline{S}^{\mvec{p},(\ell)}_{123,t,j} 
\right)_{\beta}^{b} (y) 
&:= 
	\pha^{-\mvec{p}}(z)\sum_{\sigma} 
	\left[ 
		{}_\sigma Y_{t,j}^{T,\mvec{p},(\ell)}(y) E^c  
	\right]^{cb}_{\gamma\beta} 
	\left(P_+\right)_{\gamma\sigma}\,,\nonumber \\
\left( 
	\overline{S}^{\mvec{p},(\ell)}_{213,t,j} 
\right)_{\alpha}^{a} (y) 
&:= 
	\pha^{-\mvec{p}}(z)\sum_{\sigma} 
	\left[ 
		E^c {}_\sigma Y_{t,j}^{\mvec{p},(\ell)}(y) 
	\right]^{ac}_{\alpha\gamma} 
	\left(P_+\right)_{\gamma\sigma}\,,\nonumber \\
\left( 
	\overline{S}^{\mvec{p},(\ell)}_{231,t,j} 
\right)_{\gamma}^{c} (y) 
&:= 
	\pha^{-\mvec{p}}(z)\sum_{\sigma} 
	\tr\left\{ 
		{}_\sigma Y_{t,j}^{\mvec{p},(\ell)}(y) E^c 
	\right\} 
	\left(P_+\right)_{\gamma\sigma}\,, \nonumber \\
\left( 
	\overline{S}^{\mvec{p},(\ell)}_{132,t,j} 
\right)_{\gamma}^{c} (y) 
&:= 
	\pha^{-\mvec{p}}(z)\sum_{\sigma} 
	\tr\left\{ 
		E^c {}_\sigma Y_{t,j}^{T,\mvec{p},(\ell)}(y) 
	\right\} 
	\left(P_+\right)_{\gamma\sigma}\,,\nonumber \\
\left( 
	\overline{S}^{\mvec{p},(\ell)}_{312,t,j} 
\right)_{\alpha}^{a} (y) 
&:= 
	\pha^{-\mvec{p}}(z)\sum_{\sigma} 
	\left[ 
		E^c {}_\sigma Y_{t,j}^{T,\mvec{p},(\ell)}(y) 
	\right]^{a c}_{\alpha\gamma} 
	\left(P_+\right)_{\gamma\sigma}\,,\nonumber \\
\left( 
	\overline{S}^{\mvec{p},(\ell)}_{321,t,j} 
\right)_{\beta}^{b} (y) 
&:= 
	\pha^{-\mvec{p}}(z)\sum_{\sigma} 
	\left[ 
		{}_\sigma Y_{t,j}^{\mvec{p},(\ell)}(y) E^c 
	\right]^{cb}_{\gamma\beta} 
	\left(P_+\right)_{\gamma\sigma}\,.
\label{eq:c1_src}
\end{align}
Like for the sequential sources discussed before, $q_1$ is obtained by summing over a subset of these terms, which is specific to the flavor combinations. More details and the cases needed for flavor conserving proton-proton matrix elements shall be discussed in \appx\ref{sec:c1_contr}.

\paragraph*{Sequential sources for $G_{3\mathrm{pt}}$:} 

For the disconnected three-point contractions we re-use the sequential sources that appear in three-point functions. Depending on the flavor of the quark line, they can be written as:\footnote{In contrast to the sequential source \eqref{eq:seq_c1_src} used for the $C_1$ contraction, the three-point sources $S^{\mvec{p}}_{3\mathrm{pt}}$ are defined without $\gamma_5$, which in this case is included in \eqref{eq:seq_3pt_inv}}

\begin{align}
\label{eq:3pt_seq}
\left(
	S^{\mvec{p}}_{3\mathrm{pt},u}
\right)_{\alpha\beta}^{ab}(z^\prime) 
&= 
	\pha^{\mvec{p}}(z^\prime) 
	\left[ 
		P_+ \Phi^{\mvec{p}} M_z^{\Phi,\mvec{p}}(z^\prime) 
		E^a 
		\left(
			E^b \Phi^{\mvec{p}} M_z^{\Phi,\mvec{p}} (z^\prime)
		\right)^T 
	\right]_{\alpha\beta}^{cc} \nonumber\\
&\quad + 
	\pha^{\mvec{p}}(z^\prime) 
	\left(P_+\right)_{\alpha\beta} 
	\tr \left\{ 
		\Phi^{\mvec{p}} M_z^{\Phi,\mvec{p}}(z^\prime) E^a 
		\left( 
			E^b \Phi^{\mvec{p}} M_z^{\Phi,\mvec{p}}(z^\prime) 
		\right)^T  
	\right\} \nonumber\\
&\quad + 
	\pha^{\mvec{p}}(z^\prime) 
	\left[ 
		\left( 
			E^b \Phi^{\mvec{p}} M_z^{\Phi,\mvec{p}}(z^\prime) E^{a} 
		\right)^T 
		\Phi^{\mvec{p}} M_z^{\Phi,\mvec{p}}(z^\prime) P_+ 
	\right]_{\alpha\beta}^{cc} \nonumber \\
&\quad + 
	\pha^{\mvec{p}}(z^\prime) 
	\left( 
		\Phi^{\mvec{p}} M_z^{\Phi,\mvec{p}}(z^\prime) P_+ 
	\right)_{\gamma\gamma}^{cd} 
	\left[ 
		\left( 
			E^b \Phi^{\mvec{p}} M_z^{\Phi,\mvec{p}}(z^\prime) 
			E^{a} 
		\right)^T 
	\right]_{\alpha\beta}^{dc}\,,\nonumber \\
\left(
	S^{\mvec{p}}_{3\mathrm{pt},d}
\right)_{\alpha\beta}^{ab}(z^\prime) 
&= 
	\pha^{\mvec{p}}(z^\prime) 
	\left( 
		\Phi^{\mvec{p}} M_z^{\Phi,\mvec{p}} (z^\prime) P_+ 
	\right)_{\gamma\gamma}^{cd} 
	\left[ 
		\left( 
			E^b \Phi^{\mvec{p}} M_z^{\Phi,\mvec{p}}(z^\prime) E^a 
		\right)^T 
	\right]_{\alpha\beta}^{dc}\nonumber \\
&\quad - 
	\pha^{\mvec{p}}(z^\prime) 
	\left( 
		E^b \Phi^{\mvec{p}} M_z^{\Phi,\mvec{p}}(z^\prime) 
	\right)_{\beta\gamma}^{cd} 
	\left( 
		P_+ \Phi^{\mvec{p}} M_z^{\Phi,\mvec{p}}(z^\prime) E^a 
	\right)_{\gamma\alpha}^{cd}\,.
\end{align}

\subsection{$C_1$ contractions}
\label{sec:c1_contr}
\begin{figure}[ht]
\includegraphics[scale=1]{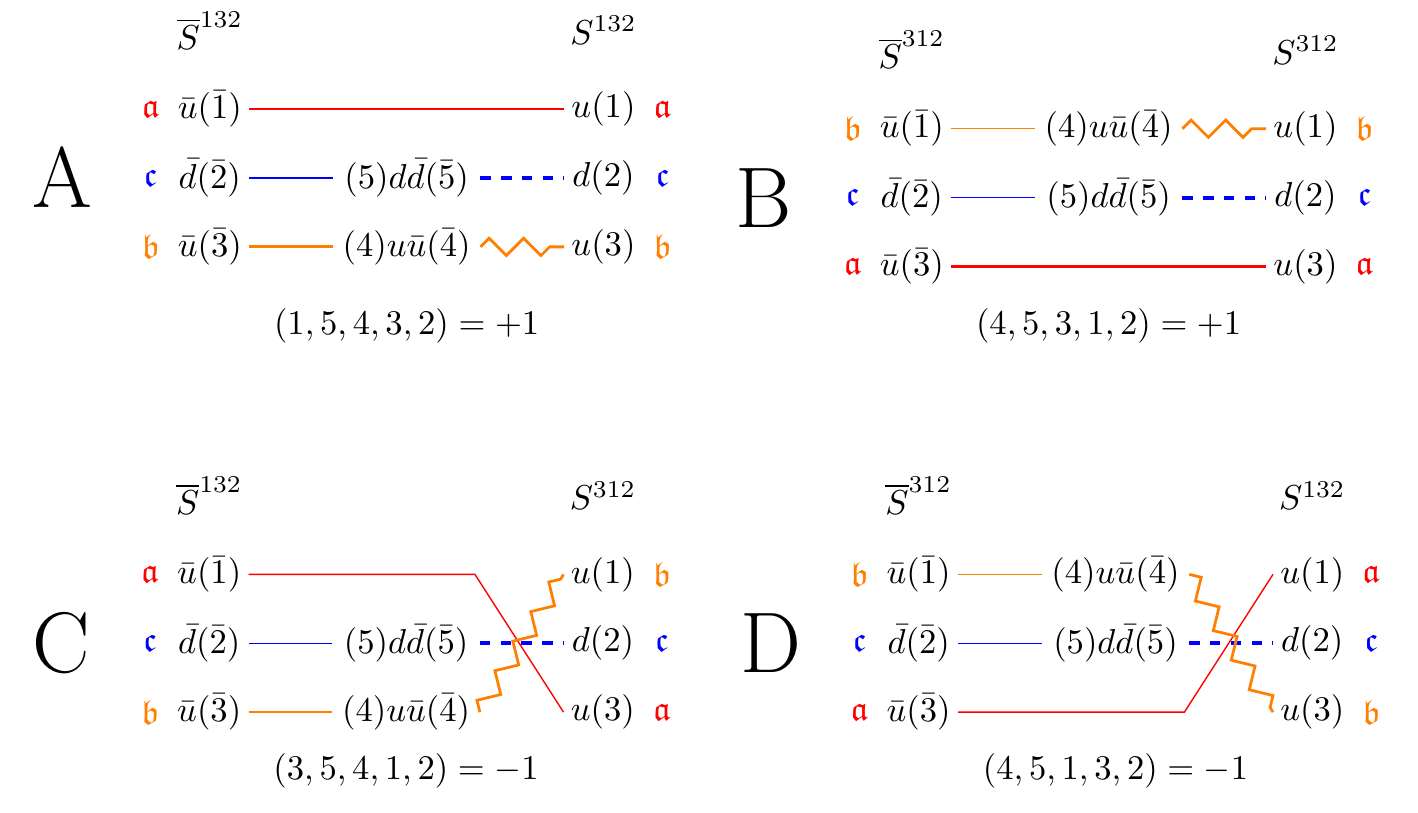}
\caption{Contributions of $C_1$ type for the combination $\Op_i^{uu}(y)\, \Op_j^{dd}(0)$ ($C_{1,uudd}$). Depending on the evaluation method, we use different symbols to depict the propagators: The forward propagator $M_z$ is represented by a simple line, the stochastic propagator $\psi$ by a zigzag line, and the sequential propagator $X$ (without the incorporated forward propagator and the stochastic source) by a dashed line. The colors indicate the quark lines: Red corresponds to $\mathfrak{a}$, orange to $\mathfrak{b}$, and blue to $\mathfrak{c}$. The combination of the quark lines with the numbers $(1)$, $(2)$, $(3)$ at the sink or $(\bar{1})$, $(\bar{2})$, $(\bar{3})$ at the source determines the sequential source type $S_{n(\mathfrak{a})n(\mathfrak{b})n(\mathfrak{c})}$ (see \eqref{eq:seq_c1_src}) or the contraction $\overline{S}_{n(\mathfrak{a})n(\mathfrak{b})n(\mathfrak{c})}$ (see \eqref{eq:c1_src}), respectively. The resulting permutation is also given for each contraction. Moreover, at the bottom line of each panel we give the permutation of quark fields represented by the propagator and the corresponding sign, which enters the total contribution and hence the physical matrix elements.\label{Fig:uudd} }
\end{figure}

We now give explicit expressions for the sequential source $\mathcal{S}$ and the contraction $q_1$ needed for the calculation of the $C_1$ graph. We start with the contributions to $\Op^{uu}_{i}(y)\, \Op^{dd}_{j}(0)$. The corresponding sub-graphs are illustrated in \fig{\ref{Fig:uudd}}. 
If the last integer index of the sequential sources \eqref{eq:seq_c1_src} is equal for two or more contractions appearing in the flavor specific sum, the corresponding sequential sources can be combined before the inversion. In the case considered, we are able to combine $A$ with $C$ and $B$ with $D$, which in both cases gives:

\begin{align}
\mathcal{S}^{\mvec{p},(\ell)}(z^\prime) = 
	S^{\mvec{p},(\ell)}_{132}(z^\prime) - 
	S^{\mvec{p},(\ell)}_{312}(z^\prime)\,,
\label{eq:seq_src_uudd}
\end{align}
up to a global sign. Inserting the source \eqref{eq:seq_src_uudd}, we obtain the corresponding sequential propagator $X$ by an inversion of \eqref{eq:c1_seq_inv}. The relative signs, which can be read off from \fig{\ref{Fig:uudd}}, correspond to the permutations of fermionic fields. The two contributions $(A,C)$ and $(B,D)$ are then combined in the quantity $q_1$ by calculating the sum:

\begin{align}
\label{eq:C1_q1_uudd}
q^{\mvec{p},(\ell)}_{1,t,j}(y) &= 
	\overline{S}^{\mvec{p},(\ell)}_{132,t,j}(y) - 
	\overline{S}^{\mvec{p},(\ell)}_{312,t,j}(y)\,.
\end{align}
The quantity $Y$ appearing in the definition \eqref{eq:c1_src} of the contractions $\overline{S}$ is obtained from the sequential propagator, one current insertion, and the forward propagator, see \eqref{eq:c1_Y}. The total contribution to $C_{1,uudd}$ is then simply given by \eqref{eq:def_c1} with $q_1$ as defined in \eqref{eq:C1_q1_uudd}. 

\begin{figure}[ht]
\includegraphics[scale=1]{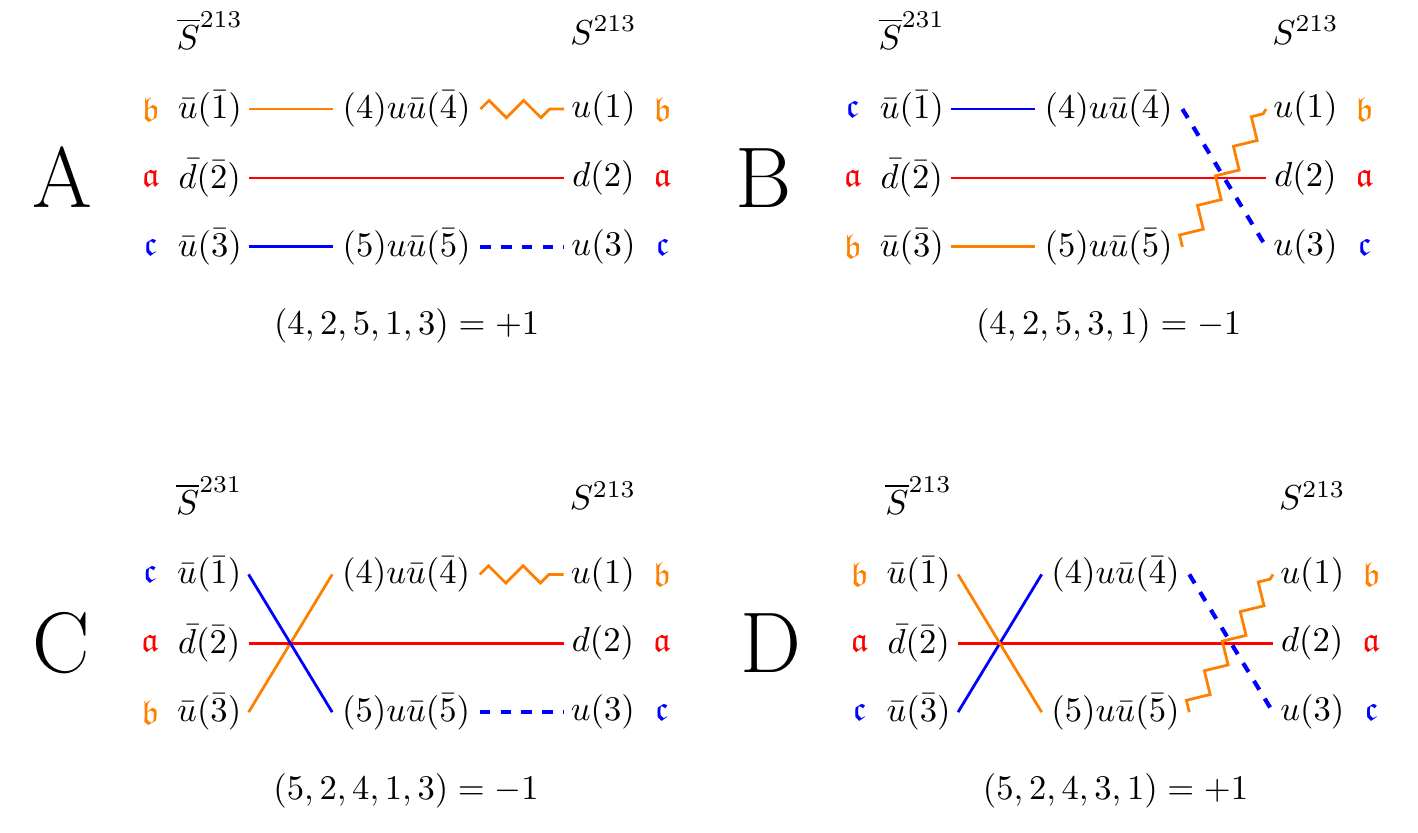}
\caption{The same as \fig{\ref{Fig:uudd}}, but for the contributions to $\Op_i^{uu}(y)\, \Op_j^{uu}(0)$ with $C_1$ topology ($C_{1,uuuu}$).\label{Fig:uuuu} }
\end{figure}

We now turn to the $C_1$ contribution for the flavor combination $\Op_i^{uu}(y)\, \Op_j^{uu}(0)$. The corresponding sub-graphs are shown in \fig\ref{Fig:uuuu}.
In this case, we only need the expression $S^{213}$ for the construction of the sequential source, \ie :

\begin{align}
\label{eq:seq_src_uuuu}
\mathcal{S}^{\mvec{p},(\ell)}(z^\prime) = 
	S^{\mvec{p},(\ell)}_{213}(z^\prime)\,.
\end{align}
Like for $C_{1,uudd}$, we calculate the sequential propagator by inverting \eqref{eq:c1_seq_inv} with the source \eqref{eq:seq_src_uuuu}, and calculate $Y$ (see \eqref{eq:c1_Y}), which is then contracted with the sources \eqref{eq:c1_src} according to the permutation that can be read off in \fig\ref{Fig:uuuu}. It is possible to combine $A$ with $C$ and $B$ with $D$ before doing the spatial correlation, since each current insertion is connected to the same quark line within these pairs. In contrast to the $C_{1,uudd}$ case, we have two terms contributing to $C_{1,uuuu}$ consisting of $q_1 q_2$ products, where $q_1$ and $q_2$ are given by:

\begin{align}
q_{1,AC,t,j}^{\mvec{p},(\ell)}(y) &= 
	\overline{S}^{\mvec{p},(\ell)}_{213,t,j}(y) 
	- \overline{S}^{\mvec{p},(\ell)}_{231,j}(y)\,,\nonumber \\
q_{1,BD,t,i}^{\mvec{p},(\ell)}(y) &= 
	-\overline{S}^{\mvec{p},(\ell)}_{231,t,i}(y) 
	+ \overline{S}^{\mvec{p},(\ell)}_{213,i}(y)\ ,\nonumber \\
q_{2,AC,t,i}^{\mvec{p},(\ell)}(x) &= 
	\psi_t^{\dagger,(\ell)}(x)\ \gamma_5 
	\Gamma_i\ M^{\Phi ,\mvec{p}}_z(x)\,, \nonumber  \\
q_{2,BD,t,j}^{\mvec{p},(\ell)}(x) &= 
	\psi_t^{\dagger,(\ell)}(x)\ \gamma_5 
	\Gamma_j\ M^{\Phi ,\mvec{p}}_z(x)\,.
\end{align}
Notice that in the $BD$ case the current insertion indices $i,j$ are exchanged compared to the $AC$ case. Putting everything together, the total $C_{1,uuuu}$ contribution reads:

\begin{align}
C_{1,uuuu}^{ij,\mvec{p}}(\mvec{y},t,\tau) &= 
	\frac{a^3}{ N_{\mathrm{st}}} 
	\sum_{\mvec{x}} \sum_{\ell}^{N_{\mathrm{st}}} 
	\left\langle \left[ 
		q_{2,AC,t,i}^{T,\mvec{p},(\ell)}(x+y)\ 
		q_{1,AC,t,j}^{\mvec{p},(\ell)}(x) 
	\right] \right.\nonumber \\
&+ 
	\left. \left. \left[ 
		q_{1,BD,t,i}^{T,\mvec{p},(\ell)}(x+y)\ 
		q_{2,BD,t,j}^{\mvec{p},(\ell)}(x) 
	\right] \right\rangle \right|_{x^4 = \tau, y^4 = 0}\,.
\label{eq:def_c1uuuu}
\end{align}

\end{appendices}

%%%%%%%%%%%%%%%%%%%%%%%%%%%%%%%%%%%%%%%%%%%%%%%%%%%%%%%%%%%%%%%%%%%%%%%%%%

\FloatBarrier

\phantomsection
\addcontentsline{toc}{section}{References}

\bibliographystyle{JHEP}
\bibliography{dpd-nucleon-paper}

\end{document}